\documentclass[10pt,twocolumn]{article}

\usepackage[T1]{fontenc}
\usepackage{lmodern}
\usepackage[margin=0.75in]{geometry}
\usepackage{amsmath}
\usepackage{booktabs}
\usepackage{graphicx}
\usepackage[hidelinks]{hyperref}
\usepackage{xcolor}
\usepackage{pgfplots}
\pgfplotsset{compat=1.18}
\usepgfplotslibrary{groupplots}
\hypersetup{
  pdftitle={Global Pass Barriers Without Per-Resource RHI Tracking:
    A Cross-Vendor Study with Blade},
  pdfauthor={Dzmitry Malyshau}
}

\newcommand{\blade}{\textsc{Blade}}
\newcommand{\wgpu}{\textsc{wgpu}}

\makeatletter
\newcommand{\bladenum}[1]{%
  \ifcsname blade@num@#1\endcsname
    \csname blade@num@#1\endcsname
  \else
    \errmessage{No generated number for `#1'; run paper/build-tables.py}%
  \fi}
\makeatother
\newcommand{\bpct}[3]{\bladenum{#1/#2/#3/pct}\%}
\newcommand{\bmag}[3]{\bladenum{#1/#2/#3/mag}\%}
\newcommand{\bpctci}[3]{\bpct{#1}{#2}{#3}\,[\bladenum{#1/#2/#3/lo},\,\bladenum{#1/#2/#3/hi}]}
\newcommand{\bmagci}[3]{\bmag{#1}{#2}{#3}\,[\bladenum{#1/#2/#3/maglo},\,\bladenum{#1/#2/#3/maghi}]}
\newcommand{\bfloor}[2]{\bladenum{#1/#2/floor}\%}
\newcommand{\bus}[3]{\bladenum{#1/#2/#3us}\,$\mu$s}

\newcommand{\hostcaveat}{}
\newcommand{\hostcaveatinline}{}
\expandafter\gdef\csname blade@num@applem3/compute-chain/hostautous\endcsname{68.3}
\expandafter\gdef\csname blade@num@applem3/compute-chain/hostwgpu/hi\endcsname{+42.9}
\expandafter\gdef\csname blade@num@applem3/compute-chain/hostwgpu/lo\endcsname{+24.8}
\expandafter\gdef\csname blade@num@applem3/compute-chain/hostwgpu/mag\endcsname{33.0}
\expandafter\gdef\csname blade@num@applem3/compute-chain/hostwgpu/maghi\endcsname{42.9}
\expandafter\gdef\csname blade@num@applem3/compute-chain/hostwgpu/maglo\endcsname{24.8}
\expandafter\gdef\csname blade@num@applem3/compute-chain/hostwgpu/pct\endcsname{+33.0}
\expandafter\gdef\csname blade@num@applem3/compute-chain/hostwgpuus\endcsname{91.1}
\expandafter\gdef\csname blade@num@applem3/compute-chain/waitautous\endcsname{1082.3}
\expandafter\gdef\csname blade@num@applem3/compute-chain/waitwgpuus\endcsname{1110.6}
\expandafter\gdef\csname blade@num@applem3/compute-independent/hostautous\endcsname{68.3}
\expandafter\gdef\csname blade@num@applem3/compute-independent/hostwgpu/hi\endcsname{+57.9}
\expandafter\gdef\csname blade@num@applem3/compute-independent/hostwgpu/lo\endcsname{+24.5}
\expandafter\gdef\csname blade@num@applem3/compute-independent/hostwgpu/mag\endcsname{33.7}
\expandafter\gdef\csname blade@num@applem3/compute-independent/hostwgpu/maghi\endcsname{57.9}
\expandafter\gdef\csname blade@num@applem3/compute-independent/hostwgpu/maglo\endcsname{24.5}
\expandafter\gdef\csname blade@num@applem3/compute-independent/hostwgpu/pct\endcsname{+33.7}
\expandafter\gdef\csname blade@num@applem3/compute-independent/hostwgpuus\endcsname{92.8}
\expandafter\gdef\csname blade@num@applem3/compute-independent/waitautous\endcsname{1449.1}
\expandafter\gdef\csname blade@num@applem3/compute-independent/waitwgpuus\endcsname{1449.2}
\expandafter\gdef\csname blade@num@applem3/graphics-chain/hostautous\endcsname{83.8}
\expandafter\gdef\csname blade@num@applem3/graphics-chain/hostwgpu/hi\endcsname{+86.0}
\expandafter\gdef\csname blade@num@applem3/graphics-chain/hostwgpu/lo\endcsname{+71.0}
\expandafter\gdef\csname blade@num@applem3/graphics-chain/hostwgpu/mag\endcsname{79.3}
\expandafter\gdef\csname blade@num@applem3/graphics-chain/hostwgpu/maghi\endcsname{86.0}
\expandafter\gdef\csname blade@num@applem3/graphics-chain/hostwgpu/maglo\endcsname{71.0}
\expandafter\gdef\csname blade@num@applem3/graphics-chain/hostwgpu/pct\endcsname{+79.3}
\expandafter\gdef\csname blade@num@applem3/graphics-chain/hostwgpuus\endcsname{149.9}
\expandafter\gdef\csname blade@num@applem3/graphics-chain/waitautous\endcsname{1421.8}
\expandafter\gdef\csname blade@num@applem3/graphics-chain/waitwgpuus\endcsname{1409.4}
\expandafter\gdef\csname blade@num@applem3/graphics-independent/hostautous\endcsname{85.2}
\expandafter\gdef\csname blade@num@applem3/graphics-independent/hostwgpu/hi\endcsname{+98.2}
\expandafter\gdef\csname blade@num@applem3/graphics-independent/hostwgpu/lo\endcsname{+64.3}
\expandafter\gdef\csname blade@num@applem3/graphics-independent/hostwgpu/mag\endcsname{77.6}
\expandafter\gdef\csname blade@num@applem3/graphics-independent/hostwgpu/maghi\endcsname{98.2}
\expandafter\gdef\csname blade@num@applem3/graphics-independent/hostwgpu/maglo\endcsname{64.3}
\expandafter\gdef\csname blade@num@applem3/graphics-independent/hostwgpu/pct\endcsname{+77.6}
\expandafter\gdef\csname blade@num@applem3/graphics-independent/hostwgpuus\endcsname{150.0}
\expandafter\gdef\csname blade@num@applem3/graphics-independent/waitautous\endcsname{1302.6}
\expandafter\gdef\csname blade@num@applem3/graphics-independent/waitwgpuus\endcsname{1247.9}
\expandafter\gdef\csname blade@num@barriers/auto/chain\endcsname{5}
\expandafter\gdef\csname blade@num@barriers/auto/chain/final\endcsname{1}
\expandafter\gdef\csname blade@num@barriers/auto/chain/initial\endcsname{1}
\expandafter\gdef\csname blade@num@barriers/auto/chain/interpass\endcsname{3}
\expandafter\gdef\csname blade@num@barriers/auto/independent\endcsname{5}
\expandafter\gdef\csname blade@num@barriers/auto/independent/final\endcsname{1}
\expandafter\gdef\csname blade@num@barriers/auto/independent/initial\endcsname{1}
\expandafter\gdef\csname blade@num@barriers/auto/independent/interpass\endcsname{3}
\expandafter\gdef\csname blade@num@barriers/hazard/chain\endcsname{4}
\expandafter\gdef\csname blade@num@barriers/hazard/chain/final\endcsname{1}
\expandafter\gdef\csname blade@num@barriers/hazard/chain/initial\endcsname{0}
\expandafter\gdef\csname blade@num@barriers/hazard/chain/interpass\endcsname{3}
\expandafter\gdef\csname blade@num@barriers/hazard/independent\endcsname{1}
\expandafter\gdef\csname blade@num@barriers/hazard/independent/final\endcsname{1}
\expandafter\gdef\csname blade@num@barriers/hazard/independent/initial\endcsname{0}
\expandafter\gdef\csname blade@num@barriers/hazard/independent/interpass\endcsname{0}
\expandafter\gdef\csname blade@num@barriers/wgpu/chain\endcsname{4}
\expandafter\gdef\csname blade@num@barriers/wgpu/chain/final\endcsname{0}
\expandafter\gdef\csname blade@num@barriers/wgpu/chain/initial\endcsname{1}
\expandafter\gdef\csname blade@num@barriers/wgpu/chain/interpass\endcsname{3}
\expandafter\gdef\csname blade@num@barriers/wgpu/independent\endcsname{1}
\expandafter\gdef\csname blade@num@barriers/wgpu/independent/final\endcsname{0}
\expandafter\gdef\csname blade@num@barriers/wgpu/independent/initial\endcsname{1}
\expandafter\gdef\csname blade@num@barriers/wgpu/independent/interpass\endcsname{0}
\expandafter\gdef\csname blade@num@bimodal/radeon780m/blade/launches\endcsname{420}
\expandafter\gdef\csname blade@num@bimodal/radeon780m/blade/slow\endcsname{67}
\expandafter\gdef\csname blade@num@bimodal/radeon780m/fastplacement\endcsname{+41.6}
\expandafter\gdef\csname blade@num@bimodal/radeon780m/ratio\endcsname{2.04}
\expandafter\gdef\csname blade@num@bimodal/radeon780m/wgpu/launches\endcsname{70}
\expandafter\gdef\csname blade@num@bimodal/radeon780m/wgpu/slow\endcsname{1}
\expandafter\gdef\csname blade@num@captures/distinct\endcsname{1}
\expandafter\gdef\csname blade@num@captures/machines\endcsname{4}
\expandafter\gdef\csname blade@num@captures/rows\endcsname{160}
\expandafter\gdef\csname blade@num@captures/sha\endcsname{f2a0b85141e1}
\expandafter\gdef\csname blade@num@chain/resolved\endcsname{0}
\expandafter\gdef\csname blade@num@depth/radeon780m/graphics-independent/p1/control/hi\endcsname{+0.2}
\expandafter\gdef\csname blade@num@depth/radeon780m/graphics-independent/p1/control/lo\endcsname{-2.7}
\expandafter\gdef\csname blade@num@depth/radeon780m/graphics-independent/p1/control/pct\endcsname{-0.0}
\expandafter\gdef\csname blade@num@depth/radeon780m/graphics-independent/p1/floor\endcsname{2.7}
\expandafter\gdef\csname blade@num@depth/radeon780m/graphics-independent/p1/placement/hi\endcsname{+0.2}
\expandafter\gdef\csname blade@num@depth/radeon780m/graphics-independent/p1/placement/lo\endcsname{-2.5}
\expandafter\gdef\csname blade@num@depth/radeon780m/graphics-independent/p1/placement/pct\endcsname{-0.3}
\expandafter\gdef\csname blade@num@depth/radeon780m/graphics-independent/p1/wgpu/hi\endcsname{-21.5}
\expandafter\gdef\csname blade@num@depth/radeon780m/graphics-independent/p1/wgpu/lo\endcsname{-23.8}
\expandafter\gdef\csname blade@num@depth/radeon780m/graphics-independent/p1/wgpu/mag\endcsname{21.7}
\expandafter\gdef\csname blade@num@depth/radeon780m/graphics-independent/p1/wgpu/maghi\endcsname{23.8}
\expandafter\gdef\csname blade@num@depth/radeon780m/graphics-independent/p1/wgpu/maglo\endcsname{21.5}
\expandafter\gdef\csname blade@num@depth/radeon780m/graphics-independent/p1/wgpu/pct\endcsname{-21.7}
\expandafter\gdef\csname blade@num@depth/radeon780m/graphics-independent/p16/control/hi\endcsname{+106.7}
\expandafter\gdef\csname blade@num@depth/radeon780m/graphics-independent/p16/control/lo\endcsname{-14.5}
\expandafter\gdef\csname blade@num@depth/radeon780m/graphics-independent/p16/control/pct\endcsname{+51.9}
\expandafter\gdef\csname blade@num@depth/radeon780m/graphics-independent/p16/floor\endcsname{106.7}
\expandafter\gdef\csname blade@num@depth/radeon780m/graphics-independent/p16/placement/hi\endcsname{+184.6}
\expandafter\gdef\csname blade@num@depth/radeon780m/graphics-independent/p16/placement/lo\endcsname{+22.5}
\expandafter\gdef\csname blade@num@depth/radeon780m/graphics-independent/p16/placement/mag\endcsname{42.0}
\expandafter\gdef\csname blade@num@depth/radeon780m/graphics-independent/p16/placement/maghi\endcsname{184.6}
\expandafter\gdef\csname blade@num@depth/radeon780m/graphics-independent/p16/placement/maglo\endcsname{22.5}
\expandafter\gdef\csname blade@num@depth/radeon780m/graphics-independent/p16/placement/pct\endcsname{+42.0}
\expandafter\gdef\csname blade@num@depth/radeon780m/graphics-independent/p16/wgpu/hi\endcsname{-10.7}
\expandafter\gdef\csname blade@num@depth/radeon780m/graphics-independent/p16/wgpu/lo\endcsname{-28.4}
\expandafter\gdef\csname blade@num@depth/radeon780m/graphics-independent/p16/wgpu/mag\endcsname{12.2}
\expandafter\gdef\csname blade@num@depth/radeon780m/graphics-independent/p16/wgpu/maghi\endcsname{28.4}
\expandafter\gdef\csname blade@num@depth/radeon780m/graphics-independent/p16/wgpu/maglo\endcsname{10.7}
\expandafter\gdef\csname blade@num@depth/radeon780m/graphics-independent/p16/wgpu/pct\endcsname{-12.2}
\expandafter\gdef\csname blade@num@depth/radeon780m/graphics-independent/p2/control/hi\endcsname{+2.2}
\expandafter\gdef\csname blade@num@depth/radeon780m/graphics-independent/p2/control/lo\endcsname{-2.0}
\expandafter\gdef\csname blade@num@depth/radeon780m/graphics-independent/p2/control/pct\endcsname{-0.2}
\expandafter\gdef\csname blade@num@depth/radeon780m/graphics-independent/p2/floor\endcsname{2.2}
\expandafter\gdef\csname blade@num@depth/radeon780m/graphics-independent/p2/placement/hi\endcsname{+148.7}
\expandafter\gdef\csname blade@num@depth/radeon780m/graphics-independent/p2/placement/lo\endcsname{+19.7}
\expandafter\gdef\csname blade@num@depth/radeon780m/graphics-independent/p2/placement/mag\endcsname{22.7}
\expandafter\gdef\csname blade@num@depth/radeon780m/graphics-independent/p2/placement/maghi\endcsname{148.7}
\expandafter\gdef\csname blade@num@depth/radeon780m/graphics-independent/p2/placement/maglo\endcsname{19.7}
\expandafter\gdef\csname blade@num@depth/radeon780m/graphics-independent/p2/placement/pct\endcsname{+22.7}
\expandafter\gdef\csname blade@num@depth/radeon780m/graphics-independent/p2/wgpu/hi\endcsname{-21.3}
\expandafter\gdef\csname blade@num@depth/radeon780m/graphics-independent/p2/wgpu/lo\endcsname{-23.2}
\expandafter\gdef\csname blade@num@depth/radeon780m/graphics-independent/p2/wgpu/mag\endcsname{21.5}
\expandafter\gdef\csname blade@num@depth/radeon780m/graphics-independent/p2/wgpu/maghi\endcsname{23.2}
\expandafter\gdef\csname blade@num@depth/radeon780m/graphics-independent/p2/wgpu/maglo\endcsname{21.3}
\expandafter\gdef\csname blade@num@depth/radeon780m/graphics-independent/p2/wgpu/pct\endcsname{-21.5}
\expandafter\gdef\csname blade@num@depth/radeon780m/graphics-independent/p32/control/hi\endcsname{+0.7}
\expandafter\gdef\csname blade@num@depth/radeon780m/graphics-independent/p32/control/lo\endcsname{-12.3}
\expandafter\gdef\csname blade@num@depth/radeon780m/graphics-independent/p32/control/pct\endcsname{-0.3}
\expandafter\gdef\csname blade@num@depth/radeon780m/graphics-independent/p32/floor\endcsname{12.3}
\expandafter\gdef\csname blade@num@depth/radeon780m/graphics-independent/p32/placement/hi\endcsname{+59.2}
\expandafter\gdef\csname blade@num@depth/radeon780m/graphics-independent/p32/placement/lo\endcsname{+40.5}
\expandafter\gdef\csname blade@num@depth/radeon780m/graphics-independent/p32/placement/mag\endcsname{42.4}
\expandafter\gdef\csname blade@num@depth/radeon780m/graphics-independent/p32/placement/maghi\endcsname{59.2}
\expandafter\gdef\csname blade@num@depth/radeon780m/graphics-independent/p32/placement/maglo\endcsname{40.5}
\expandafter\gdef\csname blade@num@depth/radeon780m/graphics-independent/p32/placement/pct\endcsname{+42.4}
\expandafter\gdef\csname blade@num@depth/radeon780m/graphics-independent/p32/wgpu/hi\endcsname{-10.9}
\expandafter\gdef\csname blade@num@depth/radeon780m/graphics-independent/p32/wgpu/lo\endcsname{-23.4}
\expandafter\gdef\csname blade@num@depth/radeon780m/graphics-independent/p32/wgpu/mag\endcsname{11.3}
\expandafter\gdef\csname blade@num@depth/radeon780m/graphics-independent/p32/wgpu/maghi\endcsname{23.4}
\expandafter\gdef\csname blade@num@depth/radeon780m/graphics-independent/p32/wgpu/maglo\endcsname{10.9}
\expandafter\gdef\csname blade@num@depth/radeon780m/graphics-independent/p32/wgpu/pct\endcsname{-11.3}
\expandafter\gdef\csname blade@num@depth/radeon780m/graphics-independent/p4/control/hi\endcsname{+0.1}
\expandafter\gdef\csname blade@num@depth/radeon780m/graphics-independent/p4/control/lo\endcsname{-51.1}
\expandafter\gdef\csname blade@num@depth/radeon780m/graphics-independent/p4/control/pct\endcsname{-0.1}
\expandafter\gdef\csname blade@num@depth/radeon780m/graphics-independent/p4/floor\endcsname{51.1}
\expandafter\gdef\csname blade@num@depth/radeon780m/graphics-independent/p4/placement/hi\endcsname{+32.0}
\expandafter\gdef\csname blade@num@depth/radeon780m/graphics-independent/p4/placement/lo\endcsname{-36.1}
\expandafter\gdef\csname blade@num@depth/radeon780m/graphics-independent/p4/placement/pct\endcsname{+30.1}
\expandafter\gdef\csname blade@num@depth/radeon780m/graphics-independent/p4/wgpu/hi\endcsname{-21.7}
\expandafter\gdef\csname blade@num@depth/radeon780m/graphics-independent/p4/wgpu/lo\endcsname{-62.1}
\expandafter\gdef\csname blade@num@depth/radeon780m/graphics-independent/p4/wgpu/mag\endcsname{23.1}
\expandafter\gdef\csname blade@num@depth/radeon780m/graphics-independent/p4/wgpu/maghi\endcsname{62.1}
\expandafter\gdef\csname blade@num@depth/radeon780m/graphics-independent/p4/wgpu/maglo\endcsname{21.7}
\expandafter\gdef\csname blade@num@depth/radeon780m/graphics-independent/p4/wgpu/pct\endcsname{-23.1}
\expandafter\gdef\csname blade@num@depth/radeon780m/graphics-independent/p64/control/hi\endcsname{+18.6}
\expandafter\gdef\csname blade@num@depth/radeon780m/graphics-independent/p64/control/lo\endcsname{-7.9}
\expandafter\gdef\csname blade@num@depth/radeon780m/graphics-independent/p64/control/pct\endcsname{+0.1}
\expandafter\gdef\csname blade@num@depth/radeon780m/graphics-independent/p64/floor\endcsname{18.6}
\expandafter\gdef\csname blade@num@depth/radeon780m/graphics-independent/p64/placement/hi\endcsname{+100.6}
\expandafter\gdef\csname blade@num@depth/radeon780m/graphics-independent/p64/placement/lo\endcsname{+30.2}
\expandafter\gdef\csname blade@num@depth/radeon780m/graphics-independent/p64/placement/mag\endcsname{42.9}
\expandafter\gdef\csname blade@num@depth/radeon780m/graphics-independent/p64/placement/maghi\endcsname{100.6}
\expandafter\gdef\csname blade@num@depth/radeon780m/graphics-independent/p64/placement/maglo\endcsname{30.2}
\expandafter\gdef\csname blade@num@depth/radeon780m/graphics-independent/p64/placement/pct\endcsname{+42.9}
\expandafter\gdef\csname blade@num@depth/radeon780m/graphics-independent/p64/wgpu/hi\endcsname{-11.3}
\expandafter\gdef\csname blade@num@depth/radeon780m/graphics-independent/p64/wgpu/lo\endcsname{-26.6}
\expandafter\gdef\csname blade@num@depth/radeon780m/graphics-independent/p64/wgpu/mag\endcsname{11.5}
\expandafter\gdef\csname blade@num@depth/radeon780m/graphics-independent/p64/wgpu/maghi\endcsname{26.6}
\expandafter\gdef\csname blade@num@depth/radeon780m/graphics-independent/p64/wgpu/maglo\endcsname{11.3}
\expandafter\gdef\csname blade@num@depth/radeon780m/graphics-independent/p64/wgpu/pct\endcsname{-11.5}
\expandafter\gdef\csname blade@num@depth/radeon780m/graphics-independent/p8/control/hi\endcsname{-0.2}
\expandafter\gdef\csname blade@num@depth/radeon780m/graphics-independent/p8/control/lo\endcsname{-50.6}
\expandafter\gdef\csname blade@num@depth/radeon780m/graphics-independent/p8/control/mag\endcsname{1.4}
\expandafter\gdef\csname blade@num@depth/radeon780m/graphics-independent/p8/control/maghi\endcsname{50.6}
\expandafter\gdef\csname blade@num@depth/radeon780m/graphics-independent/p8/control/maglo\endcsname{0.2}
\expandafter\gdef\csname blade@num@depth/radeon780m/graphics-independent/p8/control/pct\endcsname{-1.4}
\expandafter\gdef\csname blade@num@depth/radeon780m/graphics-independent/p8/floor\endcsname{50.6}
\expandafter\gdef\csname blade@num@depth/radeon780m/graphics-independent/p8/placement/hi\endcsname{+106.0}
\expandafter\gdef\csname blade@num@depth/radeon780m/graphics-independent/p8/placement/lo\endcsname{-33.0}
\expandafter\gdef\csname blade@num@depth/radeon780m/graphics-independent/p8/placement/pct\endcsname{+36.4}
\expandafter\gdef\csname blade@num@depth/radeon780m/graphics-independent/p8/wgpu/hi\endcsname{-22.2}
\expandafter\gdef\csname blade@num@depth/radeon780m/graphics-independent/p8/wgpu/lo\endcsname{-61.9}
\expandafter\gdef\csname blade@num@depth/radeon780m/graphics-independent/p8/wgpu/mag\endcsname{23.3}
\expandafter\gdef\csname blade@num@depth/radeon780m/graphics-independent/p8/wgpu/maghi\endcsname{61.9}
\expandafter\gdef\csname blade@num@depth/radeon780m/graphics-independent/p8/wgpu/maglo\endcsname{22.2}
\expandafter\gdef\csname blade@num@depth/radeon780m/graphics-independent/p8/wgpu/pct\endcsname{-23.3}
\expandafter\gdef\csname blade@num@depth/radeon780m/graphics-independent/resolved-placement-counts\endcsname{3}
\expandafter\gdef\csname blade@num@depth/rtx5070/graphics-independent/p1/control/hi\endcsname{+0.0}
\expandafter\gdef\csname blade@num@depth/rtx5070/graphics-independent/p1/control/lo\endcsname{-0.5}
\expandafter\gdef\csname blade@num@depth/rtx5070/graphics-independent/p1/control/pct\endcsname{+0.0}
\expandafter\gdef\csname blade@num@depth/rtx5070/graphics-independent/p1/floor\endcsname{0.5}
\expandafter\gdef\csname blade@num@depth/rtx5070/graphics-independent/p1/placement/hi\endcsname{+0.0}
\expandafter\gdef\csname blade@num@depth/rtx5070/graphics-independent/p1/placement/lo\endcsname{-0.4}
\expandafter\gdef\csname blade@num@depth/rtx5070/graphics-independent/p1/placement/pct\endcsname{+0.0}
\expandafter\gdef\csname blade@num@depth/rtx5070/graphics-independent/p1/wgpu/hi\endcsname{-26.7}
\expandafter\gdef\csname blade@num@depth/rtx5070/graphics-independent/p1/wgpu/lo\endcsname{-26.9}
\expandafter\gdef\csname blade@num@depth/rtx5070/graphics-independent/p1/wgpu/mag\endcsname{26.7}
\expandafter\gdef\csname blade@num@depth/rtx5070/graphics-independent/p1/wgpu/maghi\endcsname{26.9}
\expandafter\gdef\csname blade@num@depth/rtx5070/graphics-independent/p1/wgpu/maglo\endcsname{26.7}
\expandafter\gdef\csname blade@num@depth/rtx5070/graphics-independent/p1/wgpu/pct\endcsname{-26.7}
\expandafter\gdef\csname blade@num@depth/rtx5070/graphics-independent/p16/control/hi\endcsname{+0.1}
\expandafter\gdef\csname blade@num@depth/rtx5070/graphics-independent/p16/control/lo\endcsname{-0.5}
\expandafter\gdef\csname blade@num@depth/rtx5070/graphics-independent/p16/control/pct\endcsname{-0.1}
\expandafter\gdef\csname blade@num@depth/rtx5070/graphics-independent/p16/floor\endcsname{0.5}
\expandafter\gdef\csname blade@num@depth/rtx5070/graphics-independent/p16/placement/hi\endcsname{-30.9}
\expandafter\gdef\csname blade@num@depth/rtx5070/graphics-independent/p16/placement/lo\endcsname{-32.2}
\expandafter\gdef\csname blade@num@depth/rtx5070/graphics-independent/p16/placement/mag\endcsname{31.5}
\expandafter\gdef\csname blade@num@depth/rtx5070/graphics-independent/p16/placement/maghi\endcsname{32.2}
\expandafter\gdef\csname blade@num@depth/rtx5070/graphics-independent/p16/placement/maglo\endcsname{30.9}
\expandafter\gdef\csname blade@num@depth/rtx5070/graphics-independent/p16/placement/pct\endcsname{-31.5}
\expandafter\gdef\csname blade@num@depth/rtx5070/graphics-independent/p16/wgpu/hi\endcsname{-33.2}
\expandafter\gdef\csname blade@num@depth/rtx5070/graphics-independent/p16/wgpu/lo\endcsname{-35.2}
\expandafter\gdef\csname blade@num@depth/rtx5070/graphics-independent/p16/wgpu/mag\endcsname{34.7}
\expandafter\gdef\csname blade@num@depth/rtx5070/graphics-independent/p16/wgpu/maghi\endcsname{35.2}
\expandafter\gdef\csname blade@num@depth/rtx5070/graphics-independent/p16/wgpu/maglo\endcsname{33.2}
\expandafter\gdef\csname blade@num@depth/rtx5070/graphics-independent/p16/wgpu/pct\endcsname{-34.7}
\expandafter\gdef\csname blade@num@depth/rtx5070/graphics-independent/p2/control/hi\endcsname{+0.0}
\expandafter\gdef\csname blade@num@depth/rtx5070/graphics-independent/p2/control/lo\endcsname{-0.2}
\expandafter\gdef\csname blade@num@depth/rtx5070/graphics-independent/p2/control/pct\endcsname{+0.0}
\expandafter\gdef\csname blade@num@depth/rtx5070/graphics-independent/p2/floor\endcsname{0.2}
\expandafter\gdef\csname blade@num@depth/rtx5070/graphics-independent/p2/placement/hi\endcsname{-16.3}
\expandafter\gdef\csname blade@num@depth/rtx5070/graphics-independent/p2/placement/lo\endcsname{-16.5}
\expandafter\gdef\csname blade@num@depth/rtx5070/graphics-independent/p2/placement/mag\endcsname{16.4}
\expandafter\gdef\csname blade@num@depth/rtx5070/graphics-independent/p2/placement/maghi\endcsname{16.5}
\expandafter\gdef\csname blade@num@depth/rtx5070/graphics-independent/p2/placement/maglo\endcsname{16.3}
\expandafter\gdef\csname blade@num@depth/rtx5070/graphics-independent/p2/placement/pct\endcsname{-16.4}
\expandafter\gdef\csname blade@num@depth/rtx5070/graphics-independent/p2/wgpu/hi\endcsname{-30.0}
\expandafter\gdef\csname blade@num@depth/rtx5070/graphics-independent/p2/wgpu/lo\endcsname{-30.1}
\expandafter\gdef\csname blade@num@depth/rtx5070/graphics-independent/p2/wgpu/mag\endcsname{30.1}
\expandafter\gdef\csname blade@num@depth/rtx5070/graphics-independent/p2/wgpu/maghi\endcsname{30.1}
\expandafter\gdef\csname blade@num@depth/rtx5070/graphics-independent/p2/wgpu/maglo\endcsname{30.0}
\expandafter\gdef\csname blade@num@depth/rtx5070/graphics-independent/p2/wgpu/pct\endcsname{-30.1}
\expandafter\gdef\csname blade@num@depth/rtx5070/graphics-independent/p32/control/hi\endcsname{+1.0}
\expandafter\gdef\csname blade@num@depth/rtx5070/graphics-independent/p32/control/lo\endcsname{-0.1}
\expandafter\gdef\csname blade@num@depth/rtx5070/graphics-independent/p32/control/pct\endcsname{+0.1}
\expandafter\gdef\csname blade@num@depth/rtx5070/graphics-independent/p32/floor\endcsname{1.0}
\expandafter\gdef\csname blade@num@depth/rtx5070/graphics-independent/p32/placement/hi\endcsname{-30.3}
\expandafter\gdef\csname blade@num@depth/rtx5070/graphics-independent/p32/placement/lo\endcsname{-31.2}
\expandafter\gdef\csname blade@num@depth/rtx5070/graphics-independent/p32/placement/mag\endcsname{30.6}
\expandafter\gdef\csname blade@num@depth/rtx5070/graphics-independent/p32/placement/maghi\endcsname{31.2}
\expandafter\gdef\csname blade@num@depth/rtx5070/graphics-independent/p32/placement/maglo\endcsname{30.3}
\expandafter\gdef\csname blade@num@depth/rtx5070/graphics-independent/p32/placement/pct\endcsname{-30.6}
\expandafter\gdef\csname blade@num@depth/rtx5070/graphics-independent/p32/wgpu/hi\endcsname{-31.2}
\expandafter\gdef\csname blade@num@depth/rtx5070/graphics-independent/p32/wgpu/lo\endcsname{-32.3}
\expandafter\gdef\csname blade@num@depth/rtx5070/graphics-independent/p32/wgpu/mag\endcsname{31.8}
\expandafter\gdef\csname blade@num@depth/rtx5070/graphics-independent/p32/wgpu/maghi\endcsname{32.3}
\expandafter\gdef\csname blade@num@depth/rtx5070/graphics-independent/p32/wgpu/maglo\endcsname{31.2}
\expandafter\gdef\csname blade@num@depth/rtx5070/graphics-independent/p32/wgpu/pct\endcsname{-31.8}
\expandafter\gdef\csname blade@num@depth/rtx5070/graphics-independent/p4/control/hi\endcsname{+0.1}
\expandafter\gdef\csname blade@num@depth/rtx5070/graphics-independent/p4/control/lo\endcsname{-0.1}
\expandafter\gdef\csname blade@num@depth/rtx5070/graphics-independent/p4/control/pct\endcsname{+0.0}
\expandafter\gdef\csname blade@num@depth/rtx5070/graphics-independent/p4/floor\endcsname{0.1}
\expandafter\gdef\csname blade@num@depth/rtx5070/graphics-independent/p4/placement/hi\endcsname{-25.6}
\expandafter\gdef\csname blade@num@depth/rtx5070/graphics-independent/p4/placement/lo\endcsname{-25.8}
\expandafter\gdef\csname blade@num@depth/rtx5070/graphics-independent/p4/placement/mag\endcsname{25.7}
\expandafter\gdef\csname blade@num@depth/rtx5070/graphics-independent/p4/placement/maghi\endcsname{25.8}
\expandafter\gdef\csname blade@num@depth/rtx5070/graphics-independent/p4/placement/maglo\endcsname{25.6}
\expandafter\gdef\csname blade@num@depth/rtx5070/graphics-independent/p4/placement/pct\endcsname{-25.7}
\expandafter\gdef\csname blade@num@depth/rtx5070/graphics-independent/p4/wgpu/hi\endcsname{-32.7}
\expandafter\gdef\csname blade@num@depth/rtx5070/graphics-independent/p4/wgpu/lo\endcsname{-32.8}
\expandafter\gdef\csname blade@num@depth/rtx5070/graphics-independent/p4/wgpu/mag\endcsname{32.8}
\expandafter\gdef\csname blade@num@depth/rtx5070/graphics-independent/p4/wgpu/maghi\endcsname{32.8}
\expandafter\gdef\csname blade@num@depth/rtx5070/graphics-independent/p4/wgpu/maglo\endcsname{32.7}
\expandafter\gdef\csname blade@num@depth/rtx5070/graphics-independent/p4/wgpu/pct\endcsname{-32.8}
\expandafter\gdef\csname blade@num@depth/rtx5070/graphics-independent/p64/control/hi\endcsname{+0.0}
\expandafter\gdef\csname blade@num@depth/rtx5070/graphics-independent/p64/control/lo\endcsname{-0.1}
\expandafter\gdef\csname blade@num@depth/rtx5070/graphics-independent/p64/control/pct\endcsname{+0.0}
\expandafter\gdef\csname blade@num@depth/rtx5070/graphics-independent/p64/floor\endcsname{0.1}
\expandafter\gdef\csname blade@num@depth/rtx5070/graphics-independent/p64/placement/hi\endcsname{-30.0}
\expandafter\gdef\csname blade@num@depth/rtx5070/graphics-independent/p64/placement/lo\endcsname{-30.6}
\expandafter\gdef\csname blade@num@depth/rtx5070/graphics-independent/p64/placement/mag\endcsname{30.4}
\expandafter\gdef\csname blade@num@depth/rtx5070/graphics-independent/p64/placement/maghi\endcsname{30.6}
\expandafter\gdef\csname blade@num@depth/rtx5070/graphics-independent/p64/placement/maglo\endcsname{30.0}
\expandafter\gdef\csname blade@num@depth/rtx5070/graphics-independent/p64/placement/pct\endcsname{-30.4}
\expandafter\gdef\csname blade@num@depth/rtx5070/graphics-independent/p64/wgpu/hi\endcsname{-30.5}
\expandafter\gdef\csname blade@num@depth/rtx5070/graphics-independent/p64/wgpu/lo\endcsname{-31.6}
\expandafter\gdef\csname blade@num@depth/rtx5070/graphics-independent/p64/wgpu/mag\endcsname{31.4}
\expandafter\gdef\csname blade@num@depth/rtx5070/graphics-independent/p64/wgpu/maghi\endcsname{31.6}
\expandafter\gdef\csname blade@num@depth/rtx5070/graphics-independent/p64/wgpu/maglo\endcsname{30.5}
\expandafter\gdef\csname blade@num@depth/rtx5070/graphics-independent/p64/wgpu/pct\endcsname{-31.4}
\expandafter\gdef\csname blade@num@depth/rtx5070/graphics-independent/p8/control/hi\endcsname{+0.0}
\expandafter\gdef\csname blade@num@depth/rtx5070/graphics-independent/p8/control/lo\endcsname{-0.1}
\expandafter\gdef\csname blade@num@depth/rtx5070/graphics-independent/p8/control/pct\endcsname{-0.0}
\expandafter\gdef\csname blade@num@depth/rtx5070/graphics-independent/p8/floor\endcsname{0.1}
\expandafter\gdef\csname blade@num@depth/rtx5070/graphics-independent/p8/placement/hi\endcsname{-30.8}
\expandafter\gdef\csname blade@num@depth/rtx5070/graphics-independent/p8/placement/lo\endcsname{-31.0}
\expandafter\gdef\csname blade@num@depth/rtx5070/graphics-independent/p8/placement/mag\endcsname{30.9}
\expandafter\gdef\csname blade@num@depth/rtx5070/graphics-independent/p8/placement/maghi\endcsname{31.0}
\expandafter\gdef\csname blade@num@depth/rtx5070/graphics-independent/p8/placement/maglo\endcsname{30.8}
\expandafter\gdef\csname blade@num@depth/rtx5070/graphics-independent/p8/placement/pct\endcsname{-30.9}
\expandafter\gdef\csname blade@num@depth/rtx5070/graphics-independent/p8/wgpu/hi\endcsname{-34.3}
\expandafter\gdef\csname blade@num@depth/rtx5070/graphics-independent/p8/wgpu/lo\endcsname{-34.5}
\expandafter\gdef\csname blade@num@depth/rtx5070/graphics-independent/p8/wgpu/mag\endcsname{34.4}
\expandafter\gdef\csname blade@num@depth/rtx5070/graphics-independent/p8/wgpu/maghi\endcsname{34.5}
\expandafter\gdef\csname blade@num@depth/rtx5070/graphics-independent/p8/wgpu/maglo\endcsname{34.3}
\expandafter\gdef\csname blade@num@depth/rtx5070/graphics-independent/p8/wgpu/pct\endcsname{-34.4}
\expandafter\gdef\csname blade@num@depth/rtx5070/graphics-independent/resolved-placement-counts\endcsname{6}
\expandafter\gdef\csname blade@num@dispersion/applem3/gpu\endcsname{4}
\expandafter\gdef\csname blade@num@dispersion/applem3/host\endcsname{6}
\expandafter\gdef\csname blade@num@dispersion/intelxe/gpu\endcsname{1}
\expandafter\gdef\csname blade@num@dispersion/intelxe/host\endcsname{7}
\expandafter\gdef\csname blade@num@dispersion/radeon780m/gpu\endcsname{1}
\expandafter\gdef\csname blade@num@dispersion/radeon780m/host\endcsname{7}
\expandafter\gdef\csname blade@num@dispersion/raphael/gpu\endcsname{0}
\expandafter\gdef\csname blade@num@dispersion/raphael/host\endcsname{2}
\expandafter\gdef\csname blade@num@dispersion/rtx5070/gpu\endcsname{0}
\expandafter\gdef\csname blade@num@dispersion/rtx5070/host\endcsname{16}
\expandafter\gdef\csname blade@num@dispersion/rx7900xt/gpu\endcsname{0}
\expandafter\gdef\csname blade@num@dispersion/rx7900xt/host\endcsname{1}
\expandafter\gdef\csname blade@num@drift/applem3/median\endcsname{-0.1}
\expandafter\gdef\csname blade@num@drift/applem3/worst\endcsname{-100.0}
\expandafter\gdef\csname blade@num@drift/intelxe/median\endcsname{+0.0}
\expandafter\gdef\csname blade@num@drift/intelxe/worst\endcsname{-9.9}
\expandafter\gdef\csname blade@num@drift/radeon780m/median\endcsname{-0.0}
\expandafter\gdef\csname blade@num@drift/radeon780m/worst\endcsname{+114.7}
\expandafter\gdef\csname blade@num@drift/raphael/median\endcsname{-0.0}
\expandafter\gdef\csname blade@num@drift/raphael/worst\endcsname{-0.3}
\expandafter\gdef\csname blade@num@drift/rtx5070/median\endcsname{+0.1}
\expandafter\gdef\csname blade@num@drift/rtx5070/worst\endcsname{+8.6}
\expandafter\gdef\csname blade@num@drift/rx7900xt/median\endcsname{+0.0}
\expandafter\gdef\csname blade@num@drift/rx7900xt/worst\endcsname{+9.9}
\expandafter\gdef\csname blade@num@floor/above2\endcsname{10}
\expandafter\gdef\csname blade@num@floor/above30\endcsname{2}
\expandafter\gdef\csname blade@num@floor/below2max\endcsname{0.8}
\expandafter\gdef\csname blade@num@floor/count\endcsname{30}
\expandafter\gdef\csname blade@num@floor/max\endcsname{106.7}
\expandafter\gdef\csname blade@num@hazard/dependent/event/encode/p100/hi\endcsname{8.7}
\expandafter\gdef\csname blade@num@hazard/dependent/event/encode/p100/lo\endcsname{8.4}
\expandafter\gdef\csname blade@num@hazard/dependent/event/encode/p100/med\endcsname{8.5}
\expandafter\gdef\csname blade@num@hazard/dependent/event/encode/p500/hi\endcsname{8.7}
\expandafter\gdef\csname blade@num@hazard/dependent/event/encode/p500/lo\endcsname{8.4}
\expandafter\gdef\csname blade@num@hazard/dependent/event/encode/p500/med\endcsname{8.5}
\expandafter\gdef\csname blade@num@hazard/dependent/event/gpu/p100/hi\endcsname{10.3}
\expandafter\gdef\csname blade@num@hazard/dependent/event/gpu/p100/lo\endcsname{9.9}
\expandafter\gdef\csname blade@num@hazard/dependent/event/gpu/p100/med\endcsname{10.1}
\expandafter\gdef\csname blade@num@hazard/dependent/event/gpu/p500/hi\endcsname{11.3}
\expandafter\gdef\csname blade@num@hazard/dependent/event/gpu/p500/lo\endcsname{11.0}
\expandafter\gdef\csname blade@num@hazard/dependent/event/gpu/p500/med\endcsname{11.1}
\expandafter\gdef\csname blade@num@hazard/dependent/fence/encode/p100/hi\endcsname{9.1}
\expandafter\gdef\csname blade@num@hazard/dependent/fence/encode/p100/lo\endcsname{8.8}
\expandafter\gdef\csname blade@num@hazard/dependent/fence/encode/p100/med\endcsname{8.9}
\expandafter\gdef\csname blade@num@hazard/dependent/fence/encode/p500/hi\endcsname{9.0}
\expandafter\gdef\csname blade@num@hazard/dependent/fence/encode/p500/lo\endcsname{8.8}
\expandafter\gdef\csname blade@num@hazard/dependent/fence/encode/p500/med\endcsname{8.8}
\expandafter\gdef\csname blade@num@hazard/dependent/fence/gpu/p100/hi\endcsname{10.3}
\expandafter\gdef\csname blade@num@hazard/dependent/fence/gpu/p100/lo\endcsname{9.9}
\expandafter\gdef\csname blade@num@hazard/dependent/fence/gpu/p100/med\endcsname{10.1}
\expandafter\gdef\csname blade@num@hazard/dependent/fence/gpu/p500/hi\endcsname{11.4}
\expandafter\gdef\csname blade@num@hazard/dependent/fence/gpu/p500/lo\endcsname{11.0}
\expandafter\gdef\csname blade@num@hazard/dependent/fence/gpu/p500/med\endcsname{11.1}
\expandafter\gdef\csname blade@num@hazard/independent/encode/hi\endcsname{+5.3}
\expandafter\gdef\csname blade@num@hazard/independent/encode/lo\endcsname{-2.0}
\expandafter\gdef\csname blade@num@hazard/independent/gpu/hi\endcsname{+2.1}
\expandafter\gdef\csname blade@num@hazard/independent/gpu/lo\endcsname{-3.5}
\expandafter\gdef\csname blade@num@hazard/rows\endcsname{40800}
\expandafter\gdef\csname blade@num@hazard/sessions\endcsname{10}
\expandafter\gdef\csname blade@num@host/ratio/all/count\endcsname{24}
\expandafter\gdef\csname blade@num@host/ratio/all/max\endcsname{5.9}
\expandafter\gdef\csname blade@num@host/ratio/all/min\endcsname{1.3}
\expandafter\gdef\csname blade@num@host/ratio/count\endcsname{24}
\expandafter\gdef\csname blade@num@host/ratio/max\endcsname{5.9}
\expandafter\gdef\csname blade@num@host/ratio/min\endcsname{1.3}
\expandafter\gdef\csname blade@num@host/usable/dispersion\endcsname{16}
\expandafter\gdef\csname blade@num@intelxe/compute-chain/autous\endcsname{3594.0}
\expandafter\gdef\csname blade@num@intelxe/compute-chain/both/hi\endcsname{+2.6}
\expandafter\gdef\csname blade@num@intelxe/compute-chain/both/lo\endcsname{-2.5}
\expandafter\gdef\csname blade@num@intelxe/compute-chain/both/pct\endcsname{+0.0}
\expandafter\gdef\csname blade@num@intelxe/compute-chain/bothus\endcsname{3506.2}
\expandafter\gdef\csname blade@num@intelxe/compute-chain/control/hi\endcsname{+0.0}
\expandafter\gdef\csname blade@num@intelxe/compute-chain/control/lo\endcsname{-4.9}
\expandafter\gdef\csname blade@num@intelxe/compute-chain/control/pct\endcsname{-0.0}
\expandafter\gdef\csname blade@num@intelxe/compute-chain/controlus\endcsname{3505.4}
\expandafter\gdef\csname blade@num@intelxe/compute-chain/floor\endcsname{4.9}
\expandafter\gdef\csname blade@num@intelxe/compute-chain/hostautous\endcsname{83.7}
\expandafter\gdef\csname blade@num@intelxe/compute-chain/hostboth/hi\endcsname{+7.2}
\expandafter\gdef\csname blade@num@intelxe/compute-chain/hostboth/lo\endcsname{-8.4}
\expandafter\gdef\csname blade@num@intelxe/compute-chain/hostboth/pct\endcsname{-4.4}
\expandafter\gdef\csname blade@num@intelxe/compute-chain/hostbothus\endcsname{80.9}
\expandafter\gdef\csname blade@num@intelxe/compute-chain/hostcontrol/hi\endcsname{+11.8}
\expandafter\gdef\csname blade@num@intelxe/compute-chain/hostcontrol/lo\endcsname{-5.6}
\expandafter\gdef\csname blade@num@intelxe/compute-chain/hostcontrol/pct\endcsname{-0.9}
\expandafter\gdef\csname blade@num@intelxe/compute-chain/hostcontrolus\endcsname{82.2}
\expandafter\gdef\csname blade@num@intelxe/compute-chain/hostplacement/hi\endcsname{+6.8}
\expandafter\gdef\csname blade@num@intelxe/compute-chain/hostplacement/lo\endcsname{-8.9}
\expandafter\gdef\csname blade@num@intelxe/compute-chain/hostplacement/pct\endcsname{-4.4}
\expandafter\gdef\csname blade@num@intelxe/compute-chain/hostplacementus\endcsname{80.9}
\expandafter\gdef\csname blade@num@intelxe/compute-chain/hostscope/hi\endcsname{+5.8}
\expandafter\gdef\csname blade@num@intelxe/compute-chain/hostscope/lo\endcsname{-5.4}
\expandafter\gdef\csname blade@num@intelxe/compute-chain/hostscope/pct\endcsname{-3.1}
\expandafter\gdef\csname blade@num@intelxe/compute-chain/hostscopeus\endcsname{80.9}
\expandafter\gdef\csname blade@num@intelxe/compute-chain/hostwgpu/hi\endcsname{+336.7}
\expandafter\gdef\csname blade@num@intelxe/compute-chain/hostwgpu/lo\endcsname{+309.8}
\expandafter\gdef\csname blade@num@intelxe/compute-chain/hostwgpu/mag\endcsname{324.0}
\expandafter\gdef\csname blade@num@intelxe/compute-chain/hostwgpu/maghi\endcsname{336.7}
\expandafter\gdef\csname blade@num@intelxe/compute-chain/hostwgpu/maglo\endcsname{309.8}
\expandafter\gdef\csname blade@num@intelxe/compute-chain/hostwgpu/pct\endcsname{+324.0}
\expandafter\gdef\csname blade@num@intelxe/compute-chain/hostwgpuus\endcsname{354.7}
\expandafter\gdef\csname blade@num@intelxe/compute-chain/manualscope/hi\endcsname{+1.3}
\expandafter\gdef\csname blade@num@intelxe/compute-chain/manualscope/lo\endcsname{-2.9}
\expandafter\gdef\csname blade@num@intelxe/compute-chain/manualscope/pct\endcsname{-0.0}
\expandafter\gdef\csname blade@num@intelxe/compute-chain/placement/hi\endcsname{+0.8}
\expandafter\gdef\csname blade@num@intelxe/compute-chain/placement/lo\endcsname{-0.0}
\expandafter\gdef\csname blade@num@intelxe/compute-chain/placement/pct\endcsname{+0.0}
\expandafter\gdef\csname blade@num@intelxe/compute-chain/placementus\endcsname{3594.4}
\expandafter\gdef\csname blade@num@intelxe/compute-chain/scope/hi\endcsname{+5.0}
\expandafter\gdef\csname blade@num@intelxe/compute-chain/scope/lo\endcsname{-1.6}
\expandafter\gdef\csname blade@num@intelxe/compute-chain/scope/pct\endcsname{-0.1}
\expandafter\gdef\csname blade@num@intelxe/compute-chain/scopeus\endcsname{3588.0}
\expandafter\gdef\csname blade@num@intelxe/compute-chain/waitautous\endcsname{3883.0}
\expandafter\gdef\csname blade@num@intelxe/compute-chain/waitbothus\endcsname{3799.1}
\expandafter\gdef\csname blade@num@intelxe/compute-chain/waitcontrolus\endcsname{3797.6}
\expandafter\gdef\csname blade@num@intelxe/compute-chain/waitplacementus\endcsname{3800.4}
\expandafter\gdef\csname blade@num@intelxe/compute-chain/waitscopeus\endcsname{3841.0}
\expandafter\gdef\csname blade@num@intelxe/compute-chain/waitwgpuus\endcsname{4433.5}
\expandafter\gdef\csname blade@num@intelxe/compute-chain/wgpu/hi\endcsname{+13.7}
\expandafter\gdef\csname blade@num@intelxe/compute-chain/wgpu/lo\endcsname{+10.2}
\expandafter\gdef\csname blade@num@intelxe/compute-chain/wgpu/mag\endcsname{12.9}
\expandafter\gdef\csname blade@num@intelxe/compute-chain/wgpu/maghi\endcsname{13.7}
\expandafter\gdef\csname blade@num@intelxe/compute-chain/wgpu/maglo\endcsname{10.2}
\expandafter\gdef\csname blade@num@intelxe/compute-chain/wgpu/pct\endcsname{+12.9}
\expandafter\gdef\csname blade@num@intelxe/compute-chain/wgpuus\endcsname{4057.8}
\expandafter\gdef\csname blade@num@intelxe/compute-independent/autous\endcsname{4017.0}
\expandafter\gdef\csname blade@num@intelxe/compute-independent/both/hi\endcsname{-1.8}
\expandafter\gdef\csname blade@num@intelxe/compute-independent/both/lo\endcsname{-9.0}
\expandafter\gdef\csname blade@num@intelxe/compute-independent/both/mag\endcsname{5.7}
\expandafter\gdef\csname blade@num@intelxe/compute-independent/both/maghi\endcsname{9.0}
\expandafter\gdef\csname blade@num@intelxe/compute-independent/both/maglo\endcsname{1.8}
\expandafter\gdef\csname blade@num@intelxe/compute-independent/both/pct\endcsname{-5.7}
\expandafter\gdef\csname blade@num@intelxe/compute-independent/bothus\endcsname{3772.5}
\expandafter\gdef\csname blade@num@intelxe/compute-independent/control/hi\endcsname{+2.4}
\expandafter\gdef\csname blade@num@intelxe/compute-independent/control/lo\endcsname{-0.6}
\expandafter\gdef\csname blade@num@intelxe/compute-independent/control/pct\endcsname{+0.1}
\expandafter\gdef\csname blade@num@intelxe/compute-independent/controlus\endcsname{4020.5}
\expandafter\gdef\csname blade@num@intelxe/compute-independent/floor\endcsname{2.4}
\expandafter\gdef\csname blade@num@intelxe/compute-independent/hostautous\endcsname{95.8}
\expandafter\gdef\csname blade@num@intelxe/compute-independent/hostboth/hi\endcsname{+17.8}
\expandafter\gdef\csname blade@num@intelxe/compute-independent/hostboth/lo\endcsname{-0.3}
\expandafter\gdef\csname blade@num@intelxe/compute-independent/hostboth/pct\endcsname{+6.5}
\expandafter\gdef\csname blade@num@intelxe/compute-independent/hostbothus\endcsname{106.4}
\expandafter\gdef\csname blade@num@intelxe/compute-independent/hostcontrol/hi\endcsname{+7.2}
\expandafter\gdef\csname blade@num@intelxe/compute-independent/hostcontrol/lo\endcsname{-6.2}
\expandafter\gdef\csname blade@num@intelxe/compute-independent/hostcontrol/pct\endcsname{+1.0}
\expandafter\gdef\csname blade@num@intelxe/compute-independent/hostcontrolus\endcsname{97.5}
\expandafter\gdef\csname blade@num@intelxe/compute-independent/hostplacement/hi\endcsname{+16.9}
\expandafter\gdef\csname blade@num@intelxe/compute-independent/hostplacement/lo\endcsname{-1.6}
\expandafter\gdef\csname blade@num@intelxe/compute-independent/hostplacement/pct\endcsname{+0.2}
\expandafter\gdef\csname blade@num@intelxe/compute-independent/hostplacementus\endcsname{97.5}
\expandafter\gdef\csname blade@num@intelxe/compute-independent/hostscope/hi\endcsname{+4.5}
\expandafter\gdef\csname blade@num@intelxe/compute-independent/hostscope/lo\endcsname{-7.1}
\expandafter\gdef\csname blade@num@intelxe/compute-independent/hostscope/pct\endcsname{+2.1}
\expandafter\gdef\csname blade@num@intelxe/compute-independent/hostscopeus\endcsname{98.4}
\expandafter\gdef\csname blade@num@intelxe/compute-independent/hostwgpu/hi\endcsname{+308.2}
\expandafter\gdef\csname blade@num@intelxe/compute-independent/hostwgpu/lo\endcsname{+266.1}
\expandafter\gdef\csname blade@num@intelxe/compute-independent/hostwgpu/mag\endcsname{297.5}
\expandafter\gdef\csname blade@num@intelxe/compute-independent/hostwgpu/maghi\endcsname{308.2}
\expandafter\gdef\csname blade@num@intelxe/compute-independent/hostwgpu/maglo\endcsname{266.1}
\expandafter\gdef\csname blade@num@intelxe/compute-independent/hostwgpu/pct\endcsname{+297.5}
\expandafter\gdef\csname blade@num@intelxe/compute-independent/hostwgpuus\endcsname{384.4}
\expandafter\gdef\csname blade@num@intelxe/compute-independent/manualscope/hi\endcsname{+3.1}
\expandafter\gdef\csname blade@num@intelxe/compute-independent/manualscope/lo\endcsname{-1.1}
\expandafter\gdef\csname blade@num@intelxe/compute-independent/manualscope/pct\endcsname{-0.3}
\expandafter\gdef\csname blade@num@intelxe/compute-independent/placement/hi\endcsname{-5.3}
\expandafter\gdef\csname blade@num@intelxe/compute-independent/placement/lo\endcsname{-7.5}
\expandafter\gdef\csname blade@num@intelxe/compute-independent/placement/mag\endcsname{6.5}
\expandafter\gdef\csname blade@num@intelxe/compute-independent/placement/maghi\endcsname{7.5}
\expandafter\gdef\csname blade@num@intelxe/compute-independent/placement/maglo\endcsname{5.3}
\expandafter\gdef\csname blade@num@intelxe/compute-independent/placement/pct\endcsname{-6.5}
\expandafter\gdef\csname blade@num@intelxe/compute-independent/placementus\endcsname{3781.8}
\expandafter\gdef\csname blade@num@intelxe/compute-independent/scope/hi\endcsname{+1.9}
\expandafter\gdef\csname blade@num@intelxe/compute-independent/scope/lo\endcsname{-1.0}
\expandafter\gdef\csname blade@num@intelxe/compute-independent/scope/pct\endcsname{-0.2}
\expandafter\gdef\csname blade@num@intelxe/compute-independent/scopeus\endcsname{4013.2}
\expandafter\gdef\csname blade@num@intelxe/compute-independent/waitautous\endcsname{4352.3}
\expandafter\gdef\csname blade@num@intelxe/compute-independent/waitbothus\endcsname{4115.1}
\expandafter\gdef\csname blade@num@intelxe/compute-independent/waitcontrolus\endcsname{4351.0}
\expandafter\gdef\csname blade@num@intelxe/compute-independent/waitplacementus\endcsname{4118.8}
\expandafter\gdef\csname blade@num@intelxe/compute-independent/waitscopeus\endcsname{4341.0}
\expandafter\gdef\csname blade@num@intelxe/compute-independent/waitwgpuus\endcsname{4958.3}
\expandafter\gdef\csname blade@num@intelxe/compute-independent/wgpu/hi\endcsname{+12.3}
\expandafter\gdef\csname blade@num@intelxe/compute-independent/wgpu/lo\endcsname{+8.1}
\expandafter\gdef\csname blade@num@intelxe/compute-independent/wgpu/mag\endcsname{11.7}
\expandafter\gdef\csname blade@num@intelxe/compute-independent/wgpu/maghi\endcsname{12.3}
\expandafter\gdef\csname blade@num@intelxe/compute-independent/wgpu/maglo\endcsname{8.1}
\expandafter\gdef\csname blade@num@intelxe/compute-independent/wgpu/pct\endcsname{+11.7}
\expandafter\gdef\csname blade@num@intelxe/compute-independent/wgpuus\endcsname{4400.1}
\expandafter\gdef\csname blade@num@intelxe/graphics-chain/autous\endcsname{3324.9}
\expandafter\gdef\csname blade@num@intelxe/graphics-chain/both/hi\endcsname{+0.0}
\expandafter\gdef\csname blade@num@intelxe/graphics-chain/both/lo\endcsname{-0.0}
\expandafter\gdef\csname blade@num@intelxe/graphics-chain/both/pct\endcsname{+0.0}
\expandafter\gdef\csname blade@num@intelxe/graphics-chain/bothus\endcsname{3324.9}
\expandafter\gdef\csname blade@num@intelxe/graphics-chain/control/hi\endcsname{+0.0}
\expandafter\gdef\csname blade@num@intelxe/graphics-chain/control/lo\endcsname{-4.4}
\expandafter\gdef\csname blade@num@intelxe/graphics-chain/control/pct\endcsname{-0.0}
\expandafter\gdef\csname blade@num@intelxe/graphics-chain/controlus\endcsname{3324.8}
\expandafter\gdef\csname blade@num@intelxe/graphics-chain/floor\endcsname{4.4}
\expandafter\gdef\csname blade@num@intelxe/graphics-chain/hostautous\endcsname{132.1}
\expandafter\gdef\csname blade@num@intelxe/graphics-chain/hostboth/hi\endcsname{+4.5}
\expandafter\gdef\csname blade@num@intelxe/graphics-chain/hostboth/lo\endcsname{-4.4}
\expandafter\gdef\csname blade@num@intelxe/graphics-chain/hostboth/pct\endcsname{+0.5}
\expandafter\gdef\csname blade@num@intelxe/graphics-chain/hostbothus\endcsname{130.1}
\expandafter\gdef\csname blade@num@intelxe/graphics-chain/hostcontrol/hi\endcsname{+3.0}
\expandafter\gdef\csname blade@num@intelxe/graphics-chain/hostcontrol/lo\endcsname{-4.0}
\expandafter\gdef\csname blade@num@intelxe/graphics-chain/hostcontrol/pct\endcsname{-0.0}
\expandafter\gdef\csname blade@num@intelxe/graphics-chain/hostcontrolus\endcsname{131.0}
\expandafter\gdef\csname blade@num@intelxe/graphics-chain/hostplacement/hi\endcsname{+5.0}
\expandafter\gdef\csname blade@num@intelxe/graphics-chain/hostplacement/lo\endcsname{-4.4}
\expandafter\gdef\csname blade@num@intelxe/graphics-chain/hostplacement/pct\endcsname{-0.9}
\expandafter\gdef\csname blade@num@intelxe/graphics-chain/hostplacementus\endcsname{130.6}
\expandafter\gdef\csname blade@num@intelxe/graphics-chain/hostscope/hi\endcsname{+6.4}
\expandafter\gdef\csname blade@num@intelxe/graphics-chain/hostscope/lo\endcsname{-4.0}
\expandafter\gdef\csname blade@num@intelxe/graphics-chain/hostscope/pct\endcsname{+2.4}
\expandafter\gdef\csname blade@num@intelxe/graphics-chain/hostscopeus\endcsname{132.5}
\expandafter\gdef\csname blade@num@intelxe/graphics-chain/hostwgpu/hi\endcsname{+209.6}
\expandafter\gdef\csname blade@num@intelxe/graphics-chain/hostwgpu/lo\endcsname{+191.2}
\expandafter\gdef\csname blade@num@intelxe/graphics-chain/hostwgpu/mag\endcsname{200.8}
\expandafter\gdef\csname blade@num@intelxe/graphics-chain/hostwgpu/maghi\endcsname{209.6}
\expandafter\gdef\csname blade@num@intelxe/graphics-chain/hostwgpu/maglo\endcsname{191.2}
\expandafter\gdef\csname blade@num@intelxe/graphics-chain/hostwgpu/pct\endcsname{+200.8}
\expandafter\gdef\csname blade@num@intelxe/graphics-chain/hostwgpuus\endcsname{399.4}
\expandafter\gdef\csname blade@num@intelxe/graphics-chain/manualscope/hi\endcsname{+2.6}
\expandafter\gdef\csname blade@num@intelxe/graphics-chain/manualscope/lo\endcsname{-0.0}
\expandafter\gdef\csname blade@num@intelxe/graphics-chain/manualscope/pct\endcsname{+0.0}
\expandafter\gdef\csname blade@num@intelxe/graphics-chain/placement/hi\endcsname{+0.0}
\expandafter\gdef\csname blade@num@intelxe/graphics-chain/placement/lo\endcsname{-3.4}
\expandafter\gdef\csname blade@num@intelxe/graphics-chain/placement/pct\endcsname{-0.0}
\expandafter\gdef\csname blade@num@intelxe/graphics-chain/placementus\endcsname{3324.7}
\expandafter\gdef\csname blade@num@intelxe/graphics-chain/scope/hi\endcsname{+2.0}
\expandafter\gdef\csname blade@num@intelxe/graphics-chain/scope/lo\endcsname{-0.6}
\expandafter\gdef\csname blade@num@intelxe/graphics-chain/scope/pct\endcsname{-0.6}
\expandafter\gdef\csname blade@num@intelxe/graphics-chain/scopeus\endcsname{3305.0}
\expandafter\gdef\csname blade@num@intelxe/graphics-chain/waitautous\endcsname{3799.9}
\expandafter\gdef\csname blade@num@intelxe/graphics-chain/waitbothus\endcsname{3800.7}
\expandafter\gdef\csname blade@num@intelxe/graphics-chain/waitcontrolus\endcsname{3798.6}
\expandafter\gdef\csname blade@num@intelxe/graphics-chain/waitplacementus\endcsname{3787.0}
\expandafter\gdef\csname blade@num@intelxe/graphics-chain/waitscopeus\endcsname{3777.1}
\expandafter\gdef\csname blade@num@intelxe/graphics-chain/waitwgpuus\endcsname{3762.8}
\expandafter\gdef\csname blade@num@intelxe/graphics-chain/wgpu/hi\endcsname{+4.9}
\expandafter\gdef\csname blade@num@intelxe/graphics-chain/wgpu/lo\endcsname{-0.3}
\expandafter\gdef\csname blade@num@intelxe/graphics-chain/wgpu/pct\endcsname{+4.9}
\expandafter\gdef\csname blade@num@intelxe/graphics-chain/wgpuus\endcsname{3401.3}
\expandafter\gdef\csname blade@num@intelxe/graphics-independent/autous\endcsname{4256.7}
\expandafter\gdef\csname blade@num@intelxe/graphics-independent/both/hi\endcsname{+3.5}
\expandafter\gdef\csname blade@num@intelxe/graphics-independent/both/lo\endcsname{-0.5}
\expandafter\gdef\csname blade@num@intelxe/graphics-independent/both/pct\endcsname{-0.3}
\expandafter\gdef\csname blade@num@intelxe/graphics-independent/bothus\endcsname{4241.2}
\expandafter\gdef\csname blade@num@intelxe/graphics-independent/control/hi\endcsname{+3.5}
\expandafter\gdef\csname blade@num@intelxe/graphics-independent/control/lo\endcsname{-0.0}
\expandafter\gdef\csname blade@num@intelxe/graphics-independent/control/pct\endcsname{+0.1}
\expandafter\gdef\csname blade@num@intelxe/graphics-independent/controlus\endcsname{4259.9}
\expandafter\gdef\csname blade@num@intelxe/graphics-independent/floor\endcsname{3.5}
\expandafter\gdef\csname blade@num@intelxe/graphics-independent/hostautous\endcsname{240.4}
\expandafter\gdef\csname blade@num@intelxe/graphics-independent/hostboth/hi\endcsname{+8.3}
\expandafter\gdef\csname blade@num@intelxe/graphics-independent/hostboth/lo\endcsname{-4.4}
\expandafter\gdef\csname blade@num@intelxe/graphics-independent/hostboth/pct\endcsname{+0.6}
\expandafter\gdef\csname blade@num@intelxe/graphics-independent/hostbothus\endcsname{234.9}
\expandafter\gdef\csname blade@num@intelxe/graphics-independent/hostcontrol/hi\endcsname{+9.0}
\expandafter\gdef\csname blade@num@intelxe/graphics-independent/hostcontrol/lo\endcsname{-3.2}
\expandafter\gdef\csname blade@num@intelxe/graphics-independent/hostcontrol/pct\endcsname{+0.7}
\expandafter\gdef\csname blade@num@intelxe/graphics-independent/hostcontrolus\endcsname{239.7}
\expandafter\gdef\csname blade@num@intelxe/graphics-independent/hostplacement/hi\endcsname{+3.8}
\expandafter\gdef\csname blade@num@intelxe/graphics-independent/hostplacement/lo\endcsname{-4.8}
\expandafter\gdef\csname blade@num@intelxe/graphics-independent/hostplacement/pct\endcsname{-0.4}
\expandafter\gdef\csname blade@num@intelxe/graphics-independent/hostplacementus\endcsname{234.9}
\expandafter\gdef\csname blade@num@intelxe/graphics-independent/hostscope/hi\endcsname{+6.8}
\expandafter\gdef\csname blade@num@intelxe/graphics-independent/hostscope/lo\endcsname{-6.2}
\expandafter\gdef\csname blade@num@intelxe/graphics-independent/hostscope/pct\endcsname{+2.7}
\expandafter\gdef\csname blade@num@intelxe/graphics-independent/hostscopeus\endcsname{241.3}
\expandafter\gdef\csname blade@num@intelxe/graphics-independent/hostwgpu/hi\endcsname{+129.5}
\expandafter\gdef\csname blade@num@intelxe/graphics-independent/hostwgpu/lo\endcsname{+114.5}
\expandafter\gdef\csname blade@num@intelxe/graphics-independent/hostwgpu/mag\endcsname{122.5}
\expandafter\gdef\csname blade@num@intelxe/graphics-independent/hostwgpu/maghi\endcsname{129.5}
\expandafter\gdef\csname blade@num@intelxe/graphics-independent/hostwgpu/maglo\endcsname{114.5}
\expandafter\gdef\csname blade@num@intelxe/graphics-independent/hostwgpu/pct\endcsname{+122.5}
\expandafter\gdef\csname blade@num@intelxe/graphics-independent/hostwgpuus\endcsname{532.7}
\expandafter\gdef\csname blade@num@intelxe/graphics-independent/manualscope/hi\endcsname{+1.5}
\expandafter\gdef\csname blade@num@intelxe/graphics-independent/manualscope/lo\endcsname{-0.2}
\expandafter\gdef\csname blade@num@intelxe/graphics-independent/manualscope/pct\endcsname{+0.0}
\expandafter\gdef\csname blade@num@intelxe/graphics-independent/placement/hi\endcsname{+0.1}
\expandafter\gdef\csname blade@num@intelxe/graphics-independent/placement/lo\endcsname{-0.5}
\expandafter\gdef\csname blade@num@intelxe/graphics-independent/placement/pct\endcsname{-0.4}
\expandafter\gdef\csname blade@num@intelxe/graphics-independent/placementus\endcsname{4242.2}
\expandafter\gdef\csname blade@num@intelxe/graphics-independent/scope/hi\endcsname{+2.9}
\expandafter\gdef\csname blade@num@intelxe/graphics-independent/scope/lo\endcsname{-0.5}
\expandafter\gdef\csname blade@num@intelxe/graphics-independent/scope/pct\endcsname{-0.3}
\expandafter\gdef\csname blade@num@intelxe/graphics-independent/scopeus\endcsname{4241.8}
\expandafter\gdef\csname blade@num@intelxe/graphics-independent/waitautous\endcsname{4821.2}
\expandafter\gdef\csname blade@num@intelxe/graphics-independent/waitbothus\endcsname{4799.6}
\expandafter\gdef\csname blade@num@intelxe/graphics-independent/waitcontrolus\endcsname{4830.7}
\expandafter\gdef\csname blade@num@intelxe/graphics-independent/waitplacementus\endcsname{4805.9}
\expandafter\gdef\csname blade@num@intelxe/graphics-independent/waitscopeus\endcsname{4801.9}
\expandafter\gdef\csname blade@num@intelxe/graphics-independent/waitwgpuus\endcsname{4999.5}
\expandafter\gdef\csname blade@num@intelxe/graphics-independent/wgpu/hi\endcsname{+8.7}
\expandafter\gdef\csname blade@num@intelxe/graphics-independent/wgpu/lo\endcsname{+3.9}
\expandafter\gdef\csname blade@num@intelxe/graphics-independent/wgpu/mag\endcsname{7.9}
\expandafter\gdef\csname blade@num@intelxe/graphics-independent/wgpu/maghi\endcsname{8.7}
\expandafter\gdef\csname blade@num@intelxe/graphics-independent/wgpu/maglo\endcsname{3.9}
\expandafter\gdef\csname blade@num@intelxe/graphics-independent/wgpu/pct\endcsname{+7.9}
\expandafter\gdef\csname blade@num@intelxe/graphics-independent/wgpuus\endcsname{4427.1}
\expandafter\gdef\csname blade@num@intelxe/mixed-chain/autous\endcsname{3829.7}
\expandafter\gdef\csname blade@num@intelxe/mixed-chain/both/hi\endcsname{+2.9}
\expandafter\gdef\csname blade@num@intelxe/mixed-chain/both/lo\endcsname{-0.6}
\expandafter\gdef\csname blade@num@intelxe/mixed-chain/both/pct\endcsname{+0.2}
\expandafter\gdef\csname blade@num@intelxe/mixed-chain/bothus\endcsname{3837.5}
\expandafter\gdef\csname blade@num@intelxe/mixed-chain/control/hi\endcsname{+5.0}
\expandafter\gdef\csname blade@num@intelxe/mixed-chain/control/lo\endcsname{-4.6}
\expandafter\gdef\csname blade@num@intelxe/mixed-chain/control/pct\endcsname{+2.0}
\expandafter\gdef\csname blade@num@intelxe/mixed-chain/controlus\endcsname{4021.2}
\expandafter\gdef\csname blade@num@intelxe/mixed-chain/floor\endcsname{5.0}
\expandafter\gdef\csname blade@num@intelxe/mixed-chain/hostautous\endcsname{185.4}
\expandafter\gdef\csname blade@num@intelxe/mixed-chain/hostboth/hi\endcsname{+6.6}
\expandafter\gdef\csname blade@num@intelxe/mixed-chain/hostboth/lo\endcsname{-2.1}
\expandafter\gdef\csname blade@num@intelxe/mixed-chain/hostboth/pct\endcsname{+4.0}
\expandafter\gdef\csname blade@num@intelxe/mixed-chain/hostbothus\endcsname{188.1}
\expandafter\gdef\csname blade@num@intelxe/mixed-chain/hostcontrol/hi\endcsname{-0.4}
\expandafter\gdef\csname blade@num@intelxe/mixed-chain/hostcontrol/lo\endcsname{-48.7}
\expandafter\gdef\csname blade@num@intelxe/mixed-chain/hostcontrol/mag\endcsname{11.2}
\expandafter\gdef\csname blade@num@intelxe/mixed-chain/hostcontrol/maghi\endcsname{48.7}
\expandafter\gdef\csname blade@num@intelxe/mixed-chain/hostcontrol/maglo\endcsname{0.4}
\expandafter\gdef\csname blade@num@intelxe/mixed-chain/hostcontrol/pct\endcsname{-11.2}
\expandafter\gdef\csname blade@num@intelxe/mixed-chain/hostcontrolus\endcsname{164.4}
\expandafter\gdef\csname blade@num@intelxe/mixed-chain/hostplacement/hi\endcsname{+4.3}
\expandafter\gdef\csname blade@num@intelxe/mixed-chain/hostplacement/lo\endcsname{-11.7}
\expandafter\gdef\csname blade@num@intelxe/mixed-chain/hostplacement/pct\endcsname{+1.5}
\expandafter\gdef\csname blade@num@intelxe/mixed-chain/hostplacementus\endcsname{190.4}
\expandafter\gdef\csname blade@num@intelxe/mixed-chain/hostscope/hi\endcsname{+6.5}
\expandafter\gdef\csname blade@num@intelxe/mixed-chain/hostscope/lo\endcsname{-5.4}
\expandafter\gdef\csname blade@num@intelxe/mixed-chain/hostscope/pct\endcsname{+4.2}
\expandafter\gdef\csname blade@num@intelxe/mixed-chain/hostscopeus\endcsname{188.3}
\expandafter\gdef\csname blade@num@intelxe/mixed-chain/manualscope/hi\endcsname{+4.9}
\expandafter\gdef\csname blade@num@intelxe/mixed-chain/manualscope/lo\endcsname{-4.9}
\expandafter\gdef\csname blade@num@intelxe/mixed-chain/manualscope/pct\endcsname{-1.3}
\expandafter\gdef\csname blade@num@intelxe/mixed-chain/placement/hi\endcsname{+5.1}
\expandafter\gdef\csname blade@num@intelxe/mixed-chain/placement/lo\endcsname{-2.5}
\expandafter\gdef\csname blade@num@intelxe/mixed-chain/placement/pct\endcsname{+2.1}
\expandafter\gdef\csname blade@num@intelxe/mixed-chain/placementus\endcsname{4026.2}
\expandafter\gdef\csname blade@num@intelxe/mixed-chain/scope/hi\endcsname{+5.2}
\expandafter\gdef\csname blade@num@intelxe/mixed-chain/scope/lo\endcsname{-0.5}
\expandafter\gdef\csname blade@num@intelxe/mixed-chain/scope/pct\endcsname{+0.5}
\expandafter\gdef\csname blade@num@intelxe/mixed-chain/scopeus\endcsname{3920.8}
\expandafter\gdef\csname blade@num@intelxe/mixed-chain/waitautous\endcsname{4336.2}
\expandafter\gdef\csname blade@num@intelxe/mixed-chain/waitbothus\endcsname{4343.4}
\expandafter\gdef\csname blade@num@intelxe/mixed-chain/waitcontrolus\endcsname{4458.3}
\expandafter\gdef\csname blade@num@intelxe/mixed-chain/waitplacementus\endcsname{4515.1}
\expandafter\gdef\csname blade@num@intelxe/mixed-chain/waitscopeus\endcsname{4408.2}
\expandafter\gdef\csname blade@num@intelxe/mixed-independent/autous\endcsname{4098.5}
\expandafter\gdef\csname blade@num@intelxe/mixed-independent/both/hi\endcsname{-3.5}
\expandafter\gdef\csname blade@num@intelxe/mixed-independent/both/lo\endcsname{-9.5}
\expandafter\gdef\csname blade@num@intelxe/mixed-independent/both/mag\endcsname{7.4}
\expandafter\gdef\csname blade@num@intelxe/mixed-independent/both/maghi\endcsname{9.5}
\expandafter\gdef\csname blade@num@intelxe/mixed-independent/both/maglo\endcsname{3.5}
\expandafter\gdef\csname blade@num@intelxe/mixed-independent/both/pct\endcsname{-7.4}
\expandafter\gdef\csname blade@num@intelxe/mixed-independent/bothus\endcsname{3777.4}
\expandafter\gdef\csname blade@num@intelxe/mixed-independent/control/hi\endcsname{+4.0}
\expandafter\gdef\csname blade@num@intelxe/mixed-independent/control/lo\endcsname{-3.7}
\expandafter\gdef\csname blade@num@intelxe/mixed-independent/control/pct\endcsname{+0.1}
\expandafter\gdef\csname blade@num@intelxe/mixed-independent/controlus\endcsname{4097.0}
\expandafter\gdef\csname blade@num@intelxe/mixed-independent/floor\endcsname{4.0}
\expandafter\gdef\csname blade@num@intelxe/mixed-independent/hostautous\endcsname{220.4}
\expandafter\gdef\csname blade@num@intelxe/mixed-independent/hostboth/hi\endcsname{+11.2}
\expandafter\gdef\csname blade@num@intelxe/mixed-independent/hostboth/lo\endcsname{-6.1}
\expandafter\gdef\csname blade@num@intelxe/mixed-independent/hostboth/pct\endcsname{+4.3}
\expandafter\gdef\csname blade@num@intelxe/mixed-independent/hostbothus\endcsname{224.4}
\expandafter\gdef\csname blade@num@intelxe/mixed-independent/hostcontrol/hi\endcsname{+6.5}
\expandafter\gdef\csname blade@num@intelxe/mixed-independent/hostcontrol/lo\endcsname{-12.2}
\expandafter\gdef\csname blade@num@intelxe/mixed-independent/hostcontrol/pct\endcsname{-2.9}
\expandafter\gdef\csname blade@num@intelxe/mixed-independent/hostcontrolus\endcsname{214.0}
\expandafter\gdef\csname blade@num@intelxe/mixed-independent/hostplacement/hi\endcsname{+10.2}
\expandafter\gdef\csname blade@num@intelxe/mixed-independent/hostplacement/lo\endcsname{+0.9}
\expandafter\gdef\csname blade@num@intelxe/mixed-independent/hostplacement/mag\endcsname{3.5}
\expandafter\gdef\csname blade@num@intelxe/mixed-independent/hostplacement/maghi\endcsname{10.2}
\expandafter\gdef\csname blade@num@intelxe/mixed-independent/hostplacement/maglo\endcsname{0.9}
\expandafter\gdef\csname blade@num@intelxe/mixed-independent/hostplacement/pct\endcsname{+3.5}
\expandafter\gdef\csname blade@num@intelxe/mixed-independent/hostplacementus\endcsname{222.8}
\expandafter\gdef\csname blade@num@intelxe/mixed-independent/hostscope/hi\endcsname{+6.0}
\expandafter\gdef\csname blade@num@intelxe/mixed-independent/hostscope/lo\endcsname{-6.3}
\expandafter\gdef\csname blade@num@intelxe/mixed-independent/hostscope/pct\endcsname{+0.7}
\expandafter\gdef\csname blade@num@intelxe/mixed-independent/hostscopeus\endcsname{225.1}
\expandafter\gdef\csname blade@num@intelxe/mixed-independent/manualscope/hi\endcsname{+4.7}
\expandafter\gdef\csname blade@num@intelxe/mixed-independent/manualscope/lo\endcsname{-0.1}
\expandafter\gdef\csname blade@num@intelxe/mixed-independent/manualscope/pct\endcsname{+0.1}
\expandafter\gdef\csname blade@num@intelxe/mixed-independent/placement/hi\endcsname{-7.4}
\expandafter\gdef\csname blade@num@intelxe/mixed-independent/placement/lo\endcsname{-11.3}
\expandafter\gdef\csname blade@num@intelxe/mixed-independent/placement/mag\endcsname{8.2}
\expandafter\gdef\csname blade@num@intelxe/mixed-independent/placement/maghi\endcsname{11.3}
\expandafter\gdef\csname blade@num@intelxe/mixed-independent/placement/maglo\endcsname{7.4}
\expandafter\gdef\csname blade@num@intelxe/mixed-independent/placement/pct\endcsname{-8.2}
\expandafter\gdef\csname blade@num@intelxe/mixed-independent/placementus\endcsname{3777.5}
\expandafter\gdef\csname blade@num@intelxe/mixed-independent/scope/hi\endcsname{+1.3}
\expandafter\gdef\csname blade@num@intelxe/mixed-independent/scope/lo\endcsname{-2.5}
\expandafter\gdef\csname blade@num@intelxe/mixed-independent/scope/pct\endcsname{-0.4}
\expandafter\gdef\csname blade@num@intelxe/mixed-independent/scopeus\endcsname{4170.4}
\expandafter\gdef\csname blade@num@intelxe/mixed-independent/waitautous\endcsname{4657.7}
\expandafter\gdef\csname blade@num@intelxe/mixed-independent/waitbothus\endcsname{4354.4}
\expandafter\gdef\csname blade@num@intelxe/mixed-independent/waitcontrolus\endcsname{4663.9}
\expandafter\gdef\csname blade@num@intelxe/mixed-independent/waitplacementus\endcsname{4355.0}
\expandafter\gdef\csname blade@num@intelxe/mixed-independent/waitscopeus\endcsname{4720.7}
\expandafter\gdef\csname blade@num@metal/host/max\endcsname{85}
\expandafter\gdef\csname blade@num@metal/host/min\endcsname{68}
\expandafter\gdef\csname blade@num@metal/hostshare/max\endcsname{75}
\expandafter\gdef\csname blade@num@metal/hostshare/min\endcsname{56}
\expandafter\gdef\csname blade@num@metal/overvulkan/max\endcsname{1.2}
\expandafter\gdef\csname blade@num@metal/overvulkan/min\endcsname{1.2}
\expandafter\gdef\csname blade@num@metal/waitsaved/max\endcsname{3}
\expandafter\gdef\csname blade@num@metal/waitsaved/min\endcsname{-5}
\expandafter\gdef\csname blade@num@profile/S1/blade/compute-independent/allocator\endcsname{0.7}
\expandafter\gdef\csname blade@num@profile/S1/blade/compute-independent/ashloader\endcsname{0.3}
\expandafter\gdef\csname blade@num@profile/S1/blade/compute-independent/blade\endcsname{20.7}
\expandafter\gdef\csname blade@num@profile/S1/blade/compute-independent/driver\endcsname{29.2}
\expandafter\gdef\csname blade@num@profile/S1/blade/compute-independent/kernel\endcsname{45.5}
\expandafter\gdef\csname blade@num@profile/S1/blade/compute-independent/libcruntime\endcsname{3.7}
\expandafter\gdef\csname blade@num@profile/S1/blade/compute-independent/other\endcsname{1.0}
\expandafter\gdef\csname blade@num@profile/S1/blade/compute-independent/system\endcsname{75}
\expandafter\gdef\csname blade@num@profile/S1/blade/compute-independent/wgputotal\endcsname{0}
\expandafter\gdef\csname blade@num@profile/S1/blade/graphics-independent/allocator\endcsname{0.8}
\expandafter\gdef\csname blade@num@profile/S1/blade/graphics-independent/ashloader\endcsname{0.1}
\expandafter\gdef\csname blade@num@profile/S1/blade/graphics-independent/blade\endcsname{1.9}
\expandafter\gdef\csname blade@num@profile/S1/blade/graphics-independent/driver\endcsname{46.1}
\expandafter\gdef\csname blade@num@profile/S1/blade/graphics-independent/kernel\endcsname{41.1}
\expandafter\gdef\csname blade@num@profile/S1/blade/graphics-independent/libcruntime\endcsname{4.0}
\expandafter\gdef\csname blade@num@profile/S1/blade/graphics-independent/other\endcsname{4.0}
\expandafter\gdef\csname blade@num@profile/S1/blade/graphics-independent/system\endcsname{87}
\expandafter\gdef\csname blade@num@profile/S1/blade/graphics-independent/wgputotal\endcsname{0}
\expandafter\gdef\csname blade@num@profile/S1/wgpu/compute-independent/allocator\endcsname{3.5}
\expandafter\gdef\csname blade@num@profile/S1/wgpu/compute-independent/ashloader\endcsname{0.1}
\expandafter\gdef\csname blade@num@profile/S1/wgpu/compute-independent/blade\endcsname{0.2}
\expandafter\gdef\csname blade@num@profile/S1/wgpu/compute-independent/driver\endcsname{40.5}
\expandafter\gdef\csname blade@num@profile/S1/wgpu/compute-independent/kernel\endcsname{36.1}
\expandafter\gdef\csname blade@num@profile/S1/wgpu/compute-independent/libcruntime\endcsname{3.9}
\expandafter\gdef\csname blade@num@profile/S1/wgpu/compute-independent/other\endcsname{3.4}
\expandafter\gdef\csname blade@num@profile/S1/wgpu/compute-independent/system\endcsname{77}
\expandafter\gdef\csname blade@num@profile/S1/wgpu/compute-independent/wgpucommand\endcsname{6.2}
\expandafter\gdef\csname blade@num@profile/S1/wgpu/compute-independent/wgpudeviceresource\endcsname{1.2}
\expandafter\gdef\csname blade@num@profile/S1/wgpu/compute-independent/wgpuhal\endcsname{1.1}
\expandafter\gdef\csname blade@num@profile/S1/wgpu/compute-independent/wgpuother\endcsname{0.8}
\expandafter\gdef\csname blade@num@profile/S1/wgpu/compute-independent/wgputotal\endcsname{15}
\expandafter\gdef\csname blade@num@profile/S1/wgpu/compute-independent/wgputracker\endcsname{5.1}
\expandafter\gdef\csname blade@num@profile/S1/wgpu/compute-independent/wgpuvalidation\endcsname{0.2}
\expandafter\gdef\csname blade@num@profile/S1/wgpu/graphics-independent/allocator\endcsname{3.8}
\expandafter\gdef\csname blade@num@profile/S1/wgpu/graphics-independent/ashloader\endcsname{0.1}
\expandafter\gdef\csname blade@num@profile/S1/wgpu/graphics-independent/blade\endcsname{0.2}
\expandafter\gdef\csname blade@num@profile/S1/wgpu/graphics-independent/driver\endcsname{36.7}
\expandafter\gdef\csname blade@num@profile/S1/wgpu/graphics-independent/kernel\endcsname{30.2}
\expandafter\gdef\csname blade@num@profile/S1/wgpu/graphics-independent/libcruntime\endcsname{5.5}
\expandafter\gdef\csname blade@num@profile/S1/wgpu/graphics-independent/other\endcsname{6.0}
\expandafter\gdef\csname blade@num@profile/S1/wgpu/graphics-independent/system\endcsname{67}
\expandafter\gdef\csname blade@num@profile/S1/wgpu/graphics-independent/wgpucommand\endcsname{8.7}
\expandafter\gdef\csname blade@num@profile/S1/wgpu/graphics-independent/wgpudeviceresource\endcsname{1.0}
\expandafter\gdef\csname blade@num@profile/S1/wgpu/graphics-independent/wgpuhal\endcsname{1.9}
\expandafter\gdef\csname blade@num@profile/S1/wgpu/graphics-independent/wgpuother\endcsname{0.8}
\expandafter\gdef\csname blade@num@profile/S1/wgpu/graphics-independent/wgputotal\endcsname{16}
\expandafter\gdef\csname blade@num@profile/S1/wgpu/graphics-independent/wgputracker\endcsname{3.1}
\expandafter\gdef\csname blade@num@profile/S1/wgpu/graphics-independent/wgpuvalidation\endcsname{0.0}
\expandafter\gdef\csname blade@num@profile/S2/blade/compute-independent/allocator\endcsname{0.1}
\expandafter\gdef\csname blade@num@profile/S2/blade/compute-independent/ashloader\endcsname{0.1}
\expandafter\gdef\csname blade@num@profile/S2/blade/compute-independent/blade\endcsname{5.1}
\expandafter\gdef\csname blade@num@profile/S2/blade/compute-independent/driver\endcsname{5.9}
\expandafter\gdef\csname blade@num@profile/S2/blade/compute-independent/kernel\endcsname{87.9}
\expandafter\gdef\csname blade@num@profile/S2/blade/compute-independent/libcruntime\endcsname{0.9}
\expandafter\gdef\csname blade@num@profile/S2/blade/compute-independent/other\endcsname{0.6}
\expandafter\gdef\csname blade@num@profile/S2/blade/compute-independent/system\endcsname{94}
\expandafter\gdef\csname blade@num@profile/S2/blade/compute-independent/wgputotal\endcsname{0}
\expandafter\gdef\csname blade@num@profile/S2/blade/graphics-independent/allocator\endcsname{0.6}
\expandafter\gdef\csname blade@num@profile/S2/blade/graphics-independent/blade\endcsname{2.2}
\expandafter\gdef\csname blade@num@profile/S2/blade/graphics-independent/driver\endcsname{80.3}
\expandafter\gdef\csname blade@num@profile/S2/blade/graphics-independent/kernel\endcsname{10.1}
\expandafter\gdef\csname blade@num@profile/S2/blade/graphics-independent/libcruntime\endcsname{7.3}
\expandafter\gdef\csname blade@num@profile/S2/blade/graphics-independent/other\endcsname{1.7}
\expandafter\gdef\csname blade@num@profile/S2/blade/graphics-independent/system\endcsname{90}
\expandafter\gdef\csname blade@num@profile/S2/blade/graphics-independent/wgputotal\endcsname{0}
\expandafter\gdef\csname blade@num@profile/S2/wgpu/compute-independent/allocator\endcsname{1.3}
\expandafter\gdef\csname blade@num@profile/S2/wgpu/compute-independent/ashloader\endcsname{0.1}
\expandafter\gdef\csname blade@num@profile/S2/wgpu/compute-independent/blade\endcsname{0.3}
\expandafter\gdef\csname blade@num@profile/S2/wgpu/compute-independent/driver\endcsname{24.8}
\expandafter\gdef\csname blade@num@profile/S2/wgpu/compute-independent/kernel\endcsname{34.6}
\expandafter\gdef\csname blade@num@profile/S2/wgpu/compute-independent/libcruntime\endcsname{5.6}
\expandafter\gdef\csname blade@num@profile/S2/wgpu/compute-independent/other\endcsname{3.1}
\expandafter\gdef\csname blade@num@profile/S2/wgpu/compute-independent/system\endcsname{59}
\expandafter\gdef\csname blade@num@profile/S2/wgpu/compute-independent/wgpucommand\endcsname{7.7}
\expandafter\gdef\csname blade@num@profile/S2/wgpu/compute-independent/wgpudeviceresource\endcsname{2.0}
\expandafter\gdef\csname blade@num@profile/S2/wgpu/compute-independent/wgpuhal\endcsname{1.5}
\expandafter\gdef\csname blade@num@profile/S2/wgpu/compute-independent/wgpuother\endcsname{1.3}
\expandafter\gdef\csname blade@num@profile/S2/wgpu/compute-independent/wgputotal\endcsname{31}
\expandafter\gdef\csname blade@num@profile/S2/wgpu/compute-independent/wgputracker\endcsname{18.6}
\expandafter\gdef\csname blade@num@profile/S2/wgpu/compute-independent/wgpuvalidation\endcsname{0.1}
\expandafter\gdef\csname blade@num@profile/S2/wgpu/graphics-independent/allocator\endcsname{0.8}
\expandafter\gdef\csname blade@num@profile/S2/wgpu/graphics-independent/blade\endcsname{0.4}
\expandafter\gdef\csname blade@num@profile/S2/wgpu/graphics-independent/driver\endcsname{35.3}
\expandafter\gdef\csname blade@num@profile/S2/wgpu/graphics-independent/kernel\endcsname{24.7}
\expandafter\gdef\csname blade@num@profile/S2/wgpu/graphics-independent/libcruntime\endcsname{8.1}
\expandafter\gdef\csname blade@num@profile/S2/wgpu/graphics-independent/other\endcsname{2.7}
\expandafter\gdef\csname blade@num@profile/S2/wgpu/graphics-independent/system\endcsname{60}
\expandafter\gdef\csname blade@num@profile/S2/wgpu/graphics-independent/wgpucommand\endcsname{13.5}
\expandafter\gdef\csname blade@num@profile/S2/wgpu/graphics-independent/wgpudeviceresource\endcsname{1.3}
\expandafter\gdef\csname blade@num@profile/S2/wgpu/graphics-independent/wgpuhal\endcsname{3.3}
\expandafter\gdef\csname blade@num@profile/S2/wgpu/graphics-independent/wgpuother\endcsname{0.8}
\expandafter\gdef\csname blade@num@profile/S2/wgpu/graphics-independent/wgputotal\endcsname{26}
\expandafter\gdef\csname blade@num@profile/S2/wgpu/graphics-independent/wgputracker\endcsname{6.8}
\expandafter\gdef\csname blade@num@profile/S2/wgpu/graphics-independent/wgpuvalidation\endcsname{0.1}
\expandafter\gdef\csname blade@num@profile/S3/blade/compute-independent/allocator\endcsname{1.0}
\expandafter\gdef\csname blade@num@profile/S3/blade/compute-independent/ashloader\endcsname{0.5}
\expandafter\gdef\csname blade@num@profile/S3/blade/compute-independent/blade\endcsname{25.9}
\expandafter\gdef\csname blade@num@profile/S3/blade/compute-independent/driver\endcsname{31.0}
\expandafter\gdef\csname blade@num@profile/S3/blade/compute-independent/kernel\endcsname{33.5}
\expandafter\gdef\csname blade@num@profile/S3/blade/compute-independent/libcruntime\endcsname{6.1}
\expandafter\gdef\csname blade@num@profile/S3/blade/compute-independent/other\endcsname{1.7}
\expandafter\gdef\csname blade@num@profile/S3/blade/compute-independent/system\endcsname{65}
\expandafter\gdef\csname blade@num@profile/S3/blade/compute-independent/wgputotal\endcsname{0}
\expandafter\gdef\csname blade@num@profile/S3/blade/graphics-independent/allocator\endcsname{0.5}
\expandafter\gdef\csname blade@num@profile/S3/blade/graphics-independent/ashloader\endcsname{0.1}
\expandafter\gdef\csname blade@num@profile/S3/blade/graphics-independent/blade\endcsname{1.8}
\expandafter\gdef\csname blade@num@profile/S3/blade/graphics-independent/driver\endcsname{74.7}
\expandafter\gdef\csname blade@num@profile/S3/blade/graphics-independent/kernel\endcsname{15.0}
\expandafter\gdef\csname blade@num@profile/S3/blade/graphics-independent/libcruntime\endcsname{9.2}
\expandafter\gdef\csname blade@num@profile/S3/blade/graphics-independent/other\endcsname{1.5}
\expandafter\gdef\csname blade@num@profile/S3/blade/graphics-independent/system\endcsname{90}
\expandafter\gdef\csname blade@num@profile/S3/blade/graphics-independent/wgputotal\endcsname{0}
\expandafter\gdef\csname blade@num@profile/S3/wgpu/compute-independent/allocator\endcsname{2.1}
\expandafter\gdef\csname blade@num@profile/S3/wgpu/compute-independent/blade\endcsname{0.1}
\expandafter\gdef\csname blade@num@profile/S3/wgpu/compute-independent/driver\endcsname{21.4}
\expandafter\gdef\csname blade@num@profile/S3/wgpu/compute-independent/kernel\endcsname{52.3}
\expandafter\gdef\csname blade@num@profile/S3/wgpu/compute-independent/libcruntime\endcsname{6.3}
\expandafter\gdef\csname blade@num@profile/S3/wgpu/compute-independent/other\endcsname{3.7}
\expandafter\gdef\csname blade@num@profile/S3/wgpu/compute-independent/system\endcsname{74}
\expandafter\gdef\csname blade@num@profile/S3/wgpu/compute-independent/wgpucommand\endcsname{5.6}
\expandafter\gdef\csname blade@num@profile/S3/wgpu/compute-independent/wgpudeviceresource\endcsname{0.8}
\expandafter\gdef\csname blade@num@profile/S3/wgpu/compute-independent/wgpuhal\endcsname{1.4}
\expandafter\gdef\csname blade@num@profile/S3/wgpu/compute-independent/wgpuother\endcsname{0.6}
\expandafter\gdef\csname blade@num@profile/S3/wgpu/compute-independent/wgputotal\endcsname{13}
\expandafter\gdef\csname blade@num@profile/S3/wgpu/compute-independent/wgputracker\endcsname{4.7}
\expandafter\gdef\csname blade@num@profile/S3/wgpu/compute-independent/wgpuvalidation\endcsname{0.1}
\expandafter\gdef\csname blade@num@profile/S3/wgpu/graphics-independent/allocator\endcsname{1.6}
\expandafter\gdef\csname blade@num@profile/S3/wgpu/graphics-independent/blade\endcsname{0.1}
\expandafter\gdef\csname blade@num@profile/S3/wgpu/graphics-independent/driver\endcsname{36.6}
\expandafter\gdef\csname blade@num@profile/S3/wgpu/graphics-independent/kernel\endcsname{35.2}
\expandafter\gdef\csname blade@num@profile/S3/wgpu/graphics-independent/libcruntime\endcsname{6.8}
\expandafter\gdef\csname blade@num@profile/S3/wgpu/graphics-independent/other\endcsname{2.4}
\expandafter\gdef\csname blade@num@profile/S3/wgpu/graphics-independent/system\endcsname{72}
\expandafter\gdef\csname blade@num@profile/S3/wgpu/graphics-independent/wgpucommand\endcsname{9.4}
\expandafter\gdef\csname blade@num@profile/S3/wgpu/graphics-independent/wgpudeviceresource\endcsname{1.1}
\expandafter\gdef\csname blade@num@profile/S3/wgpu/graphics-independent/wgpuhal\endcsname{2.3}
\expandafter\gdef\csname blade@num@profile/S3/wgpu/graphics-independent/wgpuother\endcsname{0.4}
\expandafter\gdef\csname blade@num@profile/S3/wgpu/graphics-independent/wgputotal\endcsname{15}
\expandafter\gdef\csname blade@num@profile/S3/wgpu/graphics-independent/wgputracker\endcsname{1.5}
\expandafter\gdef\csname blade@num@profile/S3/wgpu/graphics-independent/wgpuvalidation\endcsname{0.2}
\expandafter\gdef\csname blade@num@profile/S4/blade/compute-independent/allocator\endcsname{1.8}
\expandafter\gdef\csname blade@num@profile/S4/blade/compute-independent/ashloader\endcsname{0.3}
\expandafter\gdef\csname blade@num@profile/S4/blade/compute-independent/blade\endcsname{3.3}
\expandafter\gdef\csname blade@num@profile/S4/blade/compute-independent/driver\endcsname{42.5}
\expandafter\gdef\csname blade@num@profile/S4/blade/compute-independent/kernel\endcsname{39.4}
\expandafter\gdef\csname blade@num@profile/S4/blade/compute-independent/libcruntime\endcsname{8.3}
\expandafter\gdef\csname blade@num@profile/S4/blade/compute-independent/other\endcsname{5.1}
\expandafter\gdef\csname blade@num@profile/S4/blade/compute-independent/system\endcsname{82}
\expandafter\gdef\csname blade@num@profile/S4/blade/compute-independent/wgputotal\endcsname{0}
\expandafter\gdef\csname blade@num@profile/S4/blade/graphics-independent/allocator\endcsname{1.1}
\expandafter\gdef\csname blade@num@profile/S4/blade/graphics-independent/ashloader\endcsname{0.0}
\expandafter\gdef\csname blade@num@profile/S4/blade/graphics-independent/blade\endcsname{1.9}
\expandafter\gdef\csname blade@num@profile/S4/blade/graphics-independent/driver\endcsname{57.2}
\expandafter\gdef\csname blade@num@profile/S4/blade/graphics-independent/kernel\endcsname{27.7}
\expandafter\gdef\csname blade@num@profile/S4/blade/graphics-independent/libcruntime\endcsname{5.8}
\expandafter\gdef\csname blade@num@profile/S4/blade/graphics-independent/other\endcsname{1.8}
\expandafter\gdef\csname blade@num@profile/S4/blade/graphics-independent/system\endcsname{85}
\expandafter\gdef\csname blade@num@profile/S4/blade/graphics-independent/wgputotal\endcsname{0}
\expandafter\gdef\csname blade@num@profile/S4/wgpu/compute-independent/allocator\endcsname{2.6}
\expandafter\gdef\csname blade@num@profile/S4/wgpu/compute-independent/ashloader\endcsname{0.0}
\expandafter\gdef\csname blade@num@profile/S4/wgpu/compute-independent/blade\endcsname{0.2}
\expandafter\gdef\csname blade@num@profile/S4/wgpu/compute-independent/driver\endcsname{29.4}
\expandafter\gdef\csname blade@num@profile/S4/wgpu/compute-independent/kernel\endcsname{7.8}
\expandafter\gdef\csname blade@num@profile/S4/wgpu/compute-independent/libcruntime\endcsname{44.8}
\expandafter\gdef\csname blade@num@profile/S4/wgpu/compute-independent/other\endcsname{2.1}
\expandafter\gdef\csname blade@num@profile/S4/wgpu/compute-independent/system\endcsname{37}
\expandafter\gdef\csname blade@num@profile/S4/wgpu/compute-independent/wgpucommand\endcsname{4.4}
\expandafter\gdef\csname blade@num@profile/S4/wgpu/compute-independent/wgpudeviceresource\endcsname{1.3}
\expandafter\gdef\csname blade@num@profile/S4/wgpu/compute-independent/wgpuhal\endcsname{0.8}
\expandafter\gdef\csname blade@num@profile/S4/wgpu/compute-independent/wgpuother\endcsname{0.6}
\expandafter\gdef\csname blade@num@profile/S4/wgpu/compute-independent/wgputotal\endcsname{11}
\expandafter\gdef\csname blade@num@profile/S4/wgpu/compute-independent/wgputracker\endcsname{4.2}
\expandafter\gdef\csname blade@num@profile/S4/wgpu/compute-independent/wgpuvalidation\endcsname{0.1}
\expandafter\gdef\csname blade@num@profile/S4/wgpu/graphics-independent/allocator\endcsname{2.7}
\expandafter\gdef\csname blade@num@profile/S4/wgpu/graphics-independent/ashloader\endcsname{0.0}
\expandafter\gdef\csname blade@num@profile/S4/wgpu/graphics-independent/blade\endcsname{0.1}
\expandafter\gdef\csname blade@num@profile/S4/wgpu/graphics-independent/driver\endcsname{36.3}
\expandafter\gdef\csname blade@num@profile/S4/wgpu/graphics-independent/kernel\endcsname{6.6}
\expandafter\gdef\csname blade@num@profile/S4/wgpu/graphics-independent/libcruntime\endcsname{39.1}
\expandafter\gdef\csname blade@num@profile/S4/wgpu/graphics-independent/other\endcsname{2.3}
\expandafter\gdef\csname blade@num@profile/S4/wgpu/graphics-independent/system\endcsname{43}
\expandafter\gdef\csname blade@num@profile/S4/wgpu/graphics-independent/wgpucommand\endcsname{5.8}
\expandafter\gdef\csname blade@num@profile/S4/wgpu/graphics-independent/wgpudeviceresource\endcsname{1.1}
\expandafter\gdef\csname blade@num@profile/S4/wgpu/graphics-independent/wgpuhal\endcsname{1.7}
\expandafter\gdef\csname blade@num@profile/S4/wgpu/graphics-independent/wgpuother\endcsname{0.6}
\expandafter\gdef\csname blade@num@profile/S4/wgpu/graphics-independent/wgputotal\endcsname{11}
\expandafter\gdef\csname blade@num@profile/S4/wgpu/graphics-independent/wgputracker\endcsname{2.0}
\expandafter\gdef\csname blade@num@profile/S4/wgpu/graphics-independent/wgpuvalidation\endcsname{0.1}
\expandafter\gdef\csname blade@num@profile/blade/system/max\endcsname{94}
\expandafter\gdef\csname blade@num@profile/blade/system/min\endcsname{65}
\expandafter\gdef\csname blade@num@profile/blade/wgputotal/max\endcsname{0}
\expandafter\gdef\csname blade@num@profile/blade/wgputotal/min\endcsname{0}
\expandafter\gdef\csname blade@num@profile/wgpu/system/max\endcsname{77}
\expandafter\gdef\csname blade@num@profile/wgpu/system/min\endcsname{37}
\expandafter\gdef\csname blade@num@profile/wgpu/tracker/max\endcsname{19}
\expandafter\gdef\csname blade@num@profile/wgpu/tracker/min\endcsname{1}
\expandafter\gdef\csname blade@num@profile/wgpu/wgputotal/max\endcsname{31}
\expandafter\gdef\csname blade@num@profile/wgpu/wgputotal/min\endcsname{11}
\expandafter\gdef\csname blade@num@radeon780m/compute-chain/autous\endcsname{3600.3}
\expandafter\gdef\csname blade@num@radeon780m/compute-chain/both/hi\endcsname{+0.3}
\expandafter\gdef\csname blade@num@radeon780m/compute-chain/both/lo\endcsname{-0.9}
\expandafter\gdef\csname blade@num@radeon780m/compute-chain/both/pct\endcsname{-0.0}
\expandafter\gdef\csname blade@num@radeon780m/compute-chain/bothus\endcsname{3595.2}
\expandafter\gdef\csname blade@num@radeon780m/compute-chain/control/hi\endcsname{+0.2}
\expandafter\gdef\csname blade@num@radeon780m/compute-chain/control/lo\endcsname{-0.6}
\expandafter\gdef\csname blade@num@radeon780m/compute-chain/control/pct\endcsname{-0.3}
\expandafter\gdef\csname blade@num@radeon780m/compute-chain/controlus\endcsname{3592.3}
\expandafter\gdef\csname blade@num@radeon780m/compute-chain/floor\endcsname{0.6}
\expandafter\gdef\csname blade@num@radeon780m/compute-chain/hostautous\endcsname{67.3}
\expandafter\gdef\csname blade@num@radeon780m/compute-chain/hostboth/hi\endcsname{+2.0}
\expandafter\gdef\csname blade@num@radeon780m/compute-chain/hostboth/lo\endcsname{-4.5}
\expandafter\gdef\csname blade@num@radeon780m/compute-chain/hostboth/pct\endcsname{-1.2}
\expandafter\gdef\csname blade@num@radeon780m/compute-chain/hostbothus\endcsname{66.5}
\expandafter\gdef\csname blade@num@radeon780m/compute-chain/hostcontrol/hi\endcsname{+3.3}
\expandafter\gdef\csname blade@num@radeon780m/compute-chain/hostcontrol/lo\endcsname{-1.6}
\expandafter\gdef\csname blade@num@radeon780m/compute-chain/hostcontrol/pct\endcsname{+1.6}
\expandafter\gdef\csname blade@num@radeon780m/compute-chain/hostcontrolus\endcsname{67.9}
\expandafter\gdef\csname blade@num@radeon780m/compute-chain/hostplacement/hi\endcsname{+1.8}
\expandafter\gdef\csname blade@num@radeon780m/compute-chain/hostplacement/lo\endcsname{-3.3}
\expandafter\gdef\csname blade@num@radeon780m/compute-chain/hostplacement/pct\endcsname{-0.9}
\expandafter\gdef\csname blade@num@radeon780m/compute-chain/hostplacementus\endcsname{67.1}
\expandafter\gdef\csname blade@num@radeon780m/compute-chain/hostscope/hi\endcsname{+2.5}
\expandafter\gdef\csname blade@num@radeon780m/compute-chain/hostscope/lo\endcsname{-2.5}
\expandafter\gdef\csname blade@num@radeon780m/compute-chain/hostscope/pct\endcsname{+1.5}
\expandafter\gdef\csname blade@num@radeon780m/compute-chain/hostscopeus\endcsname{67.2}
\expandafter\gdef\csname blade@num@radeon780m/compute-chain/hostwgpu/hi\endcsname{+257.5}
\expandafter\gdef\csname blade@num@radeon780m/compute-chain/hostwgpu/lo\endcsname{+243.3}
\expandafter\gdef\csname blade@num@radeon780m/compute-chain/hostwgpu/mag\endcsname{250.9}
\expandafter\gdef\csname blade@num@radeon780m/compute-chain/hostwgpu/maghi\endcsname{257.5}
\expandafter\gdef\csname blade@num@radeon780m/compute-chain/hostwgpu/maglo\endcsname{243.3}
\expandafter\gdef\csname blade@num@radeon780m/compute-chain/hostwgpu/pct\endcsname{+250.9}
\expandafter\gdef\csname blade@num@radeon780m/compute-chain/hostwgpuus\endcsname{237.0}
\expandafter\gdef\csname blade@num@radeon780m/compute-chain/manualscope/hi\endcsname{+0.2}
\expandafter\gdef\csname blade@num@radeon780m/compute-chain/manualscope/lo\endcsname{-0.8}
\expandafter\gdef\csname blade@num@radeon780m/compute-chain/manualscope/pct\endcsname{-0.3}
\expandafter\gdef\csname blade@num@radeon780m/compute-chain/placement/hi\endcsname{+1.5}
\expandafter\gdef\csname blade@num@radeon780m/compute-chain/placement/lo\endcsname{-0.4}
\expandafter\gdef\csname blade@num@radeon780m/compute-chain/placement/pct\endcsname{-0.0}
\expandafter\gdef\csname blade@num@radeon780m/compute-chain/placementus\endcsname{3595.8}
\expandafter\gdef\csname blade@num@radeon780m/compute-chain/scope/hi\endcsname{+0.3}
\expandafter\gdef\csname blade@num@radeon780m/compute-chain/scope/lo\endcsname{-0.5}
\expandafter\gdef\csname blade@num@radeon780m/compute-chain/scope/pct\endcsname{-0.1}
\expandafter\gdef\csname blade@num@radeon780m/compute-chain/scopeus\endcsname{3596.6}
\expandafter\gdef\csname blade@num@radeon780m/compute-chain/waitautous\endcsname{3908.3}
\expandafter\gdef\csname blade@num@radeon780m/compute-chain/waitbothus\endcsname{3899.2}
\expandafter\gdef\csname blade@num@radeon780m/compute-chain/waitcontrolus\endcsname{3902.7}
\expandafter\gdef\csname blade@num@radeon780m/compute-chain/waitplacementus\endcsname{3914.6}
\expandafter\gdef\csname blade@num@radeon780m/compute-chain/waitscopeus\endcsname{3910.6}
\expandafter\gdef\csname blade@num@radeon780m/compute-chain/waitwgpuus\endcsname{4196.4}
\expandafter\gdef\csname blade@num@radeon780m/compute-chain/wgpu/hi\endcsname{+5.4}
\expandafter\gdef\csname blade@num@radeon780m/compute-chain/wgpu/lo\endcsname{+4.9}
\expandafter\gdef\csname blade@num@radeon780m/compute-chain/wgpu/mag\endcsname{5.2}
\expandafter\gdef\csname blade@num@radeon780m/compute-chain/wgpu/maghi\endcsname{5.4}
\expandafter\gdef\csname blade@num@radeon780m/compute-chain/wgpu/maglo\endcsname{4.9}
\expandafter\gdef\csname blade@num@radeon780m/compute-chain/wgpu/pct\endcsname{+5.2}
\expandafter\gdef\csname blade@num@radeon780m/compute-chain/wgpuus\endcsname{3785.0}
\expandafter\gdef\csname blade@num@radeon780m/compute-independent/autous\endcsname{3597.8}
\expandafter\gdef\csname blade@num@radeon780m/compute-independent/both/hi\endcsname{+4.3}
\expandafter\gdef\csname blade@num@radeon780m/compute-independent/both/lo\endcsname{+3.0}
\expandafter\gdef\csname blade@num@radeon780m/compute-independent/both/mag\endcsname{3.4}
\expandafter\gdef\csname blade@num@radeon780m/compute-independent/both/maghi\endcsname{4.3}
\expandafter\gdef\csname blade@num@radeon780m/compute-independent/both/maglo\endcsname{3.0}
\expandafter\gdef\csname blade@num@radeon780m/compute-independent/both/pct\endcsname{+3.4}
\expandafter\gdef\csname blade@num@radeon780m/compute-independent/bothus\endcsname{3719.6}
\expandafter\gdef\csname blade@num@radeon780m/compute-independent/control/hi\endcsname{+0.7}
\expandafter\gdef\csname blade@num@radeon780m/compute-independent/control/lo\endcsname{-0.3}
\expandafter\gdef\csname blade@num@radeon780m/compute-independent/control/pct\endcsname{-0.1}
\expandafter\gdef\csname blade@num@radeon780m/compute-independent/controlus\endcsname{3598.6}
\expandafter\gdef\csname blade@num@radeon780m/compute-independent/floor\endcsname{0.7}
\expandafter\gdef\csname blade@num@radeon780m/compute-independent/hostautous\endcsname{68.6}
\expandafter\gdef\csname blade@num@radeon780m/compute-independent/hostboth/hi\endcsname{+6.0}
\expandafter\gdef\csname blade@num@radeon780m/compute-independent/hostboth/lo\endcsname{-4.3}
\expandafter\gdef\csname blade@num@radeon780m/compute-independent/hostboth/pct\endcsname{-1.0}
\expandafter\gdef\csname blade@num@radeon780m/compute-independent/hostbothus\endcsname{66.4}
\expandafter\gdef\csname blade@num@radeon780m/compute-independent/hostcontrol/hi\endcsname{+5.1}
\expandafter\gdef\csname blade@num@radeon780m/compute-independent/hostcontrol/lo\endcsname{-2.8}
\expandafter\gdef\csname blade@num@radeon780m/compute-independent/hostcontrol/pct\endcsname{-1.0}
\expandafter\gdef\csname blade@num@radeon780m/compute-independent/hostcontrolus\endcsname{68.3}
\expandafter\gdef\csname blade@num@radeon780m/compute-independent/hostplacement/hi\endcsname{+2.2}
\expandafter\gdef\csname blade@num@radeon780m/compute-independent/hostplacement/lo\endcsname{-6.3}
\expandafter\gdef\csname blade@num@radeon780m/compute-independent/hostplacement/pct\endcsname{-3.7}
\expandafter\gdef\csname blade@num@radeon780m/compute-independent/hostplacementus\endcsname{66.1}
\expandafter\gdef\csname blade@num@radeon780m/compute-independent/hostscope/hi\endcsname{+6.1}
\expandafter\gdef\csname blade@num@radeon780m/compute-independent/hostscope/lo\endcsname{-3.9}
\expandafter\gdef\csname blade@num@radeon780m/compute-independent/hostscope/pct\endcsname{+0.6}
\expandafter\gdef\csname blade@num@radeon780m/compute-independent/hostscopeus\endcsname{68.0}
\expandafter\gdef\csname blade@num@radeon780m/compute-independent/hostwgpu/hi\endcsname{+271.9}
\expandafter\gdef\csname blade@num@radeon780m/compute-independent/hostwgpu/lo\endcsname{+239.6}
\expandafter\gdef\csname blade@num@radeon780m/compute-independent/hostwgpu/mag\endcsname{246.2}
\expandafter\gdef\csname blade@num@radeon780m/compute-independent/hostwgpu/maghi\endcsname{271.9}
\expandafter\gdef\csname blade@num@radeon780m/compute-independent/hostwgpu/maglo\endcsname{239.6}
\expandafter\gdef\csname blade@num@radeon780m/compute-independent/hostwgpu/pct\endcsname{+246.2}
\expandafter\gdef\csname blade@num@radeon780m/compute-independent/hostwgpuus\endcsname{237.7}
\expandafter\gdef\csname blade@num@radeon780m/compute-independent/manualscope/hi\endcsname{+0.7}
\expandafter\gdef\csname blade@num@radeon780m/compute-independent/manualscope/lo\endcsname{-0.4}
\expandafter\gdef\csname blade@num@radeon780m/compute-independent/manualscope/pct\endcsname{+0.3}
\expandafter\gdef\csname blade@num@radeon780m/compute-independent/placement/hi\endcsname{+4.0}
\expandafter\gdef\csname blade@num@radeon780m/compute-independent/placement/lo\endcsname{+3.0}
\expandafter\gdef\csname blade@num@radeon780m/compute-independent/placement/mag\endcsname{3.6}
\expandafter\gdef\csname blade@num@radeon780m/compute-independent/placement/maghi\endcsname{4.0}
\expandafter\gdef\csname blade@num@radeon780m/compute-independent/placement/maglo\endcsname{3.0}
\expandafter\gdef\csname blade@num@radeon780m/compute-independent/placement/pct\endcsname{+3.6}
\expandafter\gdef\csname blade@num@radeon780m/compute-independent/placementus\endcsname{3716.7}
\expandafter\gdef\csname blade@num@radeon780m/compute-independent/scope/hi\endcsname{+0.8}
\expandafter\gdef\csname blade@num@radeon780m/compute-independent/scope/lo\endcsname{-0.1}
\expandafter\gdef\csname blade@num@radeon780m/compute-independent/scope/pct\endcsname{+0.2}
\expandafter\gdef\csname blade@num@radeon780m/compute-independent/scopeus\endcsname{3598.0}
\expandafter\gdef\csname blade@num@radeon780m/compute-independent/waitautous\endcsname{3897.6}
\expandafter\gdef\csname blade@num@radeon780m/compute-independent/waitbothus\endcsname{4034.9}
\expandafter\gdef\csname blade@num@radeon780m/compute-independent/waitcontrolus\endcsname{3903.8}
\expandafter\gdef\csname blade@num@radeon780m/compute-independent/waitplacementus\endcsname{4050.4}
\expandafter\gdef\csname blade@num@radeon780m/compute-independent/waitscopeus\endcsname{3911.7}
\expandafter\gdef\csname blade@num@radeon780m/compute-independent/waitwgpuus\endcsname{4116.1}
\expandafter\gdef\csname blade@num@radeon780m/compute-independent/wgpu/hi\endcsname{+2.7}
\expandafter\gdef\csname blade@num@radeon780m/compute-independent/wgpu/lo\endcsname{+2.1}
\expandafter\gdef\csname blade@num@radeon780m/compute-independent/wgpu/mag\endcsname{2.4}
\expandafter\gdef\csname blade@num@radeon780m/compute-independent/wgpu/maghi\endcsname{2.7}
\expandafter\gdef\csname blade@num@radeon780m/compute-independent/wgpu/maglo\endcsname{2.1}
\expandafter\gdef\csname blade@num@radeon780m/compute-independent/wgpu/pct\endcsname{+2.4}
\expandafter\gdef\csname blade@num@radeon780m/compute-independent/wgpuus\endcsname{3683.9}
\expandafter\gdef\csname blade@num@radeon780m/graphics-chain/autous\endcsname{1465.2}
\expandafter\gdef\csname blade@num@radeon780m/graphics-chain/both/hi\endcsname{+0.7}
\expandafter\gdef\csname blade@num@radeon780m/graphics-chain/both/lo\endcsname{-0.6}
\expandafter\gdef\csname blade@num@radeon780m/graphics-chain/both/pct\endcsname{+0.0}
\expandafter\gdef\csname blade@num@radeon780m/graphics-chain/bothus\endcsname{1464.7}
\expandafter\gdef\csname blade@num@radeon780m/graphics-chain/control/hi\endcsname{+0.6}
\expandafter\gdef\csname blade@num@radeon780m/graphics-chain/control/lo\endcsname{-0.8}
\expandafter\gdef\csname blade@num@radeon780m/graphics-chain/control/pct\endcsname{+0.1}
\expandafter\gdef\csname blade@num@radeon780m/graphics-chain/controlus\endcsname{1464.9}
\expandafter\gdef\csname blade@num@radeon780m/graphics-chain/floor\endcsname{0.8}
\expandafter\gdef\csname blade@num@radeon780m/graphics-chain/hostautous\endcsname{29.0}
\expandafter\gdef\csname blade@num@radeon780m/graphics-chain/hostboth/hi\endcsname{+1.7}
\expandafter\gdef\csname blade@num@radeon780m/graphics-chain/hostboth/lo\endcsname{-3.6}
\expandafter\gdef\csname blade@num@radeon780m/graphics-chain/hostboth/pct\endcsname{+0.2}
\expandafter\gdef\csname blade@num@radeon780m/graphics-chain/hostbothus\endcsname{28.7}
\expandafter\gdef\csname blade@num@radeon780m/graphics-chain/hostcontrol/hi\endcsname{+99.5}
\expandafter\gdef\csname blade@num@radeon780m/graphics-chain/hostcontrol/lo\endcsname{-1.5}
\expandafter\gdef\csname blade@num@radeon780m/graphics-chain/hostcontrol/pct\endcsname{+0.9}
\expandafter\gdef\csname blade@num@radeon780m/graphics-chain/hostcontrolus\endcsname{29.9}
\expandafter\gdef\csname blade@num@radeon780m/graphics-chain/hostplacement/hi\endcsname{+125.8}
\expandafter\gdef\csname blade@num@radeon780m/graphics-chain/hostplacement/lo\endcsname{-49.4}
\expandafter\gdef\csname blade@num@radeon780m/graphics-chain/hostplacement/pct\endcsname{+2.2}
\expandafter\gdef\csname blade@num@radeon780m/graphics-chain/hostplacementus\endcsname{39.4}
\expandafter\gdef\csname blade@num@radeon780m/graphics-chain/hostscope/hi\endcsname{+14.7}
\expandafter\gdef\csname blade@num@radeon780m/graphics-chain/hostscope/lo\endcsname{-56.2}
\expandafter\gdef\csname blade@num@radeon780m/graphics-chain/hostscope/pct\endcsname{-2.7}
\expandafter\gdef\csname blade@num@radeon780m/graphics-chain/hostscopeus\endcsname{28.5}
\expandafter\gdef\csname blade@num@radeon780m/graphics-chain/hostwgpu/hi\endcsname{+582.1}
\expandafter\gdef\csname blade@num@radeon780m/graphics-chain/hostwgpu/lo\endcsname{+58.9}
\expandafter\gdef\csname blade@num@radeon780m/graphics-chain/hostwgpu/mag\endcsname{374.9}
\expandafter\gdef\csname blade@num@radeon780m/graphics-chain/hostwgpu/maghi\endcsname{582.1}
\expandafter\gdef\csname blade@num@radeon780m/graphics-chain/hostwgpu/maglo\endcsname{58.9}
\expandafter\gdef\csname blade@num@radeon780m/graphics-chain/hostwgpu/pct\endcsname{+374.9}
\expandafter\gdef\csname blade@num@radeon780m/graphics-chain/hostwgpuus\endcsname{139.9}
\expandafter\gdef\csname blade@num@radeon780m/graphics-chain/manualscope/hi\endcsname{+1.2}
\expandafter\gdef\csname blade@num@radeon780m/graphics-chain/manualscope/lo\endcsname{-0.7}
\expandafter\gdef\csname blade@num@radeon780m/graphics-chain/manualscope/pct\endcsname{+0.1}
\expandafter\gdef\csname blade@num@radeon780m/graphics-chain/placement/hi\endcsname{+0.5}
\expandafter\gdef\csname blade@num@radeon780m/graphics-chain/placement/lo\endcsname{-1.4}
\expandafter\gdef\csname blade@num@radeon780m/graphics-chain/placement/pct\endcsname{-0.1}
\expandafter\gdef\csname blade@num@radeon780m/graphics-chain/placementus\endcsname{1456.1}
\expandafter\gdef\csname blade@num@radeon780m/graphics-chain/scope/hi\endcsname{+1.3}
\expandafter\gdef\csname blade@num@radeon780m/graphics-chain/scope/lo\endcsname{-0.9}
\expandafter\gdef\csname blade@num@radeon780m/graphics-chain/scope/pct\endcsname{+0.1}
\expandafter\gdef\csname blade@num@radeon780m/graphics-chain/scopeus\endcsname{1463.7}
\expandafter\gdef\csname blade@num@radeon780m/graphics-chain/waitautous\endcsname{1533.0}
\expandafter\gdef\csname blade@num@radeon780m/graphics-chain/waitbothus\endcsname{1531.7}
\expandafter\gdef\csname blade@num@radeon780m/graphics-chain/waitcontrolus\endcsname{1534.3}
\expandafter\gdef\csname blade@num@radeon780m/graphics-chain/waitplacementus\endcsname{1533.2}
\expandafter\gdef\csname blade@num@radeon780m/graphics-chain/waitscopeus\endcsname{1527.3}
\expandafter\gdef\csname blade@num@radeon780m/graphics-chain/waitwgpuus\endcsname{1538.8}
\expandafter\gdef\csname blade@num@radeon780m/graphics-chain/wgpu/hi\endcsname{+7.6}
\expandafter\gdef\csname blade@num@radeon780m/graphics-chain/wgpu/lo\endcsname{-2.2}
\expandafter\gdef\csname blade@num@radeon780m/graphics-chain/wgpu/pct\endcsname{-0.2}
\expandafter\gdef\csname blade@num@radeon780m/graphics-chain/wgpuus\endcsname{1448.3}
\expandafter\gdef\csname blade@num@radeon780m/graphics-independent/autous\endcsname{2322.8}
\expandafter\gdef\csname blade@num@radeon780m/graphics-independent/both/hi\endcsname{+42.3}
\expandafter\gdef\csname blade@num@radeon780m/graphics-independent/both/lo\endcsname{+39.3}
\expandafter\gdef\csname blade@num@radeon780m/graphics-independent/both/mag\endcsname{41.6}
\expandafter\gdef\csname blade@num@radeon780m/graphics-independent/both/maghi\endcsname{42.3}
\expandafter\gdef\csname blade@num@radeon780m/graphics-independent/both/maglo\endcsname{39.3}
\expandafter\gdef\csname blade@num@radeon780m/graphics-independent/both/pct\endcsname{+41.6}
\expandafter\gdef\csname blade@num@radeon780m/graphics-independent/bothus\endcsname{3293.4}
\expandafter\gdef\csname blade@num@radeon780m/graphics-independent/control/hi\endcsname{+106.7}
\expandafter\gdef\csname blade@num@radeon780m/graphics-independent/control/lo\endcsname{-14.5}
\expandafter\gdef\csname blade@num@radeon780m/graphics-independent/control/pct\endcsname{+51.9}
\expandafter\gdef\csname blade@num@radeon780m/graphics-independent/controlus\endcsname{3526.3}
\expandafter\gdef\csname blade@num@radeon780m/graphics-independent/floor\endcsname{106.7}
\expandafter\gdef\csname blade@num@radeon780m/graphics-independent/hostautous\endcsname{112.8}
\expandafter\gdef\csname blade@num@radeon780m/graphics-independent/hostboth/hi\endcsname{+45.7}
\expandafter\gdef\csname blade@num@radeon780m/graphics-independent/hostboth/lo\endcsname{-1.9}
\expandafter\gdef\csname blade@num@radeon780m/graphics-independent/hostboth/pct\endcsname{-0.2}
\expandafter\gdef\csname blade@num@radeon780m/graphics-independent/hostbothus\endcsname{112.9}
\expandafter\gdef\csname blade@num@radeon780m/graphics-independent/hostcontrol/hi\endcsname{+52.7}
\expandafter\gdef\csname blade@num@radeon780m/graphics-independent/hostcontrol/lo\endcsname{-24.5}
\expandafter\gdef\csname blade@num@radeon780m/graphics-independent/hostcontrol/pct\endcsname{+0.3}
\expandafter\gdef\csname blade@num@radeon780m/graphics-independent/hostcontrolus\endcsname{113.5}
\expandafter\gdef\csname blade@num@radeon780m/graphics-independent/hostplacement/hi\endcsname{+43.4}
\expandafter\gdef\csname blade@num@radeon780m/graphics-independent/hostplacement/lo\endcsname{-2.5}
\expandafter\gdef\csname blade@num@radeon780m/graphics-independent/hostplacement/pct\endcsname{+0.8}
\expandafter\gdef\csname blade@num@radeon780m/graphics-independent/hostplacementus\endcsname{112.7}
\expandafter\gdef\csname blade@num@radeon780m/graphics-independent/hostscope/hi\endcsname{+45.4}
\expandafter\gdef\csname blade@num@radeon780m/graphics-independent/hostscope/lo\endcsname{-0.3}
\expandafter\gdef\csname blade@num@radeon780m/graphics-independent/hostscope/pct\endcsname{+1.8}
\expandafter\gdef\csname blade@num@radeon780m/graphics-independent/hostscopeus\endcsname{113.8}
\expandafter\gdef\csname blade@num@radeon780m/graphics-independent/hostwgpu/hi\endcsname{+168.4}
\expandafter\gdef\csname blade@num@radeon780m/graphics-independent/hostwgpu/lo\endcsname{+44.0}
\expandafter\gdef\csname blade@num@radeon780m/graphics-independent/hostwgpu/mag\endcsname{162.7}
\expandafter\gdef\csname blade@num@radeon780m/graphics-independent/hostwgpu/maghi\endcsname{168.4}
\expandafter\gdef\csname blade@num@radeon780m/graphics-independent/hostwgpu/maglo\endcsname{44.0}
\expandafter\gdef\csname blade@num@radeon780m/graphics-independent/hostwgpu/pct\endcsname{+162.7}
\expandafter\gdef\csname blade@num@radeon780m/graphics-independent/hostwgpuus\endcsname{292.7}
\expandafter\gdef\csname blade@num@radeon780m/graphics-independent/manualscope/hi\endcsname{+5.7}
\expandafter\gdef\csname blade@num@radeon780m/graphics-independent/manualscope/lo\endcsname{-51.0}
\expandafter\gdef\csname blade@num@radeon780m/graphics-independent/manualscope/pct\endcsname{-0.2}
\expandafter\gdef\csname blade@num@radeon780m/graphics-independent/placement/hi\endcsname{+184.6}
\expandafter\gdef\csname blade@num@radeon780m/graphics-independent/placement/lo\endcsname{+22.5}
\expandafter\gdef\csname blade@num@radeon780m/graphics-independent/placement/mag\endcsname{42.0}
\expandafter\gdef\csname blade@num@radeon780m/graphics-independent/placement/maghi\endcsname{184.6}
\expandafter\gdef\csname blade@num@radeon780m/graphics-independent/placement/maglo\endcsname{22.5}
\expandafter\gdef\csname blade@num@radeon780m/graphics-independent/placement/pct\endcsname{+42.0}
\expandafter\gdef\csname blade@num@radeon780m/graphics-independent/placementus\endcsname{3297.6}
\expandafter\gdef\csname blade@num@radeon780m/graphics-independent/scope/hi\endcsname{+0.4}
\expandafter\gdef\csname blade@num@radeon780m/graphics-independent/scope/lo\endcsname{-1.9}
\expandafter\gdef\csname blade@num@radeon780m/graphics-independent/scope/pct\endcsname{+0.1}
\expandafter\gdef\csname blade@num@radeon780m/graphics-independent/scopeus\endcsname{2326.9}
\expandafter\gdef\csname blade@num@radeon780m/graphics-independent/waitautous\endcsname{2716.2}
\expandafter\gdef\csname blade@num@radeon780m/graphics-independent/waitbothus\endcsname{3661.4}
\expandafter\gdef\csname blade@num@radeon780m/graphics-independent/waitcontrolus\endcsname{3883.1}
\expandafter\gdef\csname blade@num@radeon780m/graphics-independent/waitplacementus\endcsname{3681.9}
\expandafter\gdef\csname blade@num@radeon780m/graphics-independent/waitscopeus\endcsname{2741.4}
\expandafter\gdef\csname blade@num@radeon780m/graphics-independent/waitwgpuus\endcsname{2493.3}
\expandafter\gdef\csname blade@num@radeon780m/graphics-independent/wgpu/hi\endcsname{-10.7}
\expandafter\gdef\csname blade@num@radeon780m/graphics-independent/wgpu/lo\endcsname{-28.4}
\expandafter\gdef\csname blade@num@radeon780m/graphics-independent/wgpu/mag\endcsname{12.2}
\expandafter\gdef\csname blade@num@radeon780m/graphics-independent/wgpu/maghi\endcsname{28.4}
\expandafter\gdef\csname blade@num@radeon780m/graphics-independent/wgpu/maglo\endcsname{10.7}
\expandafter\gdef\csname blade@num@radeon780m/graphics-independent/wgpu/pct\endcsname{-12.2}
\expandafter\gdef\csname blade@num@radeon780m/graphics-independent/wgpuus\endcsname{2072.3}
\expandafter\gdef\csname blade@num@radeon780m/mixed-chain/autous\endcsname{2890.4}
\expandafter\gdef\csname blade@num@radeon780m/mixed-chain/both/hi\endcsname{+0.3}
\expandafter\gdef\csname blade@num@radeon780m/mixed-chain/both/lo\endcsname{-0.3}
\expandafter\gdef\csname blade@num@radeon780m/mixed-chain/both/pct\endcsname{+0.0}
\expandafter\gdef\csname blade@num@radeon780m/mixed-chain/bothus\endcsname{2888.8}
\expandafter\gdef\csname blade@num@radeon780m/mixed-chain/control/hi\endcsname{+51.6}
\expandafter\gdef\csname blade@num@radeon780m/mixed-chain/control/lo\endcsname{-0.4}
\expandafter\gdef\csname blade@num@radeon780m/mixed-chain/control/pct\endcsname{+0.1}
\expandafter\gdef\csname blade@num@radeon780m/mixed-chain/controlus\endcsname{2894.2}
\expandafter\gdef\csname blade@num@radeon780m/mixed-chain/floor\endcsname{51.6}
\expandafter\gdef\csname blade@num@radeon780m/mixed-chain/hostautous\endcsname{90.5}
\expandafter\gdef\csname blade@num@radeon780m/mixed-chain/hostboth/hi\endcsname{+6.1}
\expandafter\gdef\csname blade@num@radeon780m/mixed-chain/hostboth/lo\endcsname{-3.5}
\expandafter\gdef\csname blade@num@radeon780m/mixed-chain/hostboth/pct\endcsname{-1.5}
\expandafter\gdef\csname blade@num@radeon780m/mixed-chain/hostbothus\endcsname{89.2}
\expandafter\gdef\csname blade@num@radeon780m/mixed-chain/hostcontrol/hi\endcsname{+7.5}
\expandafter\gdef\csname blade@num@radeon780m/mixed-chain/hostcontrol/lo\endcsname{-35.9}
\expandafter\gdef\csname blade@num@radeon780m/mixed-chain/hostcontrol/pct\endcsname{+0.4}
\expandafter\gdef\csname blade@num@radeon780m/mixed-chain/hostcontrolus\endcsname{90.2}
\expandafter\gdef\csname blade@num@radeon780m/mixed-chain/hostplacement/hi\endcsname{+8.8}
\expandafter\gdef\csname blade@num@radeon780m/mixed-chain/hostplacement/lo\endcsname{-1.0}
\expandafter\gdef\csname blade@num@radeon780m/mixed-chain/hostplacement/pct\endcsname{+1.8}
\expandafter\gdef\csname blade@num@radeon780m/mixed-chain/hostplacementus\endcsname{91.5}
\expandafter\gdef\csname blade@num@radeon780m/mixed-chain/hostscope/hi\endcsname{+4.9}
\expandafter\gdef\csname blade@num@radeon780m/mixed-chain/hostscope/lo\endcsname{-1.2}
\expandafter\gdef\csname blade@num@radeon780m/mixed-chain/hostscope/pct\endcsname{+1.1}
\expandafter\gdef\csname blade@num@radeon780m/mixed-chain/hostscopeus\endcsname{91.3}
\expandafter\gdef\csname blade@num@radeon780m/mixed-chain/manualscope/hi\endcsname{+0.1}
\expandafter\gdef\csname blade@num@radeon780m/mixed-chain/manualscope/lo\endcsname{-1.1}
\expandafter\gdef\csname blade@num@radeon780m/mixed-chain/manualscope/pct\endcsname{-0.2}
\expandafter\gdef\csname blade@num@radeon780m/mixed-chain/placement/hi\endcsname{+0.4}
\expandafter\gdef\csname blade@num@radeon780m/mixed-chain/placement/lo\endcsname{-0.1}
\expandafter\gdef\csname blade@num@radeon780m/mixed-chain/placement/pct\endcsname{+0.0}
\expandafter\gdef\csname blade@num@radeon780m/mixed-chain/placementus\endcsname{2894.2}
\expandafter\gdef\csname blade@num@radeon780m/mixed-chain/scope/hi\endcsname{+0.9}
\expandafter\gdef\csname blade@num@radeon780m/mixed-chain/scope/lo\endcsname{-0.0}
\expandafter\gdef\csname blade@num@radeon780m/mixed-chain/scope/pct\endcsname{+0.1}
\expandafter\gdef\csname blade@num@radeon780m/mixed-chain/scopeus\endcsname{2894.5}
\expandafter\gdef\csname blade@num@radeon780m/mixed-chain/waitautous\endcsname{3232.3}
\expandafter\gdef\csname blade@num@radeon780m/mixed-chain/waitbothus\endcsname{3235.9}
\expandafter\gdef\csname blade@num@radeon780m/mixed-chain/waitcontrolus\endcsname{3255.5}
\expandafter\gdef\csname blade@num@radeon780m/mixed-chain/waitplacementus\endcsname{3252.1}
\expandafter\gdef\csname blade@num@radeon780m/mixed-chain/waitscopeus\endcsname{3233.7}
\expandafter\gdef\csname blade@num@radeon780m/mixed-independent/autous\endcsname{2944.0}
\expandafter\gdef\csname blade@num@radeon780m/mixed-independent/both/hi\endcsname{+25.6}
\expandafter\gdef\csname blade@num@radeon780m/mixed-independent/both/lo\endcsname{+15.3}
\expandafter\gdef\csname blade@num@radeon780m/mixed-independent/both/mag\endcsname{25.0}
\expandafter\gdef\csname blade@num@radeon780m/mixed-independent/both/maghi\endcsname{25.6}
\expandafter\gdef\csname blade@num@radeon780m/mixed-independent/both/maglo\endcsname{15.3}
\expandafter\gdef\csname blade@num@radeon780m/mixed-independent/both/pct\endcsname{+25.0}
\expandafter\gdef\csname blade@num@radeon780m/mixed-independent/bothus\endcsname{3673.1}
\expandafter\gdef\csname blade@num@radeon780m/mixed-independent/control/hi\endcsname{+0.1}
\expandafter\gdef\csname blade@num@radeon780m/mixed-independent/control/lo\endcsname{-23.9}
\expandafter\gdef\csname blade@num@radeon780m/mixed-independent/control/pct\endcsname{-0.2}
\expandafter\gdef\csname blade@num@radeon780m/mixed-independent/controlus\endcsname{2935.9}
\expandafter\gdef\csname blade@num@radeon780m/mixed-independent/floor\endcsname{23.9}
\expandafter\gdef\csname blade@num@radeon780m/mixed-independent/hostautous\endcsname{101.5}
\expandafter\gdef\csname blade@num@radeon780m/mixed-independent/hostboth/hi\endcsname{+15.5}
\expandafter\gdef\csname blade@num@radeon780m/mixed-independent/hostboth/lo\endcsname{-3.9}
\expandafter\gdef\csname blade@num@radeon780m/mixed-independent/hostboth/pct\endcsname{-1.9}
\expandafter\gdef\csname blade@num@radeon780m/mixed-independent/hostbothus\endcsname{100.0}
\expandafter\gdef\csname blade@num@radeon780m/mixed-independent/hostcontrol/hi\endcsname{+22.9}
\expandafter\gdef\csname blade@num@radeon780m/mixed-independent/hostcontrol/lo\endcsname{-2.5}
\expandafter\gdef\csname blade@num@radeon780m/mixed-independent/hostcontrol/pct\endcsname{+0.1}
\expandafter\gdef\csname blade@num@radeon780m/mixed-independent/hostcontrolus\endcsname{101.5}
\expandafter\gdef\csname blade@num@radeon780m/mixed-independent/hostplacement/hi\endcsname{+22.5}
\expandafter\gdef\csname blade@num@radeon780m/mixed-independent/hostplacement/lo\endcsname{-2.8}
\expandafter\gdef\csname blade@num@radeon780m/mixed-independent/hostplacement/pct\endcsname{+0.6}
\expandafter\gdef\csname blade@num@radeon780m/mixed-independent/hostplacementus\endcsname{100.5}
\expandafter\gdef\csname blade@num@radeon780m/mixed-independent/hostscope/hi\endcsname{+5.6}
\expandafter\gdef\csname blade@num@radeon780m/mixed-independent/hostscope/lo\endcsname{-31.6}
\expandafter\gdef\csname blade@num@radeon780m/mixed-independent/hostscope/pct\endcsname{-0.8}
\expandafter\gdef\csname blade@num@radeon780m/mixed-independent/hostscopeus\endcsname{101.1}
\expandafter\gdef\csname blade@num@radeon780m/mixed-independent/manualscope/hi\endcsname{-0.1}
\expandafter\gdef\csname blade@num@radeon780m/mixed-independent/manualscope/lo\endcsname{-43.6}
\expandafter\gdef\csname blade@num@radeon780m/mixed-independent/manualscope/mag\endcsname{0.8}
\expandafter\gdef\csname blade@num@radeon780m/mixed-independent/manualscope/maghi\endcsname{43.6}
\expandafter\gdef\csname blade@num@radeon780m/mixed-independent/manualscope/maglo\endcsname{0.1}
\expandafter\gdef\csname blade@num@radeon780m/mixed-independent/manualscope/pct\endcsname{-0.8}
\expandafter\gdef\csname blade@num@radeon780m/mixed-independent/placement/hi\endcsname{+57.4}
\expandafter\gdef\csname blade@num@radeon780m/mixed-independent/placement/lo\endcsname{+24.8}
\expandafter\gdef\csname blade@num@radeon780m/mixed-independent/placement/mag\endcsname{25.7}
\expandafter\gdef\csname blade@num@radeon780m/mixed-independent/placement/maghi\endcsname{57.4}
\expandafter\gdef\csname blade@num@radeon780m/mixed-independent/placement/maglo\endcsname{24.8}
\expandafter\gdef\csname blade@num@radeon780m/mixed-independent/placement/pct\endcsname{+25.7}
\expandafter\gdef\csname blade@num@radeon780m/mixed-independent/placementus\endcsname{3697.7}
\expandafter\gdef\csname blade@num@radeon780m/mixed-independent/scope/hi\endcsname{+0.1}
\expandafter\gdef\csname blade@num@radeon780m/mixed-independent/scope/lo\endcsname{-24.2}
\expandafter\gdef\csname blade@num@radeon780m/mixed-independent/scope/pct\endcsname{-0.6}
\expandafter\gdef\csname blade@num@radeon780m/mixed-independent/scopeus\endcsname{2935.1}
\expandafter\gdef\csname blade@num@radeon780m/mixed-independent/waitautous\endcsname{3303.4}
\expandafter\gdef\csname blade@num@radeon780m/mixed-independent/waitbothus\endcsname{4013.0}
\expandafter\gdef\csname blade@num@radeon780m/mixed-independent/waitcontrolus\endcsname{3278.9}
\expandafter\gdef\csname blade@num@radeon780m/mixed-independent/waitplacementus\endcsname{4149.8}
\expandafter\gdef\csname blade@num@radeon780m/mixed-independent/waitscopeus\endcsname{3254.0}
\expandafter\gdef\csname blade@num@raphael/compute-chain/autous\endcsname{3833.6}
\expandafter\gdef\csname blade@num@raphael/compute-chain/both/hi\endcsname{+0.0}
\expandafter\gdef\csname blade@num@raphael/compute-chain/both/lo\endcsname{-0.0}
\expandafter\gdef\csname blade@num@raphael/compute-chain/both/pct\endcsname{-0.0}
\expandafter\gdef\csname blade@num@raphael/compute-chain/bothus\endcsname{3833.8}
\expandafter\gdef\csname blade@num@raphael/compute-chain/control/hi\endcsname{+0.0}
\expandafter\gdef\csname blade@num@raphael/compute-chain/control/lo\endcsname{-0.0}
\expandafter\gdef\csname blade@num@raphael/compute-chain/control/pct\endcsname{+0.0}
\expandafter\gdef\csname blade@num@raphael/compute-chain/controlus\endcsname{3834.5}
\expandafter\gdef\csname blade@num@raphael/compute-chain/floor\endcsname{0.0}
\expandafter\gdef\csname blade@num@raphael/compute-chain/hostautous\endcsname{8.7}
\expandafter\gdef\csname blade@num@raphael/compute-chain/hostboth/hi\endcsname{+7.3}
\expandafter\gdef\csname blade@num@raphael/compute-chain/hostboth/lo\endcsname{-6.1}
\expandafter\gdef\csname blade@num@raphael/compute-chain/hostboth/pct\endcsname{+0.5}
\expandafter\gdef\csname blade@num@raphael/compute-chain/hostbothus\endcsname{9.0}
\expandafter\gdef\csname blade@num@raphael/compute-chain/hostcontrol/hi\endcsname{+6.6}
\expandafter\gdef\csname blade@num@raphael/compute-chain/hostcontrol/lo\endcsname{-1.9}
\expandafter\gdef\csname blade@num@raphael/compute-chain/hostcontrol/pct\endcsname{+0.1}
\expandafter\gdef\csname blade@num@raphael/compute-chain/hostcontrolus\endcsname{8.9}
\expandafter\gdef\csname blade@num@raphael/compute-chain/hostplacement/hi\endcsname{+3.7}
\expandafter\gdef\csname blade@num@raphael/compute-chain/hostplacement/lo\endcsname{-9.0}
\expandafter\gdef\csname blade@num@raphael/compute-chain/hostplacement/pct\endcsname{-3.5}
\expandafter\gdef\csname blade@num@raphael/compute-chain/hostplacementus\endcsname{8.5}
\expandafter\gdef\csname blade@num@raphael/compute-chain/hostscope/hi\endcsname{+2.6}
\expandafter\gdef\csname blade@num@raphael/compute-chain/hostscope/lo\endcsname{-9.3}
\expandafter\gdef\csname blade@num@raphael/compute-chain/hostscope/pct\endcsname{-4.1}
\expandafter\gdef\csname blade@num@raphael/compute-chain/hostscopeus\endcsname{8.5}
\expandafter\gdef\csname blade@num@raphael/compute-chain/hostwgpu/hi\endcsname{+502.9}
\expandafter\gdef\csname blade@num@raphael/compute-chain/hostwgpu/lo\endcsname{+456.8}
\expandafter\gdef\csname blade@num@raphael/compute-chain/hostwgpu/mag\endcsname{490.7}
\expandafter\gdef\csname blade@num@raphael/compute-chain/hostwgpu/maghi\endcsname{502.9}
\expandafter\gdef\csname blade@num@raphael/compute-chain/hostwgpu/maglo\endcsname{456.8}
\expandafter\gdef\csname blade@num@raphael/compute-chain/hostwgpu/pct\endcsname{+490.7}
\expandafter\gdef\csname blade@num@raphael/compute-chain/hostwgpuus\endcsname{51.1}
\expandafter\gdef\csname blade@num@raphael/compute-chain/manualscope/hi\endcsname{+0.0}
\expandafter\gdef\csname blade@num@raphael/compute-chain/manualscope/lo\endcsname{-0.1}
\expandafter\gdef\csname blade@num@raphael/compute-chain/manualscope/pct\endcsname{-0.0}
\expandafter\gdef\csname blade@num@raphael/compute-chain/placement/hi\endcsname{+0.1}
\expandafter\gdef\csname blade@num@raphael/compute-chain/placement/lo\endcsname{-0.0}
\expandafter\gdef\csname blade@num@raphael/compute-chain/placement/pct\endcsname{+0.0}
\expandafter\gdef\csname blade@num@raphael/compute-chain/placementus\endcsname{3834.2}
\expandafter\gdef\csname blade@num@raphael/compute-chain/scope/hi\endcsname{-0.1}
\expandafter\gdef\csname blade@num@raphael/compute-chain/scope/lo\endcsname{-0.2}
\expandafter\gdef\csname blade@num@raphael/compute-chain/scope/mag\endcsname{0.2}
\expandafter\gdef\csname blade@num@raphael/compute-chain/scope/maghi\endcsname{0.2}
\expandafter\gdef\csname blade@num@raphael/compute-chain/scope/maglo\endcsname{0.1}
\expandafter\gdef\csname blade@num@raphael/compute-chain/scope/pct\endcsname{-0.2}
\expandafter\gdef\csname blade@num@raphael/compute-chain/scopeus\endcsname{3826.9}
\expandafter\gdef\csname blade@num@raphael/compute-chain/waitautous\endcsname{3883.2}
\expandafter\gdef\csname blade@num@raphael/compute-chain/waitbothus\endcsname{3882.9}
\expandafter\gdef\csname blade@num@raphael/compute-chain/waitcontrolus\endcsname{3883.7}
\expandafter\gdef\csname blade@num@raphael/compute-chain/waitplacementus\endcsname{3876.6}
\expandafter\gdef\csname blade@num@raphael/compute-chain/waitscopeus\endcsname{3875.3}
\expandafter\gdef\csname blade@num@raphael/compute-chain/waitwgpuus\endcsname{4177.1}
\expandafter\gdef\csname blade@num@raphael/compute-chain/wgpu/hi\endcsname{+7.5}
\expandafter\gdef\csname blade@num@raphael/compute-chain/wgpu/lo\endcsname{+7.4}
\expandafter\gdef\csname blade@num@raphael/compute-chain/wgpu/mag\endcsname{7.4}
\expandafter\gdef\csname blade@num@raphael/compute-chain/wgpu/maghi\endcsname{7.5}
\expandafter\gdef\csname blade@num@raphael/compute-chain/wgpu/maglo\endcsname{7.4}
\expandafter\gdef\csname blade@num@raphael/compute-chain/wgpu/pct\endcsname{+7.4}
\expandafter\gdef\csname blade@num@raphael/compute-chain/wgpuus\endcsname{4117.7}
\expandafter\gdef\csname blade@num@raphael/compute-independent/autous\endcsname{3854.7}
\expandafter\gdef\csname blade@num@raphael/compute-independent/both/hi\endcsname{-0.8}
\expandafter\gdef\csname blade@num@raphael/compute-independent/both/lo\endcsname{-1.0}
\expandafter\gdef\csname blade@num@raphael/compute-independent/both/mag\endcsname{0.9}
\expandafter\gdef\csname blade@num@raphael/compute-independent/both/maghi\endcsname{1.0}
\expandafter\gdef\csname blade@num@raphael/compute-independent/both/maglo\endcsname{0.8}
\expandafter\gdef\csname blade@num@raphael/compute-independent/both/pct\endcsname{-0.9}
\expandafter\gdef\csname blade@num@raphael/compute-independent/bothus\endcsname{3820.3}
\expandafter\gdef\csname blade@num@raphael/compute-independent/control/hi\endcsname{+0.1}
\expandafter\gdef\csname blade@num@raphael/compute-independent/control/lo\endcsname{-0.0}
\expandafter\gdef\csname blade@num@raphael/compute-independent/control/pct\endcsname{-0.0}
\expandafter\gdef\csname blade@num@raphael/compute-independent/controlus\endcsname{3854.6}
\expandafter\gdef\csname blade@num@raphael/compute-independent/floor\endcsname{0.1}
\expandafter\gdef\csname blade@num@raphael/compute-independent/hostautous\endcsname{8.8}
\expandafter\gdef\csname blade@num@raphael/compute-independent/hostboth/hi\endcsname{-2.1}
\expandafter\gdef\csname blade@num@raphael/compute-independent/hostboth/lo\endcsname{-17.8}
\expandafter\gdef\csname blade@num@raphael/compute-independent/hostboth/mag\endcsname{14.3}
\expandafter\gdef\csname blade@num@raphael/compute-independent/hostboth/maghi\endcsname{17.8}
\expandafter\gdef\csname blade@num@raphael/compute-independent/hostboth/maglo\endcsname{2.1}
\expandafter\gdef\csname blade@num@raphael/compute-independent/hostboth/pct\endcsname{-14.3}
\expandafter\gdef\csname blade@num@raphael/compute-independent/hostbothus\endcsname{7.4}
\expandafter\gdef\csname blade@num@raphael/compute-independent/hostcontrol/hi\endcsname{-1.1}
\expandafter\gdef\csname blade@num@raphael/compute-independent/hostcontrol/lo\endcsname{-5.8}
\expandafter\gdef\csname blade@num@raphael/compute-independent/hostcontrol/mag\endcsname{3.5}
\expandafter\gdef\csname blade@num@raphael/compute-independent/hostcontrol/maghi\endcsname{5.8}
\expandafter\gdef\csname blade@num@raphael/compute-independent/hostcontrol/maglo\endcsname{1.1}
\expandafter\gdef\csname blade@num@raphael/compute-independent/hostcontrol/pct\endcsname{-3.5}
\expandafter\gdef\csname blade@num@raphael/compute-independent/hostcontrolus\endcsname{8.4}
\expandafter\gdef\csname blade@num@raphael/compute-independent/hostplacement/hi\endcsname{-12.1}
\expandafter\gdef\csname blade@num@raphael/compute-independent/hostplacement/lo\endcsname{-18.9}
\expandafter\gdef\csname blade@num@raphael/compute-independent/hostplacement/mag\endcsname{14.9}
\expandafter\gdef\csname blade@num@raphael/compute-independent/hostplacement/maghi\endcsname{18.9}
\expandafter\gdef\csname blade@num@raphael/compute-independent/hostplacement/maglo\endcsname{12.1}
\expandafter\gdef\csname blade@num@raphael/compute-independent/hostplacement/pct\endcsname{-14.9}
\expandafter\gdef\csname blade@num@raphael/compute-independent/hostplacementus\endcsname{7.6}
\expandafter\gdef\csname blade@num@raphael/compute-independent/hostscope/hi\endcsname{-1.1}
\expandafter\gdef\csname blade@num@raphael/compute-independent/hostscope/lo\endcsname{-8.2}
\expandafter\gdef\csname blade@num@raphael/compute-independent/hostscope/mag\endcsname{5.6}
\expandafter\gdef\csname blade@num@raphael/compute-independent/hostscope/maghi\endcsname{8.2}
\expandafter\gdef\csname blade@num@raphael/compute-independent/hostscope/maglo\endcsname{1.1}
\expandafter\gdef\csname blade@num@raphael/compute-independent/hostscope/pct\endcsname{-5.6}
\expandafter\gdef\csname blade@num@raphael/compute-independent/hostscopeus\endcsname{8.3}
\expandafter\gdef\csname blade@num@raphael/compute-independent/hostwgpu/hi\endcsname{+499.3}
\expandafter\gdef\csname blade@num@raphael/compute-independent/hostwgpu/lo\endcsname{+458.3}
\expandafter\gdef\csname blade@num@raphael/compute-independent/hostwgpu/mag\endcsname{486.0}
\expandafter\gdef\csname blade@num@raphael/compute-independent/hostwgpu/maghi\endcsname{499.3}
\expandafter\gdef\csname blade@num@raphael/compute-independent/hostwgpu/maglo\endcsname{458.3}
\expandafter\gdef\csname blade@num@raphael/compute-independent/hostwgpu/pct\endcsname{+486.0}
\expandafter\gdef\csname blade@num@raphael/compute-independent/hostwgpuus\endcsname{51.4}
\expandafter\gdef\csname blade@num@raphael/compute-independent/manualscope/hi\endcsname{+0.0}
\expandafter\gdef\csname blade@num@raphael/compute-independent/manualscope/lo\endcsname{-0.1}
\expandafter\gdef\csname blade@num@raphael/compute-independent/manualscope/pct\endcsname{-0.0}
\expandafter\gdef\csname blade@num@raphael/compute-independent/placement/hi\endcsname{-0.8}
\expandafter\gdef\csname blade@num@raphael/compute-independent/placement/lo\endcsname{-0.9}
\expandafter\gdef\csname blade@num@raphael/compute-independent/placement/mag\endcsname{0.9}
\expandafter\gdef\csname blade@num@raphael/compute-independent/placement/maghi\endcsname{0.9}
\expandafter\gdef\csname blade@num@raphael/compute-independent/placement/maglo\endcsname{0.8}
\expandafter\gdef\csname blade@num@raphael/compute-independent/placement/pct\endcsname{-0.9}
\expandafter\gdef\csname blade@num@raphael/compute-independent/placementus\endcsname{3821.1}
\expandafter\gdef\csname blade@num@raphael/compute-independent/scope/hi\endcsname{-0.1}
\expandafter\gdef\csname blade@num@raphael/compute-independent/scope/lo\endcsname{-0.3}
\expandafter\gdef\csname blade@num@raphael/compute-independent/scope/mag\endcsname{0.2}
\expandafter\gdef\csname blade@num@raphael/compute-independent/scope/maghi\endcsname{0.3}
\expandafter\gdef\csname blade@num@raphael/compute-independent/scope/maglo\endcsname{0.1}
\expandafter\gdef\csname blade@num@raphael/compute-independent/scope/pct\endcsname{-0.2}
\expandafter\gdef\csname blade@num@raphael/compute-independent/scopeus\endcsname{3847.8}
\expandafter\gdef\csname blade@num@raphael/compute-independent/waitautous\endcsname{3904.2}
\expandafter\gdef\csname blade@num@raphael/compute-independent/waitbothus\endcsname{3869.3}
\expandafter\gdef\csname blade@num@raphael/compute-independent/waitcontrolus\endcsname{3896.0}
\expandafter\gdef\csname blade@num@raphael/compute-independent/waitplacementus\endcsname{3871.2}
\expandafter\gdef\csname blade@num@raphael/compute-independent/waitscopeus\endcsname{3895.6}
\expandafter\gdef\csname blade@num@raphael/compute-independent/waitwgpuus\endcsname{4151.2}
\expandafter\gdef\csname blade@num@raphael/compute-independent/wgpu/hi\endcsname{+6.2}
\expandafter\gdef\csname blade@num@raphael/compute-independent/wgpu/lo\endcsname{+6.1}
\expandafter\gdef\csname blade@num@raphael/compute-independent/wgpu/mag\endcsname{6.1}
\expandafter\gdef\csname blade@num@raphael/compute-independent/wgpu/maghi\endcsname{6.2}
\expandafter\gdef\csname blade@num@raphael/compute-independent/wgpu/maglo\endcsname{6.1}
\expandafter\gdef\csname blade@num@raphael/compute-independent/wgpu/pct\endcsname{+6.1}
\expandafter\gdef\csname blade@num@raphael/compute-independent/wgpuus\endcsname{4092.3}
\expandafter\gdef\csname blade@num@raphael/graphics-chain/autous\endcsname{4829.8}
\expandafter\gdef\csname blade@num@raphael/graphics-chain/both/hi\endcsname{+0.0}
\expandafter\gdef\csname blade@num@raphael/graphics-chain/both/lo\endcsname{-0.0}
\expandafter\gdef\csname blade@num@raphael/graphics-chain/both/pct\endcsname{-0.0}
\expandafter\gdef\csname blade@num@raphael/graphics-chain/bothus\endcsname{4829.6}
\expandafter\gdef\csname blade@num@raphael/graphics-chain/control/hi\endcsname{+0.0}
\expandafter\gdef\csname blade@num@raphael/graphics-chain/control/lo\endcsname{-0.0}
\expandafter\gdef\csname blade@num@raphael/graphics-chain/control/pct\endcsname{-0.0}
\expandafter\gdef\csname blade@num@raphael/graphics-chain/controlus\endcsname{4829.8}
\expandafter\gdef\csname blade@num@raphael/graphics-chain/floor\endcsname{0.0}
\expandafter\gdef\csname blade@num@raphael/graphics-chain/hostautous\endcsname{17.7}
\expandafter\gdef\csname blade@num@raphael/graphics-chain/hostboth/hi\endcsname{+0.7}
\expandafter\gdef\csname blade@num@raphael/graphics-chain/hostboth/lo\endcsname{-2.0}
\expandafter\gdef\csname blade@num@raphael/graphics-chain/hostboth/pct\endcsname{-0.4}
\expandafter\gdef\csname blade@num@raphael/graphics-chain/hostbothus\endcsname{17.6}
\expandafter\gdef\csname blade@num@raphael/graphics-chain/hostcontrol/hi\endcsname{+8.1}
\expandafter\gdef\csname blade@num@raphael/graphics-chain/hostcontrol/lo\endcsname{-1.5}
\expandafter\gdef\csname blade@num@raphael/graphics-chain/hostcontrol/pct\endcsname{-0.8}
\expandafter\gdef\csname blade@num@raphael/graphics-chain/hostcontrolus\endcsname{17.5}
\expandafter\gdef\csname blade@num@raphael/graphics-chain/hostplacement/hi\endcsname{+0.5}
\expandafter\gdef\csname blade@num@raphael/graphics-chain/hostplacement/lo\endcsname{-2.6}
\expandafter\gdef\csname blade@num@raphael/graphics-chain/hostplacement/pct\endcsname{-0.6}
\expandafter\gdef\csname blade@num@raphael/graphics-chain/hostplacementus\endcsname{17.6}
\expandafter\gdef\csname blade@num@raphael/graphics-chain/hostscope/hi\endcsname{-0.2}
\expandafter\gdef\csname blade@num@raphael/graphics-chain/hostscope/lo\endcsname{-2.5}
\expandafter\gdef\csname blade@num@raphael/graphics-chain/hostscope/mag\endcsname{1.1}
\expandafter\gdef\csname blade@num@raphael/graphics-chain/hostscope/maghi\endcsname{2.5}
\expandafter\gdef\csname blade@num@raphael/graphics-chain/hostscope/maglo\endcsname{0.2}
\expandafter\gdef\csname blade@num@raphael/graphics-chain/hostscope/pct\endcsname{-1.1}
\expandafter\gdef\csname blade@num@raphael/graphics-chain/hostscopeus\endcsname{17.4}
\expandafter\gdef\csname blade@num@raphael/graphics-chain/hostwgpu/hi\endcsname{+318.6}
\expandafter\gdef\csname blade@num@raphael/graphics-chain/hostwgpu/lo\endcsname{+307.6}
\expandafter\gdef\csname blade@num@raphael/graphics-chain/hostwgpu/mag\endcsname{312.8}
\expandafter\gdef\csname blade@num@raphael/graphics-chain/hostwgpu/maghi\endcsname{318.6}
\expandafter\gdef\csname blade@num@raphael/graphics-chain/hostwgpu/maglo\endcsname{307.6}
\expandafter\gdef\csname blade@num@raphael/graphics-chain/hostwgpu/pct\endcsname{+312.8}
\expandafter\gdef\csname blade@num@raphael/graphics-chain/hostwgpuus\endcsname{72.9}
\expandafter\gdef\csname blade@num@raphael/graphics-chain/manualscope/hi\endcsname{+0.0}
\expandafter\gdef\csname blade@num@raphael/graphics-chain/manualscope/lo\endcsname{-0.0}
\expandafter\gdef\csname blade@num@raphael/graphics-chain/manualscope/pct\endcsname{-0.0}
\expandafter\gdef\csname blade@num@raphael/graphics-chain/placement/hi\endcsname{+0.0}
\expandafter\gdef\csname blade@num@raphael/graphics-chain/placement/lo\endcsname{-0.0}
\expandafter\gdef\csname blade@num@raphael/graphics-chain/placement/pct\endcsname{-0.0}
\expandafter\gdef\csname blade@num@raphael/graphics-chain/placementus\endcsname{4829.8}
\expandafter\gdef\csname blade@num@raphael/graphics-chain/scope/hi\endcsname{+0.0}
\expandafter\gdef\csname blade@num@raphael/graphics-chain/scope/lo\endcsname{-0.0}
\expandafter\gdef\csname blade@num@raphael/graphics-chain/scope/pct\endcsname{-0.0}
\expandafter\gdef\csname blade@num@raphael/graphics-chain/scopeus\endcsname{4828.9}
\expandafter\gdef\csname blade@num@raphael/graphics-chain/waitautous\endcsname{4880.0}
\expandafter\gdef\csname blade@num@raphael/graphics-chain/waitbothus\endcsname{4878.9}
\expandafter\gdef\csname blade@num@raphael/graphics-chain/waitcontrolus\endcsname{4879.6}
\expandafter\gdef\csname blade@num@raphael/graphics-chain/waitplacementus\endcsname{4879.1}
\expandafter\gdef\csname blade@num@raphael/graphics-chain/waitscopeus\endcsname{4879.9}
\expandafter\gdef\csname blade@num@raphael/graphics-chain/waitwgpuus\endcsname{5017.0}
\expandafter\gdef\csname blade@num@raphael/graphics-chain/wgpu/hi\endcsname{+2.4}
\expandafter\gdef\csname blade@num@raphael/graphics-chain/wgpu/lo\endcsname{+2.4}
\expandafter\gdef\csname blade@num@raphael/graphics-chain/wgpu/mag\endcsname{2.4}
\expandafter\gdef\csname blade@num@raphael/graphics-chain/wgpu/maghi\endcsname{2.4}
\expandafter\gdef\csname blade@num@raphael/graphics-chain/wgpu/maglo\endcsname{2.4}
\expandafter\gdef\csname blade@num@raphael/graphics-chain/wgpu/pct\endcsname{+2.4}
\expandafter\gdef\csname blade@num@raphael/graphics-chain/wgpuus\endcsname{4944.0}
\expandafter\gdef\csname blade@num@raphael/graphics-independent/autous\endcsname{5724.6}
\expandafter\gdef\csname blade@num@raphael/graphics-independent/both/hi\endcsname{-1.1}
\expandafter\gdef\csname blade@num@raphael/graphics-independent/both/lo\endcsname{-1.1}
\expandafter\gdef\csname blade@num@raphael/graphics-independent/both/mag\endcsname{1.1}
\expandafter\gdef\csname blade@num@raphael/graphics-independent/both/maghi\endcsname{1.1}
\expandafter\gdef\csname blade@num@raphael/graphics-independent/both/maglo\endcsname{1.1}
\expandafter\gdef\csname blade@num@raphael/graphics-independent/both/pct\endcsname{-1.1}
\expandafter\gdef\csname blade@num@raphael/graphics-independent/bothus\endcsname{5663.0}
\expandafter\gdef\csname blade@num@raphael/graphics-independent/control/hi\endcsname{+0.0}
\expandafter\gdef\csname blade@num@raphael/graphics-independent/control/lo\endcsname{-0.1}
\expandafter\gdef\csname blade@num@raphael/graphics-independent/control/pct\endcsname{-0.0}
\expandafter\gdef\csname blade@num@raphael/graphics-independent/controlus\endcsname{5725.1}
\expandafter\gdef\csname blade@num@raphael/graphics-independent/floor\endcsname{0.1}
\expandafter\gdef\csname blade@num@raphael/graphics-independent/hostautous\endcsname{30.5}
\expandafter\gdef\csname blade@num@raphael/graphics-independent/hostboth/hi\endcsname{-3.2}
\expandafter\gdef\csname blade@num@raphael/graphics-independent/hostboth/lo\endcsname{-5.0}
\expandafter\gdef\csname blade@num@raphael/graphics-independent/hostboth/mag\endcsname{4.1}
\expandafter\gdef\csname blade@num@raphael/graphics-independent/hostboth/maghi\endcsname{5.0}
\expandafter\gdef\csname blade@num@raphael/graphics-independent/hostboth/maglo\endcsname{3.2}
\expandafter\gdef\csname blade@num@raphael/graphics-independent/hostboth/pct\endcsname{-4.1}
\expandafter\gdef\csname blade@num@raphael/graphics-independent/hostbothus\endcsname{29.0}
\expandafter\gdef\csname blade@num@raphael/graphics-independent/hostcontrol/hi\endcsname{+0.5}
\expandafter\gdef\csname blade@num@raphael/graphics-independent/hostcontrol/lo\endcsname{-0.6}
\expandafter\gdef\csname blade@num@raphael/graphics-independent/hostcontrol/pct\endcsname{-0.2}
\expandafter\gdef\csname blade@num@raphael/graphics-independent/hostcontrolus\endcsname{30.5}
\expandafter\gdef\csname blade@num@raphael/graphics-independent/hostplacement/hi\endcsname{-3.8}
\expandafter\gdef\csname blade@num@raphael/graphics-independent/hostplacement/lo\endcsname{-5.6}
\expandafter\gdef\csname blade@num@raphael/graphics-independent/hostplacement/mag\endcsname{4.4}
\expandafter\gdef\csname blade@num@raphael/graphics-independent/hostplacement/maghi\endcsname{5.6}
\expandafter\gdef\csname blade@num@raphael/graphics-independent/hostplacement/maglo\endcsname{3.8}
\expandafter\gdef\csname blade@num@raphael/graphics-independent/hostplacement/pct\endcsname{-4.4}
\expandafter\gdef\csname blade@num@raphael/graphics-independent/hostplacementus\endcsname{29.1}
\expandafter\gdef\csname blade@num@raphael/graphics-independent/hostscope/hi\endcsname{-0.1}
\expandafter\gdef\csname blade@num@raphael/graphics-independent/hostscope/lo\endcsname{-1.7}
\expandafter\gdef\csname blade@num@raphael/graphics-independent/hostscope/mag\endcsname{1.1}
\expandafter\gdef\csname blade@num@raphael/graphics-independent/hostscope/maghi\endcsname{1.7}
\expandafter\gdef\csname blade@num@raphael/graphics-independent/hostscope/maglo\endcsname{0.1}
\expandafter\gdef\csname blade@num@raphael/graphics-independent/hostscope/pct\endcsname{-1.1}
\expandafter\gdef\csname blade@num@raphael/graphics-independent/hostscopeus\endcsname{30.2}
\expandafter\gdef\csname blade@num@raphael/graphics-independent/hostwgpu/hi\endcsname{+168.0}
\expandafter\gdef\csname blade@num@raphael/graphics-independent/hostwgpu/lo\endcsname{+161.2}
\expandafter\gdef\csname blade@num@raphael/graphics-independent/hostwgpu/mag\endcsname{163.7}
\expandafter\gdef\csname blade@num@raphael/graphics-independent/hostwgpu/maghi\endcsname{168.0}
\expandafter\gdef\csname blade@num@raphael/graphics-independent/hostwgpu/maglo\endcsname{161.2}
\expandafter\gdef\csname blade@num@raphael/graphics-independent/hostwgpu/pct\endcsname{+163.7}
\expandafter\gdef\csname blade@num@raphael/graphics-independent/hostwgpuus\endcsname{80.4}
\expandafter\gdef\csname blade@num@raphael/graphics-independent/manualscope/hi\endcsname{+0.0}
\expandafter\gdef\csname blade@num@raphael/graphics-independent/manualscope/lo\endcsname{-0.0}
\expandafter\gdef\csname blade@num@raphael/graphics-independent/manualscope/pct\endcsname{+0.0}
\expandafter\gdef\csname blade@num@raphael/graphics-independent/placement/hi\endcsname{-1.1}
\expandafter\gdef\csname blade@num@raphael/graphics-independent/placement/lo\endcsname{-1.1}
\expandafter\gdef\csname blade@num@raphael/graphics-independent/placement/mag\endcsname{1.1}
\expandafter\gdef\csname blade@num@raphael/graphics-independent/placement/maghi\endcsname{1.1}
\expandafter\gdef\csname blade@num@raphael/graphics-independent/placement/maglo\endcsname{1.1}
\expandafter\gdef\csname blade@num@raphael/graphics-independent/placement/pct\endcsname{-1.1}
\expandafter\gdef\csname blade@num@raphael/graphics-independent/placementus\endcsname{5663.1}
\expandafter\gdef\csname blade@num@raphael/graphics-independent/scope/hi\endcsname{+0.0}
\expandafter\gdef\csname blade@num@raphael/graphics-independent/scope/lo\endcsname{-0.1}
\expandafter\gdef\csname blade@num@raphael/graphics-independent/scope/pct\endcsname{+0.0}
\expandafter\gdef\csname blade@num@raphael/graphics-independent/scopeus\endcsname{5725.0}
\expandafter\gdef\csname blade@num@raphael/graphics-independent/waitautous\endcsname{5778.9}
\expandafter\gdef\csname blade@num@raphael/graphics-independent/waitbothus\endcsname{5715.4}
\expandafter\gdef\csname blade@num@raphael/graphics-independent/waitcontrolus\endcsname{5775.5}
\expandafter\gdef\csname blade@num@raphael/graphics-independent/waitplacementus\endcsname{5713.1}
\expandafter\gdef\csname blade@num@raphael/graphics-independent/waitscopeus\endcsname{5776.5}
\expandafter\gdef\csname blade@num@raphael/graphics-independent/waitwgpuus\endcsname{5034.6}
\expandafter\gdef\csname blade@num@raphael/graphics-independent/wgpu/hi\endcsname{-13.4}
\expandafter\gdef\csname blade@num@raphael/graphics-independent/wgpu/lo\endcsname{-13.4}
\expandafter\gdef\csname blade@num@raphael/graphics-independent/wgpu/mag\endcsname{13.4}
\expandafter\gdef\csname blade@num@raphael/graphics-independent/wgpu/maghi\endcsname{13.4}
\expandafter\gdef\csname blade@num@raphael/graphics-independent/wgpu/maglo\endcsname{13.4}
\expandafter\gdef\csname blade@num@raphael/graphics-independent/wgpu/pct\endcsname{-13.4}
\expandafter\gdef\csname blade@num@raphael/graphics-independent/wgpuus\endcsname{4958.5}
\expandafter\gdef\csname blade@num@raphael/mixed-chain/autous\endcsname{4375.6}
\expandafter\gdef\csname blade@num@raphael/mixed-chain/both/hi\endcsname{+0.0}
\expandafter\gdef\csname blade@num@raphael/mixed-chain/both/lo\endcsname{-0.0}
\expandafter\gdef\csname blade@num@raphael/mixed-chain/both/pct\endcsname{-0.0}
\expandafter\gdef\csname blade@num@raphael/mixed-chain/bothus\endcsname{4375.3}
\expandafter\gdef\csname blade@num@raphael/mixed-chain/control/hi\endcsname{+0.0}
\expandafter\gdef\csname blade@num@raphael/mixed-chain/control/lo\endcsname{-0.0}
\expandafter\gdef\csname blade@num@raphael/mixed-chain/control/pct\endcsname{-0.0}
\expandafter\gdef\csname blade@num@raphael/mixed-chain/controlus\endcsname{4375.1}
\expandafter\gdef\csname blade@num@raphael/mixed-chain/floor\endcsname{0.0}
\expandafter\gdef\csname blade@num@raphael/mixed-chain/hostautous\endcsname{13.7}
\expandafter\gdef\csname blade@num@raphael/mixed-chain/hostboth/hi\endcsname{+0.1}
\expandafter\gdef\csname blade@num@raphael/mixed-chain/hostboth/lo\endcsname{-1.7}
\expandafter\gdef\csname blade@num@raphael/mixed-chain/hostboth/pct\endcsname{-0.7}
\expandafter\gdef\csname blade@num@raphael/mixed-chain/hostbothus\endcsname{13.6}
\expandafter\gdef\csname blade@num@raphael/mixed-chain/hostcontrol/hi\endcsname{+2.2}
\expandafter\gdef\csname blade@num@raphael/mixed-chain/hostcontrol/lo\endcsname{-0.3}
\expandafter\gdef\csname blade@num@raphael/mixed-chain/hostcontrol/pct\endcsname{+1.2}
\expandafter\gdef\csname blade@num@raphael/mixed-chain/hostcontrolus\endcsname{13.8}
\expandafter\gdef\csname blade@num@raphael/mixed-chain/hostplacement/hi\endcsname{+1.3}
\expandafter\gdef\csname blade@num@raphael/mixed-chain/hostplacement/lo\endcsname{-0.5}
\expandafter\gdef\csname blade@num@raphael/mixed-chain/hostplacement/pct\endcsname{+0.2}
\expandafter\gdef\csname blade@num@raphael/mixed-chain/hostplacementus\endcsname{13.8}
\expandafter\gdef\csname blade@num@raphael/mixed-chain/hostscope/hi\endcsname{+1.0}
\expandafter\gdef\csname blade@num@raphael/mixed-chain/hostscope/lo\endcsname{-2.3}
\expandafter\gdef\csname blade@num@raphael/mixed-chain/hostscope/pct\endcsname{-1.2}
\expandafter\gdef\csname blade@num@raphael/mixed-chain/hostscopeus\endcsname{13.7}
\expandafter\gdef\csname blade@num@raphael/mixed-chain/manualscope/hi\endcsname{+0.0}
\expandafter\gdef\csname blade@num@raphael/mixed-chain/manualscope/lo\endcsname{-0.0}
\expandafter\gdef\csname blade@num@raphael/mixed-chain/manualscope/pct\endcsname{-0.0}
\expandafter\gdef\csname blade@num@raphael/mixed-chain/placement/hi\endcsname{+0.0}
\expandafter\gdef\csname blade@num@raphael/mixed-chain/placement/lo\endcsname{-0.0}
\expandafter\gdef\csname blade@num@raphael/mixed-chain/placement/pct\endcsname{+0.0}
\expandafter\gdef\csname blade@num@raphael/mixed-chain/placementus\endcsname{4375.8}
\expandafter\gdef\csname blade@num@raphael/mixed-chain/scope/hi\endcsname{+0.0}
\expandafter\gdef\csname blade@num@raphael/mixed-chain/scope/lo\endcsname{-0.0}
\expandafter\gdef\csname blade@num@raphael/mixed-chain/scope/pct\endcsname{+0.0}
\expandafter\gdef\csname blade@num@raphael/mixed-chain/scopeus\endcsname{4376.3}
\expandafter\gdef\csname blade@num@raphael/mixed-chain/waitautous\endcsname{4426.3}
\expandafter\gdef\csname blade@num@raphael/mixed-chain/waitbothus\endcsname{4425.6}
\expandafter\gdef\csname blade@num@raphael/mixed-chain/waitcontrolus\endcsname{4426.5}
\expandafter\gdef\csname blade@num@raphael/mixed-chain/waitplacementus\endcsname{4426.3}
\expandafter\gdef\csname blade@num@raphael/mixed-chain/waitscopeus\endcsname{4426.0}
\expandafter\gdef\csname blade@num@raphael/mixed-independent/autous\endcsname{4803.9}
\expandafter\gdef\csname blade@num@raphael/mixed-independent/both/hi\endcsname{-1.8}
\expandafter\gdef\csname blade@num@raphael/mixed-independent/both/lo\endcsname{-1.9}
\expandafter\gdef\csname blade@num@raphael/mixed-independent/both/mag\endcsname{1.8}
\expandafter\gdef\csname blade@num@raphael/mixed-independent/both/maghi\endcsname{1.9}
\expandafter\gdef\csname blade@num@raphael/mixed-independent/both/maglo\endcsname{1.8}
\expandafter\gdef\csname blade@num@raphael/mixed-independent/both/pct\endcsname{-1.8}
\expandafter\gdef\csname blade@num@raphael/mixed-independent/bothus\endcsname{4717.5}
\expandafter\gdef\csname blade@num@raphael/mixed-independent/control/hi\endcsname{+0.0}
\expandafter\gdef\csname blade@num@raphael/mixed-independent/control/lo\endcsname{-0.0}
\expandafter\gdef\csname blade@num@raphael/mixed-independent/control/pct\endcsname{-0.0}
\expandafter\gdef\csname blade@num@raphael/mixed-independent/controlus\endcsname{4803.8}
\expandafter\gdef\csname blade@num@raphael/mixed-independent/floor\endcsname{0.0}
\expandafter\gdef\csname blade@num@raphael/mixed-independent/hostautous\endcsname{20.0}
\expandafter\gdef\csname blade@num@raphael/mixed-independent/hostboth/hi\endcsname{-5.4}
\expandafter\gdef\csname blade@num@raphael/mixed-independent/hostboth/lo\endcsname{-7.6}
\expandafter\gdef\csname blade@num@raphael/mixed-independent/hostboth/mag\endcsname{6.4}
\expandafter\gdef\csname blade@num@raphael/mixed-independent/hostboth/maghi\endcsname{7.6}
\expandafter\gdef\csname blade@num@raphael/mixed-independent/hostboth/maglo\endcsname{5.4}
\expandafter\gdef\csname blade@num@raphael/mixed-independent/hostboth/pct\endcsname{-6.4}
\expandafter\gdef\csname blade@num@raphael/mixed-independent/hostbothus\endcsname{18.7}
\expandafter\gdef\csname blade@num@raphael/mixed-independent/hostcontrol/hi\endcsname{+0.2}
\expandafter\gdef\csname blade@num@raphael/mixed-independent/hostcontrol/lo\endcsname{-0.8}
\expandafter\gdef\csname blade@num@raphael/mixed-independent/hostcontrol/pct\endcsname{-0.3}
\expandafter\gdef\csname blade@num@raphael/mixed-independent/hostcontrolus\endcsname{19.9}
\expandafter\gdef\csname blade@num@raphael/mixed-independent/hostplacement/hi\endcsname{-5.9}
\expandafter\gdef\csname blade@num@raphael/mixed-independent/hostplacement/lo\endcsname{-7.5}
\expandafter\gdef\csname blade@num@raphael/mixed-independent/hostplacement/mag\endcsname{6.2}
\expandafter\gdef\csname blade@num@raphael/mixed-independent/hostplacement/maghi\endcsname{7.5}
\expandafter\gdef\csname blade@num@raphael/mixed-independent/hostplacement/maglo\endcsname{5.9}
\expandafter\gdef\csname blade@num@raphael/mixed-independent/hostplacement/pct\endcsname{-6.2}
\expandafter\gdef\csname blade@num@raphael/mixed-independent/hostplacementus\endcsname{18.7}
\expandafter\gdef\csname blade@num@raphael/mixed-independent/hostscope/hi\endcsname{+0.1}
\expandafter\gdef\csname blade@num@raphael/mixed-independent/hostscope/lo\endcsname{-1.9}
\expandafter\gdef\csname blade@num@raphael/mixed-independent/hostscope/pct\endcsname{-1.2}
\expandafter\gdef\csname blade@num@raphael/mixed-independent/hostscopeus\endcsname{19.7}
\expandafter\gdef\csname blade@num@raphael/mixed-independent/manualscope/hi\endcsname{+0.1}
\expandafter\gdef\csname blade@num@raphael/mixed-independent/manualscope/lo\endcsname{-0.0}
\expandafter\gdef\csname blade@num@raphael/mixed-independent/manualscope/pct\endcsname{+0.0}
\expandafter\gdef\csname blade@num@raphael/mixed-independent/placement/hi\endcsname{-1.8}
\expandafter\gdef\csname blade@num@raphael/mixed-independent/placement/lo\endcsname{-1.9}
\expandafter\gdef\csname blade@num@raphael/mixed-independent/placement/mag\endcsname{1.8}
\expandafter\gdef\csname blade@num@raphael/mixed-independent/placement/maghi\endcsname{1.9}
\expandafter\gdef\csname blade@num@raphael/mixed-independent/placement/maglo\endcsname{1.8}
\expandafter\gdef\csname blade@num@raphael/mixed-independent/placement/pct\endcsname{-1.8}
\expandafter\gdef\csname blade@num@raphael/mixed-independent/placementus\endcsname{4715.6}
\expandafter\gdef\csname blade@num@raphael/mixed-independent/scope/hi\endcsname{+0.0}
\expandafter\gdef\csname blade@num@raphael/mixed-independent/scope/lo\endcsname{-0.0}
\expandafter\gdef\csname blade@num@raphael/mixed-independent/scope/pct\endcsname{+0.0}
\expandafter\gdef\csname blade@num@raphael/mixed-independent/scopeus\endcsname{4803.6}
\expandafter\gdef\csname blade@num@raphael/mixed-independent/waitautous\endcsname{4854.1}
\expandafter\gdef\csname blade@num@raphael/mixed-independent/waitbothus\endcsname{4767.1}
\expandafter\gdef\csname blade@num@raphael/mixed-independent/waitcontrolus\endcsname{4855.6}
\expandafter\gdef\csname blade@num@raphael/mixed-independent/waitplacementus\endcsname{4764.8}
\expandafter\gdef\csname blade@num@raphael/mixed-independent/waitscopeus\endcsname{4853.4}
\expandafter\gdef\csname blade@num@rtx5070/compute-chain/autous\endcsname{195.4}
\expandafter\gdef\csname blade@num@rtx5070/compute-chain/both/hi\endcsname{+0.0}
\expandafter\gdef\csname blade@num@rtx5070/compute-chain/both/lo\endcsname{-0.1}
\expandafter\gdef\csname blade@num@rtx5070/compute-chain/both/pct\endcsname{-0.1}
\expandafter\gdef\csname blade@num@rtx5070/compute-chain/bothus\endcsname{195.3}
\expandafter\gdef\csname blade@num@rtx5070/compute-chain/control/hi\endcsname{+0.0}
\expandafter\gdef\csname blade@num@rtx5070/compute-chain/control/lo\endcsname{-0.2}
\expandafter\gdef\csname blade@num@rtx5070/compute-chain/control/pct\endcsname{-0.1}
\expandafter\gdef\csname blade@num@rtx5070/compute-chain/controlus\endcsname{195.3}
\expandafter\gdef\csname blade@num@rtx5070/compute-chain/floor\endcsname{0.2}
\expandafter\gdef\csname blade@num@rtx5070/compute-chain/hostautous\endcsname{58.3}
\expandafter\gdef\csname blade@num@rtx5070/compute-chain/hostboth/hi\endcsname{+3.0}
\expandafter\gdef\csname blade@num@rtx5070/compute-chain/hostboth/lo\endcsname{-15.7}
\expandafter\gdef\csname blade@num@rtx5070/compute-chain/hostboth/pct\endcsname{-6.7}
\expandafter\gdef\csname blade@num@rtx5070/compute-chain/hostbothus\endcsname{55.2}
\expandafter\gdef\csname blade@num@rtx5070/compute-chain/hostcontrol/hi\endcsname{+6.5}
\expandafter\gdef\csname blade@num@rtx5070/compute-chain/hostcontrol/lo\endcsname{-19.0}
\expandafter\gdef\csname blade@num@rtx5070/compute-chain/hostcontrol/pct\endcsname{-5.2}
\expandafter\gdef\csname blade@num@rtx5070/compute-chain/hostcontrolus\endcsname{54.9}
\expandafter\gdef\csname blade@num@rtx5070/compute-chain/hostplacement/hi\endcsname{-2.6}
\expandafter\gdef\csname blade@num@rtx5070/compute-chain/hostplacement/lo\endcsname{-22.1}
\expandafter\gdef\csname blade@num@rtx5070/compute-chain/hostplacement/mag\endcsname{11.3}
\expandafter\gdef\csname blade@num@rtx5070/compute-chain/hostplacement/maghi\endcsname{22.1}
\expandafter\gdef\csname blade@num@rtx5070/compute-chain/hostplacement/maglo\endcsname{2.6}
\expandafter\gdef\csname blade@num@rtx5070/compute-chain/hostplacement/pct\endcsname{-11.3}
\expandafter\gdef\csname blade@num@rtx5070/compute-chain/hostplacementus\endcsname{53.3}
\expandafter\gdef\csname blade@num@rtx5070/compute-chain/hostscope/hi\endcsname{+8.5}
\expandafter\gdef\csname blade@num@rtx5070/compute-chain/hostscope/lo\endcsname{-10.9}
\expandafter\gdef\csname blade@num@rtx5070/compute-chain/hostscope/pct\endcsname{-0.4}
\expandafter\gdef\csname blade@num@rtx5070/compute-chain/hostscopeus\endcsname{57.6}
\expandafter\gdef\csname blade@num@rtx5070/compute-chain/hostwgpu/hi\endcsname{+322.2}
\expandafter\gdef\csname blade@num@rtx5070/compute-chain/hostwgpu/lo\endcsname{+224.2}
\expandafter\gdef\csname blade@num@rtx5070/compute-chain/hostwgpu/mag\endcsname{270.8}
\expandafter\gdef\csname blade@num@rtx5070/compute-chain/hostwgpu/maghi\endcsname{322.2}
\expandafter\gdef\csname blade@num@rtx5070/compute-chain/hostwgpu/maglo\endcsname{224.2}
\expandafter\gdef\csname blade@num@rtx5070/compute-chain/hostwgpu/pct\endcsname{+270.8}
\expandafter\gdef\csname blade@num@rtx5070/compute-chain/hostwgpuus\endcsname{220.0}
\expandafter\gdef\csname blade@num@rtx5070/compute-chain/manualscope/hi\endcsname{+0.1}
\expandafter\gdef\csname blade@num@rtx5070/compute-chain/manualscope/lo\endcsname{-0.1}
\expandafter\gdef\csname blade@num@rtx5070/compute-chain/manualscope/pct\endcsname{-0.0}
\expandafter\gdef\csname blade@num@rtx5070/compute-chain/placement/hi\endcsname{+0.0}
\expandafter\gdef\csname blade@num@rtx5070/compute-chain/placement/lo\endcsname{-0.1}
\expandafter\gdef\csname blade@num@rtx5070/compute-chain/placement/pct\endcsname{-0.1}
\expandafter\gdef\csname blade@num@rtx5070/compute-chain/placementus\endcsname{195.3}
\expandafter\gdef\csname blade@num@rtx5070/compute-chain/scope/hi\endcsname{+0.0}
\expandafter\gdef\csname blade@num@rtx5070/compute-chain/scope/lo\endcsname{-0.1}
\expandafter\gdef\csname blade@num@rtx5070/compute-chain/scope/pct\endcsname{-0.0}
\expandafter\gdef\csname blade@num@rtx5070/compute-chain/scopeus\endcsname{195.3}
\expandafter\gdef\csname blade@num@rtx5070/compute-chain/waitautous\endcsname{286.9}
\expandafter\gdef\csname blade@num@rtx5070/compute-chain/waitbothus\endcsname{280.7}
\expandafter\gdef\csname blade@num@rtx5070/compute-chain/waitcontrolus\endcsname{283.9}
\expandafter\gdef\csname blade@num@rtx5070/compute-chain/waitplacementus\endcsname{279.4}
\expandafter\gdef\csname blade@num@rtx5070/compute-chain/waitscopeus\endcsname{286.4}
\expandafter\gdef\csname blade@num@rtx5070/compute-chain/waitwgpuus\endcsname{326.4}
\expandafter\gdef\csname blade@num@rtx5070/compute-chain/wgpu/hi\endcsname{-0.3}
\expandafter\gdef\csname blade@num@rtx5070/compute-chain/wgpu/lo\endcsname{-0.5}
\expandafter\gdef\csname blade@num@rtx5070/compute-chain/wgpu/mag\endcsname{0.4}
\expandafter\gdef\csname blade@num@rtx5070/compute-chain/wgpu/maghi\endcsname{0.5}
\expandafter\gdef\csname blade@num@rtx5070/compute-chain/wgpu/maglo\endcsname{0.3}
\expandafter\gdef\csname blade@num@rtx5070/compute-chain/wgpu/pct\endcsname{-0.4}
\expandafter\gdef\csname blade@num@rtx5070/compute-chain/wgpuus\endcsname{194.6}
\expandafter\gdef\csname blade@num@rtx5070/compute-independent/autous\endcsname{195.4}
\expandafter\gdef\csname blade@num@rtx5070/compute-independent/both/hi\endcsname{-29.1}
\expandafter\gdef\csname blade@num@rtx5070/compute-independent/both/lo\endcsname{-29.3}
\expandafter\gdef\csname blade@num@rtx5070/compute-independent/both/mag\endcsname{29.3}
\expandafter\gdef\csname blade@num@rtx5070/compute-independent/both/maghi\endcsname{29.3}
\expandafter\gdef\csname blade@num@rtx5070/compute-independent/both/maglo\endcsname{29.1}
\expandafter\gdef\csname blade@num@rtx5070/compute-independent/both/pct\endcsname{-29.3}
\expandafter\gdef\csname blade@num@rtx5070/compute-independent/bothus\endcsname{138.2}
\expandafter\gdef\csname blade@num@rtx5070/compute-independent/control/hi\endcsname{+0.1}
\expandafter\gdef\csname blade@num@rtx5070/compute-independent/control/lo\endcsname{-0.1}
\expandafter\gdef\csname blade@num@rtx5070/compute-independent/control/pct\endcsname{+0.0}
\expandafter\gdef\csname blade@num@rtx5070/compute-independent/controlus\endcsname{195.4}
\expandafter\gdef\csname blade@num@rtx5070/compute-independent/floor\endcsname{0.1}
\expandafter\gdef\csname blade@num@rtx5070/compute-independent/hostautous\endcsname{56.0}
\expandafter\gdef\csname blade@num@rtx5070/compute-independent/hostboth/hi\endcsname{-58.0}
\expandafter\gdef\csname blade@num@rtx5070/compute-independent/hostboth/lo\endcsname{-62.5}
\expandafter\gdef\csname blade@num@rtx5070/compute-independent/hostboth/mag\endcsname{61.0}
\expandafter\gdef\csname blade@num@rtx5070/compute-independent/hostboth/maghi\endcsname{62.5}
\expandafter\gdef\csname blade@num@rtx5070/compute-independent/hostboth/maglo\endcsname{58.0}
\expandafter\gdef\csname blade@num@rtx5070/compute-independent/hostboth/pct\endcsname{-61.0}
\expandafter\gdef\csname blade@num@rtx5070/compute-independent/hostbothus\endcsname{21.8}
\expandafter\gdef\csname blade@num@rtx5070/compute-independent/hostcontrol/hi\endcsname{+11.8}
\expandafter\gdef\csname blade@num@rtx5070/compute-independent/hostcontrol/lo\endcsname{-10.8}
\expandafter\gdef\csname blade@num@rtx5070/compute-independent/hostcontrol/pct\endcsname{+0.4}
\expandafter\gdef\csname blade@num@rtx5070/compute-independent/hostcontrolus\endcsname{55.9}
\expandafter\gdef\csname blade@num@rtx5070/compute-independent/hostplacement/hi\endcsname{-58.2}
\expandafter\gdef\csname blade@num@rtx5070/compute-independent/hostplacement/lo\endcsname{-63.2}
\expandafter\gdef\csname blade@num@rtx5070/compute-independent/hostplacement/mag\endcsname{61.5}
\expandafter\gdef\csname blade@num@rtx5070/compute-independent/hostplacement/maghi\endcsname{63.2}
\expandafter\gdef\csname blade@num@rtx5070/compute-independent/hostplacement/maglo\endcsname{58.2}
\expandafter\gdef\csname blade@num@rtx5070/compute-independent/hostplacement/pct\endcsname{-61.5}
\expandafter\gdef\csname blade@num@rtx5070/compute-independent/hostplacementus\endcsname{21.3}
\expandafter\gdef\csname blade@num@rtx5070/compute-independent/hostscope/hi\endcsname{+9.5}
\expandafter\gdef\csname blade@num@rtx5070/compute-independent/hostscope/lo\endcsname{-11.6}
\expandafter\gdef\csname blade@num@rtx5070/compute-independent/hostscope/pct\endcsname{-1.1}
\expandafter\gdef\csname blade@num@rtx5070/compute-independent/hostscopeus\endcsname{55.9}
\expandafter\gdef\csname blade@num@rtx5070/compute-independent/hostwgpu/hi\endcsname{+339.4}
\expandafter\gdef\csname blade@num@rtx5070/compute-independent/hostwgpu/lo\endcsname{+116.2}
\expandafter\gdef\csname blade@num@rtx5070/compute-independent/hostwgpu/mag\endcsname{282.3}
\expandafter\gdef\csname blade@num@rtx5070/compute-independent/hostwgpu/maghi\endcsname{339.4}
\expandafter\gdef\csname blade@num@rtx5070/compute-independent/hostwgpu/maglo\endcsname{116.2}
\expandafter\gdef\csname blade@num@rtx5070/compute-independent/hostwgpu/pct\endcsname{+282.3}
\expandafter\gdef\csname blade@num@rtx5070/compute-independent/hostwgpuus\endcsname{220.3}
\expandafter\gdef\csname blade@num@rtx5070/compute-independent/manualscope/hi\endcsname{+0.2}
\expandafter\gdef\csname blade@num@rtx5070/compute-independent/manualscope/lo\endcsname{-0.0}
\expandafter\gdef\csname blade@num@rtx5070/compute-independent/manualscope/pct\endcsname{+0.1}
\expandafter\gdef\csname blade@num@rtx5070/compute-independent/placement/hi\endcsname{-29.2}
\expandafter\gdef\csname blade@num@rtx5070/compute-independent/placement/lo\endcsname{-29.4}
\expandafter\gdef\csname blade@num@rtx5070/compute-independent/placement/mag\endcsname{29.3}
\expandafter\gdef\csname blade@num@rtx5070/compute-independent/placement/maghi\endcsname{29.4}
\expandafter\gdef\csname blade@num@rtx5070/compute-independent/placement/maglo\endcsname{29.2}
\expandafter\gdef\csname blade@num@rtx5070/compute-independent/placement/pct\endcsname{-29.3}
\expandafter\gdef\csname blade@num@rtx5070/compute-independent/placementus\endcsname{138.0}
\expandafter\gdef\csname blade@num@rtx5070/compute-independent/scope/hi\endcsname{+0.0}
\expandafter\gdef\csname blade@num@rtx5070/compute-independent/scope/lo\endcsname{-0.1}
\expandafter\gdef\csname blade@num@rtx5070/compute-independent/scope/pct\endcsname{-0.0}
\expandafter\gdef\csname blade@num@rtx5070/compute-independent/scopeus\endcsname{195.3}
\expandafter\gdef\csname blade@num@rtx5070/compute-independent/waitautous\endcsname{284.3}
\expandafter\gdef\csname blade@num@rtx5070/compute-independent/waitbothus\endcsname{199.7}
\expandafter\gdef\csname blade@num@rtx5070/compute-independent/waitcontrolus\endcsname{289.6}
\expandafter\gdef\csname blade@num@rtx5070/compute-independent/waitplacementus\endcsname{199.7}
\expandafter\gdef\csname blade@num@rtx5070/compute-independent/waitscopeus\endcsname{283.7}
\expandafter\gdef\csname blade@num@rtx5070/compute-independent/waitwgpuus\endcsname{342.1}
\expandafter\gdef\csname blade@num@rtx5070/compute-independent/wgpu/hi\endcsname{+1.2}
\expandafter\gdef\csname blade@num@rtx5070/compute-independent/wgpu/lo\endcsname{+0.2}
\expandafter\gdef\csname blade@num@rtx5070/compute-independent/wgpu/mag\endcsname{0.9}
\expandafter\gdef\csname blade@num@rtx5070/compute-independent/wgpu/maghi\endcsname{1.2}
\expandafter\gdef\csname blade@num@rtx5070/compute-independent/wgpu/maglo\endcsname{0.2}
\expandafter\gdef\csname blade@num@rtx5070/compute-independent/wgpu/pct\endcsname{+0.9}
\expandafter\gdef\csname blade@num@rtx5070/compute-independent/wgpuus\endcsname{197.2}
\expandafter\gdef\csname blade@num@rtx5070/graphics-chain/autous\endcsname{150.4}
\expandafter\gdef\csname blade@num@rtx5070/graphics-chain/both/hi\endcsname{+0.1}
\expandafter\gdef\csname blade@num@rtx5070/graphics-chain/both/lo\endcsname{-0.4}
\expandafter\gdef\csname blade@num@rtx5070/graphics-chain/both/pct\endcsname{-0.0}
\expandafter\gdef\csname blade@num@rtx5070/graphics-chain/bothus\endcsname{150.4}
\expandafter\gdef\csname blade@num@rtx5070/graphics-chain/control/hi\endcsname{+0.1}
\expandafter\gdef\csname blade@num@rtx5070/graphics-chain/control/lo\endcsname{-0.2}
\expandafter\gdef\csname blade@num@rtx5070/graphics-chain/control/pct\endcsname{-0.1}
\expandafter\gdef\csname blade@num@rtx5070/graphics-chain/controlus\endcsname{150.4}
\expandafter\gdef\csname blade@num@rtx5070/graphics-chain/floor\endcsname{0.2}
\expandafter\gdef\csname blade@num@rtx5070/graphics-chain/hostautous\endcsname{60.3}
\expandafter\gdef\csname blade@num@rtx5070/graphics-chain/hostboth/hi\endcsname{+5.0}
\expandafter\gdef\csname blade@num@rtx5070/graphics-chain/hostboth/lo\endcsname{-21.8}
\expandafter\gdef\csname blade@num@rtx5070/graphics-chain/hostboth/pct\endcsname{-5.4}
\expandafter\gdef\csname blade@num@rtx5070/graphics-chain/hostbothus\endcsname{57.0}
\expandafter\gdef\csname blade@num@rtx5070/graphics-chain/hostcontrol/hi\endcsname{+13.0}
\expandafter\gdef\csname blade@num@rtx5070/graphics-chain/hostcontrol/lo\endcsname{-9.6}
\expandafter\gdef\csname blade@num@rtx5070/graphics-chain/hostcontrol/pct\endcsname{+0.5}
\expandafter\gdef\csname blade@num@rtx5070/graphics-chain/hostcontrolus\endcsname{59.1}
\expandafter\gdef\csname blade@num@rtx5070/graphics-chain/hostplacement/hi\endcsname{-1.7}
\expandafter\gdef\csname blade@num@rtx5070/graphics-chain/hostplacement/lo\endcsname{-36.8}
\expandafter\gdef\csname blade@num@rtx5070/graphics-chain/hostplacement/mag\endcsname{11.3}
\expandafter\gdef\csname blade@num@rtx5070/graphics-chain/hostplacement/maghi\endcsname{36.8}
\expandafter\gdef\csname blade@num@rtx5070/graphics-chain/hostplacement/maglo\endcsname{1.7}
\expandafter\gdef\csname blade@num@rtx5070/graphics-chain/hostplacement/pct\endcsname{-11.3}
\expandafter\gdef\csname blade@num@rtx5070/graphics-chain/hostplacementus\endcsname{52.1}
\expandafter\gdef\csname blade@num@rtx5070/graphics-chain/hostscope/hi\endcsname{-0.7}
\expandafter\gdef\csname blade@num@rtx5070/graphics-chain/hostscope/lo\endcsname{-12.0}
\expandafter\gdef\csname blade@num@rtx5070/graphics-chain/hostscope/mag\endcsname{6.8}
\expandafter\gdef\csname blade@num@rtx5070/graphics-chain/hostscope/maghi\endcsname{12.0}
\expandafter\gdef\csname blade@num@rtx5070/graphics-chain/hostscope/maglo\endcsname{0.7}
\expandafter\gdef\csname blade@num@rtx5070/graphics-chain/hostscope/pct\endcsname{-6.8}
\expandafter\gdef\csname blade@num@rtx5070/graphics-chain/hostscopeus\endcsname{56.1}
\expandafter\gdef\csname blade@num@rtx5070/graphics-chain/hostwgpu/hi\endcsname{+144.9}
\expandafter\gdef\csname blade@num@rtx5070/graphics-chain/hostwgpu/lo\endcsname{+115.5}
\expandafter\gdef\csname blade@num@rtx5070/graphics-chain/hostwgpu/mag\endcsname{128.7}
\expandafter\gdef\csname blade@num@rtx5070/graphics-chain/hostwgpu/maghi\endcsname{144.9}
\expandafter\gdef\csname blade@num@rtx5070/graphics-chain/hostwgpu/maglo\endcsname{115.5}
\expandafter\gdef\csname blade@num@rtx5070/graphics-chain/hostwgpu/pct\endcsname{+128.7}
\expandafter\gdef\csname blade@num@rtx5070/graphics-chain/hostwgpuus\endcsname{137.0}
\expandafter\gdef\csname blade@num@rtx5070/graphics-chain/manualscope/hi\endcsname{+0.3}
\expandafter\gdef\csname blade@num@rtx5070/graphics-chain/manualscope/lo\endcsname{+0.0}
\expandafter\gdef\csname blade@num@rtx5070/graphics-chain/manualscope/mag\endcsname{0.1}
\expandafter\gdef\csname blade@num@rtx5070/graphics-chain/manualscope/maghi\endcsname{0.3}
\expandafter\gdef\csname blade@num@rtx5070/graphics-chain/manualscope/maglo\endcsname{0.0}
\expandafter\gdef\csname blade@num@rtx5070/graphics-chain/manualscope/pct\endcsname{+0.1}
\expandafter\gdef\csname blade@num@rtx5070/graphics-chain/placement/hi\endcsname{-0.1}
\expandafter\gdef\csname blade@num@rtx5070/graphics-chain/placement/lo\endcsname{-0.7}
\expandafter\gdef\csname blade@num@rtx5070/graphics-chain/placement/mag\endcsname{0.1}
\expandafter\gdef\csname blade@num@rtx5070/graphics-chain/placement/maghi\endcsname{0.7}
\expandafter\gdef\csname blade@num@rtx5070/graphics-chain/placement/maglo\endcsname{0.1}
\expandafter\gdef\csname blade@num@rtx5070/graphics-chain/placement/pct\endcsname{-0.1}
\expandafter\gdef\csname blade@num@rtx5070/graphics-chain/placementus\endcsname{150.0}
\expandafter\gdef\csname blade@num@rtx5070/graphics-chain/scope/hi\endcsname{-4.8}
\expandafter\gdef\csname blade@num@rtx5070/graphics-chain/scope/lo\endcsname{-5.0}
\expandafter\gdef\csname blade@num@rtx5070/graphics-chain/scope/mag\endcsname{5.0}
\expandafter\gdef\csname blade@num@rtx5070/graphics-chain/scope/maghi\endcsname{5.0}
\expandafter\gdef\csname blade@num@rtx5070/graphics-chain/scope/maglo\endcsname{4.8}
\expandafter\gdef\csname blade@num@rtx5070/graphics-chain/scope/pct\endcsname{-5.0}
\expandafter\gdef\csname blade@num@rtx5070/graphics-chain/scopeus\endcsname{143.2}
\expandafter\gdef\csname blade@num@rtx5070/graphics-chain/waitautous\endcsname{225.2}
\expandafter\gdef\csname blade@num@rtx5070/graphics-chain/waitbothus\endcsname{225.6}
\expandafter\gdef\csname blade@num@rtx5070/graphics-chain/waitcontrolus\endcsname{224.9}
\expandafter\gdef\csname blade@num@rtx5070/graphics-chain/waitplacementus\endcsname{235.9}
\expandafter\gdef\csname blade@num@rtx5070/graphics-chain/waitscopeus\endcsname{214.5}
\expandafter\gdef\csname blade@num@rtx5070/graphics-chain/waitwgpuus\endcsname{197.1}
\expandafter\gdef\csname blade@num@rtx5070/graphics-chain/wgpu/hi\endcsname{-17.4}
\expandafter\gdef\csname blade@num@rtx5070/graphics-chain/wgpu/lo\endcsname{-17.6}
\expandafter\gdef\csname blade@num@rtx5070/graphics-chain/wgpu/mag\endcsname{17.5}
\expandafter\gdef\csname blade@num@rtx5070/graphics-chain/wgpu/maghi\endcsname{17.6}
\expandafter\gdef\csname blade@num@rtx5070/graphics-chain/wgpu/maglo\endcsname{17.4}
\expandafter\gdef\csname blade@num@rtx5070/graphics-chain/wgpu/pct\endcsname{-17.5}
\expandafter\gdef\csname blade@num@rtx5070/graphics-chain/wgpuus\endcsname{124.1}
\expandafter\gdef\csname blade@num@rtx5070/graphics-independent/autous\endcsname{152.1}
\expandafter\gdef\csname blade@num@rtx5070/graphics-independent/both/hi\endcsname{-31.8}
\expandafter\gdef\csname blade@num@rtx5070/graphics-independent/both/lo\endcsname{-32.6}
\expandafter\gdef\csname blade@num@rtx5070/graphics-independent/both/mag\endcsname{32.1}
\expandafter\gdef\csname blade@num@rtx5070/graphics-independent/both/maghi\endcsname{32.6}
\expandafter\gdef\csname blade@num@rtx5070/graphics-independent/both/maglo\endcsname{31.8}
\expandafter\gdef\csname blade@num@rtx5070/graphics-independent/both/pct\endcsname{-32.1}
\expandafter\gdef\csname blade@num@rtx5070/graphics-independent/bothus\endcsname{103.3}
\expandafter\gdef\csname blade@num@rtx5070/graphics-independent/control/hi\endcsname{+0.1}
\expandafter\gdef\csname blade@num@rtx5070/graphics-independent/control/lo\endcsname{-0.2}
\expandafter\gdef\csname blade@num@rtx5070/graphics-independent/control/pct\endcsname{-0.1}
\expandafter\gdef\csname blade@num@rtx5070/graphics-independent/controlus\endcsname{152.0}
\expandafter\gdef\csname blade@num@rtx5070/graphics-independent/floor\endcsname{0.2}
\expandafter\gdef\csname blade@num@rtx5070/graphics-independent/hostautous\endcsname{70.0}
\expandafter\gdef\csname blade@num@rtx5070/graphics-independent/hostboth/hi\endcsname{-51.4}
\expandafter\gdef\csname blade@num@rtx5070/graphics-independent/hostboth/lo\endcsname{-57.7}
\expandafter\gdef\csname blade@num@rtx5070/graphics-independent/hostboth/mag\endcsname{55.3}
\expandafter\gdef\csname blade@num@rtx5070/graphics-independent/hostboth/maghi\endcsname{57.7}
\expandafter\gdef\csname blade@num@rtx5070/graphics-independent/hostboth/maglo\endcsname{51.4}
\expandafter\gdef\csname blade@num@rtx5070/graphics-independent/hostboth/pct\endcsname{-55.3}
\expandafter\gdef\csname blade@num@rtx5070/graphics-independent/hostbothus\endcsname{31.3}
\expandafter\gdef\csname blade@num@rtx5070/graphics-independent/hostcontrol/hi\endcsname{+10.8}
\expandafter\gdef\csname blade@num@rtx5070/graphics-independent/hostcontrol/lo\endcsname{-8.0}
\expandafter\gdef\csname blade@num@rtx5070/graphics-independent/hostcontrol/pct\endcsname{+1.9}
\expandafter\gdef\csname blade@num@rtx5070/graphics-independent/hostcontrolus\endcsname{68.3}
\expandafter\gdef\csname blade@num@rtx5070/graphics-independent/hostplacement/hi\endcsname{-50.2}
\expandafter\gdef\csname blade@num@rtx5070/graphics-independent/hostplacement/lo\endcsname{-58.5}
\expandafter\gdef\csname blade@num@rtx5070/graphics-independent/hostplacement/mag\endcsname{54.5}
\expandafter\gdef\csname blade@num@rtx5070/graphics-independent/hostplacement/maghi\endcsname{58.5}
\expandafter\gdef\csname blade@num@rtx5070/graphics-independent/hostplacement/maglo\endcsname{50.2}
\expandafter\gdef\csname blade@num@rtx5070/graphics-independent/hostplacement/pct\endcsname{-54.5}
\expandafter\gdef\csname blade@num@rtx5070/graphics-independent/hostplacementus\endcsname{31.5}
\expandafter\gdef\csname blade@num@rtx5070/graphics-independent/hostscope/hi\endcsname{+3.5}
\expandafter\gdef\csname blade@num@rtx5070/graphics-independent/hostscope/lo\endcsname{-14.6}
\expandafter\gdef\csname blade@num@rtx5070/graphics-independent/hostscope/pct\endcsname{-4.5}
\expandafter\gdef\csname blade@num@rtx5070/graphics-independent/hostscopeus\endcsname{65.4}
\expandafter\gdef\csname blade@num@rtx5070/graphics-independent/hostwgpu/hi\endcsname{+73.6}
\expandafter\gdef\csname blade@num@rtx5070/graphics-independent/hostwgpu/lo\endcsname{+46.2}
\expandafter\gdef\csname blade@num@rtx5070/graphics-independent/hostwgpu/mag\endcsname{59.4}
\expandafter\gdef\csname blade@num@rtx5070/graphics-independent/hostwgpu/maghi\endcsname{73.6}
\expandafter\gdef\csname blade@num@rtx5070/graphics-independent/hostwgpu/maglo\endcsname{46.2}
\expandafter\gdef\csname blade@num@rtx5070/graphics-independent/hostwgpu/pct\endcsname{+59.4}
\expandafter\gdef\csname blade@num@rtx5070/graphics-independent/hostwgpuus\endcsname{112.1}
\expandafter\gdef\csname blade@num@rtx5070/graphics-independent/manualscope/hi\endcsname{+1.3}
\expandafter\gdef\csname blade@num@rtx5070/graphics-independent/manualscope/lo\endcsname{-0.5}
\expandafter\gdef\csname blade@num@rtx5070/graphics-independent/manualscope/pct\endcsname{-0.0}
\expandafter\gdef\csname blade@num@rtx5070/graphics-independent/placement/hi\endcsname{-31.8}
\expandafter\gdef\csname blade@num@rtx5070/graphics-independent/placement/lo\endcsname{-32.9}
\expandafter\gdef\csname blade@num@rtx5070/graphics-independent/placement/mag\endcsname{32.4}
\expandafter\gdef\csname blade@num@rtx5070/graphics-independent/placement/maghi\endcsname{32.9}
\expandafter\gdef\csname blade@num@rtx5070/graphics-independent/placement/maglo\endcsname{31.8}
\expandafter\gdef\csname blade@num@rtx5070/graphics-independent/placement/pct\endcsname{-32.4}
\expandafter\gdef\csname blade@num@rtx5070/graphics-independent/placementus\endcsname{102.8}
\expandafter\gdef\csname blade@num@rtx5070/graphics-independent/scope/hi\endcsname{-4.7}
\expandafter\gdef\csname blade@num@rtx5070/graphics-independent/scope/lo\endcsname{-5.0}
\expandafter\gdef\csname blade@num@rtx5070/graphics-independent/scope/mag\endcsname{4.9}
\expandafter\gdef\csname blade@num@rtx5070/graphics-independent/scope/maghi\endcsname{5.0}
\expandafter\gdef\csname blade@num@rtx5070/graphics-independent/scope/maglo\endcsname{4.7}
\expandafter\gdef\csname blade@num@rtx5070/graphics-independent/scope/pct\endcsname{-4.9}
\expandafter\gdef\csname blade@num@rtx5070/graphics-independent/scopeus\endcsname{144.8}
\expandafter\gdef\csname blade@num@rtx5070/graphics-independent/waitautous\endcsname{232.7}
\expandafter\gdef\csname blade@num@rtx5070/graphics-independent/waitbothus\endcsname{156.6}
\expandafter\gdef\csname blade@num@rtx5070/graphics-independent/waitcontrolus\endcsname{229.9}
\expandafter\gdef\csname blade@num@rtx5070/graphics-independent/waitplacementus\endcsname{156.5}
\expandafter\gdef\csname blade@num@rtx5070/graphics-independent/waitscopeus\endcsname{221.4}
\expandafter\gdef\csname blade@num@rtx5070/graphics-independent/waitwgpuus\endcsname{166.6}
\expandafter\gdef\csname blade@num@rtx5070/graphics-independent/wgpu/hi\endcsname{-33.6}
\expandafter\gdef\csname blade@num@rtx5070/graphics-independent/wgpu/lo\endcsname{-35.3}
\expandafter\gdef\csname blade@num@rtx5070/graphics-independent/wgpu/mag\endcsname{34.5}
\expandafter\gdef\csname blade@num@rtx5070/graphics-independent/wgpu/maghi\endcsname{35.3}
\expandafter\gdef\csname blade@num@rtx5070/graphics-independent/wgpu/maglo\endcsname{33.6}
\expandafter\gdef\csname blade@num@rtx5070/graphics-independent/wgpu/pct\endcsname{-34.5}
\expandafter\gdef\csname blade@num@rtx5070/graphics-independent/wgpuus\endcsname{99.7}
\expandafter\gdef\csname blade@num@rtx5070/mixed-chain/autous\endcsname{171.8}
\expandafter\gdef\csname blade@num@rtx5070/mixed-chain/both/hi\endcsname{+0.1}
\expandafter\gdef\csname blade@num@rtx5070/mixed-chain/both/lo\endcsname{-0.1}
\expandafter\gdef\csname blade@num@rtx5070/mixed-chain/both/pct\endcsname{-0.0}
\expandafter\gdef\csname blade@num@rtx5070/mixed-chain/bothus\endcsname{171.7}
\expandafter\gdef\csname blade@num@rtx5070/mixed-chain/control/hi\endcsname{+0.1}
\expandafter\gdef\csname blade@num@rtx5070/mixed-chain/control/lo\endcsname{-0.1}
\expandafter\gdef\csname blade@num@rtx5070/mixed-chain/control/pct\endcsname{+0.0}
\expandafter\gdef\csname blade@num@rtx5070/mixed-chain/controlus\endcsname{171.7}
\expandafter\gdef\csname blade@num@rtx5070/mixed-chain/floor\endcsname{0.1}
\expandafter\gdef\csname blade@num@rtx5070/mixed-chain/hostautous\endcsname{67.1}
\expandafter\gdef\csname blade@num@rtx5070/mixed-chain/hostboth/hi\endcsname{+8.1}
\expandafter\gdef\csname blade@num@rtx5070/mixed-chain/hostboth/lo\endcsname{-9.8}
\expandafter\gdef\csname blade@num@rtx5070/mixed-chain/hostboth/pct\endcsname{-0.1}
\expandafter\gdef\csname blade@num@rtx5070/mixed-chain/hostbothus\endcsname{65.7}
\expandafter\gdef\csname blade@num@rtx5070/mixed-chain/hostcontrol/hi\endcsname{+9.8}
\expandafter\gdef\csname blade@num@rtx5070/mixed-chain/hostcontrol/lo\endcsname{-12.0}
\expandafter\gdef\csname blade@num@rtx5070/mixed-chain/hostcontrol/pct\endcsname{-2.1}
\expandafter\gdef\csname blade@num@rtx5070/mixed-chain/hostcontrolus\endcsname{63.0}
\expandafter\gdef\csname blade@num@rtx5070/mixed-chain/hostplacement/hi\endcsname{+10.2}
\expandafter\gdef\csname blade@num@rtx5070/mixed-chain/hostplacement/lo\endcsname{-13.8}
\expandafter\gdef\csname blade@num@rtx5070/mixed-chain/hostplacement/pct\endcsname{-0.9}
\expandafter\gdef\csname blade@num@rtx5070/mixed-chain/hostplacementus\endcsname{65.5}
\expandafter\gdef\csname blade@num@rtx5070/mixed-chain/hostscope/hi\endcsname{+11.7}
\expandafter\gdef\csname blade@num@rtx5070/mixed-chain/hostscope/lo\endcsname{-12.6}
\expandafter\gdef\csname blade@num@rtx5070/mixed-chain/hostscope/pct\endcsname{-0.1}
\expandafter\gdef\csname blade@num@rtx5070/mixed-chain/hostscopeus\endcsname{70.1}
\expandafter\gdef\csname blade@num@rtx5070/mixed-chain/manualscope/hi\endcsname{+0.1}
\expandafter\gdef\csname blade@num@rtx5070/mixed-chain/manualscope/lo\endcsname{-0.1}
\expandafter\gdef\csname blade@num@rtx5070/mixed-chain/manualscope/pct\endcsname{-0.0}
\expandafter\gdef\csname blade@num@rtx5070/mixed-chain/placement/hi\endcsname{+0.1}
\expandafter\gdef\csname blade@num@rtx5070/mixed-chain/placement/lo\endcsname{-0.1}
\expandafter\gdef\csname blade@num@rtx5070/mixed-chain/placement/pct\endcsname{-0.0}
\expandafter\gdef\csname blade@num@rtx5070/mixed-chain/placementus\endcsname{171.8}
\expandafter\gdef\csname blade@num@rtx5070/mixed-chain/scope/hi\endcsname{-0.3}
\expandafter\gdef\csname blade@num@rtx5070/mixed-chain/scope/lo\endcsname{-3.4}
\expandafter\gdef\csname blade@num@rtx5070/mixed-chain/scope/mag\endcsname{1.2}
\expandafter\gdef\csname blade@num@rtx5070/mixed-chain/scope/maghi\endcsname{3.4}
\expandafter\gdef\csname blade@num@rtx5070/mixed-chain/scope/maglo\endcsname{0.3}
\expandafter\gdef\csname blade@num@rtx5070/mixed-chain/scope/pct\endcsname{-1.2}
\expandafter\gdef\csname blade@num@rtx5070/mixed-chain/scopeus\endcsname{169.8}
\expandafter\gdef\csname blade@num@rtx5070/mixed-chain/waitautous\endcsname{265.3}
\expandafter\gdef\csname blade@num@rtx5070/mixed-chain/waitbothus\endcsname{260.5}
\expandafter\gdef\csname blade@num@rtx5070/mixed-chain/waitcontrolus\endcsname{266.7}
\expandafter\gdef\csname blade@num@rtx5070/mixed-chain/waitplacementus\endcsname{259.8}
\expandafter\gdef\csname blade@num@rtx5070/mixed-chain/waitscopeus\endcsname{262.6}
\expandafter\gdef\csname blade@num@rtx5070/mixed-independent/autous\endcsname{172.1}
\expandafter\gdef\csname blade@num@rtx5070/mixed-independent/both/hi\endcsname{-17.3}
\expandafter\gdef\csname blade@num@rtx5070/mixed-independent/both/lo\endcsname{-17.6}
\expandafter\gdef\csname blade@num@rtx5070/mixed-independent/both/mag\endcsname{17.5}
\expandafter\gdef\csname blade@num@rtx5070/mixed-independent/both/maghi\endcsname{17.6}
\expandafter\gdef\csname blade@num@rtx5070/mixed-independent/both/maglo\endcsname{17.3}
\expandafter\gdef\csname blade@num@rtx5070/mixed-independent/both/pct\endcsname{-17.5}
\expandafter\gdef\csname blade@num@rtx5070/mixed-independent/bothus\endcsname{141.9}
\expandafter\gdef\csname blade@num@rtx5070/mixed-independent/control/hi\endcsname{+0.2}
\expandafter\gdef\csname blade@num@rtx5070/mixed-independent/control/lo\endcsname{-0.2}
\expandafter\gdef\csname blade@num@rtx5070/mixed-independent/control/pct\endcsname{+0.0}
\expandafter\gdef\csname blade@num@rtx5070/mixed-independent/controlus\endcsname{172.0}
\expandafter\gdef\csname blade@num@rtx5070/mixed-independent/floor\endcsname{0.2}
\expandafter\gdef\csname blade@num@rtx5070/mixed-independent/hostautous\endcsname{69.3}
\expandafter\gdef\csname blade@num@rtx5070/mixed-independent/hostboth/hi\endcsname{-31.4}
\expandafter\gdef\csname blade@num@rtx5070/mixed-independent/hostboth/lo\endcsname{-51.8}
\expandafter\gdef\csname blade@num@rtx5070/mixed-independent/hostboth/mag\endcsname{47.6}
\expandafter\gdef\csname blade@num@rtx5070/mixed-independent/hostboth/maghi\endcsname{51.8}
\expandafter\gdef\csname blade@num@rtx5070/mixed-independent/hostboth/maglo\endcsname{31.4}
\expandafter\gdef\csname blade@num@rtx5070/mixed-independent/hostboth/pct\endcsname{-47.6}
\expandafter\gdef\csname blade@num@rtx5070/mixed-independent/hostbothus\endcsname{35.7}
\expandafter\gdef\csname blade@num@rtx5070/mixed-independent/hostcontrol/hi\endcsname{+26.9}
\expandafter\gdef\csname blade@num@rtx5070/mixed-independent/hostcontrol/lo\endcsname{-0.2}
\expandafter\gdef\csname blade@num@rtx5070/mixed-independent/hostcontrol/pct\endcsname{+7.0}
\expandafter\gdef\csname blade@num@rtx5070/mixed-independent/hostcontrolus\endcsname{70.2}
\expandafter\gdef\csname blade@num@rtx5070/mixed-independent/hostplacement/hi\endcsname{-42.1}
\expandafter\gdef\csname blade@num@rtx5070/mixed-independent/hostplacement/lo\endcsname{-51.2}
\expandafter\gdef\csname blade@num@rtx5070/mixed-independent/hostplacement/mag\endcsname{46.7}
\expandafter\gdef\csname blade@num@rtx5070/mixed-independent/hostplacement/maghi\endcsname{51.2}
\expandafter\gdef\csname blade@num@rtx5070/mixed-independent/hostplacement/maglo\endcsname{42.1}
\expandafter\gdef\csname blade@num@rtx5070/mixed-independent/hostplacement/pct\endcsname{-46.7}
\expandafter\gdef\csname blade@num@rtx5070/mixed-independent/hostplacementus\endcsname{36.8}
\expandafter\gdef\csname blade@num@rtx5070/mixed-independent/hostscope/hi\endcsname{+29.0}
\expandafter\gdef\csname blade@num@rtx5070/mixed-independent/hostscope/lo\endcsname{-11.8}
\expandafter\gdef\csname blade@num@rtx5070/mixed-independent/hostscope/pct\endcsname{+2.8}
\expandafter\gdef\csname blade@num@rtx5070/mixed-independent/hostscopeus\endcsname{69.0}
\expandafter\gdef\csname blade@num@rtx5070/mixed-independent/manualscope/hi\endcsname{+0.1}
\expandafter\gdef\csname blade@num@rtx5070/mixed-independent/manualscope/lo\endcsname{-0.0}
\expandafter\gdef\csname blade@num@rtx5070/mixed-independent/manualscope/pct\endcsname{+0.0}
\expandafter\gdef\csname blade@num@rtx5070/mixed-independent/placement/hi\endcsname{-17.4}
\expandafter\gdef\csname blade@num@rtx5070/mixed-independent/placement/lo\endcsname{-17.6}
\expandafter\gdef\csname blade@num@rtx5070/mixed-independent/placement/mag\endcsname{17.5}
\expandafter\gdef\csname blade@num@rtx5070/mixed-independent/placement/maghi\endcsname{17.6}
\expandafter\gdef\csname blade@num@rtx5070/mixed-independent/placement/maglo\endcsname{17.4}
\expandafter\gdef\csname blade@num@rtx5070/mixed-independent/placement/pct\endcsname{-17.5}
\expandafter\gdef\csname blade@num@rtx5070/mixed-independent/placementus\endcsname{141.8}
\expandafter\gdef\csname blade@num@rtx5070/mixed-independent/scope/hi\endcsname{-0.3}
\expandafter\gdef\csname blade@num@rtx5070/mixed-independent/scope/lo\endcsname{-1.0}
\expandafter\gdef\csname blade@num@rtx5070/mixed-independent/scope/mag\endcsname{0.4}
\expandafter\gdef\csname blade@num@rtx5070/mixed-independent/scope/maghi\endcsname{1.0}
\expandafter\gdef\csname blade@num@rtx5070/mixed-independent/scope/maglo\endcsname{0.3}
\expandafter\gdef\csname blade@num@rtx5070/mixed-independent/scope/pct\endcsname{-0.4}
\expandafter\gdef\csname blade@num@rtx5070/mixed-independent/scopeus\endcsname{171.4}
\expandafter\gdef\csname blade@num@rtx5070/mixed-independent/waitautous\endcsname{261.4}
\expandafter\gdef\csname blade@num@rtx5070/mixed-independent/waitbothus\endcsname{210.1}
\expandafter\gdef\csname blade@num@rtx5070/mixed-independent/waitcontrolus\endcsname{261.1}
\expandafter\gdef\csname blade@num@rtx5070/mixed-independent/waitplacementus\endcsname{210.7}
\expandafter\gdef\csname blade@num@rtx5070/mixed-independent/waitscopeus\endcsname{261.4}
\expandafter\gdef\csname blade@num@runs/hashconflicts\endcsname{0}
\expandafter\gdef\csname blade@num@runs/hashgroups\endcsname{154}
\expandafter\gdef\csname blade@num@runs/matrix\endcsname{1680}
\expandafter\gdef\csname blade@num@runs/sweep\endcsname{8400}
\expandafter\gdef\csname blade@num@rx7900xt/compute-chain/autous\endcsname{126.2}
\expandafter\gdef\csname blade@num@rx7900xt/compute-chain/both/hi\endcsname{+0.1}
\expandafter\gdef\csname blade@num@rx7900xt/compute-chain/both/lo\endcsname{-2.1}
\expandafter\gdef\csname blade@num@rx7900xt/compute-chain/both/pct\endcsname{-1.0}
\expandafter\gdef\csname blade@num@rx7900xt/compute-chain/bothus\endcsname{124.8}
\expandafter\gdef\csname blade@num@rx7900xt/compute-chain/control/hi\endcsname{+0.8}
\expandafter\gdef\csname blade@num@rx7900xt/compute-chain/control/lo\endcsname{-2.3}
\expandafter\gdef\csname blade@num@rx7900xt/compute-chain/control/pct\endcsname{-1.1}
\expandafter\gdef\csname blade@num@rx7900xt/compute-chain/controlus\endcsname{124.2}
\expandafter\gdef\csname blade@num@rx7900xt/compute-chain/floor\endcsname{2.3}
\expandafter\gdef\csname blade@num@rx7900xt/compute-chain/hostautous\endcsname{10.6}
\expandafter\gdef\csname blade@num@rx7900xt/compute-chain/hostboth/hi\endcsname{+1.4}
\expandafter\gdef\csname blade@num@rx7900xt/compute-chain/hostboth/lo\endcsname{-0.0}
\expandafter\gdef\csname blade@num@rx7900xt/compute-chain/hostboth/pct\endcsname{+1.0}
\expandafter\gdef\csname blade@num@rx7900xt/compute-chain/hostbothus\endcsname{10.7}
\expandafter\gdef\csname blade@num@rx7900xt/compute-chain/hostcontrol/hi\endcsname{-0.1}
\expandafter\gdef\csname blade@num@rx7900xt/compute-chain/hostcontrol/lo\endcsname{-1.3}
\expandafter\gdef\csname blade@num@rx7900xt/compute-chain/hostcontrol/mag\endcsname{0.8}
\expandafter\gdef\csname blade@num@rx7900xt/compute-chain/hostcontrol/maghi\endcsname{1.3}
\expandafter\gdef\csname blade@num@rx7900xt/compute-chain/hostcontrol/maglo\endcsname{0.1}
\expandafter\gdef\csname blade@num@rx7900xt/compute-chain/hostcontrol/pct\endcsname{-0.8}
\expandafter\gdef\csname blade@num@rx7900xt/compute-chain/hostcontrolus\endcsname{10.5}
\expandafter\gdef\csname blade@num@rx7900xt/compute-chain/hostplacement/hi\endcsname{+0.7}
\expandafter\gdef\csname blade@num@rx7900xt/compute-chain/hostplacement/lo\endcsname{-1.0}
\expandafter\gdef\csname blade@num@rx7900xt/compute-chain/hostplacement/pct\endcsname{-0.5}
\expandafter\gdef\csname blade@num@rx7900xt/compute-chain/hostplacementus\endcsname{10.6}
\expandafter\gdef\csname blade@num@rx7900xt/compute-chain/hostscope/hi\endcsname{+1.7}
\expandafter\gdef\csname blade@num@rx7900xt/compute-chain/hostscope/lo\endcsname{-0.7}
\expandafter\gdef\csname blade@num@rx7900xt/compute-chain/hostscope/pct\endcsname{+0.8}
\expandafter\gdef\csname blade@num@rx7900xt/compute-chain/hostscopeus\endcsname{10.7}
\expandafter\gdef\csname blade@num@rx7900xt/compute-chain/hostwgpu/hi\endcsname{+420.5}
\expandafter\gdef\csname blade@num@rx7900xt/compute-chain/hostwgpu/lo\endcsname{+410.1}
\expandafter\gdef\csname blade@num@rx7900xt/compute-chain/hostwgpu/mag\endcsname{412.7}
\expandafter\gdef\csname blade@num@rx7900xt/compute-chain/hostwgpu/maghi\endcsname{420.5}
\expandafter\gdef\csname blade@num@rx7900xt/compute-chain/hostwgpu/maglo\endcsname{410.1}
\expandafter\gdef\csname blade@num@rx7900xt/compute-chain/hostwgpu/pct\endcsname{+412.7}
\expandafter\gdef\csname blade@num@rx7900xt/compute-chain/hostwgpuus\endcsname{54.7}
\expandafter\gdef\csname blade@num@rx7900xt/compute-chain/manualscope/hi\endcsname{+1.4}
\expandafter\gdef\csname blade@num@rx7900xt/compute-chain/manualscope/lo\endcsname{-1.8}
\expandafter\gdef\csname blade@num@rx7900xt/compute-chain/manualscope/pct\endcsname{-0.6}
\expandafter\gdef\csname blade@num@rx7900xt/compute-chain/placement/hi\endcsname{+0.6}
\expandafter\gdef\csname blade@num@rx7900xt/compute-chain/placement/lo\endcsname{-2.3}
\expandafter\gdef\csname blade@num@rx7900xt/compute-chain/placement/pct\endcsname{-0.5}
\expandafter\gdef\csname blade@num@rx7900xt/compute-chain/placementus\endcsname{124.8}
\expandafter\gdef\csname blade@num@rx7900xt/compute-chain/scope/hi\endcsname{-4.5}
\expandafter\gdef\csname blade@num@rx7900xt/compute-chain/scope/lo\endcsname{-6.9}
\expandafter\gdef\csname blade@num@rx7900xt/compute-chain/scope/mag\endcsname{6.7}
\expandafter\gdef\csname blade@num@rx7900xt/compute-chain/scope/maghi\endcsname{6.9}
\expandafter\gdef\csname blade@num@rx7900xt/compute-chain/scope/maglo\endcsname{4.5}
\expandafter\gdef\csname blade@num@rx7900xt/compute-chain/scope/pct\endcsname{-6.7}
\expandafter\gdef\csname blade@num@rx7900xt/compute-chain/scopeus\endcsname{117.7}
\expandafter\gdef\csname blade@num@rx7900xt/compute-chain/waitautous\endcsname{144.8}
\expandafter\gdef\csname blade@num@rx7900xt/compute-chain/waitbothus\endcsname{143.4}
\expandafter\gdef\csname blade@num@rx7900xt/compute-chain/waitcontrolus\endcsname{144.7}
\expandafter\gdef\csname blade@num@rx7900xt/compute-chain/waitplacementus\endcsname{143.5}
\expandafter\gdef\csname blade@num@rx7900xt/compute-chain/waitscopeus\endcsname{136.9}
\expandafter\gdef\csname blade@num@rx7900xt/compute-chain/waitwgpuus\endcsname{157.7}
\expandafter\gdef\csname blade@num@rx7900xt/compute-chain/wgpu/hi\endcsname{+1.8}
\expandafter\gdef\csname blade@num@rx7900xt/compute-chain/wgpu/lo\endcsname{-0.1}
\expandafter\gdef\csname blade@num@rx7900xt/compute-chain/wgpu/pct\endcsname{+0.0}
\expandafter\gdef\csname blade@num@rx7900xt/compute-chain/wgpuus\endcsname{126.2}
\expandafter\gdef\csname blade@num@rx7900xt/compute-independent/autous\endcsname{120.6}
\expandafter\gdef\csname blade@num@rx7900xt/compute-independent/both/hi\endcsname{-32.1}
\expandafter\gdef\csname blade@num@rx7900xt/compute-independent/both/lo\endcsname{-32.8}
\expandafter\gdef\csname blade@num@rx7900xt/compute-independent/both/mag\endcsname{32.6}
\expandafter\gdef\csname blade@num@rx7900xt/compute-independent/both/maghi\endcsname{32.8}
\expandafter\gdef\csname blade@num@rx7900xt/compute-independent/both/maglo\endcsname{32.1}
\expandafter\gdef\csname blade@num@rx7900xt/compute-independent/both/pct\endcsname{-32.6}
\expandafter\gdef\csname blade@num@rx7900xt/compute-independent/bothus\endcsname{81.4}
\expandafter\gdef\csname blade@num@rx7900xt/compute-independent/control/hi\endcsname{+0.5}
\expandafter\gdef\csname blade@num@rx7900xt/compute-independent/control/lo\endcsname{-0.2}
\expandafter\gdef\csname blade@num@rx7900xt/compute-independent/control/pct\endcsname{+0.3}
\expandafter\gdef\csname blade@num@rx7900xt/compute-independent/controlus\endcsname{121.0}
\expandafter\gdef\csname blade@num@rx7900xt/compute-independent/floor\endcsname{0.5}
\expandafter\gdef\csname blade@num@rx7900xt/compute-independent/hostautous\endcsname{10.6}
\expandafter\gdef\csname blade@num@rx7900xt/compute-independent/hostboth/hi\endcsname{-13.2}
\expandafter\gdef\csname blade@num@rx7900xt/compute-independent/hostboth/lo\endcsname{-16.3}
\expandafter\gdef\csname blade@num@rx7900xt/compute-independent/hostboth/mag\endcsname{14.0}
\expandafter\gdef\csname blade@num@rx7900xt/compute-independent/hostboth/maghi\endcsname{16.3}
\expandafter\gdef\csname blade@num@rx7900xt/compute-independent/hostboth/maglo\endcsname{13.2}
\expandafter\gdef\csname blade@num@rx7900xt/compute-independent/hostboth/pct\endcsname{-14.0}
\expandafter\gdef\csname blade@num@rx7900xt/compute-independent/hostbothus\endcsname{9.1}
\expandafter\gdef\csname blade@num@rx7900xt/compute-independent/hostcontrol/hi\endcsname{+0.2}
\expandafter\gdef\csname blade@num@rx7900xt/compute-independent/hostcontrol/lo\endcsname{-3.5}
\expandafter\gdef\csname blade@num@rx7900xt/compute-independent/hostcontrol/pct\endcsname{-1.1}
\expandafter\gdef\csname blade@num@rx7900xt/compute-independent/hostcontrolus\endcsname{10.5}
\expandafter\gdef\csname blade@num@rx7900xt/compute-independent/hostplacement/hi\endcsname{-13.5}
\expandafter\gdef\csname blade@num@rx7900xt/compute-independent/hostplacement/lo\endcsname{-16.2}
\expandafter\gdef\csname blade@num@rx7900xt/compute-independent/hostplacement/mag\endcsname{14.8}
\expandafter\gdef\csname blade@num@rx7900xt/compute-independent/hostplacement/maghi\endcsname{16.2}
\expandafter\gdef\csname blade@num@rx7900xt/compute-independent/hostplacement/maglo\endcsname{13.5}
\expandafter\gdef\csname blade@num@rx7900xt/compute-independent/hostplacement/pct\endcsname{-14.8}
\expandafter\gdef\csname blade@num@rx7900xt/compute-independent/hostplacementus\endcsname{9.1}
\expandafter\gdef\csname blade@num@rx7900xt/compute-independent/hostscope/hi\endcsname{+2.7}
\expandafter\gdef\csname blade@num@rx7900xt/compute-independent/hostscope/lo\endcsname{-1.5}
\expandafter\gdef\csname blade@num@rx7900xt/compute-independent/hostscope/pct\endcsname{+1.9}
\expandafter\gdef\csname blade@num@rx7900xt/compute-independent/hostscopeus\endcsname{10.7}
\expandafter\gdef\csname blade@num@rx7900xt/compute-independent/hostwgpu/hi\endcsname{+431.9}
\expandafter\gdef\csname blade@num@rx7900xt/compute-independent/hostwgpu/lo\endcsname{+417.0}
\expandafter\gdef\csname blade@num@rx7900xt/compute-independent/hostwgpu/mag\endcsname{429.1}
\expandafter\gdef\csname blade@num@rx7900xt/compute-independent/hostwgpu/maghi\endcsname{431.9}
\expandafter\gdef\csname blade@num@rx7900xt/compute-independent/hostwgpu/maglo\endcsname{417.0}
\expandafter\gdef\csname blade@num@rx7900xt/compute-independent/hostwgpu/pct\endcsname{+429.1}
\expandafter\gdef\csname blade@num@rx7900xt/compute-independent/hostwgpuus\endcsname{56.2}
\expandafter\gdef\csname blade@num@rx7900xt/compute-independent/manualscope/hi\endcsname{+0.3}
\expandafter\gdef\csname blade@num@rx7900xt/compute-independent/manualscope/lo\endcsname{-0.7}
\expandafter\gdef\csname blade@num@rx7900xt/compute-independent/manualscope/pct\endcsname{-0.3}
\expandafter\gdef\csname blade@num@rx7900xt/compute-independent/placement/hi\endcsname{-32.1}
\expandafter\gdef\csname blade@num@rx7900xt/compute-independent/placement/lo\endcsname{-32.6}
\expandafter\gdef\csname blade@num@rx7900xt/compute-independent/placement/mag\endcsname{32.3}
\expandafter\gdef\csname blade@num@rx7900xt/compute-independent/placement/maghi\endcsname{32.6}
\expandafter\gdef\csname blade@num@rx7900xt/compute-independent/placement/maglo\endcsname{32.1}
\expandafter\gdef\csname blade@num@rx7900xt/compute-independent/placement/pct\endcsname{-32.3}
\expandafter\gdef\csname blade@num@rx7900xt/compute-independent/placementus\endcsname{81.7}
\expandafter\gdef\csname blade@num@rx7900xt/compute-independent/scope/hi\endcsname{-4.5}
\expandafter\gdef\csname blade@num@rx7900xt/compute-independent/scope/lo\endcsname{-5.2}
\expandafter\gdef\csname blade@num@rx7900xt/compute-independent/scope/mag\endcsname{4.7}
\expandafter\gdef\csname blade@num@rx7900xt/compute-independent/scope/maghi\endcsname{5.2}
\expandafter\gdef\csname blade@num@rx7900xt/compute-independent/scope/maglo\endcsname{4.5}
\expandafter\gdef\csname blade@num@rx7900xt/compute-independent/scope/pct\endcsname{-4.7}
\expandafter\gdef\csname blade@num@rx7900xt/compute-independent/scopeus\endcsname{115.0}
\expandafter\gdef\csname blade@num@rx7900xt/compute-independent/waitautous\endcsname{139.3}
\expandafter\gdef\csname blade@num@rx7900xt/compute-independent/waitbothus\endcsname{105.0}
\expandafter\gdef\csname blade@num@rx7900xt/compute-independent/waitcontrolus\endcsname{139.5}
\expandafter\gdef\csname blade@num@rx7900xt/compute-independent/waitplacementus\endcsname{105.6}
\expandafter\gdef\csname blade@num@rx7900xt/compute-independent/waitscopeus\endcsname{134.6}
\expandafter\gdef\csname blade@num@rx7900xt/compute-independent/waitwgpuus\endcsname{115.7}
\expandafter\gdef\csname blade@num@rx7900xt/compute-independent/wgpu/hi\endcsname{-29.8}
\expandafter\gdef\csname blade@num@rx7900xt/compute-independent/wgpu/lo\endcsname{-30.2}
\expandafter\gdef\csname blade@num@rx7900xt/compute-independent/wgpu/mag\endcsname{29.9}
\expandafter\gdef\csname blade@num@rx7900xt/compute-independent/wgpu/maghi\endcsname{30.2}
\expandafter\gdef\csname blade@num@rx7900xt/compute-independent/wgpu/maglo\endcsname{29.8}
\expandafter\gdef\csname blade@num@rx7900xt/compute-independent/wgpu/pct\endcsname{-29.9}
\expandafter\gdef\csname blade@num@rx7900xt/compute-independent/wgpuus\endcsname{84.6}
\expandafter\gdef\csname blade@num@rx7900xt/graphics-chain/autous\endcsname{150.8}
\expandafter\gdef\csname blade@num@rx7900xt/graphics-chain/both/hi\endcsname{+0.2}
\expandafter\gdef\csname blade@num@rx7900xt/graphics-chain/both/lo\endcsname{-0.1}
\expandafter\gdef\csname blade@num@rx7900xt/graphics-chain/both/pct\endcsname{+0.1}
\expandafter\gdef\csname blade@num@rx7900xt/graphics-chain/bothus\endcsname{150.8}
\expandafter\gdef\csname blade@num@rx7900xt/graphics-chain/control/hi\endcsname{+0.0}
\expandafter\gdef\csname blade@num@rx7900xt/graphics-chain/control/lo\endcsname{-0.1}
\expandafter\gdef\csname blade@num@rx7900xt/graphics-chain/control/pct\endcsname{-0.1}
\expandafter\gdef\csname blade@num@rx7900xt/graphics-chain/controlus\endcsname{150.7}
\expandafter\gdef\csname blade@num@rx7900xt/graphics-chain/floor\endcsname{0.1}
\expandafter\gdef\csname blade@num@rx7900xt/graphics-chain/hostautous\endcsname{19.7}
\expandafter\gdef\csname blade@num@rx7900xt/graphics-chain/hostboth/hi\endcsname{-0.3}
\expandafter\gdef\csname blade@num@rx7900xt/graphics-chain/hostboth/lo\endcsname{-2.0}
\expandafter\gdef\csname blade@num@rx7900xt/graphics-chain/hostboth/mag\endcsname{1.0}
\expandafter\gdef\csname blade@num@rx7900xt/graphics-chain/hostboth/maghi\endcsname{2.0}
\expandafter\gdef\csname blade@num@rx7900xt/graphics-chain/hostboth/maglo\endcsname{0.3}
\expandafter\gdef\csname blade@num@rx7900xt/graphics-chain/hostboth/pct\endcsname{-1.0}
\expandafter\gdef\csname blade@num@rx7900xt/graphics-chain/hostbothus\endcsname{19.5}
\expandafter\gdef\csname blade@num@rx7900xt/graphics-chain/hostcontrol/hi\endcsname{+0.1}
\expandafter\gdef\csname blade@num@rx7900xt/graphics-chain/hostcontrol/lo\endcsname{-0.9}
\expandafter\gdef\csname blade@num@rx7900xt/graphics-chain/hostcontrol/pct\endcsname{-0.7}
\expandafter\gdef\csname blade@num@rx7900xt/graphics-chain/hostcontrolus\endcsname{19.6}
\expandafter\gdef\csname blade@num@rx7900xt/graphics-chain/hostplacement/hi\endcsname{+0.2}
\expandafter\gdef\csname blade@num@rx7900xt/graphics-chain/hostplacement/lo\endcsname{-1.6}
\expandafter\gdef\csname blade@num@rx7900xt/graphics-chain/hostplacement/pct\endcsname{-0.6}
\expandafter\gdef\csname blade@num@rx7900xt/graphics-chain/hostplacementus\endcsname{19.7}
\expandafter\gdef\csname blade@num@rx7900xt/graphics-chain/hostscope/hi\endcsname{+1.1}
\expandafter\gdef\csname blade@num@rx7900xt/graphics-chain/hostscope/lo\endcsname{-1.2}
\expandafter\gdef\csname blade@num@rx7900xt/graphics-chain/hostscope/pct\endcsname{+0.2}
\expandafter\gdef\csname blade@num@rx7900xt/graphics-chain/hostscopeus\endcsname{19.8}
\expandafter\gdef\csname blade@num@rx7900xt/graphics-chain/hostwgpu/hi\endcsname{+290.4}
\expandafter\gdef\csname blade@num@rx7900xt/graphics-chain/hostwgpu/lo\endcsname{+285.1}
\expandafter\gdef\csname blade@num@rx7900xt/graphics-chain/hostwgpu/mag\endcsname{288.0}
\expandafter\gdef\csname blade@num@rx7900xt/graphics-chain/hostwgpu/maghi\endcsname{290.4}
\expandafter\gdef\csname blade@num@rx7900xt/graphics-chain/hostwgpu/maglo\endcsname{285.1}
\expandafter\gdef\csname blade@num@rx7900xt/graphics-chain/hostwgpu/pct\endcsname{+288.0}
\expandafter\gdef\csname blade@num@rx7900xt/graphics-chain/hostwgpuus\endcsname{76.4}
\expandafter\gdef\csname blade@num@rx7900xt/graphics-chain/manualscope/hi\endcsname{+0.1}
\expandafter\gdef\csname blade@num@rx7900xt/graphics-chain/manualscope/lo\endcsname{-0.2}
\expandafter\gdef\csname blade@num@rx7900xt/graphics-chain/manualscope/pct\endcsname{+0.0}
\expandafter\gdef\csname blade@num@rx7900xt/graphics-chain/placement/hi\endcsname{+0.2}
\expandafter\gdef\csname blade@num@rx7900xt/graphics-chain/placement/lo\endcsname{-0.0}
\expandafter\gdef\csname blade@num@rx7900xt/graphics-chain/placement/pct\endcsname{+0.1}
\expandafter\gdef\csname blade@num@rx7900xt/graphics-chain/placementus\endcsname{151.0}
\expandafter\gdef\csname blade@num@rx7900xt/graphics-chain/scope/hi\endcsname{+0.2}
\expandafter\gdef\csname blade@num@rx7900xt/graphics-chain/scope/lo\endcsname{-0.1}
\expandafter\gdef\csname blade@num@rx7900xt/graphics-chain/scope/pct\endcsname{+0.1}
\expandafter\gdef\csname blade@num@rx7900xt/graphics-chain/scopeus\endcsname{150.9}
\expandafter\gdef\csname blade@num@rx7900xt/graphics-chain/waitautous\endcsname{172.3}
\expandafter\gdef\csname blade@num@rx7900xt/graphics-chain/waitbothus\endcsname{171.1}
\expandafter\gdef\csname blade@num@rx7900xt/graphics-chain/waitcontrolus\endcsname{172.2}
\expandafter\gdef\csname blade@num@rx7900xt/graphics-chain/waitplacementus\endcsname{171.6}
\expandafter\gdef\csname blade@num@rx7900xt/graphics-chain/waitscopeus\endcsname{172.4}
\expandafter\gdef\csname blade@num@rx7900xt/graphics-chain/waitwgpuus\endcsname{208.6}
\expandafter\gdef\csname blade@num@rx7900xt/graphics-chain/wgpu/hi\endcsname{+18.6}
\expandafter\gdef\csname blade@num@rx7900xt/graphics-chain/wgpu/lo\endcsname{+18.3}
\expandafter\gdef\csname blade@num@rx7900xt/graphics-chain/wgpu/mag\endcsname{18.5}
\expandafter\gdef\csname blade@num@rx7900xt/graphics-chain/wgpu/maghi\endcsname{18.6}
\expandafter\gdef\csname blade@num@rx7900xt/graphics-chain/wgpu/maglo\endcsname{18.3}
\expandafter\gdef\csname blade@num@rx7900xt/graphics-chain/wgpu/pct\endcsname{+18.5}
\expandafter\gdef\csname blade@num@rx7900xt/graphics-chain/wgpuus\endcsname{178.6}
\expandafter\gdef\csname blade@num@rx7900xt/graphics-independent/autous\endcsname{178.0}
\expandafter\gdef\csname blade@num@rx7900xt/graphics-independent/both/hi\endcsname{-7.1}
\expandafter\gdef\csname blade@num@rx7900xt/graphics-independent/both/lo\endcsname{-7.5}
\expandafter\gdef\csname blade@num@rx7900xt/graphics-independent/both/mag\endcsname{7.3}
\expandafter\gdef\csname blade@num@rx7900xt/graphics-independent/both/maghi\endcsname{7.5}
\expandafter\gdef\csname blade@num@rx7900xt/graphics-independent/both/maglo\endcsname{7.1}
\expandafter\gdef\csname blade@num@rx7900xt/graphics-independent/both/pct\endcsname{-7.3}
\expandafter\gdef\csname blade@num@rx7900xt/graphics-independent/bothus\endcsname{164.9}
\expandafter\gdef\csname blade@num@rx7900xt/graphics-independent/control/hi\endcsname{-0.0}
\expandafter\gdef\csname blade@num@rx7900xt/graphics-independent/control/lo\endcsname{-0.2}
\expandafter\gdef\csname blade@num@rx7900xt/graphics-independent/control/mag\endcsname{0.1}
\expandafter\gdef\csname blade@num@rx7900xt/graphics-independent/control/maghi\endcsname{0.2}
\expandafter\gdef\csname blade@num@rx7900xt/graphics-independent/control/maglo\endcsname{0.0}
\expandafter\gdef\csname blade@num@rx7900xt/graphics-independent/control/pct\endcsname{-0.1}
\expandafter\gdef\csname blade@num@rx7900xt/graphics-independent/controlus\endcsname{177.8}
\expandafter\gdef\csname blade@num@rx7900xt/graphics-independent/floor\endcsname{0.2}
\expandafter\gdef\csname blade@num@rx7900xt/graphics-independent/hostautous\endcsname{27.9}
\expandafter\gdef\csname blade@num@rx7900xt/graphics-independent/hostboth/hi\endcsname{-2.9}
\expandafter\gdef\csname blade@num@rx7900xt/graphics-independent/hostboth/lo\endcsname{-4.7}
\expandafter\gdef\csname blade@num@rx7900xt/graphics-independent/hostboth/mag\endcsname{3.8}
\expandafter\gdef\csname blade@num@rx7900xt/graphics-independent/hostboth/maghi\endcsname{4.7}
\expandafter\gdef\csname blade@num@rx7900xt/graphics-independent/hostboth/maglo\endcsname{2.9}
\expandafter\gdef\csname blade@num@rx7900xt/graphics-independent/hostboth/pct\endcsname{-3.8}
\expandafter\gdef\csname blade@num@rx7900xt/graphics-independent/hostbothus\endcsname{27.0}
\expandafter\gdef\csname blade@num@rx7900xt/graphics-independent/hostcontrol/hi\endcsname{+2.2}
\expandafter\gdef\csname blade@num@rx7900xt/graphics-independent/hostcontrol/lo\endcsname{-0.7}
\expandafter\gdef\csname blade@num@rx7900xt/graphics-independent/hostcontrol/pct\endcsname{+1.0}
\expandafter\gdef\csname blade@num@rx7900xt/graphics-independent/hostcontrolus\endcsname{28.2}
\expandafter\gdef\csname blade@num@rx7900xt/graphics-independent/hostplacement/hi\endcsname{-2.6}
\expandafter\gdef\csname blade@num@rx7900xt/graphics-independent/hostplacement/lo\endcsname{-4.8}
\expandafter\gdef\csname blade@num@rx7900xt/graphics-independent/hostplacement/mag\endcsname{3.0}
\expandafter\gdef\csname blade@num@rx7900xt/graphics-independent/hostplacement/maghi\endcsname{4.8}
\expandafter\gdef\csname blade@num@rx7900xt/graphics-independent/hostplacement/maglo\endcsname{2.6}
\expandafter\gdef\csname blade@num@rx7900xt/graphics-independent/hostplacement/pct\endcsname{-3.0}
\expandafter\gdef\csname blade@num@rx7900xt/graphics-independent/hostplacementus\endcsname{27.0}
\expandafter\gdef\csname blade@num@rx7900xt/graphics-independent/hostscope/hi\endcsname{-0.2}
\expandafter\gdef\csname blade@num@rx7900xt/graphics-independent/hostscope/lo\endcsname{-3.0}
\expandafter\gdef\csname blade@num@rx7900xt/graphics-independent/hostscope/mag\endcsname{1.0}
\expandafter\gdef\csname blade@num@rx7900xt/graphics-independent/hostscope/maghi\endcsname{3.0}
\expandafter\gdef\csname blade@num@rx7900xt/graphics-independent/hostscope/maglo\endcsname{0.2}
\expandafter\gdef\csname blade@num@rx7900xt/graphics-independent/hostscope/pct\endcsname{-1.0}
\expandafter\gdef\csname blade@num@rx7900xt/graphics-independent/hostscopeus\endcsname{27.6}
\expandafter\gdef\csname blade@num@rx7900xt/graphics-independent/hostwgpu/hi\endcsname{+214.3}
\expandafter\gdef\csname blade@num@rx7900xt/graphics-independent/hostwgpu/lo\endcsname{+207.2}
\expandafter\gdef\csname blade@num@rx7900xt/graphics-independent/hostwgpu/mag\endcsname{211.2}
\expandafter\gdef\csname blade@num@rx7900xt/graphics-independent/hostwgpu/maghi\endcsname{214.3}
\expandafter\gdef\csname blade@num@rx7900xt/graphics-independent/hostwgpu/maglo\endcsname{207.2}
\expandafter\gdef\csname blade@num@rx7900xt/graphics-independent/hostwgpu/pct\endcsname{+211.2}
\expandafter\gdef\csname blade@num@rx7900xt/graphics-independent/hostwgpuus\endcsname{87.0}
\expandafter\gdef\csname blade@num@rx7900xt/graphics-independent/manualscope/hi\endcsname{+0.1}
\expandafter\gdef\csname blade@num@rx7900xt/graphics-independent/manualscope/lo\endcsname{-0.1}
\expandafter\gdef\csname blade@num@rx7900xt/graphics-independent/manualscope/pct\endcsname{-0.0}
\expandafter\gdef\csname blade@num@rx7900xt/graphics-independent/placement/hi\endcsname{-7.2}
\expandafter\gdef\csname blade@num@rx7900xt/graphics-independent/placement/lo\endcsname{-7.5}
\expandafter\gdef\csname blade@num@rx7900xt/graphics-independent/placement/mag\endcsname{7.3}
\expandafter\gdef\csname blade@num@rx7900xt/graphics-independent/placement/maghi\endcsname{7.5}
\expandafter\gdef\csname blade@num@rx7900xt/graphics-independent/placement/maglo\endcsname{7.2}
\expandafter\gdef\csname blade@num@rx7900xt/graphics-independent/placement/pct\endcsname{-7.3}
\expandafter\gdef\csname blade@num@rx7900xt/graphics-independent/placementus\endcsname{165.0}
\expandafter\gdef\csname blade@num@rx7900xt/graphics-independent/scope/hi\endcsname{+0.0}
\expandafter\gdef\csname blade@num@rx7900xt/graphics-independent/scope/lo\endcsname{-0.2}
\expandafter\gdef\csname blade@num@rx7900xt/graphics-independent/scope/pct\endcsname{-0.1}
\expandafter\gdef\csname blade@num@rx7900xt/graphics-independent/scopeus\endcsname{177.9}
\expandafter\gdef\csname blade@num@rx7900xt/graphics-independent/waitautous\endcsname{199.0}
\expandafter\gdef\csname blade@num@rx7900xt/graphics-independent/waitbothus\endcsname{186.0}
\expandafter\gdef\csname blade@num@rx7900xt/graphics-independent/waitcontrolus\endcsname{198.9}
\expandafter\gdef\csname blade@num@rx7900xt/graphics-independent/waitplacementus\endcsname{186.0}
\expandafter\gdef\csname blade@num@rx7900xt/graphics-independent/waitscopeus\endcsname{198.7}
\expandafter\gdef\csname blade@num@rx7900xt/graphics-independent/waitwgpuus\endcsname{230.5}
\expandafter\gdef\csname blade@num@rx7900xt/graphics-independent/wgpu/hi\endcsname{+12.3}
\expandafter\gdef\csname blade@num@rx7900xt/graphics-independent/wgpu/lo\endcsname{+11.9}
\expandafter\gdef\csname blade@num@rx7900xt/graphics-independent/wgpu/mag\endcsname{12.0}
\expandafter\gdef\csname blade@num@rx7900xt/graphics-independent/wgpu/maghi\endcsname{12.3}
\expandafter\gdef\csname blade@num@rx7900xt/graphics-independent/wgpu/maglo\endcsname{11.9}
\expandafter\gdef\csname blade@num@rx7900xt/graphics-independent/wgpu/pct\endcsname{+12.0}
\expandafter\gdef\csname blade@num@rx7900xt/graphics-independent/wgpuus\endcsname{199.6}
\expandafter\gdef\csname blade@num@rx7900xt/mixed-chain/autous\endcsname{141.8}
\expandafter\gdef\csname blade@num@rx7900xt/mixed-chain/both/hi\endcsname{+0.1}
\expandafter\gdef\csname blade@num@rx7900xt/mixed-chain/both/lo\endcsname{-0.2}
\expandafter\gdef\csname blade@num@rx7900xt/mixed-chain/both/pct\endcsname{-0.1}
\expandafter\gdef\csname blade@num@rx7900xt/mixed-chain/bothus\endcsname{141.8}
\expandafter\gdef\csname blade@num@rx7900xt/mixed-chain/control/hi\endcsname{+0.1}
\expandafter\gdef\csname blade@num@rx7900xt/mixed-chain/control/lo\endcsname{-0.2}
\expandafter\gdef\csname blade@num@rx7900xt/mixed-chain/control/pct\endcsname{-0.1}
\expandafter\gdef\csname blade@num@rx7900xt/mixed-chain/controlus\endcsname{141.8}
\expandafter\gdef\csname blade@num@rx7900xt/mixed-chain/floor\endcsname{0.2}
\expandafter\gdef\csname blade@num@rx7900xt/mixed-chain/hostautous\endcsname{15.9}
\expandafter\gdef\csname blade@num@rx7900xt/mixed-chain/hostboth/hi\endcsname{-1.2}
\expandafter\gdef\csname blade@num@rx7900xt/mixed-chain/hostboth/lo\endcsname{-2.4}
\expandafter\gdef\csname blade@num@rx7900xt/mixed-chain/hostboth/mag\endcsname{1.8}
\expandafter\gdef\csname blade@num@rx7900xt/mixed-chain/hostboth/maghi\endcsname{2.4}
\expandafter\gdef\csname blade@num@rx7900xt/mixed-chain/hostboth/maglo\endcsname{1.2}
\expandafter\gdef\csname blade@num@rx7900xt/mixed-chain/hostboth/pct\endcsname{-1.8}
\expandafter\gdef\csname blade@num@rx7900xt/mixed-chain/hostbothus\endcsname{15.6}
\expandafter\gdef\csname blade@num@rx7900xt/mixed-chain/hostcontrol/hi\endcsname{+1.2}
\expandafter\gdef\csname blade@num@rx7900xt/mixed-chain/hostcontrol/lo\endcsname{-0.8}
\expandafter\gdef\csname blade@num@rx7900xt/mixed-chain/hostcontrol/pct\endcsname{+0.0}
\expandafter\gdef\csname blade@num@rx7900xt/mixed-chain/hostcontrolus\endcsname{15.9}
\expandafter\gdef\csname blade@num@rx7900xt/mixed-chain/hostplacement/hi\endcsname{+0.0}
\expandafter\gdef\csname blade@num@rx7900xt/mixed-chain/hostplacement/lo\endcsname{-3.1}
\expandafter\gdef\csname blade@num@rx7900xt/mixed-chain/hostplacement/pct\endcsname{-1.9}
\expandafter\gdef\csname blade@num@rx7900xt/mixed-chain/hostplacementus\endcsname{15.6}
\expandafter\gdef\csname blade@num@rx7900xt/mixed-chain/hostscope/hi\endcsname{+6.1}
\expandafter\gdef\csname blade@num@rx7900xt/mixed-chain/hostscope/lo\endcsname{-2.1}
\expandafter\gdef\csname blade@num@rx7900xt/mixed-chain/hostscope/pct\endcsname{-1.2}
\expandafter\gdef\csname blade@num@rx7900xt/mixed-chain/hostscopeus\endcsname{15.6}
\expandafter\gdef\csname blade@num@rx7900xt/mixed-chain/manualscope/hi\endcsname{+0.1}
\expandafter\gdef\csname blade@num@rx7900xt/mixed-chain/manualscope/lo\endcsname{-0.1}
\expandafter\gdef\csname blade@num@rx7900xt/mixed-chain/manualscope/pct\endcsname{-0.0}
\expandafter\gdef\csname blade@num@rx7900xt/mixed-chain/placement/hi\endcsname{+0.1}
\expandafter\gdef\csname blade@num@rx7900xt/mixed-chain/placement/lo\endcsname{-0.1}
\expandafter\gdef\csname blade@num@rx7900xt/mixed-chain/placement/pct\endcsname{-0.1}
\expandafter\gdef\csname blade@num@rx7900xt/mixed-chain/placementus\endcsname{141.9}
\expandafter\gdef\csname blade@num@rx7900xt/mixed-chain/scope/hi\endcsname{+0.2}
\expandafter\gdef\csname blade@num@rx7900xt/mixed-chain/scope/lo\endcsname{-0.1}
\expandafter\gdef\csname blade@num@rx7900xt/mixed-chain/scope/pct\endcsname{+0.0}
\expandafter\gdef\csname blade@num@rx7900xt/mixed-chain/scopeus\endcsname{141.9}
\expandafter\gdef\csname blade@num@rx7900xt/mixed-chain/waitautous\endcsname{161.7}
\expandafter\gdef\csname blade@num@rx7900xt/mixed-chain/waitbothus\endcsname{160.8}
\expandafter\gdef\csname blade@num@rx7900xt/mixed-chain/waitcontrolus\endcsname{161.6}
\expandafter\gdef\csname blade@num@rx7900xt/mixed-chain/waitplacementus\endcsname{160.6}
\expandafter\gdef\csname blade@num@rx7900xt/mixed-chain/waitscopeus\endcsname{161.7}
\expandafter\gdef\csname blade@num@rx7900xt/mixed-independent/autous\endcsname{155.9}
\expandafter\gdef\csname blade@num@rx7900xt/mixed-independent/both/hi\endcsname{-17.6}
\expandafter\gdef\csname blade@num@rx7900xt/mixed-independent/both/lo\endcsname{-17.8}
\expandafter\gdef\csname blade@num@rx7900xt/mixed-independent/both/mag\endcsname{17.7}
\expandafter\gdef\csname blade@num@rx7900xt/mixed-independent/both/maghi\endcsname{17.8}
\expandafter\gdef\csname blade@num@rx7900xt/mixed-independent/both/maglo\endcsname{17.6}
\expandafter\gdef\csname blade@num@rx7900xt/mixed-independent/both/pct\endcsname{-17.7}
\expandafter\gdef\csname blade@num@rx7900xt/mixed-independent/bothus\endcsname{128.3}
\expandafter\gdef\csname blade@num@rx7900xt/mixed-independent/control/hi\endcsname{+0.1}
\expandafter\gdef\csname blade@num@rx7900xt/mixed-independent/control/lo\endcsname{-0.1}
\expandafter\gdef\csname blade@num@rx7900xt/mixed-independent/control/pct\endcsname{-0.0}
\expandafter\gdef\csname blade@num@rx7900xt/mixed-independent/controlus\endcsname{155.8}
\expandafter\gdef\csname blade@num@rx7900xt/mixed-independent/floor\endcsname{0.1}
\expandafter\gdef\csname blade@num@rx7900xt/mixed-independent/hostautous\endcsname{20.0}
\expandafter\gdef\csname blade@num@rx7900xt/mixed-independent/hostboth/hi\endcsname{-5.7}
\expandafter\gdef\csname blade@num@rx7900xt/mixed-independent/hostboth/lo\endcsname{-7.7}
\expandafter\gdef\csname blade@num@rx7900xt/mixed-independent/hostboth/mag\endcsname{6.5}
\expandafter\gdef\csname blade@num@rx7900xt/mixed-independent/hostboth/maghi\endcsname{7.7}
\expandafter\gdef\csname blade@num@rx7900xt/mixed-independent/hostboth/maglo\endcsname{5.7}
\expandafter\gdef\csname blade@num@rx7900xt/mixed-independent/hostboth/pct\endcsname{-6.5}
\expandafter\gdef\csname blade@num@rx7900xt/mixed-independent/hostbothus\endcsname{18.6}
\expandafter\gdef\csname blade@num@rx7900xt/mixed-independent/hostcontrol/hi\endcsname{+0.3}
\expandafter\gdef\csname blade@num@rx7900xt/mixed-independent/hostcontrol/lo\endcsname{-2.5}
\expandafter\gdef\csname blade@num@rx7900xt/mixed-independent/hostcontrol/pct\endcsname{-0.8}
\expandafter\gdef\csname blade@num@rx7900xt/mixed-independent/hostcontrolus\endcsname{19.7}
\expandafter\gdef\csname blade@num@rx7900xt/mixed-independent/hostplacement/hi\endcsname{-5.9}
\expandafter\gdef\csname blade@num@rx7900xt/mixed-independent/hostplacement/lo\endcsname{-8.0}
\expandafter\gdef\csname blade@num@rx7900xt/mixed-independent/hostplacement/mag\endcsname{6.6}
\expandafter\gdef\csname blade@num@rx7900xt/mixed-independent/hostplacement/maghi\endcsname{8.0}
\expandafter\gdef\csname blade@num@rx7900xt/mixed-independent/hostplacement/maglo\endcsname{5.9}
\expandafter\gdef\csname blade@num@rx7900xt/mixed-independent/hostplacement/pct\endcsname{-6.6}
\expandafter\gdef\csname blade@num@rx7900xt/mixed-independent/hostplacementus\endcsname{18.6}
\expandafter\gdef\csname blade@num@rx7900xt/mixed-independent/hostscope/hi\endcsname{-2.1}
\expandafter\gdef\csname blade@num@rx7900xt/mixed-independent/hostscope/lo\endcsname{-4.6}
\expandafter\gdef\csname blade@num@rx7900xt/mixed-independent/hostscope/mag\endcsname{2.9}
\expandafter\gdef\csname blade@num@rx7900xt/mixed-independent/hostscope/maghi\endcsname{4.6}
\expandafter\gdef\csname blade@num@rx7900xt/mixed-independent/hostscope/maglo\endcsname{2.1}
\expandafter\gdef\csname blade@num@rx7900xt/mixed-independent/hostscope/pct\endcsname{-2.9}
\expandafter\gdef\csname blade@num@rx7900xt/mixed-independent/hostscopeus\endcsname{19.3}
\expandafter\gdef\csname blade@num@rx7900xt/mixed-independent/manualscope/hi\endcsname{+0.1}
\expandafter\gdef\csname blade@num@rx7900xt/mixed-independent/manualscope/lo\endcsname{-0.1}
\expandafter\gdef\csname blade@num@rx7900xt/mixed-independent/manualscope/pct\endcsname{+0.0}
\expandafter\gdef\csname blade@num@rx7900xt/mixed-independent/placement/hi\endcsname{-17.7}
\expandafter\gdef\csname blade@num@rx7900xt/mixed-independent/placement/lo\endcsname{-17.8}
\expandafter\gdef\csname blade@num@rx7900xt/mixed-independent/placement/mag\endcsname{17.7}
\expandafter\gdef\csname blade@num@rx7900xt/mixed-independent/placement/maghi\endcsname{17.8}
\expandafter\gdef\csname blade@num@rx7900xt/mixed-independent/placement/maglo\endcsname{17.7}
\expandafter\gdef\csname blade@num@rx7900xt/mixed-independent/placement/pct\endcsname{-17.7}
\expandafter\gdef\csname blade@num@rx7900xt/mixed-independent/placementus\endcsname{128.2}
\expandafter\gdef\csname blade@num@rx7900xt/mixed-independent/scope/hi\endcsname{+0.1}
\expandafter\gdef\csname blade@num@rx7900xt/mixed-independent/scope/lo\endcsname{-0.1}
\expandafter\gdef\csname blade@num@rx7900xt/mixed-independent/scope/pct\endcsname{-0.0}
\expandafter\gdef\csname blade@num@rx7900xt/mixed-independent/scopeus\endcsname{155.8}
\expandafter\gdef\csname blade@num@rx7900xt/mixed-independent/waitautous\endcsname{177.1}
\expandafter\gdef\csname blade@num@rx7900xt/mixed-independent/waitbothus\endcsname{154.8}
\expandafter\gdef\csname blade@num@rx7900xt/mixed-independent/waitcontrolus\endcsname{177.0}
\expandafter\gdef\csname blade@num@rx7900xt/mixed-independent/waitplacementus\endcsname{155.0}
\expandafter\gdef\csname blade@num@rx7900xt/mixed-independent/waitscopeus\endcsname{176.8}
\expandafter\gdef\csname blade@num@sweep/gpu/compute-independent/auto/at16\endcsname{195.3}
\expandafter\gdef\csname blade@num@sweep/gpu/compute-independent/auto/marginal\endcsname{12.29}
\expandafter\gdef\csname blade@num@sweep/gpu/compute-independent/barriercost\endcsname{3.77}
\expandafter\gdef\csname blade@num@sweep/gpu/compute-independent/hazard/at16\endcsname{138.2}
\expandafter\gdef\csname blade@num@sweep/gpu/compute-independent/hazard/marginal\endcsname{8.52}
\expandafter\gdef\csname blade@num@sweep/gpu/compute-independent/wgpu/at16\endcsname{196.5}
\expandafter\gdef\csname blade@num@sweep/gpu/compute-independent/wgpu/marginal\endcsname{12.34}
\expandafter\gdef\csname blade@num@sweep/gpu/graphics-independent/auto/at16\endcsname{151.6}
\expandafter\gdef\csname blade@num@sweep/gpu/graphics-independent/auto/marginal\endcsname{9.46}
\expandafter\gdef\csname blade@num@sweep/gpu/graphics-independent/barriercost\endcsname{2.94}
\expandafter\gdef\csname blade@num@sweep/gpu/graphics-independent/hazard/at16\endcsname{103.8}
\expandafter\gdef\csname blade@num@sweep/gpu/graphics-independent/hazard/marginal\endcsname{6.52}
\expandafter\gdef\csname blade@num@sweep/gpu/graphics-independent/wgpu/at16\endcsname{99.3}
\expandafter\gdef\csname blade@num@sweep/gpu/graphics-independent/wgpu/marginal\endcsname{6.48}
\expandafter\gdef\csname blade@num@sweep/host/compute-independent/auto/at16\endcsname{54.5}
\expandafter\gdef\csname blade@num@sweep/host/compute-independent/auto/marginal\endcsname{1.10}
\expandafter\gdef\csname blade@num@sweep/host/compute-independent/barriercost\endcsname{0.74}
\expandafter\gdef\csname blade@num@sweep/host/compute-independent/hazard/at16\endcsname{16.0}
\expandafter\gdef\csname blade@num@sweep/host/compute-independent/hazard/marginal\endcsname{0.36}
\expandafter\gdef\csname blade@num@sweep/host/compute-independent/wgpu/at16\endcsname{230.2}
\expandafter\gdef\csname blade@num@sweep/host/compute-independent/wgpu/marginal\endcsname{5.59}
\expandafter\gdef\csname blade@num@sweep/host/compute-independent/wgpuoverauto\endcsname{5.1}
\expandafter\gdef\csname blade@num@sweep/host/compute-independent/wgpuoverhazard\endcsname{15.5}
\expandafter\gdef\csname blade@num@sweep/host/graphics-independent/auto/at16\endcsname{62.2}
\expandafter\gdef\csname blade@num@sweep/host/graphics-independent/auto/marginal\endcsname{1.48}
\expandafter\gdef\csname blade@num@sweep/host/graphics-independent/barriercost\endcsname{0.58}
\expandafter\gdef\csname blade@num@sweep/host/graphics-independent/hazard/at16\endcsname{27.7}
\expandafter\gdef\csname blade@num@sweep/host/graphics-independent/hazard/marginal\endcsname{0.90}
\expandafter\gdef\csname blade@num@sweep/host/graphics-independent/wgpu/at16\endcsname{115.0}
\expandafter\gdef\csname blade@num@sweep/host/graphics-independent/wgpu/marginal\endcsname{3.75}
\expandafter\gdef\csname blade@num@sweep/host/graphics-independent/wgpuoverauto\endcsname{2.5}
\expandafter\gdef\csname blade@num@sweep/host/graphics-independent/wgpuoverhazard\endcsname{4.2}

\title{Global Pass Barriers Without Per-Resource RHI Tracking:\\
A Cross-Vendor Study with \blade}
\author{Dzmitry Malyshau}
\date{July 2026}

\begin{document}
\maketitle

\begin{abstract}
Explicit graphics APIs expose memory dependencies, per-resource accesses, and
image layouts. \wgpu{} reconstructs and validates this state; \blade{} keeps
Vulkan images in \texttt{GENERAL}, tracks no per-resource state, and issues
global pass-boundary barriers. We isolate barrier placement and stage/access
scope within \blade{}, compare matched \wgpu{} programs end-to-end, and measure
six GPUs from four vendors, including exploratory Apple/Metal results.

Removing fifteen redundant barriers from sixteen independent compute passes
reduces GPU span by $\bmagci{rtx5070}{compute-independent}{placement}$ on an
RTX~5070 and $\bmagci{rx7900xt}{compute-independent}{placement}$ on an
RX~7900~XT. It reduces independent-render span by
$\bmag{rtx5070}{graphics-independent}{placement}$ and
$\bmag{rx7900xt}{graphics-independent}{placement}$ on the same discrete parts,
but increases it by
$\bladenum{depth/radeon780m/graphics-independent/p32/placement/mag}\%$ on a
Radeon~780M at 32 passes, beyond that count's stability floor. No
dependent-chain placement effect clears the study's cell-specific stability
criterion. Deriving global barrier scope from the pass kinds around each
boundary, without tracking any resource, saves
$\bmagci{rtx5070}{graphics-chain}{scope}$ on an NVIDIA graphics chain and
$\bmagci{rx7900xt}{compute-chain}{scope}$ on an AMD compute chain; no AMD
render-involving scope cell resolves. \wgpu's record-and-submit cost is higher
in every measured cell, but this end-to-end difference cannot be attributed
to tracking alone.

RADV source shows why command counts do not predict these costs: broad global
dependencies expand to several flush and invalidate requests that the driver
may partly elide. It also shows that persistent \texttt{GENERAL} retains DCC
under the measured RDNA conditions but disables FMASK. The resulting \blade{}
direction is lightweight aggregate pass-kind state in a tracking-free RHI: an
upstream render graph selects global dependency cuts, while aliasing,
cross-queue use, exceptional layouts, and arbitrary per-resource DAG edges
remain resource-aware engine responsibilities.
\end{abstract}

\section{Introduction}

Explicit GPU APIs expose synchronization so applications can preserve
parallelism and avoid unnecessary cache operations. That control has a cost:
somebody has to know how every buffer and image subresource is used, maintain
that state across command buffers, validate conflicts, and generate
transitions. This machinery sits on a CPU-critical command recording path. The
\wgpu{} implementation describes its trackers as some of the hottest code in
its codebase~\cite{wgpu-trackers}.

Production engines show both the scale and why this work is normally
centralized. Activision reports that the first deployment of its Task Graph
contained 236 tasks, 258 resources, and 92 barrier calls with 328 barrier
elements; the graph later grew to roughly 350 tasks~\cite{activision-task-graph}.
Unreal's Render Dependency Graph likewise uses
pass resource declarations to derive subresource transitions, asynchronous
compute fences, transient aliasing, validation, and parallel
recording~\cite{unreal-rdg}. At that scale the useful question is more precise than
``track or do not track'': must the low-level graphics abstraction duplicate
per-resource knowledge, or can an engine-level graph remain the source of
truth and give the abstraction a smaller execution contract?

\blade{}~\cite{blade} tests the latter engineering hypothesis. The
distinction is not that its API exposes fewer state objects --- \wgpu's does
not expose resource states either, and tracking is an internal matter for both
--- but that \blade{} does not maintain the state at all. On Vulkan, ordinary
images stay in the \texttt{GENERAL} image layout and a broad memory barrier
orders writes before later reads and writes at pass boundaries, so nothing has
to be remembered between one pass and the next. What surfaces in the API is a
consequence rather than the point: no per-resource work at record time, and an
application that wants overlap has to say so itself. A render graph and
\blade{} are therefore composable rather than competing designs: the graph can
choose the dependency boundaries while the RHI declines to rediscover them.
An application without such a graph can retain \blade's conservative automatic
boundaries.

The question is newly relevant because
\texttt{VK\_KHR\_unified\_image\_layouts} guarantees, when enabled, that
\texttt{GENERAL} is as efficient as specialized layouts in most
uses~\cite{vulkan-unified-layouts}. The extension removes one reason for
tracking image state, but explicitly leaves barrier granularity as a separate
performance question. It is not universal in the measured set: four of the
five Vulkan device-driver pairs advertise it (Table~\ref{tab:platforms}), and
these current 2025--2026 drivers are not a representative deployment sample.
This paper therefore measures \texttt{GENERAL} on drivers with and without the
extension rather than assuming the guarantee.

The immediate motivation is concrete. A \blade{} user reported that their
renderer's planar-reflection pass and main pass, which share no resources, were
being serialized on AMD hardware purely because \blade{} placed a barrier
between them~\cite{blade-issue-343}. That report, the fix it produced, and the
question of how far it generalizes are the subject of Section~\ref{sec:field}.

This paper is organized around the following research questions:

\begin{description}
  \item[RQ1:] How does host cost scale with passes and dependency density for
  tracking-free and tracked record-and-submit paths?
  \item[RQ2:] When does removing redundant global barriers reduce device span,
  and when is it neutral or harmful?
  \item[RQ3:] How do matched \blade{} and \wgpu{} implementations differ
  end-to-end?
  \item[RQ4:] How do the answers differ between compute and graphics workloads,
  and across architectures?
\end{description}

The contributions are:

\begin{itemize}
  \item a cost model that distinguishes tracking, barrier precision, barrier
  placement, and image layouts without claiming that every factor is
  independently controlled;
  \item a controlled within-\blade{} evaluation of automatic versus
  application-declared barrier placement on five Vulkan GPUs, including a negative
  result where coarse placement wins;
  \item a barrier that narrows stage and access scope from the kinds of the
  surrounding passes, keeping the no-resource-state contract, with its
  correctness argument, its measured effect at fixed placement, and the
  two-sided misprediction of the driver-source reading about where it would
  pay;
  \item matched \blade{} and \wgpu{} workloads with matching post-run output
  hashes, plus captures that compare their emitted barrier requests, for
  end-to-end comparison;
  \item a reading of the AMD open-source driver that identifies, in driver
  operations rather than folklore, what a global barrier and a persistent
  \texttt{GENERAL} layout ask for --- and, from the same driver, conditions
  under which part of that request can be elided;
  \item practical guidance identifying the workload envelope in which
  tracking-free synchronization is appropriate.
\end{itemize}

\section{Background}

\subsection{Execution and memory dependencies}

A Vulkan barrier describes both an execution dependency between pipeline
stages and a memory dependency between access types. Resource-specific
barriers can additionally identify the affected buffer or image range.
Narrow scopes may preserve unrelated work, whereas overly broad scopes can
drain the pipeline~\cite{amd-barriers,nvidia-vulkan-tips}. A global memory
barrier is nevertheless attractive when many resources share the same
dependency because it combines them in one command and often maps to hardware
operations whose granularity is already coarse.

These dimensions must not be collapsed into a single ``barrier quality''
variable. A system may:

\begin{enumerate}
  \item place barriers only at true hazards or at every pass boundary;
  \item use global or resource-specific memory scopes;
  \item use broad or precise pipeline stages and access masks; and
  \item transition images between specialized layouts or retain one layout.
\end{enumerate}

This decomposition is not specific to Vulkan. Direct3D~12 Enhanced Barriers
independently expose synchronization scope, access scope, and texture layout,
and distinguish global, buffer, and texture barriers. A global barrier affects
the indicated resource-memory accesses without identifying a resource or
changing a texture layout~\cite{d3d12-enhanced-barriers}. \blade's policy
selects the analogous global point and separately chooses a persistent image
layout. We do not benchmark Direct3D~12; its API is evidence that the axes are
real interface choices, not performance evidence for \blade.

\subsection{What a coarse barrier costs a driver}
\label{sec:radv-barriers}

Vendor guidance states that broad barriers are expensive but rarely says what
work they request. The AMD open-source Vulkan driver answers that mechanics
question directly, so we read it rather than infer it. All statements below refer to
Mesa \texttt{main} at revision \texttt{d18d598e275d}~\cite{mesa-radv}.

RADV converts a dependency into a set of cache-operation flags in
\texttt{radv\_src\_access\_flush} and \texttt{radv\_dst\_access\_flush}. Before
either runs, the shared Vulkan runtime expands the access
mask~\cite{mesa-vk-synchronization}: \texttt{VK\_ACCESS\_2\_MEMORY\_WRITE\_BIT}
becomes \emph{every} write access reachable from the given stage mask.
\blade's barrier names \texttt{ALL\_COMMANDS}, so the expansion is total, and
consequently every branch of \texttt{radv\_src\_access\_flush} fires:

\begin{itemize}
  \item \texttt{COLOR\_ATTACHMENT\_WRITE} triggers
  \texttt{FLUSH\_AND\_INV\_CB}, plus colour metadata handling;
  \item \texttt{DEPTH\_STENCIL\_ATTACHMENT\_WRITE} triggers
  \texttt{FLUSH\_AND\_INV\_DB}, plus depth metadata handling;
  \item \texttt{TRANSFER\_WRITE} requests both attachment-cache paths and their
  metadata handling;
  \item \texttt{TRANSFORM\_FEEDBACK\_WRITE} requests an L2 write-back; and
  \item command-preprocess writes request an L2 invalidation; shader-storage
  and acceleration-structure writes do so where the relevant memory is not
  coherent.
\end{itemize}

Two details make the global form more pessimistic than an \emph{image}-scoped
barrier carrying the same access masks. (A buffer barrier supplies no image
either.) First, RADV initializes
\texttt{has\_CB\_meta} and \texttt{has\_DB\_meta} to true and only clears them
when an image argument is supplied; a global memory barrier therefore always
adds \texttt{FLUSH\_AND\_INV\_CB\_META} and \texttt{FLUSH\_AND\_INV\_DB\_META}
even when no compressed image is involved. The destination side is pessimistic
in the same way and for the same reason. Second, the L2-coherence check
\texttt{radv\_image\_is\_l2\_coherent} requires an image, so a global barrier
cannot use it; the source comment notes that for memory barriers without an
image the state at rest ``often devolves to just VRAM/GTT anyway''.

The execution dependency is separate and equally blunt. RADV maps source
stages to partial flushes: \texttt{ALL\_COMMANDS} appears in both the list that
sets \texttt{CS\_PARTIAL\_FLUSH} and the list that sets
\texttt{PS\_PARTIAL\_FLUSH}. A partial flush waits for all in-flight waves of
that type to retire. Placing \blade's barrier at every pass boundary therefore
requests both compute and pixel partial flushes between every pair of passes,
whether or not they share data; either request can be cheap when no matching
work is in flight. This source-level mechanism is consistent with the user
report in Section~\ref{sec:field}, but does not by itself establish its cost.

\subsubsection{What the same driver takes back}
\label{sec:radv-elision}

The paragraphs above read the request. The driver's answer to it is a few lines
further down the same function, and we record it here because
Section~\ref{sec:scope-results} needed it and did not have it.

\texttt{radv\_dst\_access\_flush} does not use the barrier's masks alone. It
first widens its notion of coherence:
\texttt{image\_is\_coherent} is set whenever
\texttt{can\_skip\_buffer\_l2\_flushes} holds and no render-backend-incoherent
image is currently dirty. The first condition is a property of the part ---
GFX9, or GFX10 and later with a coherent TCC --- and the second is tracked
across the command buffer, set only when a render pass binds an attachment for
which \texttt{radv\_image\_is\_l2\_coherent} is false. On GFX10 and later that
predicate reduces to whether the image is pipe-misaligned.

The consequence is conditional but broad: on the measured RDNA parts, while no
pipe-misaligned attachment is dirty, every destination-side
\texttt{INV\_L2} the coarse barrier asks for is dropped. The
driver has already made the argument that motivated our narrower barrier, and
made it without needing the application to say anything. What a narrower scope
can still remove is the source-side flushes and the pixel drain --- operations
whose cost is proportional to work in flight and to dirty cache state, both of
which are absent in exactly the workloads where the narrowing is cleanest to
measure. Section~\ref{sec:scope-results} reports the measurement, including
the one component this account does not predict; the explanation here arrived
after the numbers, and it is a reading of source rather than a second
measurement.

\subsection{Image layouts}

Vulkan image layouts communicate usage information that may select compression
metadata, attachment paths, or transfer representations. Vendor guidance has
historically recommended specialized layouts, particularly on
AMD~\cite{amd-rdna-guide}. The unified-image-layouts extension reflects a
hardware and driver trend in which most layouts are physically identical. It
allows \texttt{GENERAL} in most positions and promises optimal performance,
while retaining exceptions such as initialization from
\texttt{UNDEFINED}, presentation, and some video
interactions~\cite{vulkan-unified-layouts}.

RADV lets us be more specific than the extension proposal, because the layout
policy is three readable predicates~\cite{mesa-radv}:

\begin{itemize}
  \item \texttt{radv\_layout\_dcc\_compressed} ends by returning true when the
  GFX level is at least GFX10 \emph{or} the layout is not \texttt{GENERAL}.
  Delta colour compression therefore survives \texttt{GENERAL} on RDNA and
  later, and is lost on GFX9 and earlier. One earlier branch qualifies this:
  an image whose queue-family mask includes the compute family loses DCC in
  \texttt{GENERAL} on hardware that cannot compress DCC through image stores.
  \blade{} creates every image \texttt{EXCLUSIVE} on one queue family, so the
  qualification does not apply to anything measured here, but it does apply to
  an application that shares images with an async-compute queue. Subject to
  the function's other transfer-queue and feedback-loop cases, the relevant
  conclusion is narrower: for the measured single-queue colour attachments on
  GFX10 and later, \texttt{GENERAL} by itself does not disable DCC.
  \item \texttt{radv\_layout\_is\_htile\_compressed} returns true for
  \texttt{GENERAL} when TC-compatible HTILE is enabled for that level, the
  image can be used on the graphics queue, and a driver-configuration override
  has not disabled the case. The accompanying comment states that it exists
  specifically to help ``apps that use GENERAL for the main depth pass'', and
  gives ``the image doesn't have the storage bit set'' as part of its
  justification. That condition is enforced earlier when RADV decides whether
  the image may use HTILE, rather than retested by this layout predicate.
  \item \texttt{radv\_layout\_fmask\_compression} returns
  \texttt{RADV\_FMASK\_COMPRESSION\_NONE} for \texttt{GENERAL}
  unconditionally, before any of its other cases. A persistent
  \texttt{GENERAL} layout therefore forgoes FMASK compression for multisampled
  colour; the bandwidth cost is expected from that source reading but is not
  measured here.
\end{itemize}

Consequently, this work uses ``no layout tracking'' to mean no steady-state
tracking for ordinary application images. It does not claim that all layout
transitions or image-specific synchronization disappear, and it identifies MSAA
colour as the case where the claim is weakest.

\subsection{WebGPU and \wgpu}

WebGPU defines usage scopes that reject conflicting uses and provide a safe,
portable programming contract~\cite{webgpu-spec}. \wgpu{} represents buffer
and texture states in dense trackers, merges per-scope usage, and generates
transitions as usage scopes enter a command buffer and between submitted
command buffers~\cite{wgpu-trackers}. Its Vulkan backend
emits \texttt{VkBufferMemoryBarrier} and \texttt{VkImageMemoryBarrier}
structures whose stages, accesses, and image layouts are derived from the
recorded usage. That makes the requests resource-specific; it does not prove
minimum stages, minimum calls, or optimal placement. This work does not treat
that tracking as accidental overhead: it implements guarantees that \blade's
unsafe API intentionally does not provide.

The comparison instead asks what that end-to-end design costs,
and whether a narrower native contract is useful for applications that already
know their dependency structure.

\section{\blade{} Synchronization Model}

\blade{} exposes transfer, compute, and render passes within a command encoder.
It does not attach a current state to resource handles. The Vulkan backend
places the following dependency before each pass, and once more when the
encoder is finished:

\begin{verbatim}
srcStage  = ALL_COMMANDS
dstStage  = ALL_COMMANDS
srcAccess = MEMORY_WRITE
dstAccess = MEMORY_READ | MEMORY_WRITE
\end{verbatim}

Images are initialized from \texttt{UNDEFINED} to \texttt{GENERAL} and remain
in \texttt{GENERAL} for ordinary rendering, sampling, storage, and copies.
Presentation images still transition to the presentation layout. \blade{}
enables \texttt{VK\_KHR\_unified\_image\_layouts} on drivers that advertise it.

Since the resolution of the field report, \texttt{CommandEncoderDesc} carries a
\texttt{manual\_barriers} flag~\cite{blade-pr-355}. When set, the automatic pass-boundary barrier
is suppressed and the application calls \texttt{encoder.barrier()} where its
dependency graph requires one. The emitted barrier is identical; only the
placement changes. This is the mechanism the evaluation controls, and its
design history is Section~\ref{sec:field}.

\subsection{Narrowing the barrier without tracking resources}
\label{sec:scoped}

Section~\ref{sec:radv-barriers} shows that most of what the barrier above
\emph{asks} of RADV comes from the two words \texttt{ALL\_COMMANDS} and
\texttt{MEMORY\_WRITE}, not from the barrier being global. That suggests a
cheaper barrier that still tracks nothing: an encoder always knows what kind of
pass it just closed and what kind it is about to open, and a pass kind is one
enum per encoder, not a table indexed by resource. The distinction between what
a barrier asks for and what it costs is the one this configuration went on to
measure, in both directions; we keep the original reasoning here because the
configuration is worth having anyway, and report where it held and where it
did not in Section~\ref{sec:scope-results}.

We therefore added a second scope, and the first question it raises is how
much the application has to say for it to work. The answer is: nothing, for
the barriers the encoder places itself. Two quantities are needed, and the
encoder already holds both.

The \emph{source} scope must cover every write the barrier has to make
available. That is the union of the kinds of the passes recorded since the
previous barrier, which the encoder accumulates in one bitmask. Taking the
union rather than just the last pass is what makes this safe when the
application suppressed some boundaries: whichever passes went unseparated,
their writes are still in the set. The \emph{destination} scope is the kind of
the pass being opened, and the encoder is inside \texttt{compute()},
\texttt{render()}, \texttt{transfer()}, or
\texttt{acceleration\_structure()} when it places the barrier, so the kind is
the method that was called. Nothing is added to the pass declaration, and no
resource is consulted.

The one case that is not derivable is an explicitly placed barrier, because
the consumer has not been declared when \texttt{barrier()} is called. That one
takes its scope as an argument, and is emitted where it is written: its source
scope narrows like any other, but its destination stays \texttt{ALL\_COMMANDS}.
Holding the request back until the next pass opened would recover the
destination as well, and we measured that it does, but a synchronization
primitive that does not appear where the application put it is a poor trade for
a few percent. The asymmetry is instead reported as a result: it is the
measurable cost of not knowing what comes next.

The derivation itself is a table over pass kinds:

\begin{center}
\footnotesize
\begin{tabular}{@{}l@{\ \ }l@{\ \ }l@{}}
\toprule
Pass kind & Source stage & Source access \\
\midrule
transfer & \texttt{TRANSFER} & \texttt{TRANSFER\_WRITE} \\
compute & \texttt{COMPUTE\_SHADER} & \texttt{SHADER\_WRITE} \\
render & \texttt{ALL\_GRAPHICS} & attachment, \texttt{SHADER\_WRITE} \\
accel.\ struct. & \texttt{AS\_BUILD\,|\,TRANSFER} &
  \texttt{AS\_WRITE\,|\,TRANSFER\_WRITE} \\
\bottomrule
\end{tabular}
\end{center}

The destination masks mirror this with the read accesses each kind can perform.
The first barrier of an encoder starts with the source set seeded to
``unknown'', so it is as wide as before; anything recorded outside the pass
model, such as image initialization, adds ``unknown'' to the set for the same
reason. The barrier emitted when the encoder finishes keeps a wide
\emph{destination}, since the next queue consumer is unknown, but its source is
derived like any other. Host visibility separately relies on queue completion
and the memory's host-coherency or invalidation rules.

Correctness rests on the standard availability-and-visibility argument. An
automatically placed barrier names every producer accumulated since the
previous barrier on its source side and the pass being opened on its
destination side. An explicit barrier names the same accumulated producers and
keeps an \texttt{ALL\_COMMANDS} destination, which covers any later pass.
Thus every write ordered by the policy is made available by a source scope
that includes its producing pass kind and visible to a destination scope that
includes its consumer.

Synchronization validation was necessary rather than decorative here. It found
that Blade applies two driver workarounds which add transfer accesses to every
barrier; those are legal alongside \texttt{ALL\_COMMANDS} and illegal alongside
a narrowed stage mask, and they are also unnecessary in the narrowed case,
because a narrowed scope names the accesses of the passes that actually ran.
During development, the final implementation reported no Khronos
synchronization hazards or validation errors across all thirty-six workload
and policy combinations, and every policy produced the same output hash. The
output hashes are retained with the timing collections; the development-time
synchronization-validation logs are not. The validation result is therefore
not yet archival evidence and must be repeated and published before
submission.

On RADV the difference is concrete. A compute-to-compute boundary drops from
\{compute and pixel partial flush, colour-block flush, depth-block flush, both
metadata flushes, L2 invalidate, L2 write-back\} to \{compute partial flush, L2
invalidate\}. A render-to-compute boundary keeps the colour and depth flushes,
which are real, but loses the compute drain. What it does \emph{not} drop is
anything on the destination's cache side, because on these parts the driver had
already dropped it (Section~\ref{sec:radv-elision}) --- the first hint that
what this configuration removes and what it would save were not going to line
up.

The evaluated contract is deliberately limited:

\begin{itemize}
  \item synchronization within one Vulkan queue and command stream;
  \item explicit host and cross-queue synchronization outside the pass model;
  \item explicit initialization and presentation transitions; and
  \item native validation during correctness testing, but no safe-API
  guarantee in production.
\end{itemize}

The Metal backend is a different animal and should not be described as
tracking-free. Metal tracks hazards for resources created in the default
\texttt{MTLHazardTrackingModeTracked} mode~\cite{metal-hazard-tracking}, and
\blade{} relies on that: \texttt{manual\_barriers} has no effect there, and the
backend issues no inter-pass synchronization of its own. On Metal the tracking
has moved into the framework, not disappeared.

\section{Field Report: Barrier-Induced Serialization on AMD}
\label{sec:field}

The controlled experiments in this paper were prompted by a user
report~\cite{blade-issue-343}. A renderer built on \blade{} drew a planar
reflection and a main view into separate colour and depth targets, resolved
each from MSAA, and then drew reflective and refractive geometry that depended
on both. Conceptually the reflection and main passes are independent. In
practice \blade{} emitted this sequence:

\begin{verbatim}
full_barrier(); raytrace()
full_barrier(); reflection_depth_prepass()
pipeline_barrier(); render_reflection()
full_barrier(); resolve_reflection_msaa()
full_barrier(); main_depth_prepass()
pipeline_barrier(); render_main()
full_barrier(); resolve_main_msaa()
full_barrier(); render_reflective_refractive()
\end{verbatim}

\begin{figure*}[t]
\centering
\begin{tikzpicture}[
  x=1cm, y=1cm,
  pass/.style={draw=black!55, fill=black!7, rounded corners=1pt,
    minimum height=6.5mm, text width=15mm, align=center,
    font=\scriptsize, inner sep=1.5pt, anchor=south west},
  needed/.style={black!75, line width=0.9pt},
  redundant/.style={black!55, line width=0.9pt,
    dash pattern=on 1.6pt off 1.4pt},
  rowlabel/.style={font=\small, anchor=south west},
  note/.style={font=\scriptsize, anchor=west, black!70},
]

\node[rowlabel] at (0.1,2.75) {\textsc{automatic}};
\foreach \i/\name in {0/{ray trace}, 1/{refl.\ depth}, 2/{refl.\ draw},
                     3/{refl.\ resolve}, 4/{main depth}, 5/{main draw},
                     6/{main resolve}, 7/{refract}} {
  \node[pass] at ({0.2 + \i*1.95}, 1.85) {\name};
}
\foreach \i in {0,2,3,5,6,7} {
  \draw[needed] ({0.1 + \i*1.95}, 1.75) -- ({0.1 + \i*1.95}, 2.6);
}
\foreach \i in {1,4} {
  \draw[redundant] ({0.1 + \i*1.95}, 1.75) -- ({0.1 + \i*1.95}, 2.6);
}

\node[rowlabel] at (0.1,1.05) {\texttt{manual\_barriers}};
\node[pass] at (0.2, 0.15) {ray trace};
\foreach \i/\name in {0/{refl.\ depth}, 1/{refl.\ draw}, 2/{refl.\ resolve}} {
  \node[pass] at ({0.2 + \i*1.95}, -0.6) {\name};
}
\foreach \i/\name in {0/{main depth}, 1/{main draw}, 2/{main resolve}} {
  \node[pass] at ({0.2 + \i*1.95}, -1.35) {\name};
}
\foreach \i in {1,2} {
  \draw[needed] ({0.1 + \i*1.95}, -0.7) -- ({0.1 + \i*1.95}, -0.05);
  \draw[needed] ({0.1 + \i*1.95}, -1.45) -- ({0.1 + \i*1.95}, -0.8);
}
\draw[needed] (0.1, -1.45) -- (0.1, 0.9);
\draw[needed] (6.0, -1.45) -- (6.0, 0.9);
\node[pass] at (6.1, -0.6) {refract};

\draw[needed] (8.7,0.15) -- (8.7,0.6);
\node[note] at (8.95,0.375) {barrier the dependency graph needs};
\draw[redundant] (8.7,-0.7) -- (8.7,-0.25);
\node[note] at (8.95,-0.475) {barrier nothing needs: two of the eight};

\draw[->, black!45] (0.1, -1.95) -- (15.9, -1.95)
  node[midway, below, font=\scriptsize, black!70]
  {time, schematic --- every pass is drawn the same width because none was measured};

\end{tikzpicture}
\caption{The field report's frame, as \blade{} emitted it and as it was
restructured. The figure shows what the barriers \emph{permit}, not what the
hardware achieved. Two of the eight barriers separate work with no resource in
common --- ray tracing from the reflection view, and the reflection view from
the main one --- and on RADV each requests both compute and pixel partial
flushes without regard to whether the passes share a resource
(Section~\ref{sec:radv-barriers}). Removing those two is what let the
reporter's profiler show ray tracing and reflection running together. The
remaining six are real edges in the graph and stay, which is why this is a
change to where \texttt{barrier()} is called and not to any resource
declaration.}
\label{fig:field-report}
\end{figure*}
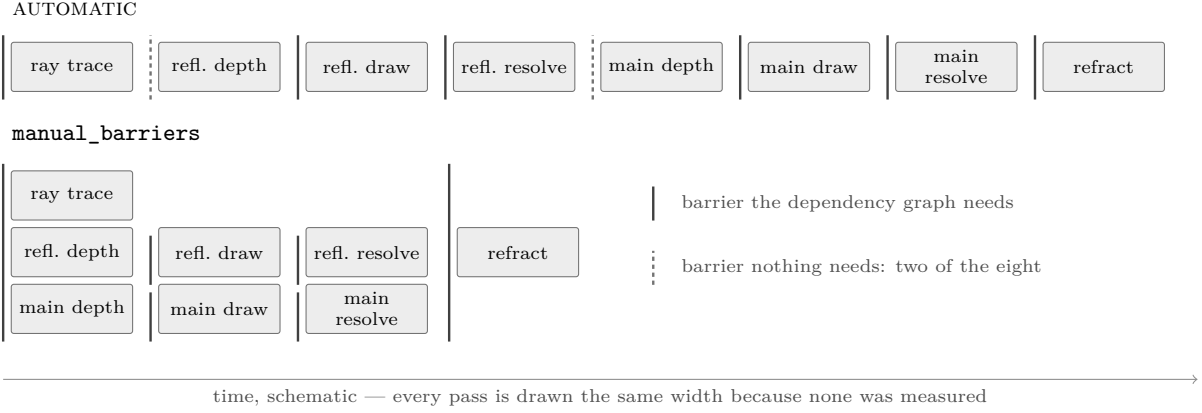

Figure~\ref{fig:field-report} draws that sequence against the dependency
graph it came from. The reporter observed in a GPU profiler that the barriers
forced the AMD scheduler to run the reflection work to completion before
starting the main view, and asked whether the hardware could overlap them at all if the barriers
were removed. Section~\ref{sec:radv-barriers} answers the mechanism question:
on RADV each \texttt{full\_barrier} sets both \texttt{CS\_PARTIAL\_FLUSH} and
\texttt{PS\_PARTIAL\_FLUSH}. The execution dependency therefore does not
depend on the driver noticing that the two passes are unrelated, although the
latency of either flush still depends on matching work being in flight. The
driver cannot discard the dependency merely because it sees no shared
resource: the application requested \texttt{ALL\_COMMANDS} ordering.

The reporter prototyped a flag to disable automatic barriers, removed the
barriers between the ray-tracing, reflection, and main passes, kept one before
the dependent refraction pass, and reported that the ray-tracing and reflection
passes then executed concurrently. Two API shapes were debated: a scoped
``group'' object whose passes share one barrier, and a mode flag on the
encoder. The group form was rejected as too much API surface for the
benefit, and per-call enable/disable was rejected because the scope of the
switch is ambiguous at the call site --- a concern the reporter later confirmed
by forgetting to re-enable it. The resolution~\cite{blade-pr-355} is a flag
fixed at encoder creation, which makes the mode a property of a whole encoder
and lets an application put its serial work in one encoder and its independent
work in another.

This paper's \texttt{B-hazard} configuration is exactly that mechanism applied
to a controlled workload, so the study can ask how far the field report
generalizes. Section~\ref{sec:placement} shows that it does not generalize
unconditionally, and Section~\ref{sec:discussion} proposes the boundary.

\section{Experimental Design}

\subsection{Configurations}

Table~\ref{tab:configs} separates the controlled \blade{} configurations from
the realistic tracked baseline. \texttt{B-explicit-all} is a same-request
insertion control: the two source paths issue the same timed global barrier
calls by construction. The retained captures do not include this policy, so
we do not present record equality as a capture-backed observation.
\texttt{W-wgpu} uses the same shaders, resources, pass graph, and output
checks, but remains an end-to-end comparison because the APIs and runtime
guarantees differ.

\begin{table}[t]
\centering
\small
\begin{tabular}{@{}l@{\ \ }l@{\ \ }l@{\ \ }l@{}}
\toprule
ID & Tracking & Placement & Scope (src/dst) \\
\midrule
B-auto & none & every pass & broad/broad \\
B-hazard & none & app-declared & broad/broad \\
B-explicit-all & none & every pass & broad/broad \\
\addlinespace
B-auto-scoped & none & every pass & kind/kind \\
B-hazard-scoped & none & app-declared & kind/broad \\
B-exp-all-scoped & none & every pass & kind/broad \\
\addlinespace
W-wgpu & tracked & derived & per resource \\
\bottomrule
\end{tabular}
\caption{Evaluation configurations. Every \texttt{B-*} row uses a global
memory barrier that names no resource and keeps images in \texttt{GENERAL};
automatic placement can derive both sides of a pass-kind scope, while an
explicit barrier can derive only its source because the consumer is not yet
known.}
\label{tab:configs}
\end{table}

\wgpu{} is reported as an end-to-end tracked baseline because its validation,
lifetime, binding, and command models differ from \blade. Direct
\texttt{wgpu-hal} measurements are optional diagnostics for separating frontend
work; they are not required for the main comparison and are not presented as
another tracking implementation.

\subsection{Workloads}
\label{sec:workloads}

The headless \texttt{sync-bench} artifact provides six workloads that vary
dependency structure across compute, render, and alternating pass streams.

\begin{itemize}
  \item \texttt{compute-independent}: one shared read-only input buffer, a
  distinct output buffer per pass. No inter-pass hazard.
  \item \texttt{compute-chain}: two ping-pong storage buffers. Every pass
  depends on its predecessor.
  \item \texttt{graphics-independent}: a distinct \texttt{RGBA8} colour target
  per pass, cleared and stored. No inter-pass hazard.
  \item \texttt{graphics-chain}: one colour target loaded and stored with
  additive blending. Every pass depends on its predecessor.
  \item \texttt{mixed-independent} and \texttt{mixed-chain}: compute and render
  passes alternating, with the same resource patterns as the corresponding
  single-kind workloads applied within each family. Every pass boundary
  therefore joins two different pass kinds, which is where the scope axis has
  the most to give. In \texttt{mixed-chain} each pass depends on the pass of
  the same kind two positions earlier. From the third pass onward each incoming
  pass therefore has a real dependency; the collector's hazard policy places a
  barrier before every pass except the first, so it differs from automatic
  placement by only the initial barrier, as the single-kind chains do. The
  barrier before the second pass is conservative---those first compute and
  render passes are independent---so ``hazard'' is a workload-level label for
  this mixed case rather than an edge-minimal schedule.
\end{itemize}

The two mixed families exist only in \blade's benchmark; the matched \wgpu{}
program has no equivalent, and they are used for the within-\blade{}
comparison rather than the end-to-end one.

Unless stated otherwise, each command buffer contains 16 passes; compute
buffers hold $2^{20}$ \texttt{u32} elements, render targets are
$1024\times1024$, and each invocation performs 8 mixing rounds. Every
configuration is warmed for 1000 iterations---chosen after the pilot's ten
warm-ups left severe within-block drift---and measured for 30, ten times, in
an order randomized jointly across implementations. Residual drift is reported
in Section~\ref{sec:deviations}; the warm-up count is not assumed to prove a
fixed clock.

The matched \wgpu{} program uses the same shader computations, resource sizes,
pass graph, and iteration schedule. The interface declarations necessarily
differ: \blade{} supplies the 16-byte per-pass parameters as an inline uniform
block, while \wgpu{} uses its native \texttt{IMMEDIATES} feature (Vulkan push
constants), so that the \wgpu{} side does not gain an extra tracked uniform
buffer. A checker in the artifact verifies that the shader bodies are textually
identical after normalizing those interface declarations and binding
decorations. After the final iteration of each process, the collector aborts
if the two implementations produce different output hashes. All
\bladenum{runs/matrix} runs of the fixed matrix
and all \bladenum{runs/sweep} runs of the pass-count sweeps agree on the output hash
within each collection/workload/pass-count group. Across the
\bladenum{runs/hashgroups} such groups from all devices, every configuration
present in a group agrees, with \bladenum{runs/hashconflicts} exceptions.

Validation is layered so that a silent failure has somewhere to be caught.
Each Vulkan machine's round begins with a retained synchronization-validation
matrix, run with the Khronos checks forced on and its logs stored beside the
timing collection. After every timed process, a slice of every independent
output is hashed with standard FNV-1a and the collector aborts on any
disagreement between policies or implementations. The graphics chain reserves
its readback row for exact two-bit increments that cannot saturate across the
64-pass sweep --- a missing late pass changes the digest instead of hiding
under clamped bytes --- while the other 1023 rows carry the full-range
measured workload, and both implementations' shaders pass SPIR-V validation.

Identical shader bodies do not guarantee identical generated code. \blade{} hands
naga the default bounds-check policies, which are unchecked throughout, and
disables integer-division checks; \wgpu's \texttt{create\_shader\_module}
keeps its own defaults, which clamp indices, guard divisions, and apply
\texttt{force\_loop\_bounding} --- a simulated 64-bit counter loaded, compared,
decremented and stored on every iteration of a loop. The benchmark therefore
uses \wgpu's unsafe trusted-shader entry point to select the unchecked policy
that \blade{} uses. This removes a shader-code confound, but it also means
\texttt{W-wgpu} is a native tracked-\wgpu{} comparison with one WebGPU safety
feature class deliberately disabled, not a browser-WebGPU baseline. A separate
\texttt{--shader-checks} diagnostic exists, but its earlier numerical results
are not reported because that diagnostic collection is not present in the
archival data.

These six are the whole workload set, deliberately. They contrast independent
resource sets with chain dependency patterns across the pass kinds a
synchronization policy can distinguish, and that is what the study is about.
Application frames would add heterogeneity of a kind these do not have, and
Section~\ref{sec:threats} states what that costs the conclusions; they would
also add scene, content, and driver-heuristic variation that this design exists
to exclude. We report a bounded claim about a controlled workload rather than
an unbounded one about renderers.

They are also deliberately barrier-dense stress tests, and the two things
they stress scale differently in a real application. A pass here performs
only microseconds of work, so every boundary cost lands on the smallest
denominator the design allows --- sixteen boundaries against roughly a
millisecond of total work. Fixed per-boundary costs --- host-side barrier
recording, the fraction-of-a-microsecond device-side scope constant ---
amortize in a production frame whose passes run hundreds of microseconds to
milliseconds each: those results should be read as microseconds per boundary
and divided by real pass durations, which shrinks them by one to two orders
of magnitude. Overlap effects are structural instead: what a redundant
barrier forfeits is proportional to how much genuinely independent work sits
on both sides of it, which sixteen equal independent passes maximize and few
real frames approach. On both axes a real application suffers less than
these numbers; the field report is what the exception looks like when it
occurs.

\subsection{Platforms}

Table~\ref{tab:platforms} lists the machines. The matrix covers the minimum
viable study of experiments.md --- one recent AMD and one recent NVIDIA
device-driver pair on Linux --- and adds an Intel Vulkan implementation and an
Apple/Metal case study. Results are reported per device-driver pair and never
pooled by vendor.

One class is absent by construction: mobile GPUs. The matrix contains no
Android-class tiler --- no Adreno, Mali, or PowerVR --- and while the AMD
iGPUs and the Apple M3 are integrated parts, only the M3 is a tile-based
deferred renderer, and it is measured through Metal's tracker rather than
through explicit barriers. On a Vulkan tiler, a barrier that forces a
render-pass break costs a tile flush and reload --- a mechanism none of the
measured architectures has --- so nothing in this paper should be
extrapolated to that class; it is the most consequential gap in the platform
set.

\begin{table*}[tp]
\centering
\small
\begin{tabular}{@{}llllllll@{}}
\toprule
Machine & Device & Backend & Driver & API & Unified & Platform & Revision \\
\midrule
\texttt{S1} & RTX 5070 & vulkan & NVIDIA 595.71.05 & 1.4.329 & yes & Linux 7.0.0-28 & \texttt{87ed067} \\
\texttt{S2} & RX 7900 XT & vulkan & radv 26.0.3-1ubuntu1 & 1.4.335 & yes & Linux 7.0.0-15 & \texttt{87ed067} \\
\texttt{S2} & Raphael iGPU & vulkan & radv 26.0.3-1ubuntu1 & 1.4.335 & yes & Linux 7.0.0-15 & \texttt{87ed067} \\
\texttt{S3} & Radeon 780M & vulkan & radv 25.2.8-0ubuntu0.24.04.1 & 1.4.318 & yes & Linux 6.14.0-37 & \texttt{87ed067} \\
\texttt{S4} & Intel Xe (RPL-U) & vulkan & anv 25.2.8-0ubuntu0.25.10.2 & 1.4.318 & no & Linux 6.17.0-22 & \texttt{87ed067} \\
\texttt{S5} & Apple M3 & metal & Metal --- & --- & --- & macOS 15.7.3 & \texttt{87ed067} \\
\bottomrule
\end{tabular}
\\[2pt]\footnotesize The Vulkan rows were all collected on one day from one benchmark build: where their revisions differ, they differ only by commits under \texttt{paper/}, which the benchmark does not compile, and every row used the same \wgpu{} revision. The Metal harness has only its backend's automatic policy; the scope variants in this study alter Vulkan barriers and therefore do not apply to that row.
\caption{Machines in the fixed 16-pass matrix. Every row is one collection under \texttt{paper/data/raw/}; results are never pooled across rows. ``Unified'' reports whether the driver advertises \texttt{VK\_KHR\_unified\_image\_layouts}, which Blade enables when present.}
\label{tab:platforms}
\end{table*}

\subsection{Metrics and statistics}

We measure command-buffer start, pass recording, queue submission, and host
wait separately, and report \emph{host cost} as recording plus submission,
because \wgpu{} defers command translation and tracker resolution to
\texttt{finish()} and \texttt{submit()} while \blade{} performs it during
recording. Charging only recording would misattribute the difference. The
separately reported \texttt{start\_ns} field is not included: it resets and
reuses a \blade{} backend encoder but creates a fresh \wgpu{} frontend encoder,
so it is a different lifecycle comparison rather than part of the
record-and-submit estimand used here.

Device time is the GPU span of one command buffer. On Vulkan, \blade{} writes a
\texttt{TOP\_OF\_PIPE} timestamp before each pass and one after the final
barrier; consecutive differences telescope, so their sum is exactly the span
from the first pass to the end of the command buffer, and it remains
well-defined when passes overlap. The \wgpu{} program uses one
beginning-of-first-pass and one end-of-last-pass timestamp. Thus \blade{} emits
17 timestamp writes for a 16-pass run and includes its final barrier, whereas
\wgpu{} emits two writes and ends its window with the last pass. Both
differences give \blade{} more instrumented work, but they may also perturb
scheduling; the instrumentation and windows are not identical. Only
within-\blade{} device comparisons are single-factor measurements.

The experimental unit for one configuration is a process launch, not one of
the 30 samples inside it. One outer matrix repetition launches every
configuration once in randomized order; for each of the ten outer
repetitions we pair the contender launch with the \texttt{B-auto} launch from
the same randomized block and compute their percent difference. The reported
point is the median of those ten block effects; quoted absolute timings are
likewise the median of the ten process medians. The 95\%
hierarchical-bootstrap percentile interval resamples outer blocks first and
observations within each selected process second, using a deterministic
seed~\cite{davison-hinkley-bootstrap,loy-korobova-bootstrap}. Quartiles
describe the pooled observations, but no interval treats all 300 as
independent. With ten process launches per configuration, and possible serial
correlation among observations inside a launch, these intervals describe
uncertainty across the observed sessions rather than a sampled population ---
where launches disagree with one another, the interval says so by widening.
We do not report multiplicity-adjusted hypothesis tests; the
device-by-workload cells and their control floors are exploratory effect
estimates. Validation layers are enabled for correctness runs and disabled for
performance runs, as recommended by vendor guidance~\cite{nvidia-vulkan-tips}.

We did not preregister a practical-equivalence region, and in the end we do not
use one. An earlier version of this analysis derived a single $\pm3\%$ band
from the worst \texttt{B-explicit-all} control deviation; with all
\bladenum{floor/count} cells collected, control floors vary by cell and reach
$\bladenum{floor/max}\%$, so no single band describes them. Every directional
result is therefore compared against the control floor of the cell and pass
count it comes from. Figures~\ref{fig:placement} and~\ref{fig:scope} draw the
fixed-matrix floors beside the effects; Section~\ref{sec:placement} states the
corresponding sweep floors. A band wide enough to cover the worst cell would
discard the best ones, and a band fitted to the best would license the worst.
This post-hoc floor is a conservative stability diagnostic, not a confidence
bound or a preregistered equivalence test. Curve shapes and post-hoc
state-conditioned summaries are explicitly labelled descriptive and are not
counted as directional results.

\subsection{Deviations from the protocol}
\label{sec:deviations}

These deviations affect how the results may be read, and are stated here
rather than in a footnote.

\textbf{Metal timestamps.} \blade's Metal backend reports per-pass durations
from counter sample buffers rather than a span, so summing them double-counts
overlapping passes; the matched \wgpu{} program returned zero for three of the
four workloads. Metal device time is therefore not reported at all, and
Section~\ref{sec:metal} uses host wall time. The machine also had macOS Low
Power Mode enabled, which is recorded in the collection but not controlled.

\textbf{Clocks that move during a measurement.} Figures~\ref{fig:placement} and~\ref{fig:scope} carry a
control floor for every cell: the disagreement between two configurations that
issue the same global barrier calls by construction. Of the \bladenum{floor/count} cells,
\bladenum{floor/above2} have floors above $2\%$; the maximum is
$\bladenum{floor/max}\%$, and the largest at or below two is
$\bladenum{floor/below2max}\%$. A directional effect is treated as a result
only when its entire interval lies beyond the same-side floor; otherwise it is
not credited.

The pilot collections that preceded these showed severe monotone drift in
their time-ordered samples. Those pilot rows are not part of the retained
artifact, so they motivate the protocol but do not support a result. Each
block --- one workload, one policy, one repetition --- is a separate process
launch, so the device idles between blocks and ramps its clock again inside
each one. Ten warm-ups were visibly insufficient on the pilot RX~7900~XT even
with \texttt{power\_dpm\_force\_performance\_level} set to \texttt{high};
increasing that count is a protocol parameter, not a benchmark change.

The retained collections use 1000 warm-ups, and comparing the median of each
block's first third with its last third checks that it worked: at the median
block, drift is $\bladenum{drift/rtx5070/median}\%$ on the RTX~5070,
$\bladenum{drift/rx7900xt/median}\%$ on the RX~7900~XT,
$\bladenum{drift/raphael/median}\%$ on the Raphael iGPU,
$\bladenum{drift/radeon780m/median}\%$ on the Radeon~780M, and
$\bladenum{drift/intelxe/median}\%$ on the Intel part. The worst single blocks
still move --- $\bladenum{drift/rtx5070/worst}\%$ on the RTX~5070 and
$\bladenum{drift/intelxe/worst}\%$ on the Intel part --- which is the residue
that the post-hoc control diagnostic is intended to flag.

\textbf{The process repetitions expose what pooling hides.} At three blocks
the Intel part's controls swung across signs and disqualified its cells; at
ten its floors settle to
$\bfloor{intelxe}{compute-independent}$--$\bfloor{intelxe}{mixed-chain}$ of
run-to-run variation and its compute-placement cells resolve
(Section~\ref{sec:placement}). The instability that remains in this round is
a different machine and a different kind.

\textbf{One host's render cells vary launch to launch.} On the Radeon~780M's
\texttt{graphics-independent} workload, paired block effects disperse far
more between process launches than samples do within one launch, which is
visible directly in the block marks of Figure~\ref{fig:overlap-depth}. The
host ran unrelated services during its multi-hour sweep, and that kind of
interference is the mundane explanation we can neither exclude nor need to:
the identical-request controls price the dispersion per pass count, the
16-pass control floor reaches
$\bfloor{radeon780m}{graphics-independent}$, and no conclusion is drawn from
a cell whose interval fails its own floor. The directional support comes from
the sweep counts whose full intervals clear their own controls, reported in
Section~\ref{sec:placement}.

\hostcaveat

\textbf{Collections are one day, one benchmark build.} Every machine,
including the Apple M3, was measured on 2026-07-28 from the same corrected
revisions of both repositories; where Table~\ref{tab:platforms} lists a
different \blade{} revision for the study host's sweeps, it differs only by
commits under \texttt{paper/}, which the benchmark does not compile. The
Metal collection has no scoped configurations because those variants alter
Vulkan barrier masks, not Metal's framework-managed resource hazards.

\section{Results}

\subsection{Control validity}

The \texttt{B-exp-all} column of Table~\ref{tab:gpu-matrix} is the
instrumentation control: manual barriers before every pass reproduce the same
global barrier calls by construction. A device-time difference is therefore
measurement instability under an unchanged timed request. Its interval defines
the paper's conservative, post-hoc stability threshold for that cell, and it is
drawn as a pair of ticks in
Figures~\ref{fig:placement} and~\ref{fig:scope}.

Two choices about that column matter. It is read per cell, because noise varies
by workload even on one device. It is also the far end of the control's
hierarchical interval rather than its point estimate: the point can be near
zero while different process repetitions occupy different regimes. Of the
\bladenum{floor/count} cells, \bladenum{floor/above2} have floors above $2\%$.
That does not disqualify a whole device; it means the entire interval of a
claimed effect must lie beyond the same-side floor of its own cell.
The overlap-depth sweep applies the same construction separately at each pass
count.

\subsection{Barrier placement and dependency structure}
\label{sec:placement}

\begin{table*}[tp]
\centering
\small
\begin{tabular}{@{}llrr@{\,}lr@{\,}lr@{\,}l@{}}
\toprule
Device & Workload & B-auto & \multicolumn{2}{c}{B-hazard \%} & \multicolumn{2}{c}{B-exp-all \%} & \multicolumn{2}{c}{W-wgpu \%} \\
\midrule
RTX 5070 & c-ind & 195.4 & -29.3 & [-29.4,\,-29.2] & +0.0 & [-0.1,\,+0.1] & +0.9 & [+0.2,\,+1.2] \\
 & c-chain & 195.4 & -0.1 & [-0.1,\,+0.0] & -0.1 & [-0.2,\,+0.0] & -0.4 & [-0.5,\,-0.3] \\
 & g-ind & 152.1 & -32.4 & [-32.9,\,-31.8] & -0.1 & [-0.2,\,+0.1] & -34.5 & [-35.3,\,-33.6] \\
 & g-chain & 150.4 & -0.1 & [-0.7,\,-0.1] & -0.1 & [-0.2,\,+0.1] & -17.5 & [-17.6,\,-17.4] \\
 & m-ind & 172.1 & -17.5 & [-17.6,\,-17.4] & +0.0 & [-0.2,\,+0.2] & --- &  \\
 & m-chain & 171.8 & -0.0 & [-0.1,\,+0.1] & +0.0 & [-0.1,\,+0.1] & --- &  \\
\addlinespace
RX 7900 XT & c-ind & 120.6 & -32.3 & [-32.6,\,-32.1] & +0.3 & [-0.2,\,+0.5] & -29.9 & [-30.2,\,-29.8] \\
 & c-chain & 126.2 & -0.5 & [-2.3,\,+0.6] & -1.1 & [-2.3,\,+0.8] & +0.0 & [-0.1,\,+1.8] \\
 & g-ind & 178.0 & -7.3 & [-7.5,\,-7.2] & -0.1 & [-0.2,\,-0.0] & +12.0 & [+11.9,\,+12.3] \\
 & g-chain & 150.8 & +0.1 & [-0.0,\,+0.2] & -0.1 & [-0.1,\,+0.0] & +18.5 & [+18.3,\,+18.6] \\
 & m-ind & 155.9 & -17.7 & [-17.8,\,-17.7] & -0.0 & [-0.1,\,+0.1] & --- &  \\
 & m-chain & 141.8 & -0.1 & [-0.1,\,+0.1] & -0.1 & [-0.2,\,+0.1] & --- &  \\
\addlinespace
Raphael iGPU & c-ind & 3854.7 & -0.9 & [-0.9,\,-0.8] & -0.0 & [-0.0,\,+0.1] & +6.1 & [+6.1,\,+6.2] \\
 & c-chain & 3833.6 & +0.0 & [-0.0,\,+0.1] & +0.0 & [-0.0,\,+0.0] & +7.4 & [+7.4,\,+7.5] \\
 & g-ind & 5724.6 & -1.1 & [-1.1,\,-1.1] & -0.0 & [-0.1,\,+0.0] & -13.4 & [-13.4,\,-13.4] \\
 & g-chain & 4829.8 & -0.0 & [-0.0,\,+0.0] & -0.0 & [-0.0,\,+0.0] & +2.4 & [+2.4,\,+2.4] \\
 & m-ind & 4803.9 & -1.8 & [-1.9,\,-1.8] & -0.0 & [-0.0,\,+0.0] & --- &  \\
 & m-chain & 4375.6 & +0.0 & [-0.0,\,+0.0] & -0.0 & [-0.0,\,+0.0] & --- &  \\
\addlinespace
Radeon 780M & c-ind & 3597.8 & +3.6 & [+3.0,\,+4.0] & -0.1 & [-0.3,\,+0.7] & +2.4 & [+2.1,\,+2.7] \\
 & c-chain & 3600.3 & -0.0 & [-0.4,\,+1.5] & -0.3 & [-0.6,\,+0.2] & +5.2 & [+4.9,\,+5.4] \\
 & g-ind & 2322.8 & +42.0 & [+22.5,\,+184.6] & +51.9 & [-14.5,\,+106.7] & -12.2 & [-28.4,\,-10.7] \\
 & g-chain & 1465.2 & -0.1 & [-1.4,\,+0.5] & +0.1 & [-0.8,\,+0.6] & -0.2 & [-2.2,\,+7.6] \\
 & m-ind & 2944.0 & +25.7 & [+24.8,\,+57.4] & -0.2 & [-23.9,\,+0.1] & --- &  \\
 & m-chain & 2890.4 & +0.0 & [-0.1,\,+0.4] & +0.1 & [-0.4,\,+51.6] & --- &  \\
\addlinespace
Intel Xe (RPL-U) & c-ind & 4017.0 & -6.5 & [-7.5,\,-5.3] & +0.1 & [-0.6,\,+2.4] & +11.7 & [+8.1,\,+12.3] \\
 & c-chain & 3594.0 & +0.0 & [-0.0,\,+0.8] & -0.0 & [-4.9,\,+0.0] & +12.9 & [+10.2,\,+13.7] \\
 & g-ind & 4256.7 & -0.4 & [-0.5,\,+0.1] & +0.1 & [-0.0,\,+3.5] & +7.9 & [+3.9,\,+8.7] \\
 & g-chain & 3324.9 & -0.0 & [-3.4,\,+0.0] & -0.0 & [-4.4,\,+0.0] & +4.9 & [-0.3,\,+4.9] \\
 & m-ind & 4098.5 & -8.2 & [-11.3,\,-7.4] & +0.1 & [-3.7,\,+4.0] & --- &  \\
 & m-chain & 3829.7 & +2.1 & [-2.5,\,+5.1] & +2.0 & [-4.6,\,+5.0] & --- &  \\
\bottomrule
\end{tabular}
\\[2pt]\footnotesize Every cell compares the two implementations on the same physical device; the collector aborts otherwise (Section~\ref{sec:deviations}).
\caption{GPU span of one 16-pass command buffer on Vulkan: \texttt{B-auto} median of process medians in microseconds, then the median paired process-level percent differences with 95\% hierarchical-bootstrap intervals. \texttt{B-explicit-all} is the instrumentation control: it issues the same timed global barrier calls as \texttt{B-auto} by construction. The far endpoint of its interval is used as a conservative, post-hoc stability threshold (Section~\ref{sec:deviations}). \texttt{W-wgpu} is an end-to-end comparison with two timestamps ending at the final pass, whereas \blade{} writes one per pass and includes its final barrier in the span.}
\label{tab:gpu-matrix}
\end{table*}

\begin{figure*}[t]
\centering
\begin{tikzpicture}
\begin{groupplot}[
  group style={group size=2 by 1, horizontal sep=1.4cm},
  width=8.0cm, height=5.4cm,
  symbolic x coords={1,2,4,8,16,32,64},
  xtick=data,
  xlabel={passes per command buffer},
  ylabel={\% of \texttt{B-auto}},
  xlabel style={font=\scriptsize},
  ylabel style={font=\scriptsize},
  tick label style={font=\scriptsize},
  title style={font=\scriptsize},
  legend style={font=\scriptsize, draw=none, at={(0.03,0.97)},
    anchor=north west},
  grid=major, grid style={black!10},
]
\nextgroupplot[title={RTX 5070}]
\addplot[only marks, mark=*, mark size=0.8pt, black!35, forget plot] coordinates {(1,-0.3) (1,0.0) (1,0.0) (1,-0.2) (1,0.0) (1,-0.6) (1,0.0) (1,0.0) (1,-1.2) (1,0.0) (2,-16.4) (2,-16.4) (2,-16.2) (2,-16.4) (2,-16.2) (2,-16.4) (2,-16.5) (2,-16.5) (2,-16.4) (2,-16.5) (4,-25.8) (4,-25.7) (4,-25.8) (4,-25.7) (4,-25.7) (4,-25.8) (4,-25.8) (4,-25.4) (4,-25.7) (4,-25.7) (8,-31.1) (8,-30.9) (8,-30.6) (8,-30.7) (8,-30.8) (8,-30.8) (8,-31.0) (8,-30.9) (8,-31.0) (8,-30.8) (16,-31.6) (16,-31.6) (16,-31.3) (16,-32.2) (16,-31.1) (16,-30.5) (16,-32.6) (16,-32.3) (16,-31.2) (16,-30.6) (32,-30.7) (32,-31.1) (32,-30.5) (32,-31.5) (32,-29.5) (32,-30.5) (32,-31.4) (32,-31.2) (32,-29.9) (32,-30.2) (64,-30.5) (64,-31.0) (64,-30.4) (64,-30.6) (64,-29.5) (64,-29.4) (64,-30.2) (64,-30.4) (64,-30.7) (64,-30.3)};
\addplot[black, mark=*, mark size=1.4pt] coordinates {(1,0.0) (2,-16.4) (4,-25.7) (8,-30.9) (16,-31.5) (32,-30.6) (64,-30.4)};
\addlegendentry{\texttt{B-hazard}}
\addplot[only marks, mark=*, mark size=0.8pt, black!20, forget plot] coordinates {(1,-26.9) (1,-26.7) (1,-26.7) (1,-26.7) (1,-26.7) (1,-27.1) (1,-26.7) (1,-26.7) (1,-27.6) (1,-26.7) (2,-30.0) (2,-30.0) (2,-30.0) (2,-30.2) (2,-30.1) (2,-30.0) (2,-30.3) (2,-30.1) (2,-30.1) (2,-30.1) (4,-32.8) (4,-32.7) (4,-32.8) (4,-32.7) (4,-32.7) (4,-32.8) (4,-32.8) (4,-32.8) (4,-32.7) (4,-32.8) (8,-34.3) (8,-34.5) (8,-34.5) (8,-34.5) (8,-34.3) (8,-34.4) (8,-34.4) (8,-34.4) (8,-34.7) (8,-34.3) (16,-35.2) (16,-33.9) (16,-35.2) (16,-35.3) (16,-32.5) (16,-33.3) (16,-35.1) (16,-35.3) (16,-34.3) (16,-32.5) (32,-32.5) (32,-32.5) (32,-31.9) (32,-32.0) (32,-30.6) (32,-30.9) (32,-32.4) (32,-31.6) (32,-31.8) (32,-31.4) (64,-30.6) (64,-31.8) (64,-31.0) (64,-31.8) (64,-30.0) (64,-29.6) (64,-31.4) (64,-31.5) (64,-31.5) (64,-31.5)};
\addplot[black!55, densely dashed, mark=o, mark size=1.4pt] coordinates {(1,-26.7) (2,-30.1) (4,-32.8) (8,-34.4) (16,-34.7) (32,-31.8) (64,-31.4)};
\addlegendentry{\texttt{W-wgpu}}
\addplot[only marks, mark=*, mark size=0.8pt, black!15, forget plot] coordinates {(1,-0.6) (1,0.0) (1,0.0) (1,0.0) (1,0.0) (1,-0.6) (1,0.2) (1,0.0) (1,-1.2) (1,0.0) (2,0.0) (2,0.0) (2,0.1) (2,-0.2) (2,0.2) (2,0.1) (2,-0.2) (2,0.0) (2,-0.1) (2,-0.2) (4,0.0) (4,0.1) (4,0.0) (4,0.0) (4,-0.1) (4,-0.1) (4,0.0) (4,0.0) (4,0.1) (4,-0.1) (8,-0.2) (8,-0.0) (8,0.0) (8,-0.0) (8,-0.1) (8,0.0) (8,0.0) (8,-0.1) (8,-0.5) (8,0.1) (16,0.3) (16,-0.4) (16,-0.9) (16,-0.2) (16,0.0) (16,0.0) (16,-0.9) (16,-0.0) (16,-0.1) (16,1.2) (32,0.4) (32,-0.0) (32,-0.3) (32,-0.5) (32,0.0) (32,2.1) (32,0.1) (32,0.1) (32,1.1) (32,0.9) (64,0.4) (64,-0.0) (64,0.0) (64,0.1) (64,-0.0) (64,0.0) (64,-0.1) (64,0.0) (64,-0.1) (64,-0.0)};
\addplot[black!45, densely dotted, mark=triangle*, mark size=1.4pt] coordinates {(1,0.0) (2,0.0) (4,0.0) (8,-0.0) (16,-0.1) (32,0.1) (64,0.0)};
\addlegendentry{\texttt{B-exp-all}}
\nextgroupplot[title={Radeon 780M}]
\addplot[only marks, mark=*, mark size=0.8pt, black!35, forget plot] coordinates {(1,-0.5) (1,-0.3) (1,0.7) (1,-0.1) (1,-3.2) (1,0.5) (1,-0.3) (1,-2.6) (1,-2.6) (1,0.3) (2,22.4) (2,18.5) (2,19.8) (2,148.8) (2,143.6) (2,22.9) (2,149.2) (2,22.5) (2,18.4) (2,151.6) (4,63.6) (4,32.3) (4,32.1) (4,-36.1) (4,30.7) (4,30.3) (4,-36.2) (4,-36.0) (4,-36.1) (4,29.9) (8,175.3) (8,36.7) (8,36.7) (8,-32.9) (8,176.1) (8,39.4) (8,-33.8) (8,35.4) (8,36.1) (8,-33.6) (16,177.7) (16,40.2) (16,42.1) (16,-31.7) (16,41.9) (16,41.6) (16,190.1) (16,189.6) (16,189.7) (16,-31.7) (32,42.3) (32,79.2) (32,-17.7) (32,-16.8) (32,50.9) (32,40.4) (32,42.6) (32,42.4) (32,187.5) (32,41.6) (64,102.0) (64,52.1) (64,-5.7) (64,20.6) (64,180.5) (64,42.9) (64,20.3) (64,41.8) (64,43.0) (64,54.4)};
\addplot[black, mark=*, mark size=1.4pt] coordinates {(1,-0.3) (2,22.7) (4,30.1) (8,36.4) (16,42.0) (32,42.4) (64,42.9)};
\addlegendentry{\texttt{B-hazard}}
\addplot[only marks, mark=*, mark size=0.8pt, black!20, forget plot] coordinates {(1,-21.8) (1,-21.9) (1,-21.3) (1,-21.4) (1,-24.0) (1,-21.5) (1,-21.7) (1,-23.8) (1,-24.0) (1,-21.7) (2,-21.4) (2,-23.2) (2,-23.3) (2,-21.3) (2,-23.2) (2,-20.9) (2,-21.4) (2,-21.5) (2,-23.3) (2,-21.4) (4,-23.1) (4,-21.4) (4,-21.7) (4,-62.2) (4,60.9) (4,-22.3) (4,-62.1) (4,-61.9) (4,-62.2) (4,-23.1) (8,-23.3) (8,-23.3) (8,-23.3) (8,-62.2) (8,-12.1) (8,-21.1) (8,-62.3) (8,-23.0) (8,-23.2) (8,-61.9) (16,-15.9) (16,-11.6) (16,-10.5) (16,-57.8) (16,-10.7) (16,-10.7) (16,-12.7) (16,-10.7) (16,-21.9) (16,-56.6) (32,-11.0) (32,-10.6) (32,-53.6) (32,-47.9) (32,-11.3) (32,-11.9) (32,-11.1) (32,-11.0) (32,-11.3) (32,-11.3) (64,-11.2) (64,-11.6) (64,-41.3) (64,-42.1) (64,-11.3) (64,-11.2) (64,-24.8) (64,-11.3) (64,-11.5) (64,-11.7)};
\addplot[black!55, densely dashed, mark=o, mark size=1.4pt] coordinates {(1,-21.7) (2,-21.5) (4,-23.1) (8,-23.3) (16,-12.2) (32,-11.3) (64,-11.5)};
\addlegendentry{\texttt{W-wgpu}}
\addplot[only marks, mark=*, mark size=0.8pt, black!15, forget plot] coordinates {(1,-0.0) (1,-0.1) (1,0.4) (1,0.3) (1,-3.0) (1,0.1) (1,-0.4) (1,-2.7) (1,-2.9) (1,0.3) (2,-0.1) (2,-2.3) (2,-0.3) (2,2.3) (2,-2.0) (2,2.5) (2,2.3) (2,-0.2) (2,-2.2) (2,-0.2) (4,-0.0) (4,1.5) (4,0.1) (4,-51.5) (4,0.0) (4,0.0) (4,-51.0) (4,-50.9) (4,-51.4) (4,-0.1) (8,103.2) (8,-2.1) (8,-0.2) (8,-50.6) (8,-0.3) (8,-0.4) (8,-50.9) (8,-1.0) (8,-1.8) (8,-51.8) (16,-4.0) (16,-1.5) (16,107.3) (16,-51.7) (16,103.6) (16,107.4) (16,107.4) (16,106.6) (16,0.2) (16,-51.7) (32,0.9) (32,0.7) (32,-46.8) (32,-41.6) (32,-0.1) (32,-1.3) (32,-0.2) (32,105.9) (32,-0.9) (32,-0.4) (64,0.1) (64,-0.3) (64,1.0) (64,-34.8) (64,104.2) (64,0.1) (64,-15.1) (64,52.1) (64,0.2) (64,-0.4)};
\addplot[black!45, densely dotted, mark=triangle*, mark size=1.4pt] coordinates {(1,-0.0) (2,-0.2) (4,-0.1) (8,-1.4) (16,51.9) (32,-0.3) (64,0.1)};
\addlegendentry{\texttt{B-exp-all}}
\end{groupplot}
\end{tikzpicture}
\caption{GPU-span effect of removing the redundant barriers (\texttt{B-hazard}) and of the tracked implementation (\texttt{W-wgpu}) against \texttt{B-auto} on \texttt{graphics-independent}, as the number of overlapping passes grows. \texttt{B-exp-all} is the identical-request control. Each small mark is one paired process launch; the line joins the medians of those block effects. The scattered marks on the Radeon~780M panel are that host's launch-to-launch dispersion (Section~\ref{sec:deviations}); directional results use the count-specific control intervals stated in the text, while the joined medians describe the sweep's shape.}
\label{fig:overlap-depth}
\end{figure*}

Table~\ref{tab:gpu-matrix} answers RQ2 and RQ4 together. The pattern is not
``coarse barriers are slow''; it is conditional in a way that is stable across
devices for the dependent workloads and unstable for the independent ones.

\textbf{The chain comparison is an expected null.} For the chain workloads,
\texttt{B-hazard} places a barrier before every pass except the first, so the
emitted stream differs from \texttt{B-auto} only by the initial barrier.
Because both timing windows begin after that initial barrier, their timed
barrier calls are the same by construction. Across the five Vulkan devices,
\bladenum{chain/resolved} of the fifteen chain cells nevertheless resolve a
difference above their own stability threshold. This is a useful check on the
measurement, not evidence that a required barrier is free. The independent
workloads are the placement experiment: there \texttt{B-hazard} removes the
fifteen timed inter-pass barriers and effects reach
$\bmag{rtx5070}{graphics-independent}{placement}$.

\textbf{When the barrier is redundant, the cost depends on the device and on
the kind of work.} For independent \emph{compute} passes, removing the
redundant barriers is worth $\bmagci{rtx5070}{compute-independent}{placement}$
on the RTX~5070 and $\bmagci{rx7900xt}{compute-independent}{placement}$ on the
RX~7900~XT. The reductions are consistent with recovered overlap and/or avoided
pipeline drains and cache work; the timestamp span alone does not distinguish
those mechanisms. Both figures moved between the ramping pilot
and this warmed collection --- the AMD one grew, and its control floor fell
from a quarter of the effect to $\bfloor{rx7900xt}{compute-independent}$.

The Intel part, unreadable at three blocks, resolves at ten:
$\bpctci{intelxe}{compute-independent}{placement}$ on
\texttt{compute-independent} against a floor of
$\bfloor{intelxe}{compute-independent}$, with
$\bpctci{intelxe}{mixed-independent}{placement}$ on
\texttt{mixed-independent} against $\bfloor{intelxe}{mixed-independent}$
agreeing --- a third vendor for the direction of the compute result, at well
under the NVIDIA magnitude.

For independent \emph{render} passes the same change is worth
$\bmagci{rtx5070}{graphics-independent}{placement}$ on the RTX~5070 but only
$\bmagci{rx7900xt}{graphics-independent}{placement}$ on the RX~7900~XT ---
resolved against a floor of $\bfloor{rx7900xt}{graphics-independent}$ in the
observed sessions, and a quarter of what the same machine recovers on compute
--- and
$\bpctci{raphael}{graphics-independent}{placement}$ on the Raphael iGPU,
resolved and still nearly nothing. (The pilot could not separate the
RX~7900~XT's render cell from zero; the warmed collection yields a small
directional estimate.) The sharpest split in
the study is therefore not vendor against vendor but kind of work on the same
silicon: the RX~7900~XT frees four times more span between compute passes than
between render passes, while the RTX~5070 frees the same third either way. The
headline ``placement dominates'' is carried by both discrete parts for
compute, and at full size by NVIDIA alone for graphics.

\textbf{Removing barriers can make things worse, and the sweep shows where the
claim resolves.} On the Radeon~780M, \texttt{B-hazard} costs at
\bladenum{depth/radeon780m/graphics-independent/resolved-placement-counts}
pass counts under the same criterion used for the fixed matrix. The effect is
absent at one pass
($\bladenum{depth/radeon780m/graphics-independent/p1/placement/pct}\%$).
At two passes its full interval clears a
$\bladenum{depth/radeon780m/graphics-independent/p2/floor}\%$ control floor.
At 32 passes the effect is
$\bladenum{depth/radeon780m/graphics-independent/p32/placement/pct}\%$
[$\bladenum{depth/radeon780m/graphics-independent/p32/placement/lo}$,
$\bladenum{depth/radeon780m/graphics-independent/p32/placement/hi}$] against
a $\bladenum{depth/radeon780m/graphics-independent/p32/floor}\%$ floor, and at
64 it is
$\bladenum{depth/radeon780m/graphics-independent/p64/placement/pct}\%$
[$\bladenum{depth/radeon780m/graphics-independent/p64/placement/lo}$,
$\bladenum{depth/radeon780m/graphics-independent/p64/placement/hi}$] against
$\bladenum{depth/radeon780m/graphics-independent/p64/floor}\%$.

The medians at four, eight, and sixteen passes suggest growth and saturation
near sixteen, but their intervals do not clear their count-specific floors;
that curve shape is descriptive rather than a fourth result. In particular,
the 16-pass point
($\bladenum{depth/radeon780m/graphics-independent/p16/placement/pct}\%$)
is not promoted: that count's launch-to-launch dispersion
(Section~\ref{sec:deviations}) puts its floor past its interval.

That descriptive growth is what a shared-resource account predicts, but the
tracked comparison makes the simplest version inadequate. \wgpu{} overlaps the
same passes onto usage-matched targets (both request attachment and copy usage
only, so both are DCC-eligible by the predicates of
Section~\ref{sec:radv-barriers}), writes the same bytes, and has a faster point
estimate than serialized \texttt{B-auto} across the sweep:
$\bladenum{depth/radeon780m/graphics-independent/p2/wgpu/pct}\%$ at two
passes, $\bladenum{depth/radeon780m/graphics-independent/p64/wgpu/pct}\%$ at
sixty-four. This descriptive contrast argues against overlap depth or target
bytes alone as the explanation; it does not identify the resource involved.
What still differs between the implementations here is narrow: the layout the
targets sit in (\texttt{GENERAL} against
\texttt{COLOR\_ATTACHMENT\_OPTIMAL}) and the endpoint barrier form --- on a
driver that advertises \texttt{VK\_KHR\_unified\_image\_layouts} and
therefore promises that the layout half should not matter --- along with
parameter delivery, timestamp instrumentation, and backend-specific per-pass
work. Under hazard-only placement there is no interior barrier left to
blame. The controlled layout variant --- the same \blade{} stream with
attachment-optimal layouts --- is the experiment that would close this, and
the study does not have it. We report the shape and the eliminations rather
than the tidy explanation the numbers do not support.

\textbf{Long-running passes leave little measured headroom.} On the Raphael
iGPU the largest highlighted within-\blade{} change is
$\bmag{raphael}{mixed-independent}{placement}$, and every floor is at or below
$\bfloor{raphael}{graphics-independent}$, so the small effects are resolved
where their intervals and thresholds permit. This is
consistent with one pass saturating the small GPU and leaving no capacity for
a second, although utilization was not measured directly. What
does move on this device is the tracked implementation --- by up to
$\bmag{raphael}{graphics-independent}{wgpu}$, in both directions across
workloads --- which the saturation account alone would not explain, and is
taken up in Section~\ref{sec:endtoend}.

This suggests a boundary the field report does not reveal on its own. Overlap
can pay when adjacent passes have complementary bottlenecks and the device has
idle capacity --- ray tracing next to rasterization, in the reported case. It
need not pay when each pass already saturates the same resource. The Radeon
result also shows that removing a barrier can lose, although this experiment
does not establish memory pressure as the cause.

\subsection{Barrier scope at fixed placement}
\label{sec:scope-results}

\begin{figure*}[t]
\centering
\begin{tikzpicture}
\begin{groupplot}[
  group style={group size=3 by 2, horizontal sep=1.15cm,
    vertical sep=1.75cm, y descriptions at=edge left},
  width=5.2cm, height=4.3cm,
  xbar, /pgf/bar width=5pt,
  symbolic y coords={c-ind,c-chain,g-ind,g-chain,m-ind,m-chain},
  ytick=data, y dir=reverse,
  xlabel={\% of \texttt{B-auto}},
  xlabel style={font=\scriptsize},
  tick label style={font=\scriptsize},
  title style={font=\scriptsize},
  grid=major, grid style={black!10},
  scaled x ticks=false,
]
\nextgroupplot[title={RTX 5070}]
\addplot[xbar, fill=black!12, draw=black!55]
coordinates {(-29.3,c-ind) (-0.1,c-chain) (-32.4,g-ind) (-0.1,g-chain) (-17.5,m-ind) (-0.0,m-chain)};
\addplot[only marks, mark=|, mark size=3pt, black, forget plot]
coordinates {(-29.4,c-ind) (-29.2,c-ind) (-0.1,c-chain) (0.0,c-chain) (-32.9,g-ind) (-31.8,g-ind) (-0.7,g-chain) (-0.1,g-chain) (-17.6,m-ind) (-17.4,m-ind) (-0.1,m-chain) (0.1,m-chain)};
\addplot[only marks, mark=|, mark size=5.5pt, black!35, forget plot]
coordinates {(0.1,c-ind) (-0.1,c-ind) (0.2,c-chain) (-0.2,c-chain) (0.2,g-ind) (-0.2,g-ind) (0.2,g-chain) (-0.2,g-chain) (0.2,m-ind) (-0.2,m-ind) (0.1,m-chain) (-0.1,m-chain)};
\nextgroupplot[title={RX 7900 XT}]
\addplot[xbar, fill=black!12, draw=black!55]
coordinates {(-32.3,c-ind) (-0.5,c-chain) (-7.3,g-ind) (0.1,g-chain) (-17.7,m-ind) (-0.1,m-chain)};
\addplot[only marks, mark=|, mark size=3pt, black, forget plot]
coordinates {(-32.6,c-ind) (-32.1,c-ind) (-2.3,c-chain) (0.6,c-chain) (-7.5,g-ind) (-7.2,g-ind) (-0.0,g-chain) (0.2,g-chain) (-17.8,m-ind) (-17.7,m-ind) (-0.1,m-chain) (0.1,m-chain)};
\addplot[only marks, mark=|, mark size=5.5pt, black!35, forget plot]
coordinates {(0.5,c-ind) (-0.5,c-ind) (2.3,c-chain) (-2.3,c-chain) (0.2,g-ind) (-0.2,g-ind) (0.1,g-chain) (-0.1,g-chain) (0.1,m-ind) (-0.1,m-ind) (0.2,m-chain) (-0.2,m-chain)};
\nextgroupplot[title={Raphael iGPU}]
\addplot[xbar, fill=black!12, draw=black!55]
coordinates {(-0.9,c-ind) (0.0,c-chain) (-1.1,g-ind) (-0.0,g-chain) (-1.8,m-ind) (0.0,m-chain)};
\addplot[only marks, mark=|, mark size=3pt, black, forget plot]
coordinates {(-0.9,c-ind) (-0.8,c-ind) (-0.0,c-chain) (0.1,c-chain) (-1.1,g-ind) (-1.1,g-ind) (-0.0,g-chain) (0.0,g-chain) (-1.9,m-ind) (-1.8,m-ind) (-0.0,m-chain) (0.0,m-chain)};
\addplot[only marks, mark=|, mark size=5.5pt, black!35, forget plot]
coordinates {(0.1,c-ind) (-0.1,c-ind) (0.0,c-chain) (-0.0,c-chain) (0.1,g-ind) (-0.1,g-ind) (0.0,g-chain) (-0.0,g-chain) (0.0,m-ind) (-0.0,m-ind) (0.0,m-chain) (-0.0,m-chain)};
\nextgroupplot[title={Radeon 780M}]
\addplot[xbar, fill=black!12, draw=black!55]
coordinates {(3.6,c-ind) (-0.0,c-chain) (42.0,g-ind) (-0.1,g-chain) (25.7,m-ind) (0.0,m-chain)};
\addplot[only marks, mark=|, mark size=3pt, black, forget plot]
coordinates {(3.0,c-ind) (4.0,c-ind) (-0.4,c-chain) (1.5,c-chain) (22.5,g-ind) (184.6,g-ind) (-1.4,g-chain) (0.5,g-chain) (24.8,m-ind) (57.4,m-ind) (-0.1,m-chain) (0.4,m-chain)};
\addplot[only marks, mark=|, mark size=5.5pt, black!35, forget plot]
coordinates {(0.7,c-ind) (-0.7,c-ind) (0.6,c-chain) (-0.6,c-chain) (106.7,g-ind) (-106.7,g-ind) (0.8,g-chain) (-0.8,g-chain) (23.9,m-ind) (-23.9,m-ind) (51.6,m-chain) (-51.6,m-chain)};
\nextgroupplot[title={Intel Xe (RPL-U)}]
\addplot[xbar, fill=black!12, draw=black!55]
coordinates {(-6.5,c-ind) (0.0,c-chain) (-0.4,g-ind) (-0.0,g-chain) (-8.2,m-ind) (2.1,m-chain)};
\addplot[only marks, mark=|, mark size=3pt, black, forget plot]
coordinates {(-7.5,c-ind) (-5.3,c-ind) (-0.0,c-chain) (0.8,c-chain) (-0.5,g-ind) (0.1,g-ind) (-3.4,g-chain) (0.0,g-chain) (-11.3,m-ind) (-7.4,m-ind) (-2.5,m-chain) (5.1,m-chain)};
\addplot[only marks, mark=|, mark size=5.5pt, black!35, forget plot]
coordinates {(2.4,c-ind) (-2.4,c-ind) (4.9,c-chain) (-4.9,c-chain) (3.5,g-ind) (-3.5,g-ind) (4.4,g-chain) (-4.4,g-chain) (4.0,m-ind) (-4.0,m-ind) (5.0,m-chain) (-5.0,m-chain)};
\end{groupplot}
\end{tikzpicture}
\caption{Barrier placement: percent difference in GPU span between \texttt{B-hazard} and \texttt{B-auto} at 16 passes, by device and workload. The bar is the median paired process effect, the two black ticks its 95\% hierarchical-bootstrap interval, and the wider grey ticks plus and minus that cell's post-hoc stability threshold. A directional effect is read only when its whole interval lies beyond the corresponding grey tick. Note the per-panel horizontal scales.}
\label{fig:placement}
\end{figure*}
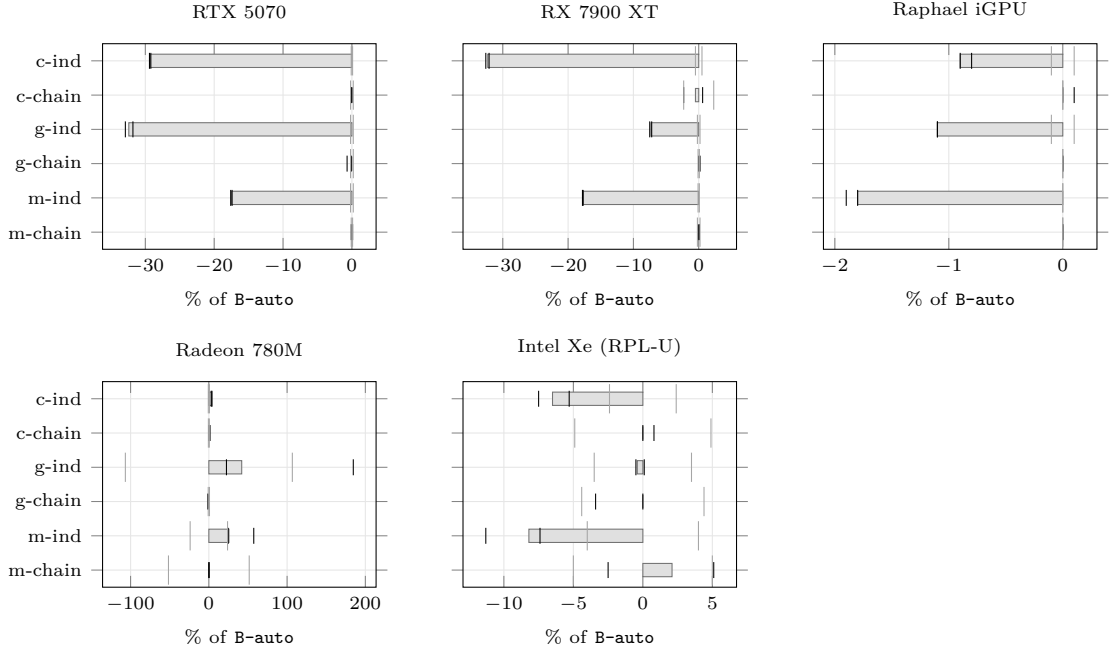
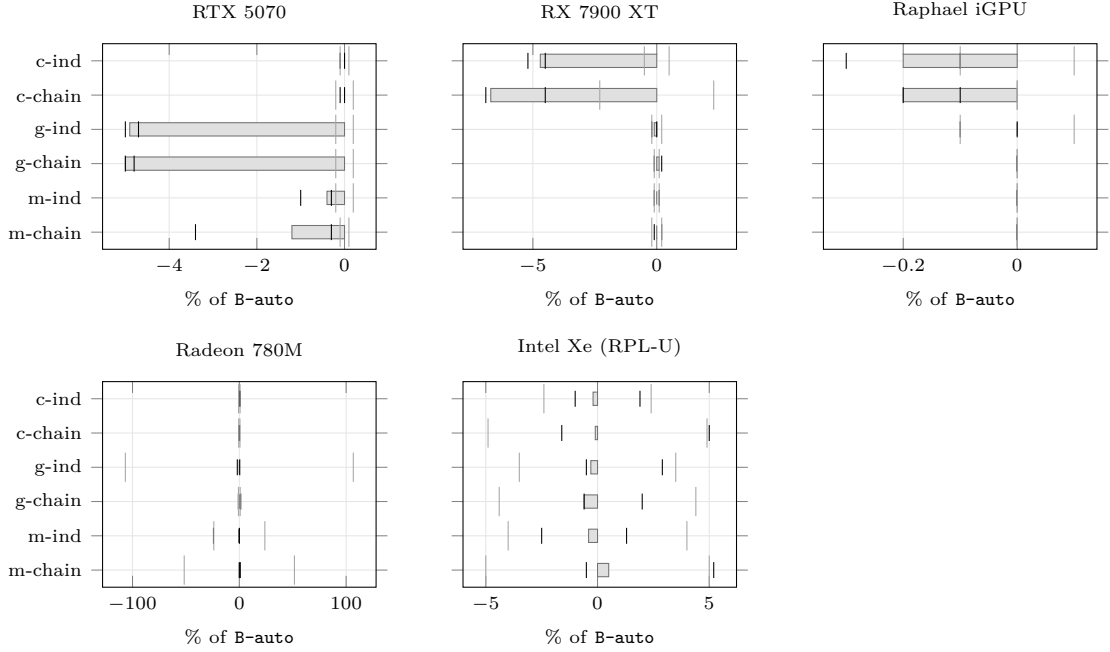
\begin{figure*}[t]
\centering
\begin{tikzpicture}
\begin{groupplot}[
  group style={group size=3 by 2, horizontal sep=1.15cm,
    vertical sep=1.75cm, y descriptions at=edge left},
  width=5.2cm, height=4.3cm,
  xbar, /pgf/bar width=5pt,
  symbolic y coords={c-ind,c-chain,g-ind,g-chain,m-ind,m-chain},
  ytick=data, y dir=reverse,
  xlabel={\% of \texttt{B-auto}},
  xlabel style={font=\scriptsize},
  tick label style={font=\scriptsize},
  title style={font=\scriptsize},
  grid=major, grid style={black!10},
  scaled x ticks=false,
]
\nextgroupplot[title={RTX 5070}]
\addplot[xbar, fill=black!12, draw=black!55]
coordinates {(-0.0,c-ind) (-0.0,c-chain) (-4.9,g-ind) (-5.0,g-chain) (-0.4,m-ind) (-1.2,m-chain)};
\addplot[only marks, mark=|, mark size=3pt, black, forget plot]
coordinates {(-0.1,c-ind) (0.0,c-ind) (-0.1,c-chain) (0.0,c-chain) (-5.0,g-ind) (-4.7,g-ind) (-5.0,g-chain) (-4.8,g-chain) (-1.0,m-ind) (-0.3,m-ind) (-3.4,m-chain) (-0.3,m-chain)};
\addplot[only marks, mark=|, mark size=5.5pt, black!35, forget plot]
coordinates {(0.1,c-ind) (-0.1,c-ind) (0.2,c-chain) (-0.2,c-chain) (0.2,g-ind) (-0.2,g-ind) (0.2,g-chain) (-0.2,g-chain) (0.2,m-ind) (-0.2,m-ind) (0.1,m-chain) (-0.1,m-chain)};
\nextgroupplot[title={RX 7900 XT}]
\addplot[xbar, fill=black!12, draw=black!55]
coordinates {(-4.7,c-ind) (-6.7,c-chain) (-0.1,g-ind) (0.1,g-chain) (-0.0,m-ind) (0.0,m-chain)};
\addplot[only marks, mark=|, mark size=3pt, black, forget plot]
coordinates {(-5.2,c-ind) (-4.5,c-ind) (-6.9,c-chain) (-4.5,c-chain) (-0.2,g-ind) (0.0,g-ind) (-0.1,g-chain) (0.2,g-chain) (-0.1,m-ind) (0.1,m-ind) (-0.1,m-chain) (0.2,m-chain)};
\addplot[only marks, mark=|, mark size=5.5pt, black!35, forget plot]
coordinates {(0.5,c-ind) (-0.5,c-ind) (2.3,c-chain) (-2.3,c-chain) (0.2,g-ind) (-0.2,g-ind) (0.1,g-chain) (-0.1,g-chain) (0.1,m-ind) (-0.1,m-ind) (0.2,m-chain) (-0.2,m-chain)};
\nextgroupplot[title={Raphael iGPU}]
\addplot[xbar, fill=black!12, draw=black!55]
coordinates {(-0.2,c-ind) (-0.2,c-chain) (0.0,g-ind) (-0.0,g-chain) (0.0,m-ind) (0.0,m-chain)};
\addplot[only marks, mark=|, mark size=3pt, black, forget plot]
coordinates {(-0.3,c-ind) (-0.1,c-ind) (-0.2,c-chain) (-0.1,c-chain) (-0.1,g-ind) (0.0,g-ind) (-0.0,g-chain) (0.0,g-chain) (-0.0,m-ind) (0.0,m-ind) (-0.0,m-chain) (0.0,m-chain)};
\addplot[only marks, mark=|, mark size=5.5pt, black!35, forget plot]
coordinates {(0.1,c-ind) (-0.1,c-ind) (0.0,c-chain) (-0.0,c-chain) (0.1,g-ind) (-0.1,g-ind) (0.0,g-chain) (-0.0,g-chain) (0.0,m-ind) (-0.0,m-ind) (0.0,m-chain) (-0.0,m-chain)};
\nextgroupplot[title={Radeon 780M}]
\addplot[xbar, fill=black!12, draw=black!55]
coordinates {(0.2,c-ind) (-0.1,c-chain) (0.1,g-ind) (0.1,g-chain) (-0.6,m-ind) (0.1,m-chain)};
\addplot[only marks, mark=|, mark size=3pt, black, forget plot]
coordinates {(-0.1,c-ind) (0.8,c-ind) (-0.5,c-chain) (0.3,c-chain) (-1.9,g-ind) (0.4,g-ind) (-0.9,g-chain) (1.3,g-chain) (-24.2,m-ind) (0.1,m-ind) (-0.0,m-chain) (0.9,m-chain)};
\addplot[only marks, mark=|, mark size=5.5pt, black!35, forget plot]
coordinates {(0.7,c-ind) (-0.7,c-ind) (0.6,c-chain) (-0.6,c-chain) (106.7,g-ind) (-106.7,g-ind) (0.8,g-chain) (-0.8,g-chain) (23.9,m-ind) (-23.9,m-ind) (51.6,m-chain) (-51.6,m-chain)};
\nextgroupplot[title={Intel Xe (RPL-U)}]
\addplot[xbar, fill=black!12, draw=black!55]
coordinates {(-0.2,c-ind) (-0.1,c-chain) (-0.3,g-ind) (-0.6,g-chain) (-0.4,m-ind) (0.5,m-chain)};
\addplot[only marks, mark=|, mark size=3pt, black, forget plot]
coordinates {(-1.0,c-ind) (1.9,c-ind) (-1.6,c-chain) (5.0,c-chain) (-0.5,g-ind) (2.9,g-ind) (-0.6,g-chain) (2.0,g-chain) (-2.5,m-ind) (1.3,m-ind) (-0.5,m-chain) (5.2,m-chain)};
\addplot[only marks, mark=|, mark size=5.5pt, black!35, forget plot]
coordinates {(2.4,c-ind) (-2.4,c-ind) (4.9,c-chain) (-4.9,c-chain) (3.5,g-ind) (-3.5,g-ind) (4.4,g-chain) (-4.4,g-chain) (4.0,m-ind) (-4.0,m-ind) (5.0,m-chain) (-5.0,m-chain)};
\end{groupplot}
\end{tikzpicture}
\caption{Barrier scope at fixed placement: percent difference in GPU span between \texttt{B-auto-scoped} and \texttt{B-auto}, drawn as in Figure~\ref{fig:placement}. The chain workloads are the clean test, because every timed inter-pass barrier is retained there. The grey threshold comes from the global identical-request control; it is a cell-stability diagnostic, not a scoped control.}
\label{fig:scope}
\end{figure*}

Figure~\ref{fig:scope} isolates the axis introduced in
Section~\ref{sec:scoped}. The scoped policies emit the same number of barriers
at the same boundaries as their unscoped twins; only the declared stages and
accesses change. The two chain workloads make the separation sharp, because
there every timed inter-pass barrier is retained, so any difference between
the automatic scoped and unscoped policies must come from scope.

The result splits the prediction that motivated the configuration down the
middle, and both halves are informative. It also corrects this paper's own
pilot, whose floors on the decisive AMD cells were wider than the effect and
which we initially summarized as ``nothing measurable on AMD''. That summary
was a statement about the pilot's noise, not about the device; the warmed
collection resolves what it could not.

\textbf{Where it pays.} On the RTX~5070 narrowing the scope is worth
$\bmagci{rtx5070}{graphics-chain}{scope}$ on \texttt{graphics-chain} and
$\bmagci{rtx5070}{mixed-chain}{scope}$ on \texttt{mixed-chain}, against control
floors of $\bfloor{rtx5070}{graphics-chain}$ and $\bfloor{rtx5070}{mixed-chain}$
in those cells. It is not confined to chains on that device: the same change is
worth $\bmagci{rtx5070}{graphics-independent}{scope}$ on
\texttt{graphics-independent}, floor $\bfloor{rtx5070}{graphics-independent}$,
where it is available \emph{in addition} to the much larger placement effect.
And it now pays on AMD --- on the discrete part, at compute-to-compute
boundaries only: $\bmagci{rx7900xt}{compute-chain}{scope}$ on
\texttt{compute-chain} against a floor of $\bfloor{rx7900xt}{compute-chain}$,
confirmed from a far cleaner cell by
$\bmagci{rx7900xt}{compute-independent}{scope}$ on
\texttt{compute-independent}, floor $\bfloor{rx7900xt}{compute-independent}$.
At this 16-pass point \texttt{compute-chain} falls from
\bus{rx7900xt}{compute-chain}{auto} to
\bus{rx7900xt}{compute-chain}{scope}, while
\texttt{compute-independent} falls from
\bus{rx7900xt}{compute-independent}{auto} to
\bus{rx7900xt}{compute-independent}{scope}. We did not sweep pass count on
this device, so their similar absolute savings do not establish a fixed
per-boundary law.

\textbf{Threshold-edge iGPU observations.} Raphael's two compute cells have
one-signed intervals:
$\bpctci{raphael}{compute-independent}{scope}$ and
$\bpctci{raphael}{compute-chain}{scope}$, against floors of
$\bfloor{raphael}{compute-independent}$ and
$\bfloor{raphael}{compute-chain}$. The
\texttt{compute-independent} interval still touches its stability band; the
\texttt{compute-chain} interval clears its band only at the rounding edge.
Either way these are fractions of a percent on a saturated device, not the
practical claim made for the discrete part.

\textbf{Where no direction resolves.} Every RX~7900~XT boundary that involves
a render pass on either side fails the paper's exploratory directional
criterion: $\bpctci{rx7900xt}{graphics-independent}{scope}$ on
\texttt{graphics-independent} at a floor of
$\bfloor{rx7900xt}{graphics-independent}$, with no resolved directional effect
on \texttt{graphics-chain} or either mixed workload. This is not an
equivalence result. On the Radeon~780M the two compute cells read
$\bpctci{radeon780m}{compute-independent}{scope}$ and
$\bpctci{radeon780m}{compute-chain}{scope}$ against floors of
$\bfloor{radeon780m}{compute-independent}$ and
$\bfloor{radeon780m}{compute-chain}$, and neither clears its cell's floor.
The Intel part is not credited in either direction. Its
\texttt{compute-chain} column reads
$\bpctci{intelxe}{compute-chain}{scope}$ against a floor of
$\bfloor{intelxe}{compute-chain}$
(Section~\ref{sec:deviations}). Its
\texttt{graphics-independent} estimate is
$\bpctci{intelxe}{graphics-independent}{scope}$, but that wide interval
overlaps the $\pm\bfloor{intelxe}{graphics-independent}$ stability band.
The pilot's Intel \texttt{graphics-chain} effect, once quoted here as the Intel
result, did not reproduce
($\bpctci{intelxe}{graphics-chain}{scope}$).

\textbf{The pattern points to the destination scope.} The mixed workloads
localize it.
Every \texttt{mixed} boundary joins two pass kinds, so its scoped barrier
narrows the \emph{source} exactly as a compute-to-compute barrier does ---
and no mixed AMD cell resolves a directional effect. What a mixed boundary
never has is a compute-only \emph{destination}. The crossed policies point the
same way from the other side: \texttt{B-hazard-scoped}, whose explicitly
placed barriers narrow their source but must keep a wide destination
(Section~\ref{sec:scoped}), shows little direct scope effect:
$\bpctci{rx7900xt}{compute-chain}{manualscope}$ on the AMD compute chain,
where automatic scope is worth
$\bpctci{rx7900xt}{compute-chain}{scope}$, and
$\bpctci{rtx5070}{graphics-chain}{manualscope}$ on the NVIDIA graphics chain,
where automatic scope is worth
$\bpctci{rtx5070}{graphics-chain}{scope}$. Together these comparisons are
consistent with the saving depending on the narrowed destination unavailable
to an explicit barrier. They localize the pattern to that side of the request;
they do not identify the driver operation responsible.

This is not the shape the prediction had. Section~\ref{sec:radv-barriers}
derives from RADV that narrowing sheds the most at render-involving boundaries
--- a pixel drain, two attachment flushes, two metadata flushes --- and it is
right about the commands; that part is read from the source. But those
boundaries show no resolved saving on any AMD part, while the
compute-to-compute boundary, where the shed operations run against idle pixel
hardware and clean caches and were priced accordingly, is where the discrete
part actually pays.
The drivers whose source we did not read split the same way for their own
reasons: broad savings on NVIDIA, nothing creditable on Intel.

Going back to the source explains most of it (Section~\ref{sec:radv-elision}):
on GFX10 and later with a coherent TCC, \texttt{radv\_dst\_access\_flush}
drops every destination-side L2 invalidation unless a render-backend-incoherent
attachment is dirty. If the measured targets did not set that pipe-misaligned
state, half of what the narrower barrier was designed to save was already being
discarded by the driver without the application saying anything; the
benchmark did not instrument that internal flag, so the account remains
conditional. On the source side, a flush costs time only when there is matching
work in flight or dirty state. A render boundary retains its attachment-flush
requests under either scope, while a pure compute stream has no attachment
work. This supplies a plausible source-level account of the unresolved
render-boundary effects, not a counter measurement. What it does not account
for is the compute-boundary
saving itself: the operations our reading priced there are the free-when-idle
ones, and the measurement instead follows the destination mask --- a
dependence the reading never enumerated, because everything it enumerated on
the destination side was cache work the driver had already elided. We can name
the side the cost lives on and not, from the text we read, the operation.

The general mistake is worth naming, because it is easy to repeat and we made
it in the obvious way. Counting the commands a barrier expands into records an
over-approximate request set; it is not a numerical upper bound on elapsed
cost. Here that request set is longest exactly where its timing effect is least
resolved. We read the half of the driver that translates a request and not the
half that decides what to do about it.
Reading it was still worth doing --- it produced the configuration, and it
explains most of the result --- but the explanation arrived after the
measurement, and would not have been trusted before it.

Section~\ref{sec:generated} checks the relevant request with captures: the
extracted barrier tables on all \bladenum{captures/machines} machines are
byte-identical. The spread --- $\bmag{rtx5070}{graphics-chain}{scope}$ on one
vendor's render chain, $\bmag{rx7900xt}{compute-chain}{scope}$ on another's
compute chain, no resolved direction at any AMD render boundary --- therefore
is not explained by different masks or barrier counts in the extracted
records. The captures do not establish that the complete command streams or
driver-internal work are byte-identical.

\textbf{Noise is a property of the cell, not the device.} The floor ticks in
Figures~\ref{fig:placement} and~\ref{fig:scope} make this legible. The Intel
part ranges from a floor of
$\bfloor{intelxe}{graphics-independent}$ on one cell to
$\bfloor{intelxe}{mixed-chain}$ on another in the same session. The
RX~7900~XT also varies by workload,
from $\bfloor{rx7900xt}{graphics-independent}$ on its cleanest cell to
$\bfloor{rx7900xt}{compute-chain}$ on a less stable one. A per-device maximum
would discard usable cells; a pooled sample bootstrap or the control's point
estimate would license unstable ones.

\textbf{Manual placement forfeits the scope saving.} The ``both'' column is
therefore not the sum of the first two: combining hazard-only placement with
the narrow scope resolves the placement effect but no additional scope saving,
in every cell of every vendor where scope pays, for the reason
Section~\ref{sec:scoped} builds in --- an explicitly placed barrier narrows
its source but not its destination, matching the destination-side pattern
above.
That is the opposite of the intuition that more control is strictly better. An
application that leaves placement to the encoder gets a barrier the encoder
can describe on both sides; one that takes placement into its own hands gets a
barrier that only half describes itself. The scope saving is also the cheaper
of the two gains to obtain --- no API change and no resource tracking --- so
where it exists it should be taken first, and where placement is also taken
the two should not be assumed to compose.

\subsection{Scaling with pass count}

\begin{table*}[tp]
\centering
\small
\begin{tabular}{@{}llrrrrrrrr@{}}
\toprule
 & Config & 1 & 2 & 4 & 8 & 16 & 32 & 64 & $\mu$s/pass \\
\midrule
\multicolumn{10}{l}{\emph{compute-independent}} \\
host & B-auto & 2.8 & 3.6 & 5.2 & 10.5 & 54.5 & 142.0 & 269.3 & 1.10 \\
host & B-hazard & 2.5 & 2.8 & 3.5 & 5.0 & 16.0 & 64.4 & 122.9 & 0.36 \\
host & W-wgpu & 6.3 & 9.0 & 14.2 & 45.4 & 230.2 & 383.6 & 672.6 & 5.59 \\
GPU & B-auto & 10.9 & 21.1 & 43.5 & 86.6 & 195.3 & 391.8 & 785.0 & 12.29 \\
GPU & B-hazard & 10.9 & 19.2 & 35.8 & 70.8 & 138.2 & 274.3 & 547.9 & 8.52 \\
GPU & W-wgpu & 11.0 & 22.6 & 47.0 & 96.3 & 196.5 & 394.5 & 788.3 & 12.34 \\
\multicolumn{10}{l}{\emph{graphics-independent}} \\
host & B-auto & 3.4 & 4.7 & 7.2 & 13.8 & 62.2 & 215.9 & 415.5 & 1.48 \\
host & B-hazard & 3.1 & 4.0 & 5.8 & 9.4 & 27.7 & 147.2 & 306.1 & 0.90 \\
host & W-wgpu & 7.3 & 10.8 & 17.7 & 33.6 & 115.0 & 195.6 & 356.0 & 3.75 \\
GPU & B-auto & 10.4 & 19.7 & 38.3 & 75.6 & 151.6 & 303.0 & 606.4 & 9.46 \\
GPU & B-hazard & 10.4 & 16.5 & 28.4 & 52.3 & 103.8 & 209.8 & 421.1 & 6.52 \\
GPU & W-wgpu & 7.6 & 13.8 & 25.7 & 49.6 & 99.3 & 206.3 & 416.1 & 6.48 \\
\bottomrule
\end{tabular}
\\[2pt]\footnotesize GPU span has nearly constant increments across the whole range. Host cost is linear only to 8 passes: between 8 and 16 its medians jump several-fold into a second, steeper regime, so one slope fitted over all counts would describe neither. The 1--8 average characterizes the linear regime; the larger counts are reported as measured.
\caption{Pass-count sweep on the RTX~5070, medians of process medians in microseconds. Host rows come from the timestamp-free collection; GPU rows come from the GPU-timed collection. The last column is the average increment per additional pass between the endpoint medians, taken over the whole range for the GPU rows and over 1--8 passes for the host rows.}
\label{tab:sweep}
\end{table*}

\begin{figure}[t]
\centering
\begin{tikzpicture}
\begin{axis}[
  width=\columnwidth, height=5.2cm,
  ybar, /pgf/bar width=7pt,
  symbolic x coords={host~c-ind,host~g-ind,GPU~c-ind,GPU~g-ind},
  xtick=data, ymin=0,
  ylabel={$\mu$s per additional pass},
  ylabel style={font=\scriptsize},
  tick label style={font=\scriptsize},
  legend style={font=\scriptsize, at={(0.5,1.03)}, anchor=south,
    legend columns=3, draw=none},
  grid=major, grid style={black!10},
  nodes near coords,
  every node near coord/.append style={font=\tiny},
]
\addplot+[ybar] coordinates {(host~c-ind,1.10) (host~g-ind,1.48) (GPU~c-ind,12.29) (GPU~g-ind,9.46)};
\addplot+[ybar] coordinates {(host~c-ind,0.36) (host~g-ind,0.90) (GPU~c-ind,8.52) (GPU~g-ind,6.52)};
\addplot+[ybar] coordinates {(host~c-ind,5.59) (host~g-ind,3.75) (GPU~c-ind,12.34) (GPU~g-ind,6.48)};
\legend{\texttt{B-auto},\texttt{B-hazard},\texttt{W-wgpu}}
\end{axis}
\end{tikzpicture}
\caption{Average cost of one additional pass on the RTX~5070, from the pass-count sweeps. Host figures come from the timestamp-free collection and device figures from the GPU-timed one; each is the endpoint increment over the range stated in Table~\ref{tab:sweep}. The gap between \texttt{B-auto} and \texttt{B-hazard} is what one redundant barrier costs; the gap to \texttt{W-wgpu} is end-to-end.}
\label{fig:marginal}
\end{figure}
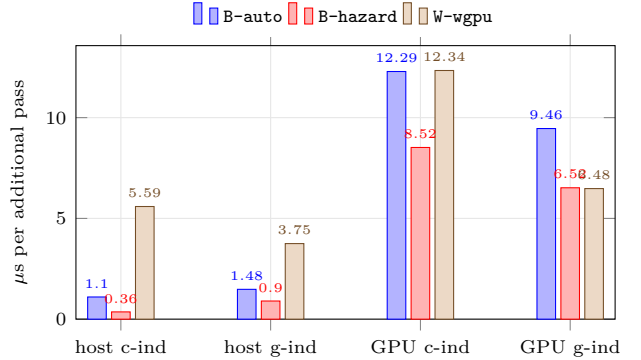

Table~\ref{tab:sweep} sweeps the pass count on the RTX~5070 and answers RQ1.
GPU span has nearly constant increments across the whole range. Host medians
are well behaved only over the 1--8-pass range used for the reported average
increment; Figure~\ref{fig:marginal} compares those endpoint averages directly.
The 16-pass values are reported as measured but lie outside that host summary
range.

On the host side, the endpoint averages imply an additional global-barrier
cost of
$\bladenum{sweep/host/compute-independent/barriercost}\,\mu$s per pass for
compute ($\bladenum{sweep/host/compute-independent/auto/marginal}$ against
$\bladenum{sweep/host/compute-independent/hazard/marginal}$) and
$\bladenum{sweep/host/graphics-independent/barriercost}\,\mu$s for graphics
($\bladenum{sweep/host/graphics-independent/auto/marginal}$ against
$\bladenum{sweep/host/graphics-independent/hazard/marginal}$). The matched
\wgpu{} program averages
$\bladenum{sweep/host/compute-independent/wgpu/marginal}$ and
$\bladenum{sweep/host/graphics-independent/wgpu/marginal}\,\mu$s per pass,
which is $\bladenum{sweep/host/compute-independent/wgpuoverauto}\times$ and
$\bladenum{sweep/host/graphics-independent/wgpuoverauto}\times$ \blade's
automatic policy and
$\bladenum{sweep/host/compute-independent/wgpuoverhazard}\times$ and
$\bladenum{sweep/host/graphics-independent/wgpuoverhazard}\times$ \blade{} with
no inter-pass barrier. At sixteen passes the absolute gap is
$\bladenum{sweep/host/compute-independent/auto/at16}$ against
$\bladenum{sweep/host/compute-independent/wgpu/at16}\,\mu$s.

Past eight passes the host cost leaves that linear regime: between the 8- and
16-pass medians it jumps several-fold on both implementations, and beyond
sixteen it grows steeply but smoothly. Something in the submission path
changes mode near that size, and its onset wandered between collections ---
the superseded three-block sweep crossed over between sixteen and thirty-two,
erratically --- so no single slope describes the full range. The averages
above are therefore restricted to 1--8 passes; the 16-pass matrix is a
directly measured point inside the second regime, not an extrapolation from
the first.

On the device side, the corresponding endpoint-average cost of each additional
redundant barrier is
$\bladenum{sweep/gpu/compute-independent/barriercost}\,\mu$s of GPU span for
compute ($\bladenum{sweep/gpu/compute-independent/auto/marginal}$ against
$\bladenum{sweep/gpu/compute-independent/hazard/marginal}$ per pass) and
$\bladenum{sweep/gpu/graphics-independent/barriercost}\,\mu$s for graphics
($\bladenum{sweep/gpu/graphics-independent/auto/marginal}$ against
$\bladenum{sweep/gpu/graphics-independent/hazard/marginal}$). Against passes
that themselves cost $\bladenum{sweep/gpu/compute-independent/hazard/marginal}$
and $\bladenum{sweep/gpu/graphics-independent/hazard/marginal}\,\mu$s, that is
more than a third of a pass lost per boundary. Within this repeated-pass sweep,
the auto--hazard gap is descriptively visible at every measured count above
one; at one pass there is no timed inter-pass barrier to remove. This is not a
claim about other graph shapes or devices.

The timestamp-free sweep exists to check whether GPU timing perturbs the host
measurement --- \blade{} records one timestamp per pass against two in total
for \wgpu{}, so any effect would also be unmatched between them. On this
driver, at ten repetitions, there is none to find: the timestamped and
timestamp-free host medians agree at every pass count for both
implementations. A superseded three-block sweep had appeared to show a gap
growing with the pass count; it did not survive the repetitions. The fixed
matrix's host columns are therefore read as host cost, with the
instrumentation concern tested on this machine and untested on the others.

\subsection{Host cost across devices}

\begin{table*}[tp]
\centering
\small
\begin{tabular}{@{}llrr@{\,}lr@{\,}l@{}}
\toprule
Device & Workload & B-auto & \multicolumn{2}{c}{B-hazard \%} & \multicolumn{2}{c}{W-wgpu \%} \\
\midrule
RTX 5070 & c-ind & 56.0 & -61.5 & [-63.2,\,-58.2] & +282.3 & [+116.2,\,+339.4] \\
 & c-chain & 58.3 & -11.3 & [-22.1,\,-2.6] & +270.8 & [+224.2,\,+322.2] \\
 & g-ind & 70.0 & -54.5 & [-58.5,\,-50.2] & +59.4 & [+46.2,\,+73.6] \\
 & g-chain & 60.3 & -11.3 & [-36.8,\,-1.7] & +128.7 & [+115.5,\,+144.9] \\
 & m-ind & 69.3 & -46.7 & [-51.2,\,-42.1] & --- &  \\
 & m-chain & 67.1 & -0.9 & [-13.8,\,+10.2] & --- &  \\
\addlinespace
RX 7900 XT & c-ind & 10.6 & -14.8 & [-16.2,\,-13.5] & +429.1 & [+417.0,\,+431.9] \\
 & c-chain & 10.6 & -0.5 & [-1.0,\,+0.7] & +412.7 & [+410.1,\,+420.5] \\
 & g-ind & 27.9 & -3.0 & [-4.8,\,-2.6] & +211.2 & [+207.2,\,+214.3] \\
 & g-chain & 19.7 & -0.6 & [-1.6,\,+0.2] & +288.0 & [+285.1,\,+290.4] \\
 & m-ind & 20.0 & -6.6 & [-8.0,\,-5.9] & --- &  \\
 & m-chain & 15.9 & -1.9 & [-3.1,\,+0.0] & --- &  \\
\addlinespace
Raphael iGPU & c-ind & 8.8 & -14.9 & [-18.9,\,-12.1] & +486.0 & [+458.3,\,+499.3] \\
 & c-chain & 8.7 & -3.5 & [-9.0,\,+3.7] & +490.7 & [+456.8,\,+502.9] \\
 & g-ind & 30.5 & -4.4 & [-5.6,\,-3.8] & +163.7 & [+161.2,\,+168.0] \\
 & g-chain & 17.7 & -0.6 & [-2.6,\,+0.5] & +312.8 & [+307.6,\,+318.6] \\
 & m-ind & 20.0 & -6.2 & [-7.5,\,-5.9] & --- &  \\
 & m-chain & 13.7 & +0.2 & [-0.5,\,+1.3] & --- &  \\
\addlinespace
Radeon 780M & c-ind & 68.6 & -3.7 & [-6.3,\,+2.2] & +246.2 & [+239.6,\,+271.9] \\
 & c-chain & 67.3 & -0.9 & [-3.3,\,+1.8] & +250.9 & [+243.3,\,+257.5] \\
 & g-ind & 112.8 & +0.8 & [-2.5,\,+43.4] & +162.7 & [+44.0,\,+168.4] \\
 & g-chain & 29.0 & +2.2 & [-49.4,\,+125.8] & +374.9 & [+58.9,\,+582.1] \\
 & m-ind & 101.5 & +0.6 & [-2.8,\,+22.5] & --- &  \\
 & m-chain & 90.5 & +1.8 & [-1.0,\,+8.8] & --- &  \\
\addlinespace
Intel Xe (RPL-U) & c-ind & 95.8 & +0.2 & [-1.6,\,+16.9] & +297.5 & [+266.1,\,+308.2] \\
 & c-chain & 83.7 & -4.4 & [-8.9,\,+6.8] & +324.0 & [+309.8,\,+336.7] \\
 & g-ind & 240.4 & -0.4 & [-4.8,\,+3.8] & +122.5 & [+114.5,\,+129.5] \\
 & g-chain & 132.1 & -0.9 & [-4.4,\,+5.0] & +200.8 & [+191.2,\,+209.6] \\
 & m-ind & 220.4 & +3.5 & [+0.9,\,+10.2] & --- &  \\
 & m-chain & 185.4 & +1.5 & [-11.7,\,+4.3] & --- &  \\
\addlinespace
Apple M3 & c-ind & 68.3 & --- &  & +33.7 & [+24.5,\,+57.9] \\
 & c-chain & 68.3 & --- &  & +33.0 & [+24.8,\,+42.9] \\
 & g-ind & 85.2 & --- &  & +77.6 & [+64.3,\,+98.2] \\
 & g-chain & 83.8 & --- &  & +79.3 & [+71.0,\,+86.0] \\
\bottomrule
\end{tabular}
\\[2pt]\footnotesize Percent differences rather than ratios: a \texttt{W-wgpu} entry of $+200\%$ is three times \texttt{B-auto}.
\caption{Host cost of one 16-pass command buffer: recording plus submission, median of process medians in microseconds, then percent differences from \texttt{B-auto} with 95\% hierarchical-bootstrap intervals over paired process repetitions. These collections have GPU timestamps enabled. Blade records one \texttt{vkCmdWriteTimestamp} per pass against two in total for wgpu, so this is an instrumented end-to-end comparison, not a matched timestamp workload.}
\label{tab:host-matrix}
\end{table*}

Table~\ref{tab:host-matrix} repeats the host comparison on every machine. Over
the \bladenum{host/ratio/count} shared cells, the matched \wgpu{} program uses
between
$\bladenum{host/ratio/min}\times$ and $\bladenum{host/ratio/max}\times$ the
host time of \blade's automatic policy for the same sixteen passes, and the
direction is the same in every one of them.\hostcaveatinline{} These collections have GPU
timestamps enabled, which charges \blade{} one \texttt{vkCmdWriteTimestamp} per
pass against two in total for \wgpu. That extra recorded work plausibly biases
against \blade{}, but the two timestamp paths differ enough that we do not
treat the comparison as a formal lower bound. The ratios describe this
instrumented end-to-end path, not an uninstrumented record-only cost; the
timestamp-free RTX~5070 sweep supplies the clean magnitude check. The sign
never changes in the fixed matrix.

The narrowest Vulkan ratios are the NVIDIA render cells, and the
\texttt{B-hazard} column says why: that driver charges heavily for recording
the barriers themselves, so removing them cuts \blade's own host cost by
$\bmag{rtx5070}{graphics-independent}{hostplacement}$ on
\texttt{graphics-independent}. Where \blade's denominator is mostly barrier
recording --- a cost the application controls --- the ratio narrows; nowhere
does it invert. The Metal rows are narrower still for the backend's own
reason: each \blade{} pass there creates a fresh command encoder
(Section~\ref{sec:metal}), which inflates the same denominator.

The difference is end-to-end and must not be attributed to tracking alone:
\wgpu{} also validates, manages lifetimes, resolves bind groups, and builds an
intermediate command representation. It is also not a barrier-count effect:
on the independent workloads Section~\ref{sec:generated} counts
\bladenum{barriers/auto/independent} barriers from \texttt{B-auto} against
\bladenum{barriers/wgpu/independent} from \wgpu, and \wgpu{} still costs
several times more host time.

\subsection{Where the host time goes}
\label{sec:profile}

\begin{table*}[tp]
\centering
\small
\setlength{\tabcolsep}{3.5pt}
\begin{tabular}{@{}lrrrrrrrrrrrrrrrr@{}}
\toprule
 & \multicolumn{4}{c}{\texttt{S1}} & \multicolumn{4}{c}{\texttt{S2}} & \multicolumn{4}{c}{\texttt{S3}} & \multicolumn{4}{c}{\texttt{S4}} \\
\cmidrule(lr){2-5}\cmidrule(lr){6-9}\cmidrule(lr){10-13}\cmidrule(lr){14-17}
 & \multicolumn{2}{c}{c-ind} & \multicolumn{2}{c}{g-ind} & \multicolumn{2}{c}{c-ind} & \multicolumn{2}{c}{g-ind} & \multicolumn{2}{c}{c-ind} & \multicolumn{2}{c}{g-ind} & \multicolumn{2}{c}{c-ind} & \multicolumn{2}{c}{g-ind} \\
\cmidrule(lr){2-3}\cmidrule(lr){4-5}\cmidrule(lr){6-7}\cmidrule(lr){8-9}\cmidrule(lr){10-11}\cmidrule(lr){12-13}\cmidrule(lr){14-15}\cmidrule(lr){16-17}
Component & \textsc{b} & \textsc{w} & \textsc{b} & \textsc{w} & \textsc{b} & \textsc{w} & \textsc{b} & \textsc{w} & \textsc{b} & \textsc{w} & \textsc{b} & \textsc{w} & \textsc{b} & \textsc{w} & \textsc{b} & \textsc{w} \\
\midrule
kernel & 45.5 & 36.1 & 41.1 & 30.2 & 87.9 & 34.6 & 10.1 & 24.7 & 33.5 & 52.3 & 15.0 & 35.2 & 39.4 & 7.8 & 27.7 & 6.6 \\
driver & 29.2 & 40.5 & 46.1 & 36.7 & 5.9 & 24.8 & 80.3 & 35.3 & 31.0 & 21.4 & 74.7 & 36.6 & 42.5 & 29.4 & 57.2 & 36.3 \\
ash / loader & 0.3 & 0.1 & 0.1 & 0.1 & 0.1 & 0.1 & --- & --- & 0.5 & --- & 0.1 & --- & 0.3 & --- & --- & --- \\
wgpu tracker & --- & 5.1 & --- & 3.1 & --- & 18.6 & --- & 6.8 & --- & 4.7 & --- & 1.5 & --- & 4.2 & --- & 2.0 \\
wgpu command & --- & 6.2 & --- & 8.7 & --- & 7.7 & --- & 13.5 & --- & 5.6 & --- & 9.4 & --- & 4.4 & --- & 5.8 \\
wgpu-hal & --- & 1.1 & --- & 1.9 & --- & 1.5 & --- & 3.3 & --- & 1.4 & --- & 2.3 & --- & 0.8 & --- & 1.7 \\
wgpu device/resource & --- & 1.2 & --- & 1.0 & --- & 2.0 & --- & 1.3 & --- & 0.8 & --- & 1.1 & --- & 1.3 & --- & 1.1 \\
wgpu validation & --- & 0.2 & --- & --- & --- & 0.1 & --- & 0.1 & --- & 0.1 & --- & 0.2 & --- & 0.1 & --- & 0.1 \\
wgpu other & --- & 0.8 & --- & 0.8 & --- & 1.3 & --- & 0.8 & --- & 0.6 & --- & 0.4 & --- & 0.6 & --- & 0.6 \\
blade & 20.7 & 0.2 & 1.9 & 0.2 & 5.1 & 0.3 & 2.2 & 0.4 & 25.9 & 0.1 & 1.8 & 0.1 & 3.3 & 0.2 & 1.9 & 0.1 \\
allocator & 0.7 & 3.5 & 0.8 & 3.8 & 0.1 & 1.3 & 0.6 & 0.8 & 1.0 & 2.1 & 0.5 & 1.6 & 1.8 & 2.6 & 1.1 & 2.7 \\
libc / runtime & 3.7 & 3.9 & 4.0 & 5.5 & 0.9 & 5.6 & 7.3 & 8.1 & 6.1 & 6.3 & 9.2 & 6.8 & 8.3 & 44.8 & 5.8 & 39.1 \\
other & 1.0 & 3.4 & 4.0 & 6.0 & 0.6 & 3.1 & 1.7 & 2.7 & 1.7 & 3.7 & 1.5 & 2.4 & 5.1 & 2.1 & 1.8 & 2.3 \\
\bottomrule
\end{tabular}
\\[2pt]\footnotesize Shares within a process, not times between processes: \texttt{task-clock} charges a blocking fence wait to the process whenever the driver spins inside it, which inflates the \emph{driver} row differently for the two implementations. Self time is attributed to the symbol a sample landed in, so inlined tracker work can be charged to its caller; the tracker-labelled share can therefore undercount, but is not a formal lower bound. Columns sum to slightly more or less than 100 because \texttt{perf} rounds each symbol's share independently. The \emph{driver} row is \texttt{S1} on \texttt{libnvidia-eglcore.so.595.71.05}, \texttt{S2} on \texttt{libvulkan\_radeon.so}, \texttt{S3} on \texttt{libvulkan\_radeon.so}, \texttt{S4} on \texttt{libvulkan\_intel.so}. The columns are labelled by machine and driver rather than by device; only shares within one process are read from them. Because waits dominate these whole-process profiles, the component shares cannot apportion the record-and-submit gap.
\caption{Share of process CPU time by component, from flat \texttt{perf} profiles. \textsc{b} is \blade{} and \textsc{w} the corresponding \wgpu{} program. The workload uses tiny dispatches, small targets, many passes, and no timestamp queries, but the profile covers the whole process, including completion waits.}
\label{tab:profile}
\end{table*}

Table~\ref{tab:profile} attributes each implementation's CPU time to the crate
or library that spent it. The workload was shaped to emphasize host work ---
tiny dispatches, small targets, many passes, no timestamp queries --- but the
profile samples the whole process, including a completion wait after every
iteration. It therefore did not isolate the recording interval as intended.

That limitation changes what the table can support. Symbols assigned directly
to \wgpu's resource tracker account for
$\bladenum{profile/wgpu/tracker/min}$--$\bladenum{profile/wgpu/tracker/max}\%$
of its process samples, and all \wgpu{} buckets together account for
$\bladenum{profile/wgpu/wgputotal/min}$--$\bladenum{profile/wgpu/wgputotal/max}\%$.
The tracker-labelled figure likely undercounts inlined tracker work, which is
charged to its caller, but it is also a share of a denominator dominated by
waits and driver activity. It cannot establish either that tracking explains the
record-and-submit gap or that it is too small to explain it.

What dominates \blade's whole process is the kernel and the driver:
$\bladenum{profile/blade/system/min}$--$\bladenum{profile/blade/system/max}\%$
in every column. The same two rows take
$\bladenum{profile/wgpu/system/min}$--$\bladenum{profile/wgpu/system/max}\%$
of \wgpu's --- its largest share everywhere except the Intel machine, where
a single symbol out-bills them: the \emph{libc / runtime} row reaches
$\bladenum{profile/S4/wgpu/compute-independent/libcruntime}\%$ of the \wgpu{}
process, almost all of it \texttt{\_\_memset\_avx2\_unaligned\_erms} ---
something in that stack clears a lot of memory on ANV, and a flat profile
cannot name the caller --- against
$\bladenum{profile/S4/blade/compute-independent/libcruntime}\%$ for all of
libc in \blade's column on the same machine. The totals are also not
comparable between implementations --- \texttt{task-clock} charges a blocking
fence wait to the process whenever the driver spins inside it, and \blade{}
and \wgpu{} wait differently --- which is why the table reports shares within
a process rather than times between processes.

Repeating one profile moves shares by around a percentage point, so small row
differences are not meaningful. The profiling invocations were not
adapter-pinned, so on a multi-GPU host the two implementations need not have
profiled the same device; only within-process shares are read from the table
for that reason. A causal decomposition
of the host gap needs a follow-up that gates sampling to record plus submit (or
batches many recordings before one wait) on the same pinned device. These flat
whole-process profiles are retained as a diagnostic, not presented as that
decomposition.

\subsection{End-to-end \blade{} versus \wgpu}
\label{sec:endtoend}

The \texttt{W-wgpu} column of Table~\ref{tab:gpu-matrix} is more interesting
than a single ratio suggests, because its sign changes --- between devices,
and on one device between kinds of work.

On the RTX~5070 the matched \wgpu{} path follows the same direction as
application-declared placement. On \texttt{graphics-independent} it is
$\bmagci{rtx5070}{graphics-independent}{wgpu}$ faster than \texttt{B-auto}:
it knows the sixteen render targets are distinct, emits no dependency between
the passes, and lands within a couple of points of \blade's own declared
placement ($\bmag{rtx5070}{graphics-independent}{placement}$) --- a residue of
the same order as the destination-scope saving that an explicitly placed
barrier forfeits (Section~\ref{sec:scope-results}). That numerical proximity is
descriptive: the barrier form, parameter interface, command machinery, and
timestamp instrumentation still differ. The host column of
Table~\ref{tab:host-matrix} separates the paths much more sharply.

On \texttt{graphics-chain} the \wgpu{} path is
$\bmagci{rtx5070}{graphics-chain}{wgpu}$ faster
than \texttt{B-auto} on a workload where every boundary is a real hazard, so
the within-\blade{} placement variants have identical timed barriers and scope
recovers only
$\bmag{rtx5070}{graphics-chain}{scope}$. The rest of the gap --- well over
twice what scope recovers --- remains bundled among the resource-scoped form of
the barrier, the specialized image layout, differing timestamp windows, and
per-pass work outside synchronization. This driver advertises
\texttt{VK\_KHR\_unified\_image\_layouts} and \blade{} enables it, which is a
reason not to expect a \texttt{GENERAL} penalty in supported uses; we observe
the gap and cannot apportion it. Neither compute cell shows a resolved
\wgpu{} advantage, so the large end-to-end difference observed here is
specific to the render path.

On the discrete AMD part the comparison runs the other way. The
\texttt{compute-independent} \wgpu{} result is close to declared placement
($\bmag{rx7900xt}{compute-independent}{wgpu}$ faster than \texttt{B-auto},
beside $\bmag{rx7900xt}{compute-independent}{placement}$ for
\texttt{B-hazard}), while its render path is
$\bmag{rx7900xt}{graphics-independent}{wgpu}$ slower on
\texttt{graphics-independent} and $\bmag{rx7900xt}{graphics-chain}{wgpu}$
slower on \texttt{graphics-chain} --- slower, that is, than a \blade{}
baseline whose global barriers prevent inter-pass overlap. The current capture
extract identifies barrier counts, kinds, masks, and layouts but does not by
itself apportion this gap; it is specific to this chip in the measured set.
On the Radeon~780M the \wgpu{} render point estimate is
$\bmag{radeon780m}{graphics-independent}{wgpu}$ faster, but neither it nor the
16-pass \blade{} placement contrast clears that cell's
$\bfloor{radeon780m}{graphics-independent}$ floor; both are descriptive. The
Raphael iGPU does resolve a $\bmag{raphael}{graphics-independent}{wgpu}$
\wgpu{} advantage, where the largest highlighted within-\blade{} change is
$\bmag{raphael}{mixed-independent}{placement}$, while both iGPUs are slower on
compute. The discrete-against-integrated contrast is reported as measured
rather than assigned to one of the bundled factors.

On the Intel device the \wgpu{} point estimate is slower on every shared
workload. Three cells have one-signed intervals, from
$\bmag{intelxe}{graphics-independent}{wgpu}$ to
$\bmag{intelxe}{compute-independent}{wgpu}$; the
\texttt{graphics-chain} interval spans both signs and is not counted as a
directional result. Since \texttt{compute-chain} is fully serialized in both
implementations, its resolved gap cannot be an overlap difference. Given this
machine's systematic control offsets (Section~\ref{sec:deviations}), we treat
even the three resolved directions as end-to-end observations rather than
mechanism estimates.

Separating the candidates that remain --- the memory scope of each barrier,
the layout the images sit in, timestamp-window differences, or per-pass setup outside synchronization ---
would need a configuration this study does not have: \blade{} emitting
resource-scoped barriers, which is the tracking it exists to avoid. What the
column does establish is that neither matched stack dominates the device side.
The \wgpu{} path wins in some cells and loses in others; because tracking,
barrier form, layouts, validation, and per-pass setup move together, the
end-to-end column cannot assign those signs to tracking alone. Its host
record-and-submit time is higher in every measured cell
(Table~\ref{tab:host-matrix}).

\subsection{Apple/Metal case study}
\label{sec:metal}

\begin{table}[tp]
\centering
\small
\begin{tabular}{@{}lrrrr@{}}
\toprule
 & \multicolumn{2}{c}{\blade{}} & \multicolumn{2}{c}{\wgpu{}} \\
\cmidrule(lr){2-3}\cmidrule(lr){4-5}
Workload & host & wait & host & wait \\
\midrule
c-ind & 68.3 & 1449.1 & 92.8 & 1449.2 \\
c-chain & 68.3 & 1082.3 & 91.1 & 1110.6 \\
g-ind & 85.2 & 1302.6 & 150.0 & 1247.9 \\
g-chain & 83.8 & 1421.8 & 149.9 & 1409.4 \\
\bottomrule
\end{tabular}
\caption{Apple M3 (Metal), 16 passes: host cost (recording plus submission) and host wall time waiting for completion, median of process medians in microseconds. Metal GPU timestamps are reported per pass rather than as a span and were unavailable for three of the four wgpu workloads, so wall time is used instead of device time.}
\label{tab:metal}
\end{table}

Metal is not pooled into the Vulkan comparison, because it hides image layouts
and because the framework performs the hazard tracking that \blade{} declines
to perform on Vulkan~\cite{metal-hazard-tracking}. On Metal, \blade{} is not a
tracking-free implementation; it is a client of Metal's tracker. This study
uses the pre-Metal-4 \texttt{MTLCommandQueue} path. Apple documents that
resource hazard-tracking mode has no effect on \texttt{MTL4CommandQueue}, whose
explicit queue barriers are a separate design point~\cite{metal-hazard-tracking}.

Table~\ref{tab:metal} therefore reports only what is comparable. \blade's host
cost is $\bladenum{metal/hostshare/min}$--$\bladenum{metal/hostshare/max}\%$ of
\wgpu's. Completion wall time is within a few percent either way on every
workload; the double-digit wait saving the pre-parity pilot appeared to show
does not survive shader parity, which is consistent with it having been the
injected-check artifact of Section~\ref{sec:workloads}. Wall time is not used
as a substitute for device span. This recollection uses the shader-check settings of
Section~\ref{sec:workloads} on both implementations.
Note that \blade's Metal host cost
($\bladenum{metal/host/min}$--$\bladenum{metal/host/max}\,\mu$s for sixteen
passes) is about $\bladenum{metal/overvulkan/max}$ times its Vulkan host cost
on the RTX~5070, because each \blade{} pass creates a new \texttt{MTL} command
encoder; the Metal backend has room to improve independently of anything in
this paper.

The question Metal actually poses --- what hazard tracking costs when the framework
rather than the application performs it --- is not answered by these workloads,
because both implementations pay it. A separate harness, listed in
Appendix~\ref{sec:metal-harness}, measures it directly by creating the same
resources with \texttt{MTLHazardTrackingModeTracked} and
\texttt{MTLHazardTrackingModeUntracked} and, in the untracked dependent case,
reconstructing the dependency with an \texttt{MTLFence} or an
\texttt{MTLEvent}. Each pass is one compute encoder with one single-thread
dispatch, so the measurement is dominated by synchronization rather than by
work. The full protocol and tables are in the repository's investigation
note~\cite{blade-metal-note}.

The replicated result differs sharply from the Vulkan placement result, in
the direction the single-session pilot indicated. Across
$\bladenum{hazard/sessions}$ retained process sessions on AC power, turning
tracking off for \emph{independent} passes moves encode time by
$\bladenum{hazard/independent/encode/lo}$ to
$\bladenum{hazard/independent/encode/hi}\%$ and GPU time by
$\bladenum{hazard/independent/gpu/lo}$ to
$\bladenum{hazard/independent/gpu/hi}\%$ across every session and pass count
--- consistent with Metal's tracker already distinguishing the resources, and
now bounded by the observed spread rather than asserted from one launch. For
\emph{dependent} passes, opting out is expensive and stable across sessions:
at 100 passes an \texttt{MTLFence} costs
$\bladenum{hazard/dependent/fence/encode/p100/med}\times$ the encode time
(session range
$[\bladenum{hazard/dependent/fence/encode/p100/lo},\,\bladenum{hazard/dependent/fence/encode/p100/hi}]$)
and $\bladenum{hazard/dependent/fence/gpu/p100/med}\times$ the GPU time of
the tracker, an \texttt{MTLEvent}
$\bladenum{hazard/dependent/event/encode/p100/med}\times$ and
$\bladenum{hazard/dependent/event/gpu/p100/med}\times$; at 500 passes the
GPU ratios rise to
$\bladenum{hazard/dependent/fence/gpu/p500/med}\times$ and
$\bladenum{hazard/dependent/event/gpu/p500/med}\times$.

Every session retains its raw observations, and each of the
$\bladenum{hazard/rows}$ rows records the resource mode the driver actually
applied and an output-correctness check; the requested mode took effect and
the check passed in all of them. The harness's deliberately unsynchronized
untracked case again produced correct values, which is still not evidence
that the mode is safe.

The API consequence is firmer still than the timings:
\texttt{manual\_barriers} should stay a Vulkan concept. Metal's tracking mode is chosen per resource at
creation, while \blade's flag belongs to a command encoder, and a resource
outlives any one encoder; moreover \blade's \texttt{barrier()} is called
between RAII pass encoders, after the producer's encoder has already ended,
which is too late to place the producer-side \texttt{updateFence}. The cost
model is consistent with the abstraction: \blade's Vulkan fallback can
serialize unrelated passes, whereas opting out of Metal tracking makes the
application reconstruct real dependencies using an API that \blade{} does not
currently expose at the right lifetime.

\subsection{Generated barriers and image layouts}
\label{sec:generated}

RenderDoc captures of all six Blade workloads under three policies, and all
four shared \wgpu{} workloads, confirm what
Sections~\ref{sec:radv-barriers} and~\ref{sec:scoped} describe from source. The
Blade counts are exactly the contract: with four passes recorded,
\texttt{B-auto} emits \bladenum{barriers/auto/independent} barriers per
configuration---one before each pass plus the one the encoder emits on
finish---
\texttt{B-hazard} emits \bladenum{barriers/hazard/independent} on the
independent workloads and \bladenum{barriers/hazard/chain} on the chains, and
the captured \texttt{B-auto-scoped} policy emits the same number as
\texttt{B-auto}. The uncaptured crossed policies follow from the same source
path and are publication capture targets, not capture-backed observations in
the retained artifact.

The masks match too. Every \texttt{B-auto} barrier in every workload is the
same shape, \texttt{ALL\_COMMANDS} to \texttt{ALL\_COMMANDS}, with
\texttt{TRANSFER\_WRITE\,|\,MEMORY\_WRITE} made available to the four reads and
writes \texttt{TRANSFER\_READ}, \texttt{TRANSFER\_WRITE},
\texttt{MEMORY\_READ}, and \texttt{MEMORY\_WRITE} ---
including the two driver-workaround accesses. Under \texttt{B-scoped} the
derivation is visible in the capture: compute-to-compute boundaries carry
\texttt{COMPUTE\_SHADER} to \texttt{DRAW\_INDIRECT\,|\,COMPUTE\_SHADER},
render-to-render carry \texttt{ALL\_GRAPHICS} on both sides, mixed workloads
show both crossings, the first barrier of each encoder still carries
\texttt{ALL\_COMMANDS} on the source side, the barrier emitted on finish still
carries it on the destination side, and the workaround accesses are absent, as
intended.

Two things the Blade captures settle that source reading could not. Every
steady-state Blade barrier is a global memory barrier: no buffer or image
barrier appears in any Blade configuration. And no image layout transition
appears in the captured measured iteration---the setup command has already
moved images to \texttt{GENERAL}, and they stay there. The capture therefore
observes the steady-state layout claim rather than the initialization
transition itself.

The matched \wgpu{} program was captured the same way, which directly observes
barrier form and layout but does not turn the end-to-end comparison into a
single-factor experiment.

\emph{Counts are equal; endpoint placement is not.} \wgpu{} emits exactly as
many barrier calls as \texttt{B-hazard}:
\bladenum{barriers/wgpu/independent} on the independent workloads and
\bladenum{barriers/wgpu/chain} on the chains, where \texttt{B-auto} emits
\bladenum{barriers/auto/independent}. The extractor replays Vulkan command
buffers in \texttt{vkQueueSubmit} order and records how many draw or dispatch
commands precede each call; using host recording order would be wrong because
\wgpu{} records pass and transition command buffers separately.

That pass-relative position exposes an endpoint shift. On the independent
compute workload, \wgpu's one call is an initial transition
(\bladenum{barriers/wgpu/independent/initial} initial,
\bladenum{barriers/wgpu/independent/interpass} inter-pass,
\bladenum{barriers/wgpu/independent/final} final), whereas
\texttt{B-hazard}'s one call is the encoder's final barrier
(\bladenum{barriers/hazard/independent/initial},
\bladenum{barriers/hazard/independent/interpass}, and
\bladenum{barriers/hazard/independent/final}, respectively). On the chain,
both have \bladenum{barriers/wgpu/chain/interpass} inter-pass calls, but
\wgpu{} again has an initial call and \texttt{B-hazard} a final one. Thus the
tracker derives the required inter-pass placement in this workload, but it
does not reproduce \texttt{B-hazard}'s complete call sequence. This also
explains why equal call counts do not make their timestamp windows equivalent:
\wgpu's initial transition lies before its first timestamp, while \blade's
final barrier lies inside its reported span.

\emph{Memory scope.} Every \blade{} barrier is a global memory barrier and
every \wgpu{} barrier is resource-scoped --- a \texttt{VkBufferMemoryBarrier}
for the compute workloads, a \texttt{VkImageMemoryBarrier} for the graphics
ones. Neither implementation emits a barrier of the other's kind in any
configuration. This is factor~2, which the experiment contract lists as
observed-only, and it is now observed rather than read from source.

\emph{Layouts.} The most useful thing the captures say is negative. \wgpu's
image barriers carry \texttt{COLOR\_ATTACHMENT\_OPTIMAL} on \emph{both} sides:
they are execution and memory dependencies, not layout transitions. \blade{}
emits no image barrier at all. So neither implementation performs a
steady-state layout transition on these workloads, and the layout difference
between them is static --- which layout the driver sees the image in --- rather
than a per-pass cost. Any layout effect in the end-to-end numbers is therefore
a property of \texttt{GENERAL} versus \texttt{COLOR\_ATTACHMENT\_OPTIMAL} as
states, not of moving between them.

The capture tables from all \bladenum{captures/machines} machines are
byte-identical --- \bladenum{captures/rows} rows, the same SHA-256 --- which is
what the design intends. These CSVs contain extracted barrier records, not the
entire command stream. They license the narrower inference needed here:
barrier counts, pass-relative positions, barrier kind, stages, accesses, and
layouts in the relevant records do not vary by capture machine. They do not
show unextracted commands or prove that every driver-internal representation
is byte-identical. The retained manifests name those hosts but did not record
the adapter selected by either implementation, so this is deliberately a
cross-host statement, not evidence from three identified GPU models. The
publication recollection pins both selectors and records the reported device
name.

One case is deliberately not covered. Every workload here is single-sampled,
and Section~\ref{sec:radv-barriers} reads from RADV that \texttt{GENERAL}
disables FMASK compression unconditionally, so multisampled colour is the one
configuration where a persistent \texttt{GENERAL} layout forgoes that
compression on AMD. We state the predicate from driver source and do not
measure its performance consequence.

The layout claim this draft makes comes from source, not a controlled layout
experiment: on RADV, a persistent \texttt{GENERAL} layout retains DCC on RDNA
and later under the stated queue conditions, conditionally retains HTILE, and
disables FMASK compression. Our graphics workloads use single-sampled
\texttt{RGBA8} colour targets: they can retain DCC and have no FMASK to lose.
They contain no depth target, so the HTILE predicate is source context rather
than a measured workload property. A study that wanted to stress the layout
dimension would need at least MSAA colour and a depth case. The Intel system is
the only one whose driver does not
advertise \texttt{VK\_KHR\_unified\_image\_layouts}, and \blade{} still
outperforms \wgpu{} there on both graphics workloads. That is
consistent with the missing extension not dominating, but it is an end-to-end
observation and not a layout experiment.

\section{Discussion}
\label{sec:discussion}

The results support a decision boundary rather than a winner.

\textbf{Placement is the big lever, and it does not require tracking.} The
largest effect among the axes controlled within \blade{} comes from barrier
\emph{placement}: on the RTX~5070's independent render passes,
redundant barriers cost a third of the span
(\bus{rtx5070}{graphics-independent}{auto} for \texttt{B-auto} against
\bus{rtx5070}{graphics-independent}{placement} for \texttt{B-hazard}). The
matched \wgpu{} path lands a couple of percentage points farther away
(\bus{rtx5070}{graphics-independent}{wgpu}), but that residue is end-to-end,
not a measurement of tracking precision. Resource knowledge lets \wgpu{} avoid
dependencies between distinct targets automatically; an application
able to say ``these passes are unrelated'' collects nearly all of it without
tracking anything, and an abstraction that can express neither leaves a third
of this command-buffer span on the table. The counterweight is the NVIDIA
render chain, where the \wgpu{} path is
$\bmag{rtx5070}{graphics-chain}{wgpu}$ faster and no declaration \blade{}
accepts reaches that end-to-end result (Section~\ref{sec:endtoend}).

\textbf{The device decides, and reading the driver does not substitute for
asking it.} The captures show that all \bladenum{captures/machines} machines
produce identical extracted barrier tables (Section~\ref{sec:generated}), so
the cross-machine \blade{} scope differences are not caused by different
application barrier masks or counts. Against that fixed request, two predictions we made from mechanism failed in
the same direction. Redundant barriers between independent render passes cost
$\bmag{rtx5070}{graphics-independent}{placement}$ on NVIDIA against
$\bmag{rx7900xt}{graphics-independent}{placement}$ on the discrete AMD part
and only $\bmag{raphael}{graphics-independent}{placement}$ on Raphael, while
the Radeon~780M sweep has threshold-cleared points in the opposite direction,
despite AMD being the architecture whose vendor guidance warns hardest about
broad barriers. Narrowing the barrier
scope removes the most commands from the RADV stream at render-involving
boundaries, where no AMD cell resolves a directional effect, and the fewest at
the compute-to-compute boundaries where the discrete part actually pays: the
request count and the measured effects do not correlate. Both times the source
told us what would be requested and we took that for what it would cost.

The second failure had an answer in the same file, and finding it afterwards is
the part worth keeping. \texttt{radv\_dst\_access\_flush} discards the
destination-side cache invalidation our narrower barrier was designed to avoid,
on the measured generations whenever no render-backend-incoherent attachment
is marked dirty
(Section~\ref{sec:radv-elision}). We had read the function that translates a
request into work and not the one that decides how much of that work to do.
Counting emitted commands describes how much the driver is \emph{asked} for;
that request set can be a very loose predictor of latency, because a drain is
free when there is nothing in flight to drain and an invalidation is free when
the driver has already proved it unnecessary. Mechanism is good for explaining
a measurement and poor for replacing one --- and a mechanism that only explains what has already been
measured is still worth having, because it says where the result should
generalize. Here it says: on a part whose TCC is not coherent with the render
backends, an AMD render-side saving is a concrete hypothesis. We did not have
one to try.

\textbf{The workload envelope for coarse barriers.} Tracking-free
synchronization with pass-boundary barriers is a candidate when passes form
chains, when each pass leaves little device headroom, or when host cost
dominates. It becomes risky when independent passes have complementary
bottlenecks and spare device capacity, which is where the field report sits. An
encoder-level opt-out covers that case at negligible API cost, which is why it
was the right resolution to the field report~\cite{blade-issue-343}: it turns
the expensive case into a two-line change without adding resource states to the
API.

\textbf{The end-to-end host difference is the most consistent result.} The
matched \wgpu{} stack takes more record-and-submit time in every fixed-matrix
cell across four vendors, and the timestamp-free RTX~5070 sweep confirms the
difference over the measured pass counts. The fixed-matrix ratios
($\bladenum{host/ratio/min}$--$\bladenum{host/ratio/max}\times$) include
timestamp instrumentation, and the sweep becomes non-linear at larger counts,
so we do not extrapolate them to hundreds of passes or whole-frame savings.

\textbf{What tracking itself costs the CPU is bracketed, not measured.} The
question this study set out from --- how much host time \wgpu{} sacrifices
specifically for resource tracking --- gets a lower and an upper bound here,
and they are far apart. Symbols directly attributable to the trackers are
$\bladenum{profile/wgpu/tracker/min}$--$\bladenum{profile/wgpu/tracker/max}\%$
of \wgpu's whole-process samples, a bound blind to inlined tracker work; the
full end-to-end gap is the upper bound, and it bundles validation, lifetime
management, bind-group resolution, and command translation with the
trackers --- machinery that exists partly \emph{because} state is tracked but
is not the tracking itself. The experiment that would separate them --- a
\wgpu{} build with tracking short-circuited, or a profile gated to the
record-and-submit interval on a pinned adapter
(Section~\ref{sec:profile}) --- was not performed. Anyone quoting this paper
as ``tracking costs $N\times$'' is quoting a bracket as a point.

\textbf{The negative results matter.} The Radeon~780M sweep contains
threshold-cleared regressions, showing that removing redundant barriers is not
a free optimization, and the AMD render boundaries show that a saving visible
in a command stream need not be visible in a timer. A layer that exposes
manual barriers must therefore treat them as a measured optimization rather
than a default, and profiling guidance belongs next to the API.

\subsection{Implications for engine architecture}
\label{sec:engine-implications}

The architectural result concerns where resource knowledge lives, not whether
it exists. An engine that already has a render or task graph need not reproduce
that graph as per-resource state inside its RHI. It can use the graph to group
independent producers, place a \blade{} barrier at a dependency cut, and then
record the consumers. The field report has exactly this shape: two independent
views form one phase and join before refraction. In a phase-structured graph,
one global barrier can represent many aligned resource edges.

A global barrier is nevertheless a \emph{cut}, not an edge. Suppose one
producer must complete before a consumer while another long producer is
independent of that consumer. A global barrier placed for the first dependency
also constrains the second producer whenever it falls on the barrier's source
side and matches its scope; a resource-specific barrier can name only the real
edge. The controlled workloads here are the two endpoints --- all inter-pass
edges or none --- plus a mixed chain whose incoming passes are almost all
hazardous. They do not show that an arbitrary partially dependent render graph
can be lowered to global cuts without losing scheduling freedom.

For a small or immediate-mode renderer without a graph, automatic pass
boundaries remain the conservative starting point. Deriving stage and access
scope from pass kind is internal, requires no resource declarations, and
caused no resolved regression in the measured cells. Manual placement is then
a profiler-guided escape hatch for a region where unrelated passes demonstrably
serialize; it is not a new default for the whole frame.

Finally, a tracking-free RHI is not a resource-management system. Transient
aliasing and lifetime planning, subresource layout transitions, queue-family
ownership and cross-queue dependencies, and the multisampled-colour exception
identified on RADV all still require resource identity somewhere. An
engine-level graph can own those responsibilities; an engine that wants the
RHI to infer them needs a tracked abstraction such as \wgpu{} or NVRHI. The
choice exposed by this study is whether that knowledge must be duplicated on
the command-recording path when the engine already has it.

\subsection{What to do on AMD}
\label{sec:amd-advice}

The short answer to ``does this abstraction work on AMD'' is that its host-side
advantage holds there as firmly as anywhere ---
$\bladenum{rx7900xt/compute-independent/hostwgpuus}\,\mu$s against
$\bladenum{rx7900xt/compute-independent/hostautous}\,\mu$s for the same sixteen
passes on the RX~7900~XT --- and its device-side cost is the widest-ranging of
any vendor in the study, from
$\bmag{rx7900xt}{compute-independent}{placement}$ recoverable to
$\bmag{radeon780m}{graphics-independent}{placement}$ lost by trying. Three
things follow, in the order an application should try them.

\emph{Take the narrow barrier, and expect it to pay only between compute
passes.} It is tempting to read Section~\ref{sec:radv-barriers} as a shopping
list of savings, and we did. What the measurement licenses is narrower: a
$\bmagci{rx7900xt}{compute-chain}{scope}$ reduction at this 16-pass shape on
the discrete part where a compute pass hands to a compute pass,
with no directional effect meeting the exploratory criterion at
render-involving boundaries or on the Radeon~780M; Raphael's two compute cells
show only the threshold-edge
$\bmag{raphael}{compute-chain}{scope}$ observation discussed in
Section~\ref{sec:scope-results}. Section~\ref{sec:radv-elision} gives a
conditional account of why much of the render-side request may already be
elided. No regression clears the stability diagnostic in these tests; because
the scope is derived internally rather than asserted by the application, it is
the lower-risk change to take first. It is not an established AMD render
optimization, and the practically meaningful saving observed here appears
only under automatic placement (below).

\emph{Restructure the passes so the remaining barriers are ones you need.}
This is the change that pays. On the RX~7900~XT, sixteen independent compute
passes run $\bmagci{rx7900xt}{compute-independent}{placement}$ faster once the
redundant barriers are gone, and the mixed equivalent
$\bmag{rx7900xt}{mixed-independent}{placement}$ faster. The field report's
renderer serialized because independent work sat on either side of a barrier.
The fix is not only to delete that barrier but to group the independent passes
so one barrier separates each group from the next --- ray tracing with the
reflection prepass, both view passes together, then the resolves, then the
dependent refraction pass. With \texttt{manual\_barriers} this is a change to
where \texttt{barrier()} is called, not a change to any resource declaration,
and the encoder-level flag makes the scope of the change explicit. Applications
that want the automatic policy for most of a frame can put the restructured
part in its own encoder and submit both to the same queue.

\emph{Then measure, because the restructuring is not always a win.} On the
Radeon~780M, the 16-pass mixed-independent cell costs
$\bmagci{radeon780m}{mixed-independent}{placement}$ against a
$\bfloor{radeon780m}{mixed-independent}$ floor. The corresponding
graphics-independent point is
$\bmag{radeon780m}{graphics-independent}{placement}$ but is not resolved
because launch-to-launch dispersion raises the floor; the graphics sweep resolves
the same direction at 2, 32, and 64 passes, with descriptive medians suggesting
growth and saturation (Figure~\ref{fig:overlap-depth}). On the RX~7900~XT the
same change on render passes instead bought
$\bmag{rx7900xt}{graphics-independent}{placement}$ --- real, but a quarter of
the $\bmag{rx7900xt}{compute-independent}{placement}$ its compute equivalent
buys. Two AMD parts of the same generation therefore disagree about the sign
on render work, which is the strongest reason in this paper not to ship the
change unmeasured. The passes worth grouping are the ones with
complementary bottlenecks --- ray tracing next to rasterization, a depth
prepass next to a compute simulation --- not several instances of the same
bottleneck. A GPU profiler showing the two passes actually overlapping, as in
the field report, is the evidence to look for; a frame-time delta on its own
will not distinguish a scheduling win from whatever the 780M is doing.

Note also that suppressing the encoder's barriers forfeits the scope narrowing
on the boundaries that remain, since an explicitly placed barrier describes
only its source. On AMD that now costs the compute-boundary saving; on NVIDIA
it is a real trade everywhere the scope pays.

Two things we looked for and did not find are worth recording, because they
bound the mechanisms examined here. The narrowed scope removes the operations
Section~\ref{sec:radv-barriers} identifies without naming a resource, and on
the discrete AMD device that changes compute boundaries by a fraction of a
microsecond. The remaining pessimism that a global barrier requests in RADV
--- including metadata-flush requests and the unavailable image-specific
coherence check --- is only avoidable by supplying an image, which is the state
\blade{} declines to keep. The remaining AMD lever is placement, and placement
is the application's knowledge, not the layer's.

The one case where AMD still argues for genuine layout tracking is
multisampled colour, where RADV disables FMASK compression in
\texttt{GENERAL}. That is orthogonal to everything above and is not addressed
by either barrier axis.

\section{Threats to Validity}
\label{sec:threats}

GPU timestamps could perturb the host measurement, and the paired RTX~5070
sweeps test for it: their timestamped and timestamp-free host medians agree
at every pass count, so on that machine the fixed matrix's host columns carry
no measurable instrumentation inflation. The other machines have no such
pair, so the concern is tested on one driver and assumed bounded, not proven
absent, on the rest. Device-side \blade{}--\wgpu{} comparisons also use different
timestamp counts and endpoints (17 writes including Blade's final barrier
versus two writes ending with wgpu's last pass), so they remain end-to-end
instrumented spans; within-\blade{} comparisons use identical instrumentation.

Driver heuristics may recognize repeated synthetic work; the benchmark varies a
per-iteration seed while preserving the graph shape, but does not vary resource
identity. Results from one driver version cannot establish permanent vendor
behaviour. The local collection directories retain metadata for comparison,
but they are git-ignored; submission therefore requires a checksummed raw-data
artifact rather than a claim that the repository alone reproduces the tables.

The controlled \blade{} variants now isolate two of the four factors ---
placement and stage/access scope --- but still not global versus
resource-specific memory scope, nor persistent \texttt{GENERAL} versus
usage-specific layouts. Those two are argued from driver sources and observed
end-to-end through \wgpu, so causal claims about them are limited by
construction. The scoped results cover five Vulkan device-driver pairs across
three vendors, but not every cell within them:
\bladenum{floor/above2} of \bladenum{floor/count} have control floors above
$2\%$, and every directional conclusion is checked against the post-hoc
stability threshold of the cell and pass count it comes from; curve shapes and
state-conditioned contrasts are labelled descriptive. Every archival timing
collection is
preceded by a retained synchronization-validation matrix, run with the
Khronos checks forced on and its logs stored beside the timing rows; the
timing rows themselves come from the validated shaders and layered output
checks of Section~\ref{sec:workloads}. Retained validation is still evidence
about the schedules that were executed rather than a proof.
The automated workload matrix exercises compute, render, and their crossings;
it does not exercise the transfer or acceleration-structure rows of the
pass-kind table. Those rows have a source-level correctness argument but need
dedicated synchronization-validation coverage before the scoped mode is
presented as validated for every \blade{} pass kind.
End-to-end \wgpu{} differences are labelled as such throughout, and the ones we
cannot account for --- the Intel gap in one direction, the RX~7900~XT gap in
the other --- are left unexplained rather than attributed.

Each configuration has 30 within-process samples and ten independent
process-level repetitions. The paired hierarchical bootstrap preserves that
nesting and reveals variation a pooled analysis hides --- at three pilot
blocks it disqualified cells that ten now resolve, and on one host it
exposes launch-to-launch dispersion rather than averaging it away. Ten
blocks still estimate tails coarsely, the inner resampling treats
time-ordered observations as exchangeable, and the matrix has no second-day
or in-process-alternation sensitivity run. A moving-block or
process-median-only re-analysis of the archived observations remains a
worthwhile check.

The workloads are microbenchmarks of sixteen passes each. Two of the six
alternate compute and render work, so heterogeneity is represented, but only
the synthetic kind: two pass types in a fixed alternation, with identical work
in every instance of each. Real frames vary the cost, the resources, and the
bottleneck from pass to pass, which is precisely the regime in which the field
report saw a benefit and our \texttt{graphics-independent} workload saw a loss
on one device. The boundary proposed in Section~\ref{sec:discussion} is
therefore a hypothesis supported by microbenchmarks --- including mixed ones ---
and one field report, not a validated rule. We chose not to close that gap with
application workloads: doing so would trade a controlled comparison for a
plausible one, since scenes, content, and driver heuristics vary in ways this
design exists to hold still. A reader who needs the rule validated on their own
frame should measure their own frame, which is the recommendation
Section~\ref{sec:amd-advice} makes anyway.

The dependency structures are also endpoints rather than representative DAGs.
The independent workloads remove every timed inter-pass barrier, while the
single-kind chains retain every one; in \texttt{mixed-chain}, every incoming
pass from the third onward has a dependency and the policy is intentionally
conservative at the second pass. No controlled workload contains two
interleaved chains whose dependency edges cross different global cuts. Such a
staggered partial DAG is the most important next microbenchmark for the engine
claim: it would measure the scheduling freedom lost when one real edge forces
a global cut across otherwise independent work.

The AMD evidence is now a discrete RDNA3 part, a mobile RDNA3 iGPU, a very
small desktop iGPU, the driver sources, and a user report. The small iGPU is
the most precisely measured device in the study and the least responsive to
barrier changes, consistent with but not proving single-pass saturation. The pilot's version of this paragraph
asked for the discrete part's placement cells to be repeated with enough
warm-up for the device to stop accelerating before being leaned on --- its
clocks were already pinned, and that was not enough. This revision is that
repeat, and the claims survived it sharpened rather than softened: the compute
effect grew while its floor fell to $\bfloor{rx7900xt}{compute-independent}$,
the render cell went from unresolvable to a resolved
$\bmag{rx7900xt}{graphics-independent}{placement}$, and the scope axis,
unreadable on the pilot, resolved to $\bmag{rx7900xt}{compute-chain}{scope}$
at discrete compute boundaries, with only threshold-edge
$\bmag{raphael}{compute-chain}{scope}$ observations on Raphael and no
render-boundary result. The third AMD part still disagrees in
sign on the render workloads, which is itself part of the result rather than a
contradiction of it.

Finally, no mobile tiler is represented at all. A pass-boundary barrier
policy would be stressed hardest exactly there --- a barrier that breaks a
render pass on a tile-based GPU forces a tile flush and reload --- so the
class where this design has the most to lose is the class the study says
nothing about.

\section{Related Work}

The Vulkan specification defines explicit execution and memory dependencies
and application-controlled image layouts~\cite{vulkan-spec}. AMD and NVIDIA
have published architecture-specific synchronization
guidance~\cite{amd-barriers,amd-rdna-guide,nvidia-vulkan-tips}. The
unified-image-layouts extension standardizes the observation that specialized
layouts no longer correspond to distinct representations on many modern
implementations~\cite{vulkan-unified-layouts}; the RADV predicates quoted in
this paper are a concrete instance of that convergence, including its
exceptions.

Direct3D~12 Enhanced Barriers separate work synchronization, memory access,
and texture layout, and provide global as well as resource-specific barrier
types~\cite{d3d12-enhanced-barriers}. This closely matches the factorization
used here. The present study differs by evaluating an RHI policy that chooses
global barriers systematically at pass boundaries and by measuring that policy
through Vulkan; it makes no Direct3D performance claim.

Render graphs take the workload declaration as their source of resource
lifetimes and pass dependencies. Frostbite's FrameGraph is a canonical
production example: it builds a graph of render passes and resources so the
engine can derive scheduling and resource-management decisions
centrally~\cite{frostbite-framegraph}. Activision's Task Graph reports the
maintenance failures that motivated the same move and its scale in a shipping
renderer~\cite{activision-task-graph}; Unreal's RDG is a current production
system deriving transitions, queue fences, aliasing, and validation from pass
parameters~\cite{unreal-rdg}. These systems are not alternatives to \blade{}.
An upstream graph may select \blade's dependency cuts. What \blade{} declines
is the resource declaration \emph{inside the RHI}, and its smaller contract
therefore cannot itself derive resource-scoped barriers, aliasing, or queue
schedules.

Academic rendering systems explore the resource-aware end of the same design
space. LuisaRender builds a subresource dependency graph and reorders
independent commands, reporting up to a 19\% rendering-time improvement from
its command scheduler~\cite{luisa-render}. RenderKernel moves resource
declarations into a compile-time DSL and generates Vulkan command recording
and barrier insertion~\cite{renderkernel}. Both demonstrate the value of rich
resource knowledge in a framework or compiler. \blade{} does not dispute that
value; it asks whether a low-level RHI must maintain the same knowledge at run
time when an upstream system already owns it.

WebGPU specifies portable usage-scope validation~\cite{webgpu-spec}; \wgpu{}
implements it using optimized dense resource trackers~\cite{wgpu-trackers}.
NVRHI is the closest prior RHI design point: it tracks resource and optional
subresource state per command list, while allowing automatic barriers to be
disabled for lower-frequency explicit state calls when an application wants
to reduce CPU overhead or control placement~\cite{nvrhi-state-tracking}.
Unlike \blade{}, that manual mode retains the per-resource state machinery and
its explicit calls still name resources.
Metal takes a third position, performing hazard tracking in the framework with
an opt-out per resource for \texttt{MTLCommandQueue}; Metal~4 queue barriers
are outside this study~\cite{metal-hazard-tracking}. \blade{} occupies the
tracking-free RHI point by exposing an unsafe native contract and relying on
application-declared pass dependencies.

\section{Artifact}

The manuscript, collectors, and \blade{} implementation are public on the
\texttt{blade-sync-study} branch, and the matched program is public on the
corresponding \wgpu{} fork branch~\cite{blade,wgpu-study}. The archival round
measures the corrected benchmark on every machine: \wgpu{} revision
\texttt{7d37a77} and \blade{} revision \texttt{87ed067}, with the study
host's sweeps at one later commit that changed only the paper directory and
therefore did not change the compiled benchmark. The bibliography records the
full identifiers, and per-collection manifests record the revisions, the clock
configuration, and device metadata.

The measurements are the irreplaceable half, and they travel with this
submission: the ancillary files hold every raw collection --- timing
matrices, sweeps, retained synchronization-validation logs, host profiles,
captures with extracted barrier tables, and the Metal hazard sessions ---
with a digest of every file. The measured code is public and
content-addressed: both repositories carry study-code tags
(\texttt{sync-study-v1} on \blade{}, \texttt{blade-sync-study-v1} on the
\wgpu{} fork), and the commit hashes recorded in the bibliography and in every
collection manifest remain the identifiers of record --- verifiable from any
clone or archive of either repository. The tags pin the code, not this article
revision; arXiv preserves the submitted article source. Extracting the
collections into \texttt{paper/data/raw/} in a study-branch checkout containing
this submission's analysis tooling and running the table builder regenerates
every table, figure, and quoted number in this paper; a missing collection is a
build error rather than a silent absence.

\paragraph{Generative-AI disclosure.}
Anthropic Claude and OpenAI Codex were used interactively for literature
discovery, manuscript editing, and development and review of benchmark and
analysis tooling. They are not authors; the human author takes responsibility
for the cited sources, code, analyses, and final text.

\section{Conclusion}

This study does not isolate a single price for fine-grained resource state:
the \wgpu{} comparison also changes validation, barrier form, layouts, and
command machinery. Within \blade{}, it does isolate the effect of placement
for the measured cells.
The dependent-chain variants have the same timed barrier calls by construction
and no chain interval clears its cell-specific stability threshold; removing
fifteen redundant inter-pass barriers reduces device span by
$\bmag{rtx5070}{compute-independent}{placement}$ and
$\bmag{rx7900xt}{compute-independent}{placement}$ on the two discrete GPUs,
while the Radeon~780M graphics sweep resolves a roughly two-fifths penalty at
32 and 64 passes, for reasons we could not establish. Its 16-pass graphics
point has the same magnitude but does not clear that noisy cell's control
floor; the 16-pass mixed-independent regression does. The matched \wgpu{} stack
takes more record-and-submit time in every measured cell, but that end-to-end
difference must not be called the cost of tracking alone.

The corresponding \blade{} development outcome is deliberately small. The
mainline Vulkan encoder can retain its existing API and add one encoder-wide
pass-kind summary: a bitmask of producers recorded since the previous barrier.
An automatic boundary can then derive both sides of its global barrier from
the accumulated producers and the pass being opened; an explicitly placed
barrier derives its source and keeps a conservative destination. This is
lightweight aggregate access tracking, not per-resource state: it stores no
resource identities, subresources, layouts, or lifetimes. The AMD measurements
also make this a scope refinement rather than a vendor heuristic. Narrowing
paid at the discrete part's compute boundaries but not at its render
boundaries, while removing barriers from independent render work hurt the
Radeon~780M. \blade{} therefore should neither select placement by vendor nor
assume that fewer barriers are always faster.

The practical conclusion is that resource-dependency knowledge is
indispensable, but per-resource state inside the RHI is not always. For the
phase-structured field report and the independent-pass endpoints measured
here, the expressive gap that matters is the ability to say ``these passes are
unrelated''. An engine-level graph can remain the source of truth and drive
\blade's dependency cuts without maintaining a second tracker in the
abstraction. An encoder-level opt-out gives an application that already knows
its dependency graph nearly the same device-span reduction that the matched
\wgpu{} path shows on the RTX~5070 independent-render cell, while retaining
\blade's lower measured host cost, without adding per-resource state surface.
That is a cell-specific end-to-end comparison, not a decomposition of tracking
cost, and the current workloads do not establish the same result for a
partially dependent DAG. The opt-out does add an
encoder-level mode and explicit barrier calls. What the current call cannot
reach is a destination scope derived from a future pass, and, on
one NVIDIA render chain, a further $\bmag{rtx5070}{graphics-chain}{wgpu}$
bundled among barrier form, layout, timestamp window, and other end-to-end differences. A global
barrier can express a narrow destination mask; \blade's current explicitly
placed call lacks the information to derive one automatically. The
opt-out should be presented as a profiled optimization rather than a default,
because on at least one real device it makes things worse.

The remaining bounds are specific: no application frames or multisampled
targets, no mechanism for the Radeon~780M overlap penalty or the discrete AMD
scope saving, and no controlled layout variant.
They bound the corresponding claims rather than supporting extrapolation
beyond the measured workloads.

\bibliographystyle{plain}
\bibliography{references}

\appendix

\section{Metal tracked-versus-untracked harness}
\label{sec:metal-harness}

The Metal question in Section~\ref{sec:metal} --- what hazard tracking costs
when the framework rather than the application performs it --- is not answered
by the main workloads, because both implementations pay it there. A standalone
Swift harness measures it and is distributed with the artifact.

It allocates the same resources twice, once with
\texttt{MTLHazardTrackingModeTracked} and once with
\texttt{MTLHazardTrackingModeUntracked}, and queries each buffer's
\texttt{hazardTrackingMode} afterwards to confirm the requested mode took
effect. Each pass is one compute encoder containing one single-thread dispatch,
so the measurement is dominated by synchronization rather than by work. Two
workloads are run at 1, 10, 100, and 500 passes: an \emph{independent} one in
which every pass increments its own four-byte buffer, and a \emph{dependent}
one in which every pass increments the same buffer. In the dependent untracked
cases the harness reconstructs the ordering explicitly, with either one
\texttt{MTLFence} updated and waited on between consecutive encoders or one
\texttt{MTLEvent} signalled and waited on between them.

Resources and synchronization objects are reused across iterations. Each case
receives ten warm-up runs, cases are interleaved in alternating order to reduce
mode-order and thermal bias, and results are medians of 200 samples, except the
500-pass cases which use 80. Encoding time excludes \texttt{commit()}, which is
measured separately; device time comes from \texttt{gpuStartTime} and
\texttt{gpuEndTime}. The archival run repeats the whole procedure as
$\bladenum{hazard/sessions}$ separate process sessions on AC power, driven by
a collection script distributed with the artifact; every session retains its
raw per-iteration observations, the resource mode the driver actually
applied, and an output-correctness check per row. The harness also contains a
deliberately unsynchronized untracked case as a canary; it produced correct
values on this machine and workload, which is not evidence that the mode is
safe.

\end{document}